\documentclass[a4paper,fleqn,sort&compress]{cas-sc}

\usepackage[numbers]{natbib}
\usepackage{graphicx}
\usepackage{soul}
\usepackage{multirow}

\newcommand{\ket}[1]{\vert #1 \rangle}
\newcommand{\bra}[1]{\langle #1 \vert}

\newcommand{\braket}[2]{\langle #1 \vert #2 \rangle}

\newcommand{\tabincell}[2]{\begin{tabular}{@{}#1@{}}#2\end{tabular}}
\renewcommand{\eqref}[1]{Eq.~(\ref{#1})} 
\newcommand{\figref}[1]{Fig.~\ref{#1}} 
\newcommand{\secref}[1]{Sec.~\ref{#1}} 

\def\tsc#1{\csdef{#1}{\textsc{\lowercase{#1}}\xspace}}
\tsc{WGM}
\tsc{QE}
\tsc{EP}
\tsc{PMS}
\tsc{BEC}
\tsc{DE}

\begin{document}
\let\WriteBookmarks\relax
\def\floatpagepagefraction{1}
\def\textpagefraction{.001}
\shorttitle{Geometric and holonomic quantum computation}
\shortauthors{J. Zhang, T.H. Kyaw, et.~al.}

\title [mode = title]{Geometric and holonomic quantum computation}                      

\author[1]{Jiang Zhang}
\ead{zhangjiang@baqis.ac.cn}
\address[1]{Beijing Academy of Quantum Information Sciences, Beijing 100193, China}

\author[2,3,4]{Thi Ha Kyaw}[orcid=0000-0002-3557-2709]
\ead{thiha.kyaw@lge.com}
\address[2]{LG Electronics Toronto AI Lab, Toronto, Ontario M5V 1M3, Canada}
\address[3]{Department  of  Chemistry,  University  of  Toronto,  Toronto,  Ontario  M5G 1Z8,  Canada}
\address[4]{Department  of  Computer Science,  University  of  Toronto,  Toronto,  Ontario  M5S 2E4,  Canada}

\author[5,6,7]{Stefan Filipp}
\ead{Stefan.Filipp@wmi.badw.de}
\address[5]{Physik-Department, Technische Universit\"{a}t M\"{u}nchen, 85748 Garching, Germany}
\address[6]{Walther-Mei{\ss}ner-Institut, Bayerische Akademie der Wissenschaften, 85748 Garching, Germany}
\address[7]{Munich Center for Quantum Science and Technology (MSQCT), Schellingstra\ss e 4, 80799 M\"{u}nchen, Germany}

\author[8,9,10,11]{Leong-Chuan Kwek}
\ead{kwekleongchuan@nus.edu.sg}
\address[8]{Centre for Quantum Technologies, National University of Singapore, 117543 Singapore} 
\address[9]{MajuLab, CNRS-UNS-NUS-NTU International Joint Research Unit, Singapore UMI 3654, Singapore}
\address[10]{National Institute of Education, Nanyang Technological University, 1 Nanyang Avenue, 637616 Singapore}
\address[11]{Quantum Science and Engineering Centre (QSEC), Nanyang Technological University, 50 Nanyang Avenue, 639798 Singapore}

\author[12]{Erik Sj\"oqvist}[orcid=0000-0002-4669-1818]
\ead{erik.sjoqvist@physics.uu.se}
\address[12]{Department of Physics and Astronomy, Uppsala University, Box 516, Se-751 20 Uppsala, Sweden}

\author[13]{Dianmin Tong}
\ead{tdm@sdu.edu.cn}
\address[13]{Department of Physics, Shandong University, China}


\begin{abstract}
Geometric and holonomic quantum computation utilizes intrinsic geometric properties of quantum-mechanical state spaces to realize quantum logic gates. Since both geometric phases and quantum holonomies are global quantities depending only on the evolution paths of quantum systems, quantum gates based on them possess built-in resilience to certain kinds of errors. This review provides an introduction to the topic as well as gives an overview of the theoretical and experimental progress for constructing geometric and holonomic quantum gates and how to combine them with other error-resistant techniques.

\end{abstract}


\maketitle

\tableofcontents

\section{Introduction}

Nature is replete with examples of  phenomena where  
holonomy constitutes the principal mechanism: interference patterns of the peacock feather, oscillations in electrical circuits, optical effects of biological systems, and the precession of the Foucault pendulum to name a few. A common feature of these phenomena is the failure of some quantity to return to its original value when transported with no rate of change around a loop in some parameter space \cite{berry90}. For instance, the plane of swing of the Foucault pendulum fails to return after Earth has made a full turn around its axis, although this plane never rotates locally with respect to the fixed stars. 

Quantum holonomies, including Abelian and non-Abelian ones, are other examples arising from the cyclic evolution of quantum systems. A quantum holonomy is a well-known analog of the rotation effect in differential geometry that occurs when a vector is parallel transported, {\it i.e.}, transported without local rotation, around a loop on a curved surface. The rotation is thus a global property caused by the curvature of the underlying space \cite{wilczek1989geometric,bohm2003the,chruscinski2004}. In quantum mechanics, such rotations are represented as phase factors (Abelian quantum holonomies) or unitary matrices (non-Abelian quantum holonomies). In existing papers, Abelian quantum holonomies are also called Abelian geometric phases or just geometric phases, while non-Abelian quantum holonomies are also referred to as non-Abelian geometric phases. To be specific, and in line with the literature,  we shall call `Abelian quantum holonomies' geometric phases and `non-Abelian quantum holonomies' quantum holonomies for short in the following.

In the quantum regime, an adiabatically evolving system driven by a nondegenerate Hamiltonian exhibits a phase factor of purely geometric origin under cyclic evolution. This was demonstrated by Berry back in 1984 \cite{berry1984quantal}. The phase factor is now commonly known as the Berry phase or the adiabatic geometric phase. Yet, although Berry's finding is 
often considered the first general theory for geometric phases, there were several 
studies on the geometric phase in specific quantum systems a few decades earlier 
\cite{longuet1975intersection,stone1976spin}. Furthermore, Rytov \cite{rytov1938transmitting}, 
Vladimirskii \cite{vladimirskii1941rotation}, and Pancharatnam \cite{pancharatnam1956,bhandari1997polarization} 
studied geometric phases in classical optics (see also the reviews in \cite{bliokh2015spin,cisowski2022geometric}). 

In 1959, Aharonov and Bohm \cite{aharonov1959significance} predicted that 
the interference pattern of free electrons is influenced by a confined magnetic flux, 
despite the fact that the magnetic field is negligible in the region of space through which
the electrons propagate. The interference pattern results 
from the emergence of the Aharonov-Bohm phase, which can be treated as a special case 
of the geometric phase \cite{berry1984quantal}. In a subsequent study, Mead and Truhlar 
\cite{mead1979on} obtained an explicit form of the geometric phase and the corresponding 
gauge potential in molecular systems by recasting the implicit formula of the geometric phases in an earlier study of a Jahn-Teller system by Longuet-Higgins {\it et al.} \cite{longuet1958studies}.

Berry's original derivation of the adiabatic geometric phase invoked the quantum adiabatic 
approximation and, therefore strictly speaking, applies only to slowly changing Hamiltonians.
For an adiabatic evolution of the quantum state, the adiabatic theorem ensures that the initial state of the system continues to track the change in the Hamiltonian, completing a cyclic evolution in parameter space. However, this situation no longer holds in nonadiabatic evolution. In 1987, Aharonov and Anandan \cite{aharonov1987phase} removed the adiabatic limitation of the Berry phase by considering closed loops of quantum states in the corresponding state space instead of loops in parameter space. This result is important because every loop in state space can be realized by an infinite set of Hamiltonians, all resulting in the same geometric phase.  Since adiabatic evolution is one of the ways to 
achieve a given quantum state evolution, the Berry phase becomes a special case of 
the nonadiabatic geometric phase (also called the Aharonov-Anandan phase, or AA phase). The latter phase was subsequently generalized to arbitrary (nonunitary or noncyclic) pure state evolution by Samuel
and Bhandari in 1988 \cite{Samuel1988general}.  

Soon after Berry's discovery, Simon realized that the geometric phase is deeply and 
profoundly connected to the holonomy of a fiber bundle in differential geometry 
\cite{simon1983holonomy}. Moreover, Berry's gauge potential plays the role of a 
connection on this fiber bundle. Thus, the adiabatic and nonadiabatic geometric 
phases are examples of Abelian holonomies in quantum physics. Non-Abelian geometric 
phase or quantum holonomy was first proposed by Wilczek and Zee for adiabatic 
evolution in \cite{wilczek1984appearance}, where they showed that a quantum state 
confined to a degenerate energy subspace of a Hamiltonian undergoing a cyclic 
evolution in the space of slow control parameters, may acquire a geometric unitary 
change, in addition to the global dynamical phase. In 1988, Anandan \cite{anadan1988non} extended 
this result to nonadiabatic evolution by considering general cyclic evolutions of a subspace of the state space. Non-Abelian holonomies for open path evolution were demonstrated in the particular context of dynamical invariants \cite{mostafazadeh99} as well as for continuously evolving subspaces \cite{kult2006noncyclic}. In fact, both geometric phases and quantum holonomies are beautifully propounded within a fiber bundle theory. 
In particular, the geometric phases are related to line bundles, while the quantum holonomies correspond to vector bundles. For reviews of the underlying geometric structures 
behind the geometric phases and quantum holonomies, we refer to Refs. \cite{wilczek1989geometric,bohm2003the,chruscinski2004,moore1991calculation,kolodrubetz2017geometry} 
and references therein.

Most quantum systems are not in pure states, where full knowledge regarding the system is available, but in mixed states. Mixed states are described by density matrices, first introduced by von Neumann in 1927 \cite{von1927wahrscheinlichkeitstheoretischer}.  Geometric phases for mixed states can also be studied since mixed states in any physical system can be mathematically described by a pure state in a larger system, known as purification. In 1986, Uhlmann \cite{uhlmann1986parallel} introduced a holonomy for sequences of density matrices through purification. In 2000, Sj\"{o}qvist {\it et al.} \cite{sjoqvist2000geometric} proposed a concept of geometric phase for nondegenerate mixed-states undergoing unitary evolution in an interferometric setting. In 2004, Tong {\it et al.} \cite{tong2004kinematic} proposed a quantum kinematic approach to the geometric phase for mixed states in nonunitary evolution. Further studies along this line of approach can be found in Refs. \cite{singh2003geometric,filipp2003off,ericsson2003mixed,chaturvedi04,marzlin2004geometric,mikael2005mixed}. When a quantum system interacts with its environment, its state is best described as a mixed state. In this direction, geometric phases in open quantum systems have been examined in the contexts of quantum jumps \cite{carollo2003geometric,fuentes2005holo}, quantum maps \cite{ericsson2003general,ramberg2019environment}, and the adiabatic approximation \cite{thunstrom2005weak,sarandy2006abelian,oreshkov2010adiabatic}.
Summary of pioneering works on different kinds of geometric phases is shown in Table \ref{tab:sec1:summary_geometric_phases}.
We remind the readers that the scope of this review is on pure states physics, and mixed states cases will not be discussed in-depth.

Nevertheless, the environment is not always a bane since it can help to generate geometric phases and quantum holonomies. Carollo {\it et al.} showed that geometric phases of a quantum system could be generated through slow variation of parameters of an engineered reservoir \cite{carollo2006coherent}. Carollo {\it et al.} also suggested a scheme to observe this effect by using a multilevel atom interacting with a broadband squeezed vacuum with an adiabatically changing squeezing parameter \cite{carollo2003geometric}. This scheme was proposed to be realized in a system where an atom is trapped in an optical cavity through engineering decay in the framework of cavity quantum electrodynamics \cite{zheng2012dissipation}. 
In parallel, the decoherence-induced geometric phases and quantum holonomies were also studied in open quantum systems undergoing cyclic adiabatic evolution \cite{sarandy2006abelian,dasgupta2007decoherence} and in quantum channels \cite{kult2008holonomy}.

In addition to the theoretical progress, geometric phases and quantum holonomies have been experimentally studied and tested in various quantum systems (for a review, see Ref. \cite{bohm2003the}). In particular, geometric phases are used for state preparation \cite{khosla2013quantum} and energy transfer \cite{fu2019geometric} in optomechanical systems (for reviews, see Refs. \cite{kippenberg2008cavity,aspelmeyer2014cavity}, and a recent advance \cite{xiong2022higher}).

\begin{table}[ht]
\caption{Summary of pioneering works on different kinds of geometric phases. To categorize the geometric phases according to their features, we use $C/NC$ to denote cyclic/noncyclic 
evolution, $P/M$ for pure/mixed states in parentheses.}
\label{tab:sec1:summary_geometric_phases}
\centering
\resizebox{\textwidth}{!}{
 \begin{tabular}{|c|c|c|} 
 \hline
  & Adiabatic & Nonadiabatic \\ 
 \hline
 Abelian & Berry  ($C,P$) \cite{berry1984quantal} & \tabincell{c}{Aharonov-Anandan ($C,P$) 
 \cite{aharonov1987phase}, Samuel-Bhandari ($NC, P$) \cite{Samuel1988general}, \\
 Uhlmann ($NC, M$) \cite{uhlmann1986parallel}, Sj\"{o}qvist {\it et al.} ($NC, M$) 
  \cite{sjoqvist2000geometric}, \\ Tong {\it et al.} ($NC, M$) \cite{tong2004kinematic}}
  \\ 
 \hline
 Non-Abelian & Wilczek-Zee ($C,P$) \cite{wilczek1984appearance} 
 & Anandan 
 ($C,P$) \cite{anadan1988non}, Kult {\it et al.} ($NC, P$) \cite{kult2006noncyclic}  \\
 \hline
\end{tabular}}
\end{table}

The rise of quantum information science has opened up a new avenue for applying geometric phase and quantum holonomy as a tool for robust quantum information processing \cite{vedral2003geometric,sjoqvist2008trend,sjoqvist15}. While large-scale fault-tolerant quantum computation with quantum error corrections is still far away, we are now in the noisy intermediate-scale quantum (NISQ) era \cite{preskill2018quantum}. Building high-fidelity quantum gates to perform quantum algorithms is essential at this stage. 
According to the literature, the approach to using geometric phases for implementing quantum gates for circuit-based quantum computation is referred to as geometric  quantum computation (GQC) \cite{jones2000geometric,ekert2000geometric}; the approach to using quantum holonomies for quantum gates is usually called holonomic quantum computation (HQC) \cite{zanardi1999holonomic}. GQC and HQC are believed to be useful in reaching the error threshold, below which quantum computation with faulty gates can be performed. The basic reasoning behind the conjectured robustness is that the geometric phase is a global feature and is resilient to errors, such as parameter noise and environment-induced decoherence, which are picked up locally along the path in state space. 

For the Abelian case, the basic idea of GQC follows directly from the fact that a quantum state $\ket{\psi}$ undergoing a cyclic evolution will acquire a geometric phase. In the qubit case, $\ket{\psi}$'s orthogonal state $\ket{\psi_\perp}$ accumulates the same geometric phase but with opposite sign. This amounts to applying a phase-shift gate to the qubit by geometric means if only the geometric phases are taken into account.
According to how the evolution is performed, the geometric gate can be adiabatic (with Berry phases) or nonadiabatic (with AA phases).
In both cases, however, dynamical phases may accumulate. Therefore, designing a special evolution path along which the unwanted dynamical phase can be canceled or avoided is essential.

Adiabatic GQC was first proposed and experimentally demonstrated by Jones {\it et al.} in 2000 \cite{jones2000geometric}.
They used a nuclear magnetic resonance (NMR) system to implement geometric gates by adiabatically varying the system Hamiltonian along loops in parameter space so that the Hamiltonian's eigenstates undergo cyclic evolution. Since each eigenstate has a different energy, the eigenstates must pick up different dynamical phases, but the effect of these phases was erased through spin echo techniques. Despite being proposed in a particular system, the idea in \cite{jones2000geometric} is general, {\it i.e.}, any system that has the same effective Hamiltonian and control capacity is suitable for adiabatic GQC. For example, setups formed by superconducting nanocircuits coupled through capacitors are another candidate system \cite{falci00detection}.

Adiabatic gates need a long evolution time, which may expose the qubits to severe decoherence. One solution to this problem is to use the AA phase to implement geometric gates so that the adiabatic constraint can be relaxed. Wang and Matsumoto put forward the first proposal of the AA phase in NMR, where the dynamics of spin qubits in an external rotating field is exactly solvable \cite{wang2001nonadiabatic}. 
Phase-shift gates based on a certain pair of bases commute with each other and cannot form universal single-qubit gates. However, Zhu and Wang demonstrated that an arbitrary geometric single-qubit gate could be constructed by combining two phase-shift gates for different bases \cite{zhu02}.
In addition to NMR, this idea has been applied to trapped ions \cite{li02non}, Faraday rotation spectroscopy \cite{li2002ultrafast}, superconducting nanocircuits \cite{zhu02geometric}, semiconductor quantum dots \cite{solinas2003nonadiabatic}, Rydberg atoms \cite{zhao17rydberg}, transmons \cite{chen2018non}, and silicon-based qubits \cite{zhang20high}. 
Moreover, geometric gates can be designed with dynamical invariant theory (DIT) \cite{teo2005geometric,shao2007implementation}. Examples are provided in NMR systems \cite{wang2009noncyclic} and in the Jaynes-Cummings model \cite{wang2009geometric}.
With DIT, geometric gates can be easily generalized to noncyclic evolution \cite{teo2005geometric,wang2009noncyclic}.

All the aforementioned proposals have to take care of dynamical phases, either by avoiding or removing them. However, when the dynamical phases have the same global feature as their geometric counterpart, for instance, when the dynamical phase is proportional to the geometric phase, it is not necessary to remove the dynamical phase. Geometric gates based on this idea are referred to as unconventional geometric gates. The idea was triggered in Ref.~\cite{leibfried2003experi} and developed in Ref.~\cite{zhu03unconventional}.

For the non-Abelian case, the development of HQC started by the work \cite{zanardi1999holonomic,pachos1999non}, using the Wilczek-Zee phase (adiabatic quantum holonomy) \cite{wilczek1989geometric} to implement quantum gates for circuit-based quantum computation.
The underlying idea is to encode a set of qubits in a set of degenerate eigenstates of a parameter-dependent Hamiltonian and to adiabatically transport these states around a loop in the corresponding parameter space. Based on this idea, Duan {\it et al.} proposed a tripod scheme to achieve a universal set of adiabatic holonomic gates for trapped ions confined in a linear Paul trap \cite{duan2001geometric}. This scheme has been adapted to many other systems \cite{faoro2003non,solinas2003semiconductor} and experimentally realized in a trapped ion system \cite{toyoda2013realization}.

Similar to adiabatic GQC, an obstacle in achieving adiabatic HQC is the long run time required for the desired adiabatic evolution. To overcome this problem, nonadiabatic HQC was proposed \cite{sjoqvist2012non,xu2012non} based on nonadiabatic holonomies \cite{anadan1988non}. The key feature of the nonadiabatic scheme is that it removes the adiabatic constraint and thus combines speed and universality. Due to its modest requirements, nonadiabatic HQC was soon experimentally demonstrated in various quantum systems \cite{abdumalikov2013experimental,feng2013experimental,arroyo2014room,zu2014experimental}. The earliest experiments on nonadiabatic HQC were performed with a superconducting transmon \cite{abdumalikov2013experimental} for single-qubit gates and with NMR \cite{feng2013experimental} for single- and two-qubit gates.

Following the original three-level setting based on the resonant $\Lambda$ model configuration \cite{sjoqvist2012non,xu2012non}, many other schemes for nonadiabatic HQC have been put forward. One important development is the single-shot scheme \cite{xu2015non,sjoqvist2016non} in which an arbitrary single-qubit gate can be realized in one loop by introducing a detuning in the $\Lambda$ model. The length of the single loop can be further shortened by separating the loop into several segments \cite{xu2018path} so that the total evolution time is compressed. When the detuning is not available, multi-pulse schemes with resonant control can be utilized for the same purpose \cite{herterich2016single}. In addition to the schemes with resonant or off-resonant $\Lambda$ model, nonadiabatic HQC can be implemented with a four-level structure \cite{mousolou2014universal} or with auxiliary systems \cite{gurkan2015realization,zhang2018holonomic,wang2020dephasing}. This development allows for the realization of holonomic gates with qubits rather than multi-level systems.

A broad spectrum of reverse-engineering approaches can be applied to HQC  for fast or robust holonomic gates. This idea has been adopted for accelerating adiabatic HQC via adiabatic shortcut \cite{zhang2015fast}. It has been realized that one can also design geometric and holonomic gates via reverse engineering directly. Explicit formulas for both Abelian and non-Abelian cases have been given \cite{liu2019plug,li2020approach,zhao2020general}, followed by many concrete examples of high-performance gates \cite{liu2021super}.

A significant feature of HQC as well as GQC is that they are compatible with many other error suppression and error-correcting techniques. The combination of GQC and HQC with such quantum technologies can obviously merge advantages from both sides. For example, similar to the traditional dynamical gates, the performance of GQC and HQC gates can be improved by applying optimal control \cite{werschnik2007quantum}. Another reason to do this is that holonomic gates are not resilient to all kinds of errors, such as environmental noise \cite{clerk2010introduction,schlosshauer2019quantum}. Broadly speaking, the environmental-noise countering strategies fall into two categories: passive error-avoiding ones, such as decoherence-free subspaces \cite{zanardi1997noiseless,duan1997preserving,lidar1998decoherence} and noiseless subsystems \cite{knill2000theory}, and active ones, such as dynamical decoupling \cite{yang2011preserving} and quantum error correcting codes \cite{terhal2015quantum,lidar2013quantum}. Geometric \cite{feng2009geometric} and holonomic \cite{wu2005holonomic,xu2012non} gates performed in decoherence-free subspaces or noiseless subsystems \cite{wu2005holonomic,zhang2014quantum} are not affected by the collective decoherence. For a general linear-independent environment, dynamical decoupling can be used to protect geometric gates \cite{xu2014protecting,wu2020universal,zhao2021dynamical}. When the errors are local and with a low enough rate, performing quantum error correction in a holonomic manner will lead to a new fault-tolerant quantum computation scheme \cite{oreshkov2009fault,oreshkov2009scheme}. On the other hand,  the environment can even help to generate holonomic gates \cite{carollo2006coherent,yin2007implementation,ramberg2019environment}.

In this review article, we aim to provide a detailed account of the theoretical and experimental development of the different forms of GQC and HQC with three main objectives: (i) to explain pedagogically the mathematical concepts in the literature by constructing a unique framework with clear physical insights into various schemes; (ii) to build bridges that connect different subfields of quantum computing, such as HQC and other quantum computing and control approaches in order to realize scalable physical qubits with higher resistance to noise and parameter fluctuations; and (iii) to point out potential future research directions, along which theorists and experimentalists can collaborate to find new realizations of quantum gates based on geometric principles.

This review is organized as follows. Section \ref{sec:intro} introduces and illustrates the basic concepts of geometric phases and quantum holonomies in both adiabatic and nonadiabatic regimes. Section \ref{sec:GQC} reviews different schemes for GQC and their experimental realizations. Section \ref{sec:nonabelianHQC} is dedicated to HQC schemes as realized in various physical settings. Section \ref{sec:combineHQC} describes how GQC and HQC can be combined with other error suppression and correcting techniques. The review ends with concluding remarks in Section \ref{sec:conclude}.

\section{Geometric phases and quantum holonomies} 
\label{sec:intro} 

\subsection{Adiabatic geometric phase}\label{sec:1.1}

\subsubsection{General theory}\label{subsec:ch1:general}
Consider a quantum system driven by a nondegenerate Hamiltonian $H(R)$ that depends on a set of time-dependent real-valued parameters $R$. The dynamics of the system is described by the Schr\"{o}dinger equation
\begin{equation}\label{eq:Schrodinger}
    i\hbar\frac{\mathrm{d}}{\mathrm{d}t}\ket{\psi(t)}=H(R(t))\ket{\psi(t)}.
\end{equation}
The Hamiltonian can be decomposed according to its eigenvalues and eigenstates as
\begin{equation}
    H(R(t))=\sum_{n}E_n(R(t))\ket{\varphi_n(R(t))}\bra{\varphi_n(R(t))}.
\end{equation}
We assume that the Hamiltonian has a discrete energy spectrum, and the initial state $\ket{\psi(0)}$ is prepared in the $n$th eigenstate:
\begin{equation}
    \ket{\psi(0)}=\ket{\varphi_n(0)},
\end{equation}
where $\ket{\varphi_n(t)}$ is the shortened expression for $\ket{\varphi_n(R(t))}$.
We adopt the same convention for $E_n$ as well: $E_n(t) \equiv E_n (R(t))$.
If the control parameters $R$ are varied slowly enough, the adiabatic theorem guarantees that the system would remain in the $n$th eigenstate of the instantaneous Hamiltonian $H(t)$ over the entire evolution (recent discussions on the adiabatic condition can be seen in Refs. \cite{marzlin2004incosistency,tong2005quantitative,tong2010quantitative,liu2005optical} and references therein).
Notice that the instantaneous eigenstate $\ket{\varphi_n(t)}$ is in general related to its time-evolving solution to the Schr\"odinger equation by
\begin{equation}\label{eq:psit}
    \ket{\psi(t)}=e^{i\gamma_n(t)}\ket{\varphi_n(t)}.
\end{equation}

The phase $\gamma_n(t)$ consists of two terms.
The first refers to the well-known dynamical phase angle $\gamma^d_n(t)$, which is the time integral over the energy eigenvalue
\begin{equation}\label{dp}
    \gamma^d_n(t)=-\frac{1}{\hbar}\int_{0}^{t}E_n(t')\mathrm{d}t',
\end{equation}
and the second is readily obtained by inserting \eqref{eq:psit} into the Schr\"{o}dinger equation, \eqref{eq:Schrodinger}.
By taking the inner product with $\bra{\varphi_n(t)}$, we obtain
\begin{equation}
    \frac{\mathrm{d}}{\mathrm{d}t}\gamma_n(t)=-\frac{1}{\hbar}E_n(t)+i\bra{\varphi_n(t)}\frac{\mathrm{d}}{\mathrm{d} t}\ket{\varphi_n(t)}.
\end{equation}
The latter phase is known as Berry's phase, and it can be expressed as
\begin{equation}\label{gp}
    \gamma^g_n(t)=i\int_0^t\bra{\varphi_n(t')}\frac{\mathrm{d}}{\mathrm{d} t'}\ket{\varphi_n(t')}\mathrm{d}t=i\int_{R(0)}^{R(t)}\bra{\varphi_n(R)}\nabla_R\ket{\varphi_n(R)}
    \cdot \mathrm{d}R,
\end{equation}
where $\nabla_R$ is the gradient operator with respect to the parameters $R$.
We note that $\bra{\varphi_n(t)}\frac{\mathrm{d}}{\mathrm{d} t}\ket{\varphi_n(t)}$ is a purely imaginary number, which follows from normalization:
\begin{eqnarray}\label{eq:zero}
    & & \braket{\varphi_n(t)}{\varphi_n(t)} = 1
    \nonumber \\ 
    & & \Rightarrow \frac{\mathrm{d}}{\mathrm{d} t}(\braket{\varphi_n(t)}{\varphi_n(t)})=\left(\frac{\mathrm{d}}{\mathrm{d} t}\bra{\varphi_n(t)}\right)\ket{\varphi_n(t)} + \bra{\varphi_n(t)}\frac{\mathrm{d}}{\mathrm{d} t}\ket{\varphi_n(t)} = 2{\rm Re} \bra{\varphi_n(t)}\frac{\mathrm{d}}{\mathrm{d} t}\ket{\varphi_n(t)} = 0 , 
\end{eqnarray}
where we have used that 
$\left(\frac{\mathrm{d}}{\mathrm{d} t}\bra{\varphi_n(t)}\right)\ket{\varphi_n(t)}=\left(\bra{\varphi_n(t)}\frac{\mathrm{d}}{\mathrm{d} t}\ket{\varphi_n(t)}\right)^*$. It thus follows that $\bra{\varphi_n(t)}\frac{\mathrm{d}}{\mathrm{d} t}\ket{\varphi_n(t)}$ must be purely imaginary, implying that $\gamma^g_n(t)$ is real.

Though Berry was not the first to notice this phase $\gamma^g_n(t)$, it is generally regarded that Berry's work highlighted its nontriviality. The alleged triviality can be seen from a reformulation of the phase as $\gamma^g_n(t)=\int A_n(R) \cdot \mathrm{d}R$, where
\begin{equation}
    A_n(R)=i\bra{\varphi_n(R)}\nabla_R\ket{\varphi_n(R)}
\end{equation}
is the Berry connection that can be treated as a gauge potential.
One can choose a different gauge by doing a gauge transformation of the eigenstates
\begin{equation}
    \ket{\varphi_n(R)}\rightarrow\ket{\varphi_n(R)}'=e^{i\xi_n(R)}\ket{\varphi_n(R)},
\end{equation}
where $\xi_n(R)$ is an arbitrary single- and real-valued function of $R$. It is interesting to note that $\ket{\varphi_n(R)}$ is not the only suitable choice of eigenstate for $H(R)$. A new set of $\ket{\varphi_n(R)}'$ also forms a set of eigenstates for the same Hamiltonian, where the corresponding new gauge potential is related to the original one via
\begin{equation}
    A_n'(R)=A_n(R)-\nabla_R\xi_n(R).
\end{equation}
Given that $\xi_n(R)$ is arbitrary, $A_n'(R)$ can always be set to zero by choosing
\begin{equation}
    \xi_n(R)=\int_{R(0)}^R A_n(R') \cdot \mathrm{d}R'.
\end{equation}
This is one of the reasons why $\gamma_n^g(t)$ was treated as trivial.  Yet it is somehow deeply connected to holonomic theory.
Evolution according to the above choice of $\xi_n(R)$ is usually referred to as `parallel transport' in geometry.

However, as Berry pointed out \cite{berry1984quantal},  the above observation is not true when the adiabatic evolution of the eigenstate is cyclic, {\it i.e.}, when $R(T)=R(0)$ with $T$ being the total evolution time.
In this case, the phase angle $\gamma_n^g(t)$ takes the form 
\begin{equation}\label{gpC}
    \gamma^g_n(C)=i\oint_C \bra{\varphi_n(R)}\nabla_R\ket{\varphi_n(R)} \cdot \mathrm{d}R,
\end{equation}
where $C$ is the closed loop traversed by $\ket{\varphi_n(R)}$ in parameter space. 
We can show that
\begin{equation}
    {\gamma^g_n}'(C)=\oint_C [A_n(R)-\nabla_R\xi_n(R)]\cdot \mathrm{d}R=\oint_C A_n(R) \cdot \mathrm{d}R=\gamma_n^g(C),
\end{equation}
where we have used the relation 
$\oint_C\nabla_R\xi_n(R)\cdot 
\mathrm{d}R=\xi_n(R_f)-\xi_n(R_i)=0$.
Note that since the Berry phase $\gamma_n^g(C)$ is a phase 
angle, the final computation should be taken modulo 
$2\pi$. The above equation exactly shows that $\gamma_n^{g}(C)$ is a $U(1)$ gauge 
invariant quantity, {\it i.e.}, a different choice of 
eigenstates of the Hamiltonian gives no change for 
$\gamma^g_n(C)$.
More importantly, this phase has a geometrical origin, and this can be seen from the latter part of Eq.~(\ref{gp}): the integration is determined by the initial and final points as well as how the curve traverses in parameter space, making the geometric phase a nonintegrable (path-dependent) phase, and not by the rate of change of control parameters or energy of the eigenstate during the evolution. 
When the cyclic condition is considered, the line integral in \eqref{gpC} can be transformed into a surface integral over any surface in parameter space with loop $C$ as its boundary (Stokes' theorem). Therefore, Berry's phase is also referred to as the geometric phase in adiabatic evolution.

\subsubsection{Berry phase of a spin-$\frac{1}{2}$ particle in an adiabatically rotating magnetic field}
\label{subsec:ch1:Berry_phase}

To elucidate the general theory outlined above, we study here a simple quantum system in some detail.
The quantum system is a spin-$\frac{1}{2}$ particle driven by an external magnetic field $\Vec{B}$ that rotates slowly around the $z$-axis with an angle $\theta$.
The angular frequency $\omega$ of the rotation is sufficiently small so that the adiabatic condition is satisfied, {\it i.e.}, the spin state of the particle would closely adhere to the direction of the magnetic field.
The magnetic field is shown in vector form as 
\begin{equation}
    \Vec{B}(t)=B_0(\sin\theta\cos(\omega t),\sin\theta\sin(\omega t),\cos\theta),
\end{equation}
where $B_0$ is the magnitude of $\Vec{B}$, and $t$ is the evolution time.
The interaction Hamiltonian for the particle in the field is given by
\begin{equation}\label{hw}
    H(t)=\mu\Vec{B}(t)\cdot\Vec{\sigma},
\end{equation}
where $\mu$ is  the gyromagnetic ratio, and $\Vec{\sigma}=(\sigma_x,\sigma_y,\sigma_z)$ are the standard Pauli operators.
In the ordered basis $\{\ket{\uparrow_z} \equiv \ket{0}, \ket{\downarrow_z} \equiv \ket{1}$, with $\sigma_z \ket{0}=\ket{0}, \sigma_z\ket{1}=-\ket{1}\}$, $H(t)$ takes the form 
\begin{equation}\label{hw2}
    H(t)=\mu B_0\left(\begin{array}{cc}
       \cos\theta  & e^{-i\omega t}\sin\theta \\
        e^{i\omega t}\sin\theta & -\cos\theta
    \end{array}\right).
\end{equation}
The corresponding eigenvalues and eigenstates of $H(t)$ are 
\begin{align}
    &E_+=\mu B_0, && \ket{\varphi_+(t)}=\cos\frac{\theta}{2}\ket{0}+e^{i\omega t}\sin\frac{\theta}{2}\ket{1} , 
    \nonumber \\
    &E_-=-\mu B_0, && \ket{\varphi_-(t)}=-\sin\frac{\theta}{2}\ket{0}+e^{i\omega t}\cos\frac{\theta}{2}\ket{1}.
\end{align}
It is clear that the Hamiltonian is parameterized by $\theta$, $\phi(t)=\omega t$, and $r=B_0$, which can also be used to define a parameter space with $\theta$ the nutation angle, $\phi$ the precession angle, and $r$ the polar radius. While the space of slow parameters is identical to the unit sphere $S^2$, the radius $r$ controls the rate at which the adiabatic limit is approached. A point on the sphere defined by the angles $\theta$ and $\phi$ can be associated with the eigenstates $\ket{\varphi_+}=\cos\frac{\theta}{2}\ket{0}+e^{i\phi}\sin\frac{\theta}{2}\ket{1}$ and $\ket{\varphi_-}=-\sin\frac{\theta}{2}\ket{0}+e^{i\phi}\cos\frac{\theta}{2}\ket{1}$.
Within the time interval $t\in(0,{2\pi/\omega})$, $H(t)$ traces a closed curve $C$ on the sphere. 
Suppose the initial state of the particle is prepared in one of the eigenstates of $H(t)$, {\it i.e.}, one of the points on the spherical shell, and the adiabatic condition is satisfied, the adiabatic theorem guarantees that the final state at $t={2\pi}/{\omega}$ returns to the original state, except for a phase factor.

The dynamical and geometric phases can be calculated with the aid of Eqs.~(\ref{dp}) and (\ref{gp}), respectively.
On the one hand, the dynamical phases for the two eigenstates $\ket{\varphi_{\pm}(t)}$ for the period $T={2\pi}/{\omega}$ are
\begin{equation}\label{dppm}
    \gamma_{\pm}^d(T)=-\frac{1}{\hbar}\int_0^T \pm\mu B_0\mathrm{d}t=\mp\frac{1}{\hbar}\mu B_0T.
\end{equation}
On the other hand, the geometric phases are shown to be
\begin{equation}\label{gppm}
    \gamma_{\pm}^g(C)=i\oint_C \bra{\varphi_{\pm}(R)}\nabla_R\ket{\varphi_{\pm}(R)} \cdot \mathrm{d}R=-\pi(1\mp \cos\theta) = \mp \pi(1 - \cos\theta), \ {\rm mod} \ 2\pi,
\end{equation}
where we have taken $\theta$ and $r$ as constant and $\phi\in[0,2\pi]$.
We recall that the solid angle surrounded by $C$ takes the form of 
\begin{equation}
    \Omega(C)=2\pi(1-\cos\theta).
\end{equation}
Therefore, the geometric phases can be expressed in terms of the solid angle as follows
\begin{equation}\label{eq:Omega}
    \gamma_{\pm}^g(C)=\mp\frac{1}{2}\Omega(C) , 
\end{equation}
which reveals a concrete dependence on geometry.
From Eqs.~(\ref{dppm}) and (\ref{gppm}), we see that the dynamical phases depend on the rotation time $T$ and the energy eigenvalues $\pm \mu B_0$, while the geometric phases depend only on the solid angles traced out by the time-dependent magnetic field.

\subsection{Nonadiabatic geometric phase}\label{sec:1.2}
Nonadiabatic generalization of the Berry phase was first proposed by Aharonov and Anandan in 1987 \cite{aharonov1987phase}, where they  relaxed the adiabatic constraint on the rate of change of the system Hamiltonian by considering general cyclic evolution of quantum states. In this scenario, the space used to describe traces of evolution is not the parameter space employed in Berry's framework, but the projective Hilbert space with each vector representing a quantum state itself.

In quantum theory, state vectors of a system differing by global phase factors correspond to the same physical state because they are not distinguishable by any measurements. 
Thus, such a set of states can be defined as an equivalence class corresponding to a certain vector in Hilbert space.
When a quantum state evolves in time according to the Schr\"{o}dinger equation, it changes from $\ket{\psi(0)}$ to another state $\ket{\psi(t)}$ and thus traces out a curve in state space.
By comparing two points on the curve (say, $\ket{\psi(t_1)}$ and $\ket{\psi(t_2)}$), one can conjecture that in addition to the state difference, there is also an accompanying phase change with the time evolution.
However, we cannot directly observe this change of the phase from the curve since the phase information has been eliminated in the projective Hilbert space.
To study the phase difference properly, we need to rely on the solution of the Schr\"{o}dinger equation.

Before doing this formally, the above analysis of the Berry phase suggests a dynamical phase of the form  
\begin{equation}\label{gammad}
    \gamma^d(t)=-\frac{1}{\hbar}\int_0^t\bra{\psi(t')}H(t')\ket{\psi(t')}\mathrm{d}t'
\end{equation}
as it reduces to $-\hbar^{-1} \int_0^t E_n (t'){\rm d} t'$ in the adiabatic limit. 
We can also expect a geometric contribution to the total phase. 
To verify this, let us study the difference between the total phase and the dynamical one, and check if the difference has a geometric origin just like the Berry phase.

In the derivation of Berry's phase, an eigenstate of the system Hamiltonian returns to its initial state when the Hamiltonian completes a cyclic change adiabatically. This cyclic evolution is depicted as a closed loop in parameter space since phase factors are irrelevant for eigenstates.
In order to describe the evolution of the state based on the Schr\"{o}dinger equation, a Hilbert space is needed so that not only the state vector but also the accompanying phase factor is represented at the same time. Therefore, a closed curve in parameter space  generally corresponds to an open curve in Hilbert space. If we set the initial state $\ket{\psi_n(0)}$ in parameter space and the initial state $\ket{\psi(0)}$ in Hilbert space to be identical, then the difference between the final state $\ket{\psi_n(T)}=\ket{\psi_n(0)}$ in the parameter space and the final state $\ket{\psi(T)}$ in the Hilbert space exhibits the acquired total phase (the sum of the dynamical and geometric phases). This same strategy is used to study the Aharonov-Anandan phase in the following general theory.

\subsubsection{General theory}
\begin{figure}
  \centering
  \includegraphics[width=1.0\textwidth]{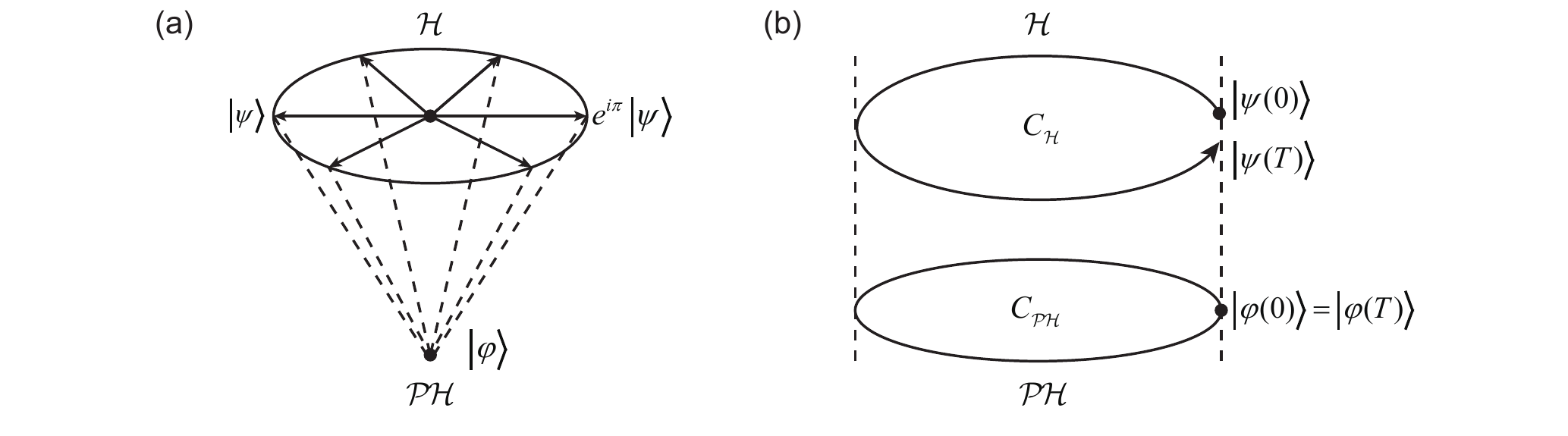}\\
  \caption{Schematic diagrams of the projection from Hilbert space ($\mathcal{H}$) to projective Hilbert space ($\mathcal{PH}$). (a) Given a quantum state $\ket{\psi}$ in $\mathcal{H}$, all vectors differing from $\ket{\psi}$ only by a global phase factor  $e^{i\gamma}$,$\gamma \in [0,2\pi)$, form a circle. All these $e^{i\gamma}\ket{\psi}$ correspond to the state $\ket{\varphi}$ in $\mathcal{PH}$ according to the projection $\Pi$. (b) The curve $[0,T] \ni t \mapsto \ket{\psi(t)}$ in $\mathcal{H}$ is projected to $\mathcal{PH}$ as $\ket{\varphi(t)}$, forming a closed path $C_\mathcal{PH}$, {\it i.e.}, $\ket{\varphi(T)}=\ket{\varphi(0)}$. The path $C_\mathcal{H}$ traced out by $\ket{\psi(t)}$ may be open, {\it i.e.}, $\ket{\psi(T)}$ may differ from $\ket{\psi(0)}$ by a phase factor.}\label{fig:phs}
\end{figure}
For an equivalence class of quantum states that differ up to global phases, we can define a projection operator $\Pi$ such that
\begin{equation}
    \Pi(e^{i\gamma}\ket{\psi})=\ket{\varphi},
\end{equation}
where $e^{i\gamma}$ is an arbitrary phase factor, $e^{i\gamma}\ket{\psi}$ are the states in Hilbert space, and $\ket{\varphi}$ is the corresponding state in the projective Hilbert space. 
The projection operator maps a Hilbert space $\mathcal{H}$ into a projective Hilbert space $\mathcal{PH}$, the latter being built up from the equivalence classes of all states in the Hilbert space [as an illustration, see Fig.~\ref{fig:phs}(a)].
This indicates that all state vectors of $\mathcal{H}$ related to one physical state are mapped to one single vector in $\mathcal{PH}$.

A cyclic evolution with period $T$ corresponds to a closed curve $C_{\mathcal{PH}}$ in projective Hilbert space, but typically to an open curve $C_\mathcal{H}$ in Hilbert space such that $\Pi(C_\mathcal{H})=C_\mathcal{PH}$ [see Fig.~\ref{fig:phs}(b)]. In Hilbert space $\mathcal{H}$, the evolution corresponds to a change of state $\ket{\psi(t)}$ according to the Schr\"{o}dinger equation; while in projective Hilbert space $\mathcal{PH}$, the projection operator $\Pi$ reduces the evolution of $\ket{\psi(t)}$ to that of $\ket{\varphi(t)}$ through $\Pi(\ket{\psi(t)})=\ket{\varphi(t)}$.
Note that, owing to the fact that $\ket{\varphi(t)}$ does not have to satisfy the Schr\"{o}dinger equation, it is always possible to find a single-valued differentiable $\ket{\varphi(t)}$, {\it i.e.}, 
$\ket{\varphi(T)}=\ket{\varphi(0)}$. 
Thus, the difference between $\ket{\psi(T)}$ and $\ket{\psi(0)}$ is only a phase factor:
\begin{equation}\label{eq:cc}
    \ket{\psi(T)} = e^{if} \ket{\psi(0)},
\end{equation}
where $f$ is the total phase acquired in the cyclic evolution. The phase factor $e^{if}$ is depicted in Fig.~\ref{fig:phs}(b) by the distance between $\ket{\psi(T)}$ and $\ket{\psi(0)}$ which are projected (shown by the dashed line) to the same state $\ket{\varphi(0)}$ in the projective Hilbert space.

The state $\ket{\psi(t)}$ and $\ket{\varphi(t)}$ can be connected by the following relation
\begin{equation}\label{gammat}
    \ket{\psi(t)}=e^{i\gamma(t)}\ket{\varphi(t)},
\end{equation}
where $\gamma(t)$ is a real function satisfying $\gamma(T) - \gamma(0) = f$.
By inserting Eq.~(\ref{gammat}) into the Schr\"{o}dinger equation, one obtains 
\begin{equation}
    \frac{\mathrm{d}}{\mathrm{d}t}\gamma(t)=-\frac{1}{\hbar}\bra{\varphi(t)}H(t)\ket{\varphi(t)}+i\bra{\varphi(t)}\frac{\mathrm{d}}{\mathrm{d} t}\ket{\varphi(t)}.
\end{equation}
Hence, the acquired total phase can be written as
\begin{equation}\label{gamma}
    \gamma(T)=\gamma^d(T)+\gamma^g(T)= -\frac{1}{\hbar}\int_0^T\bra{\varphi(t)}H(t)\ket{\varphi(t)}\mathrm{d}t+i\int_0^T \bra{\varphi(t)}\frac{\mathrm{d}}{\mathrm{d} t}\ket{\varphi(t)}\mathrm{d}t.
\end{equation}
It is evident that the total phase $\gamma(T)$ consists of two parts: the dynamical phase $\gamma^d(T)$ as defined in Eq.~(\ref{gammad}), since $\bra{\varphi}H\ket{\varphi}=\bra{\psi}H\ket{\psi}$, and $\gamma^g(T)$. 
One can show that $\gamma^g(T)$ does not depend on the details of the system Hamiltonian and is of geometric origin.
Note that, in the derivation of the Aharonov-Anandan phase, the adiabatic condition is not used so that $H(T)$ and $H(0)$ are allowed to be nontrivially different (more precisely, $H(T)$ and $H(0)$ can be noncommuting) still satisfying the cyclic condition in Eq. (\ref{eq:cc}). 
Therefore, the Aharonov-Anandan phase is the nonadiabatic geometric phase.

\subsubsection{Aharonov-Anandan phase of a spin-$\frac{1}{2}$ particle in a rotating magnetic field with arbitrary speed}\label{subsec:ch1:A-A}
As an example, we calculate the Aharonov-Anandan phase of a spin-$\frac{1}{2}$ particle in a rotating magnetic field. In contrast to the adiabatic parameter change used in the Berry phase example in  \secref{subsec:ch1:Berry_phase}, the angular frequency $\omega$ of the rotating magnetic field studied here can be arbitrarily large.
The system Hamiltonian is the same as in 
Eq.~(\ref{hw2}), and can be rewritten as 
\begin{equation}\label{eq:hAA}
    H(t)=e^{-i\frac{\omega t}{2}\sigma_z}[\mu B_0(\cos\theta \sigma_z+\sin\theta\sigma_x)]e^{i\frac{\omega t}{2}\sigma_z}=e^{-i\frac{\omega t}{2}\sigma_z}H(0) e^{i\frac{\omega t}{2}\sigma_z}.
\end{equation}
Thus, $H(t)$ represents a stationary system rotating around $\sigma_z$ with frequency $\omega$.
This means that if we transform to a rotating frame using the rotation operator $W(t)=e^{-i\frac{\omega t}{2}\sigma_z}$, the time-dependent Hamiltonian becomes time-independent.
The solution $\ket{\psi (t)}$ of the Schr\"{o}dinger equation in the laboratory frame is related to the solution $\ket{\eta (t)}$ in the rotating frame by
\begin{equation}
    \ket{\psi(t)}=W(t)\ket{\eta(t)} 
\end{equation}
with the initial condition $\ket{\psi(0)}=\ket{\eta(0)}$.
The state $\ket{\eta(t)}$ is determined by the Schr\"odinger equation 
\begin{equation}
    i\hbar\frac{\mathrm{d}}{\mathrm{d} t}\ket{\eta(t)}=\bar{H}\ket{\eta(t)},
\end{equation}
where $\bar{H}=H(0)-\frac{\hbar\omega}{2}\sigma_z$ is time-independent.
As a result, one finds 
\begin{equation}
    \ket{\eta(t)}=e^{-\frac{i}{\hbar}\bar{H}t}\ket{\eta(0)},
\end{equation}
which implies 
\begin{equation}
    \ket{\psi(t)}=e^{-i\frac{\omega t}{2}\sigma_z}e^{-\frac{i}{\hbar}\bar{H}t}\ket{\psi(0)}.
\end{equation}
When $\ket{\psi(0)}$ is an eigenstate of $\bar{H}$, $e^{-i\bar{H}t/\hbar}\ket{\psi(0)}$ equals to $e^{-i\bar{E}t/\hbar}\ket{\psi(0)}$ with $\bar{E}$ being the corresponding eigenvalue for $\ket{\psi(0)}$.
Hence, a cyclic evolution can be accomplished when $\ket{\psi(0)}$ is an eigenstate of $\bar{H}$ and the evolution time $T=\frac{2\pi}{\omega}$.
The eigenvalues $\bar{E}_{\pm}$ and eigenvectors $\ket{\eta_{\pm}}$ of $\bar{H}$ read 
\begin{equation}
    \bar{E}_\pm=\pm\mu B_0\Delta,\quad 
    \ket{\eta_+}=e^{-i\frac{\bar{\theta}}{2}\sigma_y}
    \ket{0},\quad 
    \ket{\eta_-}=e^{-i\frac{\bar{\theta}}{2}\sigma_y}
    \ket{1}
\end{equation}
with the parameters 
\begin{equation}
    \Delta=\sqrt{1-\frac{\hbar\omega}{\mu B_0}\cos\theta+\frac{\hbar^2\omega^2}{4\mu^2B_0^2}},\quad \tan\bar{\theta}=\frac{2\mu B_0\sin\theta}{2\mu B_0\cos\theta-\hbar\omega}.
\end{equation}
When the initial state is one of the eigenstates, the final state can be written as 
\begin{equation}
    \ket{\psi(T)}=e^{\mp i\pi-\frac{i}{\hbar}\bar{E}_\pm T}\ket{\eta_\pm}.
\end{equation}
The additional phase factors 
\begin{equation}
    \gamma_\pm(T)=-\frac{1}{\hbar}\bar{E}_\pm T\mp\pi 
\end{equation}
can be interpreted as the total phase factors acquired during the evolution.  
The dynamical phases read 
\begin{align}
    \gamma_\pm^d(T) &=-\frac{1}{\hbar}\int_0^T\bra{\psi(t)}H(t)\ket{\psi(t)}\mathrm{d}t=-\frac{1}{\hbar}\int_0^T\bra{\eta_\pm}H(0)\ket{\eta_{\pm}}\mathrm{d}t \nonumber \\
    &=-\frac{1}{\hbar}\bar{E}_\pm T\mp\pi\cos\bar{\theta},
\end{align}
where we have used the relation $\bra{\eta_\pm}\sigma_z\ket{\eta_\pm}=\pm\cos\bar{\theta}$.
Therefore, the nonadiabatic geometric phases are given by
\begin{equation}
    \gamma_\pm^g(T)=\gamma_\pm(T)-\gamma_\pm^d(T)=-\pi(1\mp\cos\bar{\theta}).
\end{equation}
The difference between the Berry phase in Eq.~(\ref{gppm}) and the Aharonov-Anandan phase is that the latter is determined by the modified angle $\bar{\theta}$, which tends to $\theta$ in the adiabatic limit, where $\omega / (\mu B_0) \rightarrow 0$.

\subsection{Adiabatic quantum holonomy}\label{subsec:adia_quant_holonomy}
In \secref{sec:1.1}, we introduced Berry's geometric phase by considering cyclic adiabatic evolution driven by a  slowly changing nondegenerate Hamiltonian. In this case, the geometric phases are phase factors since each eigenspace of the system Hamiltonian is one-dimensional. 
In 1984, Wilczek and Zee \cite{wilczek1984appearance} generalized Berry's formula to the case of degenerate Hamiltonians, leading to the discovery of the non-Abelian geometric phase, also known as quantum holonomy.

Consider a Hamiltonian $H(R)$ whose eigenvalue $E_n(R)$ is $k_n$-fold degenerate ($k_n \geq 1$), corresponding to a $k_n$-dimensional eigenspace $\mathcal{V}_n$. The Hilbert space spanned by $H(R)$ is the direct sum of all $\mathcal{V}_n$
\begin{equation}
    \mathcal{H}=\oplus_n \mathcal{V}_n,
\end{equation}
and thus the dimension of $\mathcal{H}$ is the sum of all the dimensions of the eigenspaces: $\dim (\mathcal{H})=\sum_n \dim(\mathcal{V}_n)$.
The degenerate eigenstates $\{\ket{\varphi_n^k(t)}, k=1,2,\cdots,k_n\}$ of a given energy $E_n$ constitute a complete set of  orthogonal normalized basis vectors of $\mathcal{V}_n$.
When the system parameters $R$ change adiabatically, the degeneracy structure of $H(R)$ is assumed to be  maintained, {\it i.e.}, no energy eigensubspaces are allowed to split or cross. It implies that a state initially in one subspace $\mathcal{V}_n$ remains in the same subspace throughout the time evolution.
If a cyclic evolution is attained in a period $T$ during which the parameters $R(t)$ accomplish a closed loop in parameter space, {\it i.e.}, $R(T)=R(0)$, then eigenspaces $\mathcal{V}_n(R)$ of the Hamiltonian $H(R)$ return to their original form at time $T$.

However, our experience in studying Berry phases suggests that a state in $\mathcal{V}_n(R(0))$ may not return to the same initial state when a cyclic subspace evolution, {\it i.e.}, $\mathcal{V}_n(R(T))=\mathcal{V}_n(R(0))$, is realized.
More precisely, a set of eigenstates $\{\ket{\varphi_n^k(0)}\}$ can be transformed into another set of eigenstates spanning  $\mathcal{V}_n(R(0))$.
This transformation between the new set of eigenstates and $\{\ket{\varphi_n^k(0)}\}$ can be described with $k_n\times k_n$ unitary matrices.
To check this, we consider the Schr\"{o}dinger evolution of a state initially prepared in a particular eigenstate $\ket{\varphi_n^m(0)}$. The adiabatic theorem guarantees that the instantaneous state continues to be a linear combination of the eigenstates spanning the instantaneous eigensubspace $\mathcal{V}_n (R(t)) \equiv \mathcal{V}_n (t)$, {\it i.e.},
\begin{equation}\label{umn}
    \ket{\psi_n^m(t)}=\sum_{p=1}^{k_n} U_n^{pm}(t)\ket{\varphi_n^p(t)} \in \mathcal{V}_n (t), \quad (\ket{\psi_n^m(0)}=\ket{\varphi_n^m(0)} \in \mathcal{V}_n (0)),
\end{equation}
where $U_n^{pm}$ are matrix elements of a unitary operator $U$.
Substituting Eq.~(\ref{umn}) into the Schr\"{o}dinger equation, we obtain
\begin{equation}\label{umnd}
\frac{\mathrm{d}}{\mathrm{d}t}U_n^{pm}(t)=i\left[-\frac{1}{\hbar}E_n(t)U_n^{pm}(t)+i\sum_k\bra{\varphi_n^p(t)}\frac{\mathrm{d}}{\mathrm{d}t}\ket{\varphi_n^k(t)}U_n^{km}(t)\right].
\end{equation}
By defining $A_n^{pk}(t)=i\bra{\varphi_n^p(t)}\frac{\mathrm{d}}{\mathrm{d}t}\ket{\varphi_n^k(t)}$, the unitary matrix can be written as
\begin{equation}
    U_n(t)=\mathcal{T}\exp \left(i\int_0^tA_n(t')\mathrm{d}t'\right) e^{-\frac{i}{\hbar}\int_0^tE_n(t')\mathrm{d}t'}=\mathcal{P}\exp\left(i\int_{R(0)}^{R(T)}A_n(R') \cdot \mathrm{d}R'\right) e^{-\frac{i}{\hbar}\int_0^tE_n(t')\mathrm{d}t'},
\end{equation}
where $\mathcal{T}$ and $\mathcal{P}$ are  time- and path-ordering, respectively. 
After a cyclic evolution, $U_n(T)$ takes the form 
\begin{equation}
    U_n(T)=U_n^g(C)e^{i\gamma_n^d}=\mathcal{P}e^{i\oint_CA_n(R) \cdot \mathrm{d}R}e^{-\frac{i}{\hbar}\int_0^tE_n(t')\mathrm{d}t'}.
\end{equation}
$\gamma_n^d$ is the dynamical phase accumulation during the evolution. Since all the eigenstates have the same eigenvalue, the acquired dynamical phase factor is a global phase factor associated with the energy of the eigensubspace. 
On the other hand, the matrix part $U_n^g(C)$ depends only on the evolution loop of the degenerate space $\mathcal{V}_n(R)$. Due to its matrix nature, $U_n^g(C)$ need not commute for different loops in parameter space; in this sense, it is a non-Abelian generalization of the Berry phase for adiabatically evolving quantum systems. 
Similar to the Berry curvature, the matrix $A_n(R)$ serves as a gauge potential. 
To see this, let us consider a unitary transformation of the basis vectors,
\begin{equation}
    \ket{\varphi_n^k(t)}'=\sum_{l=1}^{k_n}\Omega_n^{lk}(t)\ket{\varphi_n^l(t)},
\end{equation}
where $\Omega_n^{lk}(t)$ are elements of a unitary matrix $\Omega_n(t)$. 
The matrix $A_n$ transforms under the basis change as
\begin{equation}
    A'_n=\Omega_n^{-1}A_n\Omega_n +i\Omega_n^{-1}\frac{\mathrm{d}}{\mathrm{d}t}\Omega_n ,
\end{equation}
which is exactly the transformation law for a non-Abelian gauge potential.
An important distinction between the Abelian and non-Abelian geometric phases is that unlike the Abelian ones, such as the Berry phase and the Aharonov-Anandan phase, the non-Abelian geometric phase $U^g_n(T)$ obtained here is not gauge invariant, but gauge covariant under the 
gauge transformation, {\it i.e.}, ${U_n^{g}}'(t)=\Omega_n^{-1}(t)U_n^g(t)\Omega_n(t)$. 
However, if the condition $\Omega_n(T)=\Omega_n(0)=I$ ($I$ is the identity acting on $\mathcal{V}_n(0)$) is taken into account, we find ${U_n^g}'(T)=U_n^g(T)$.

\subsection{Nonadiabatic quantum holonomy}\label{subsec:non_quant_holonomy}
In 1988, Anandan proposed a nonadiabatic extension of quantum holonomy by considering cyclic continuous evolution of a subspace $\mathcal{V}(t)$ of Hilbert space $\mathcal{H}$\footnote{By `continuous', we mean that the subspace makes no jump and have fixed dimension throughout the evolution.} \cite{anadan1988non}.
Unlike the case of adiabatic quantum holonomy \cite{wilczek1984appearance}, any set of vectors spanning the subspace $\mathcal{V}(t)$ considered by Anandan are exact solutions of the time-dependent Schr\"{o}dinger equation. Furthermore, the system Hamiltonian need not be cyclic in this scenario but can take arbitrary form as long as it generates cyclic evolution of $\mathcal{V}(t)$, {\it i.e.}, $\mathcal{V}(T)=\mathcal{V}(0)$ for some $T$.

Suppose the Hilbert space $\mathcal{H}$ of the system is decomposed into an $N$-dimensional subspace $\mathcal{V}(t)$ and its $M$-dimensional orthogonal complement $\mathcal{W}(t)$, {\it i.e.},
\begin{equation}
    \mathcal{H}=\mathcal{V}(t)\oplus\mathcal{W}(t).
\end{equation}
We can choose an orthonormal basis $\{\ket{\varphi_k(t)}, k=1,2,\ldots,N\}$ for $\mathcal{V}(t)$ with $\ket{\varphi_k(T)}=\ket{\varphi_k(0)}$ for every $k$. 
We define another set of basis vectors $\{\ket{\psi_k(t)}\}$ spanning $\mathcal{V}(t)$ and  evolves according to the Schr\"{o}dinger equation with the initial condition $\ket{\psi_k(0)}=\ket{\varphi_k(0)}$. Since the two bases both span $\mathcal{V}(t)$ they are related by
\begin{equation}\label{psik}
    \ket{\psi_k(t)}=\sum_{l=1}^{N}U_{lk}(t)\ket{\varphi_l(t)},
\end{equation}
where $U_{lk}(t)$ are elements of a unitary matrix $U(t)$. 
Substituting Eq.~(\ref{psik}) into the Schr\"{o}dinger equation, we obtain
\begin{equation}
    \frac{\mathrm{d}}{\mathrm{d}t}U_{lk}(t)=i\sum_{m=1}^{N}[A_{lm}(t)-K_{lm}(t)]U_{mk}(t),
\end{equation}
where $A_{lm}(t)=i\bra{\varphi_l(t)}\frac{\mathrm{d}}{\mathrm{d}t}\ket{\varphi_m(t)}$, and $K_{lm}(t)=\frac{1}{\hbar}\bra{\varphi_l(t)}H(t)\ket{\varphi_m(t)}$.
By integration, one finds 
\begin{equation}\label{ut}
    U(T)=\mathcal{T}e^{i\int_0^T[A(t)-K(t)]\mathrm{d}t},
\end{equation}
where $\mathcal{T}$ is time-ordering and $U(T)$ acts on $\mathcal{V}(0)$. 
We see that the evolution operator has two contributions: an $N\times N$ matrix $K(t)$, depending on the Hamiltonian $H(t)$, which can be regarded as a matrix generalization of the dynamical phase $\gamma^d(t)$ considered in the Aharonov-Anandan approach discussed above, and an $N\times N$ matrix $A(t)$, which has a purely geometric origin in the sense that it is independent of the Hamiltonian, and it depends only on the loop traced by the subspace $\mathcal{V} (t)$. More precisely, for ${\rm dim} \mathcal{V} = N$ and ${\rm dim} \mathcal{H} = M$, the loop resides in the Grassmannian manifold $\mathcal{G} (M;N)$, the latter being the space of complex $N$-dimensional subspaces of the $M$-dimensional Hilbert space. 

As we have discussed for the adiabatic case, if we choose a different basis $\ket{\varphi_k(t)}'=\sum_{l=1}^{N}\Omega_{lk}(t)\ket{\varphi_l(t)}$, where the matrix $\Omega(t)$ is unitary, then the geometric matrix $A(t)$ transforms according to 
\begin{equation}
    A \mapsto A'=\Omega^{-1}A\Omega +i\Omega^{-1}\frac{\mathrm{d}}{\mathrm{d}t}\Omega , 
\end{equation}
which  is again a gauge transformation under which the geometric part 
$U^g(T)=\mathcal{T}e^{i\int_0^T A(t)\mathrm{d}t}$ transforms unitarily: $U^g(T) \mapsto \Omega^{\dagger} (0)U^g(T) \Omega (0)$ as long as $\Omega(T)=\Omega(0)$. 
Similarly, the dynamical matrix $K(t)$ transforms as $K \mapsto K'=\Omega^{\dagger}K\Omega$ under the basis change.

In a special case where $[K(t),A(t')]=0$, $\forall t, t' \in [0,T]$, 
Eq.~(\ref{ut}) can be rewritten as 
\begin{equation}\label{eq:udg}
    U(T)=\mathcal{T} e^{-i\int_0^tK(t)\mathrm{d}t}\cdot \mathcal{P} e^{i\int_C \mathcal{A}} = U^d(T)\cdot U^g(T),
\end{equation}
where $U^d(T)$ and $U^g(T)$ are the evolution operators generated by the dynamical and geometric contributions, respectively, $\mathcal{A}$ is the matrix-valued connection one-form, and $\mathcal{P}$ is path-ordering along $C$ in $\mathcal{G} (M;N)$. 
The adiabatic quantum holonomy corresponds to a case where $K(t)=E_n(t)I/\hbar$, which apparently commutes with any $A(t)$. 
Therefore, the nonadiabatic quantum holonomy can be treated as a nonadiabatic generalization of the Wilczek-Zee holonomy discussed in  \secref{subsec:adia_quant_holonomy}.
For another case, when $\mathcal{V}$ is one-dimensional, the integral of $A(t)$ and $K(t)$  reduce to $\gamma^g(t)$ and $\gamma^d(t)$ in Eq.~(\ref{gamma}), respectively. Therefore, the nonadiabatic quantum holonomy is the non-Abelian generalization of the Aharonov-Anandan phase.

\section{Geometric quantum computation}
\label{sec:GQC}

\subsection{Adiabatic geometric gates}\label{sec:adiabaticgeo}

In the adiabatic version of geometric quantum computation (GQC), Berry phases are used to implement quantum gates. To see how this works, let us consider a system consisting of a spin-$\frac{1}{2}$ particle in a magnetic field, as described by the Zeeman Hamiltonian $H(t)=\mu\Vec{B}(t)\cdot\Vec{\sigma}$. If $\Vec{B}(t)$ adiabatically traverses a loop $C$, the two eigenstates $\ket{\varphi_{\pm} (t)}$ of $H(t)$ pick up Berry phases $\gamma_{\pm}^g(C)= \mp \frac{1}{2}\Omega(C)$ upon completion of $C$. Thus, by encoding a single qubit in the initial energy eigenstates, {\it i.e.}, by putting $\ket{0}=\ket{\varphi_{+} (0)}$ and $\ket{1}=\ket{\varphi_{-} (0)}$, this results in the single-qubit phase-shift gate\footnote{We have omitted the dynamical phases for now and will address the removal of such phases later.} 
\begin{equation} \ket{k} \rightarrow e^{ik\Omega (C)} \ket{k}, \ k=0,1, 
\label{eq:1qadgate}
\end{equation} 
up to an unimportant overall phase factor $e^{-i\frac{1}{2} \Omega (C)}$. This gate is purely geometric in that it is determined by the loop $C$, but is independent of details of the evolution, such as the evolution rate. 

The above technique is not limited to spin$-\frac{1}{2}$ but can be applied to any realization of a single qubit, being exposed to a Hamiltonian of the generic form 
\begin{equation}\label{eq:h1}
    H^{(1)}(t)= h(t)[\cos\theta\sigma_z+\sin\theta(\cos\phi\sigma_x+\sin\phi\sigma_y)] . 
\end{equation}
The geometric gate in Eq.~(\ref{eq:1qadgate}) can, in this setting, be obtained by adiabatically varying the control parameters $\theta$ and $\phi$ around a loop $C$ such that $\theta (T) = \theta (0) = 0$ and $\phi (T) = \phi (0)$, provided the dynamical phases can be factored out or canceled.

The idea can be extended to the two-qubit case. Indeed, a conditional phase-shift gate that can entangle a qubit pair can be implemented through the Hamiltonian
\begin{equation}
    H^{(2)}(t)= h_1(t)[\cos\theta_1\sigma_z^1+\sin\theta_1(\cos\phi_1\sigma_x^1+\sin\phi_1\sigma_y^1)]+ h_2\sigma_z^2 + J\sigma_z^1\sigma_z^2 ,
\end{equation}
where $h_{1}$, $\theta_{1},\phi_{1},h_2$ are single-qubit parameters, $J$ is the qubit-qubit coupling strength (of Ising type), and $\sigma_{\alpha}^{1(2)}$, $\alpha = x,y,z$, are Pauli matrices of qubit $1(2)$.
By putting $\theta_1 = 0$ at the initial time $t=0$, we 
find $H^{(2)}(0)=\mathrm{diag}\{h_1(0)+h_2+J,h_1(0)-h_2-J,h_2-h_1(0)-J,-h_1(0)-h_2+J\},$ in the ordered computational basis $\{\ket{00},\ket{01},\ket{10},\ket{11}\}$. This demonstrates that the energy shift in qubit $1$  depends on the state of qubit 2. To be specific, when qubit $2$ is in the state $\ket{0}$, the transition frequency of qubit $1$ is $[h_1(0)+J]/\pi$, while if qubit $2$ is in the state $\ket{1}$, the transition frequency of qubit $1$ is $[h_1(0)-J]/\pi$.

Next, we keep qubit $2$'s Hamiltonian unchanged, but gradually add the Hamiltonian $ h_{xy}(t)(\cos\phi_1\sigma_x^1+\sin\phi_1\sigma_y^1)$ to qubit 1 and change the parameters $h_{xy}(t)$ and $\phi_1$ such that they trace out a closed path, just as in the single-qubit phase-shift gate.  
The key point is that the adiabatic path traced by qubit $1$ depends on the state of qubit $2$: when
qubit $2$ is in state $\ket{0}$, the initial Hamiltonian of qubit $1$ is $[h_1(0)+J]\sigma_z$, yielding $\theta_1=\arctan [h_{xy}/(h_1+J)]$, 
while when qubit $2$ is in state $\ket{1}$, the initial Hamiltonian of qubit $1$ reads  $[h_1(0)-J]\sigma_z$, and thereby $\theta_1=\arctan[h_{xy}/(h_1-J)]$. 
Therefore, $H^{(2)}(t)$ will go through two paths with different opening angles $\theta_1$. This means that the solid angle enclosed by qubit 1 will differ for the two states of qubit 2. By denoting these solid angles by $\Omega_1^0$ and $\Omega_1^1$, corresponding to qubit $2$ in states $\ket{0}$ and $\ket{1}$, respectively, the resulting conditional phase-shift gate can be written in the ordered computational basis as
\begin{equation}
    U_2 (C)=\mathrm{diag}\{e^{-i\Omega_1^0/2},e^{-i\Omega_1^1/2},e^{i\Omega_1^0/2},e^{i\Omega_1^1/2} \} ,
\end{equation}
which is an entangling two-qubit gate purely determined by $\Omega_1^k$, $k=0,1$, provided $\Omega_1^0 - \Omega_1^1$ is not an integer multiple of $2\pi$.

It remains to address the dynamical phases accumulated during the adiabatic process. In the single-qubit case, these phases become   $\gamma_{\pm}^d(t)=\mp\int_0^th(t')\mathrm{d}t'$, where $\gamma_+^d$ ($\gamma_-^d$) is acquired by the initial state $\ket{0}$ ($\ket{1}$). Since these phases are generally different, their effect must be eliminated in order to implement a purely geometric gate. This can be done by designing the adiabatic evolution so as to make the dynamical phase difference $\gamma_+^d - \gamma_-^d$ equal to an integer multiple of $2\pi$. However, one should be very careful in the two-qubit case since the eigenvalues of qubit $1$ are determined by the state of qubit $2$. When qubit $2$ is in state $\ket{0}$, the energy eigenvalues of qubit $1$ are $E_{\pm}^0(t)=\pm[(h_1(0)+J)^2+h_{xy}^2(t)]^{1/2}$, while they are $E_{\pm}^1(t)=\pm[(h_1(0)-J)^2+h_{xy}^2(t)]^{1/2}$ when the qubit $2$ is in the state $\ket{1}$.
In this case, we must make sure that $\int_0^TE_{\pm}^0(t')\mathrm{d}t'=2 n\pi$ and  $\int_0^TE_{\pm}^1(t')\mathrm{d}t'=2 m\pi$ ($m$ is an integer) so that the dynamical phases for the two different cases can be removed at the same time. However, this condition may not be easy to satisfy in a practical quantum system. Hence, as we will describe next, a two-loop method was proposed and implemented in an NMR system \cite{jones2000geometric,ekert2000geometric}. This technique has been used also for schemes to detect the Berry phase in superconducting nanoncircuits \cite{falci00detection} and molecular nanonmagnets \cite{mousolou2016nano}.

\subsubsection{Adiabatic geometric phase-shift gates in NMR}\label{sec:adiabaticNMR}
As described above, the basic idea of constructing geometric gates is not straightforward to implement in a practical experimental setup. For example, in NMR systems, the $\sigma_z$ term in the Hamiltonian is caused by a static magnetic field while the $\sigma_x$ and $\sigma_y$ terms are introduced by a radio-frequency (rf) field \cite{jones2000geometric,vandersypen2005NMR}. Due to the fact that the magnitude of the rf field is usually much smaller than that of the vertical field, the nutation angle $\theta$ can only be a small number, which makes the solid angle $\Omega$ small as well. In other words, realizing a large geometric phase shift is difficult.

This issue can be addressed by introducing another adjustable parameter $\phi$ in the system Hamiltonian 
according to 
\begin{equation}\label{eq:hNMR}
    H(t)=\frac{\omega_0}{2}\sigma_z+\frac{\omega_1}{2}[\cos(\omega t+\phi)\sigma_x+\sin(\omega t+\phi)\sigma_y],
\end{equation}
where $\omega_0$ is the strength of the static bias field, $\omega$ and $\omega_1$ are the frequency and amplitude, respectively, of the rf field, and $\phi$ is a variable phase shift setting the initial direction of the rf field in the $xy$-plane. By transforming to a frame that rotates with the field by means of the operator $W(t)=e^{-i\omega t\sigma_z/2}$, we find  
\begin{eqnarray}
H \mapsto \tilde{H} & = & W^{\dagger} H W + i \frac{\mathrm{d} W^{\dagger}}{\mathrm{d} t} W = \frac{1}{2}(\omega_0-\omega)\sigma_z+\frac{\omega_1}{2}(\cos\phi\sigma_x+\sin\phi\sigma_y) 
\nonumber \\ 
& = &  \frac{1}{2} \sqrt{(\omega_0 - \omega)^2 + \omega_1^2} \left[ \cos \theta \sigma_z + \sin \theta (\cos\phi\sigma_x+\sin\phi\sigma_y) \right] \equiv \tilde{H} (\theta,\phi) 
\end{eqnarray}
with $\tan\theta=\omega_1/(\omega_0-\omega)$. The two eigenstates of $\tilde{H}(\theta,\phi)$ are $\ket{\varphi_+(\theta,\phi)}=\cos(\theta/2)\ket{0}+e^{i\phi}\sin(\theta/2)\ket{1}$ and $\ket{\varphi_-(\theta,\phi)}=\sin(\theta/2)\ket{0}-e^{i\phi}\cos(\theta/2)\ket{1}$. It is clear that, by sweeping $\omega$ for a fixed nonzero $\omega_1$, the angle $\theta$ can be varied on $(0,\pi)$. 

Now, we set $\theta (0) = 0$ ({\it i.e.}, $\omega_1=0$), $\phi = \phi_0$, and initialize the spin in the state $\ket{0}$.  We thereafter turn on the rf field and increase the parameter $\omega_1$ to adiabatically move the initial state $\ket{0}$ to the eigenstate $\ket{\varphi_+(\theta,\phi_0)}$ of $\tilde{H} (\theta,\phi)$. This step tilts the Hamiltonian $\tilde{H}(\theta,\phi)$ towards the $xy$-plane along the meridian at $\phi = \phi_0$ to the latitude at $\theta$. Varying the phase from $\phi_0$ to $\phi_0+2\pi$ causes $\tilde{H}(\theta,\phi)$ to rotate around the $z$-axis forming a closed loop. The eigenstate $\ket{\varphi_+(\phi)}=\cos(\theta/2)\ket{0}+e^{i\phi}\sin(\theta/2)\ket{1}$ thereby completes a loop and acquires a phase factor, being the sum of the dynamical phase  $\gamma_+^d=-\sqrt{(\omega_0-\omega)^2+\omega_1^2} T/2$ and the geometric phase  $\gamma_+^g=-\pi(1-\cos\theta)$.

In contrast to the geometric phases, the dynamical phases depend on the experimental details. In NMR, it is difficult to calculate the exact dynamical phase for a particular spin and correct it as described above. There are two main reasons behind this. First, the strength of the rf field varies over the sample. Second, nuclei at different positions are affected by the chemical potential surrounding them. So, different nuclei would acquire slightly different dynamical phases, and averaging the dynamical phases over the entire sample would result in extensive dephasing \cite{jones2000geometric,vandersypen2005NMR}.  

\begin{figure}
  \centering
  \includegraphics[width=1.0\textwidth]{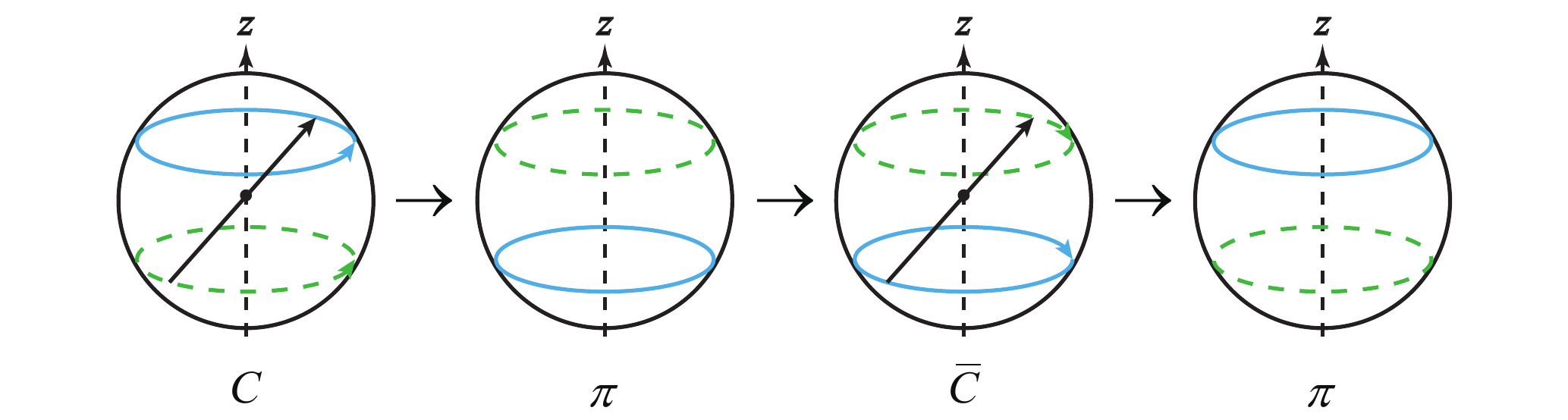}\\
  \caption{Schematic diagram for the echo control sequence to cancel out the dynamical phases. The black lines with arrows denote the choice of control parameters in the Hamiltonian $\tilde{H}(t)$ in the parameter space. The blue lines are the evolution paths for the initial state $\ket{0}$, while the green dashed lines are the paths for the initial state $\ket{1}$. The arrows on the lines indicate the rotational direction of the evolution. Step 1: Vary the phase $\phi$ to rotate the Hamiltonian $\tilde{H}(t)$ around the $z$-axis for a closed loop $C$. Step 2: Apply a $\pi$ pulse to exchange $\ket{\varphi_+}$ and $\ket{\varphi_-}$. Step 3: Vary the phase $\phi$ again to rotate the Hamiltonian $\tilde{H}(t)$ around the $z$-axis with the opposite rotational direction for a closed loop $\bar{C}$. Step 4: Apply another $\pi$ pulse to transform the eigenstates back.}\label{fig:echo}
\end{figure}

Such a problem can be overcome using spin echo, as illustrated in Fig.~\ref{fig:echo}. This technique is based on a control sequence where the spin is taken around a closed loop $C$, flipped to its orthogonal state by applying a short $\pi$ pulse, followed by an adiabatic traversal again along $C$ but in the opposite direction (denoted as $\bar{C}$), and finally flipped back by another $\pi$ pulse. 
This procedure can be summarized as follows:
\begin{align}
    \ket{\varphi_+}&\xrightarrow{C}e^{i(\gamma_+^d+\gamma_+^g)}\ket{\varphi_+}\xrightarrow{\pi}  e^{i(\gamma_+^d+\gamma_+^g)}\ket{\varphi_-}\xrightarrow{\bar{C}}e^{i(\gamma_+^d+\gamma_-^d+2\gamma_+^g)}\ket{\varphi_-}\xrightarrow{\pi}e^{i(\gamma_+^d+\gamma_-^d+2\gamma_+^g)}\ket{\varphi_+} \nonumber\\
    \ket{\varphi_-}&\xrightarrow{C}e^{i(\gamma_-^d+\gamma_-^g)}\ket{\varphi_-}\xrightarrow{\pi}
    e^{i(\gamma_-^d+\gamma_-^g)}\ket{\varphi_+}\xrightarrow{\bar{C}}e^{i(\gamma_-^d+\gamma_+^d+2\gamma_-^g)}\ket{\varphi_+}\xrightarrow{\pi}e^{i(\gamma_-^d+\gamma_+^d+2\gamma_-^g)}\ket{\varphi_-}.
\end{align}
The dynamical phases of the two eigenstates appear as a common global phase factor, leaving a state change only dependent on the geometric phases acquired by the two eigenstates. Here, it is important to ensure that the cyclic paths are implemented under the same experimental conditions, a feature that is achieved to high precision in NMR \cite{jones2000geometric}.

In addition to the dynamical phases incurred during state evolution, such phases typically arise also during state initialization and finalization. The change $\ket{0} \mapsto \ket{\varphi_+}$ induced by sweeping $\omega$ for a nonzero $\omega_1$, results in a dynamical phase $\gamma_+^i$, where the superscript $i$ stands for initialization process. Similarly, when the echo sequence is completed, the final state should be moved back to $\ket{0}$, and another dynamical phase $\gamma_+^i$ will be incurred.  We note that no geometric phase appears in the initialization process since the evolution path is along a geodesic line, part of a great circle on the sphere. The two dynamical phases can be offset by adding an `anti-initialization process' right after the $\bar{C}$ path (see Fig.~\ref{fig:echo}), {\it i.e.}, by driving the state $\ket{\varphi_-(\phi)}$ to $\ket{1}$ along the same great circle, and then back to $\ket{\varphi_-(\phi)}$. Since the energy eigenvalues of $\ket{\varphi_{\pm} (\phi)}$ differ by a sign, the corresponding dynamical phases satisfy $\gamma_-^i=-\gamma_+^i$ given that the experimental details are the same for the initialization and anti-initialization processes, eventually canceling the dynamical phases. The same applies to the initial state $\ket{1}$. Therefore, the final phase-shift gate is pure of geometric origin and takes the form
\begin{equation}
    U^g=\left(\begin{array}{cc}
       e^{i2\gamma_+^g}  & 0 \\
        0 & e^{i2\gamma_-^g}
    \end{array}\right) 
\end{equation}
in the ordered computational basis $\{\ket{0},\ket{1}\}$.

To implement a conditional phase-shift gate in NMR, we start from a coupled spin-$\frac{1}{2}$ nuclei (two-qubit) Hamiltonian 
\begin{equation}\label{eq:htNMR}
    H^{(2)}(0)=\frac{1}{2}(\omega_0^1\sigma_z^1+\omega_0^2\sigma_z^2+ J\sigma_z^1\sigma_z^2),
\end{equation}
where $\omega_0^{1(2)}$ is the Rabi frequency for spin 1(2), and $J/2$ is the spin-spin coupling strength. The first two terms arise from a static magnetic field pointing along the $z$ direction. By choosing a heteronuclear system, $\omega_0^1$ and $\omega_0^2$ can be very different, and only one of the spins can be close to resonance. The two spins interact with each other when they are close. Here, we assume $\omega_0^1 \gg \omega_0^2$ so that spin 2 can be used as a control qubit and spin 1 will undergo cyclic evolution and obtain phases depending on the state of spin 2. 

Assuming that spin 1 is initially in $\ket{0}$, an rf field $\omega_1(t)[\cos(\omega t+\phi)\sigma_x+\sin(\omega t+\phi)\sigma_y]/2$ with constant $\omega$ is gradually added to spin 1 by increasing the strength $\omega_1 (t)$ of the rf field from 0 to $\omega_1(\tau) \equiv \omega_1$. This prepares spin 1 in the state $\ket{\varphi_+}_{0(1)}=\cos(\theta_{0(1)}/2)\ket{0}+e^{i\phi}\sin(\theta_{0(1)}/2)\ket{1}$, where $\ket{\varphi_+}_{0(1)}$ and $\theta_{0(1)}$ are the state and nutation angle, respectively, when spin 2 is in state $\ket{0}$ ($\ket{1}$).
The dynamical phase obtained in the initialization can be canceled with the  method described above for the single-qubit gate case.
Changing $\phi$ to accomplish a cyclic evolution will result in two different geometric phases $\gamma_{+0(1)}^g=-\pi(1-\cos\theta_{0(1)})$ and also different dynamical phases $\gamma_{+0(1)}^d$ for spin 1, where $\gamma_{+0(1)}^g$ and $\gamma_{+0(1)}^d$ are  the geometric and dynamical phases, respectively, when spin 2 is in state $\ket{0}$ ($\ket{1}$).

In the two-qubit case, the unwanted dynamical phases can be eliminated with the following generalized spin echo sequence \cite{jones2000geometric,ekert2000geometric}:
\begin{equation}
    C\rightarrow\pi^{1}\rightarrow\bar{C}\rightarrow\pi^2\rightarrow C\rightarrow\pi^1\rightarrow\bar{C}\rightarrow\pi^2,
\end{equation}
where $C$ and $\bar{C}$ are the cyclic evolutions defined in Fig.~\ref{fig:echo}, and $\pi^{1(2)}$ is the $\pi$ pulse acting on qubit $1(2)$. If we define the difference between the geometric phases as 
\begin{equation}
    \Delta\gamma^g=\gamma_{+0}^g-\gamma_{+1}^g=\pi(\cos\theta_0-\cos\theta_1)=\pi\left(\frac{\omega_0^1-\omega+J}{\sqrt{(\omega_0^1-\omega+J)^2+\omega_1}}-\frac{\omega_0^1-\omega-J}{\sqrt{(\omega_0^1-\omega-J)^2+\omega_1}}\right),
\end{equation}
then the net conditional phase-shift gate, up to a global dynamical phase, takes the form  
\begin{equation}
    U_2(C)=\left(\begin{array}{cccc}
       e^{2i\Delta\gamma^g}  & 0 & 0 & 0 \\
        0 & e^{-2i\Delta\gamma^g} & 0 & 0 \\
        0 & 0 & e^{-2i\Delta\gamma^g} & 0 \\
        0 & 0 & 0 & e^{2i\Delta\gamma^g}
    \end{array}\right).
\end{equation}
It is evident that $\Delta\gamma^g$ depends on $\omega_0^1$, $\omega$, $\omega_1$, and $J$, but is independent of how the process is carried out.

\subsubsection{Adiabatic geometric phase-shift gates in superconducting nanocircuits}
\begin{figure}[tb]
  \centering
  \includegraphics[width=0.8\textwidth]{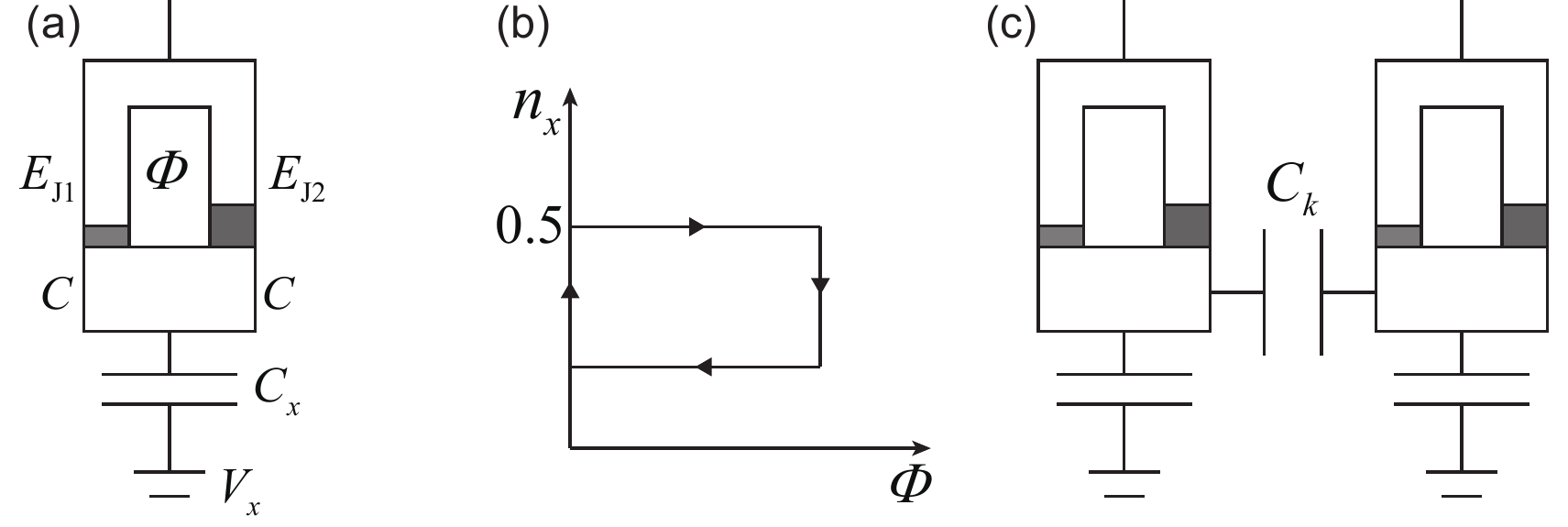}\\
  \caption{Schematic diagram of the superconducting nanocircuits. (a) A superconducting qubit is represented by the system consisting of a superconducting electron box formed by an asymmetric SQUID, controlled by a magnetic flux $\Phi$ and with a gate voltage $V_x=2en_x/C_x$. (b) A cyclic change of the Hamiltonian is realized by changing the gate voltage and the magnetic flux along the circle. (c) System design for two-qubit gates. Two qubits are coupled capacitively. The design is adopted from \cite{falci00detection}. More state-of-the-art designs can be found in Refs.~\cite{buluta2011natural,gu2017microwave,kockum2019quantum}.}\label{fig:SQUID}
\end{figure}
Geometric phases not only appear on a microscopic level but can also be observed in macroscopic systems such as superconducting devices \cite{shnirman1997quantum,makhlin1999josephson,wei2008controllable,you2011atomic,kockum2019ultrastrong}. Consider a setup in which an asymmetric superconducting quantum interference device (SQUID) forms a superconducting electron box and is controlled by a magnetic flux $\Phi$ as well as by an applied gate voltage $V_x$ [see Fig.~\ref{fig:SQUID}(a)] \cite{falci00detection}. When the Josephson coupling $E_{\mathrm{J}1(2)}$ of the junctions are much smaller than the charging energy $E_{\mathrm{ch}}$, the setup works in the charging regime. Further, when the temperature is much lower than $E_{\mathrm{J}1(2)}$, the system can be described by the Hamiltonian \cite{tinkham2004introduction,kyaw2019towards,stassi2020scalable} 
\begin{equation}\label{eq:hjose}
    H=E_{\mathrm{ch}}(n-n_x)^2-E_\mathrm{J}(\Phi)\cos(\theta-\alpha),
\end{equation}
where $n$ is the number of Cooper pairs, $2en_x$ is the offset charge, $E_\mathrm{J}(\Phi)=[(E_{\mathrm{J}1}-E_{\mathrm{J}2})^2+4E_{\mathrm{J}1}E_{\mathrm{J}2}\cos^2(\pi\Phi/\Phi_0)]^{1/2}$, $\theta$ is the phase difference across the junction, and $\tan\alpha=(E_{\mathrm{J}1}-E_{\mathrm{J}2})\tan(\pi\Phi/\Phi_0)/(E_{\mathrm{J}1}+E_{\mathrm{J}2})$ with $\Phi_0 = h/(2e)$ is the superconducting magnetic flux quantum.
The parameters $n_x$ and $E_J(\Phi)$ can be controlled by the gate voltage $V_x$ and the magnetic flux $\Phi$, respectively \cite{shnirman1997quantum,makhlin1999josephson}. The control of the phase shift $\alpha$ is allowed by the asymmetric design of SQUID \cite{falci00detection}; otherwise, $\alpha=0$ and no Berry phase will be accumulated.

If $n_x$ is set around $1/2$, the setup can be approximately treated as a two-level system since only the two charge eigenstates $n=0,1$ become relevant.
These two states define the qubit basis $\{\ket{0},\ket{1}\}$. In this approximation, the Hamiltonian in \eqref{eq:hjose} can be rewritten as 
\begin{equation}
    H_B=-\frac{1}{2}\Vec{B}\cdot\Vec{\sigma}=-\frac{1}{2}[E_\mathrm{J}(\Phi)\cos\alpha\sigma_x-E_\mathrm{J}(\Phi)\sin\alpha\sigma_y+E_{\mathrm{ch}}(1-2n_x)\sigma_z].
\end{equation}
The superconducting system thus behaves like a spin-$1/2$ in a `magnetic field´, the $z$ component of which controlled by the charging and the $xy$ components determined by the Josephson terms. By slowly changing the gate voltage $V_x$ and the flux $\Phi$, $H_B$ may trace out a cyclic path in parameter space defined by $\Vec{B}$, and thus a Berry phase is obtained for each of its eigenstates [see Fig.~\ref{fig:SQUID}(b)]. Similar to the NMR scheme discussed in Sec. \ref{sec:adiabaticNMR}, there will be a dynamical phase accompanying the geometric Berry phase. The same echo technique can remove the dynamical component as in the NMR system described above, but the $\pi$ pulse that swaps the eigenstates is now achieved by applying an a.c. gate voltage pulse \cite{falci00detection}.

The interaction Hamiltonian describing two such asymmetric SQUIDs coupled through a capacitor $C_K$ [see Fig.~\ref{fig:SQUID}(c)] reads 
\begin{equation}
    H_I=\frac{2E_{\mathrm{ch}}C_K}{C}(n_1-n_{x,1})(n_2-n_{x,2}),
\end{equation}
where $n_{1(2)}$ and $2en_{x,1(2)}$ are the number of Cooper pairs and offset charge for qubit $1(2)$, respectively. 
By truncating each SQUID to a two-level system (qubit), the $(n_1-n_{x,1})(n_2-n_{x,2})$ term represents a $\sigma_z^1\sigma_z^2$ interaction. 
As demonstrated in Sec. \ref{sec:adiabaticNMR}, this kind of interaction can be used to implement a controlled phase shift gate on the two qubits \cite{falci00detection}.
However, there is still some difference between NMR and superconducting systems: in NMR, the $xy$ components of the Hamiltonian for the control qubit can be set to zero since they are introduced by the microwaves, but it is not possible to do so for superconducting qubits because the Josephson coupling cannot be switched off completely. This leads to (small) off-diagonal phases in the gate.

\subsection{Nonadiabatic geometric gates}
By using Berry phases to realize quantum gates requires a relatively long evolution time, which may hamper their experimental implementation because quantum operations must be performed within coherence time. Thus, geometric gates will be potentially more valuable and practical if they run at a speed comparable to the usual dynamical quantum gates.  One solution to this problem is to employ the Aharonov-Anandan (AA) phase rather than the Berry phase for implementing geometric gates. The first experimental proposal of such nonadiabatic geometric gates was put forward by Wang and Keiji \cite{wang2001nonadiabatic} in NMR, where the evolution of spin qubits in an external rotating field is exactly solvable. In their proposal, the dynamical phases were shown to vanish by driving the states along geodesic paths on the Bloch sphere (see Fig.~\ref{fig:HandS}).

To elaborate on the AA approach to implementing nonadiabatic geometric gates, let us look at the generic cyclic evolution of a two-level quantum system driven by a time-dependent Hamiltonian $H(t)$. The corresponding evolution operator $U(T)$ at time $T$ can be expressed in terms of its eigenstates $\ket{\eta_\pm}$ and eigenvalues $e^{i\gamma_\pm}$. 
Suppose now that the initial state is prepared in one of these eigenstates.
At the final evolution time $T$, we have $U(T)\ket{\eta_\pm}=e^{i\gamma_\pm}\ket{\eta_\pm}$. 
In other words, the state $\ket{\eta_\pm}$ completes a cyclic evolution at $T$, with $e^{i\gamma_\pm}$ being the total phase incurred during the evolution.
This results in a phase-shift gate $\mathrm{diag}\{e^{i\gamma_+},e^{i\gamma_-}\}$ acting on the ordered basis $\{\ket{\eta_+},\ket{\eta_-}\}$. Again, if we find a way to eliminate the dynamical phases but keep only the geometric part of $\gamma_{\pm}$, a purely geometric one-qubit gate $\mathrm{diag}\{e^{i\gamma_+^g},e^{i\gamma_-^g}\}$ is obtained.

This idea can be directly applied to the entangling gate, provided one imposes a certain extra restriction on the phase eigenvalues.
Consider a two-qubit Hamiltonian $H^{(2)}(t)$ whose evolution operator is $U^{(2)}(t)$. If the final unitary $U^{(2)}(T)$ diagonalizes in a product state basis $\ket{\eta_i} \otimes \ket{\eta_j}$, $i,j=\pm$, with eigenvalues $e^{i\gamma_{ij}}$, this unitary will act as an entangling gate on the two qubits provided the condition $\gamma_{++} + \gamma_{--} \neq \gamma_{+-} + \gamma_{-+}$ mod $2\pi$ is satisfied. As before, this corresponds to a geometric gate $\mathrm{diag}\{e^{i\gamma_{++}^g},e^{i\gamma_{+-}^g},e^{i\gamma_{-+}^g},e^{i\gamma_{--}^g}\}$ provided the effect of the dynamical phases can be eliminated. Note that there must be an interaction term between the two qubits in $H^{(2)}(t)$ to satisfy the above condition. Otherwise, $U^{(2)}(T)$ can always be separated as $U^{(2)}(T)=U_1(T)\otimes U_2(T)$, where $U_{1(2)}(T)$ is the single-qubit gate acting on qubit 1 (2), implying $\gamma_{++} (T)+\gamma_{--}(T)=\gamma_{+-}(T)+\gamma_{-+}(T)$.

\subsubsection{Nonadiabatic geometric phase-shift gates}\label{sec:nonadia_ps}
\begin{figure}[tb]
  \centering
  \includegraphics[width=1.0\textwidth]{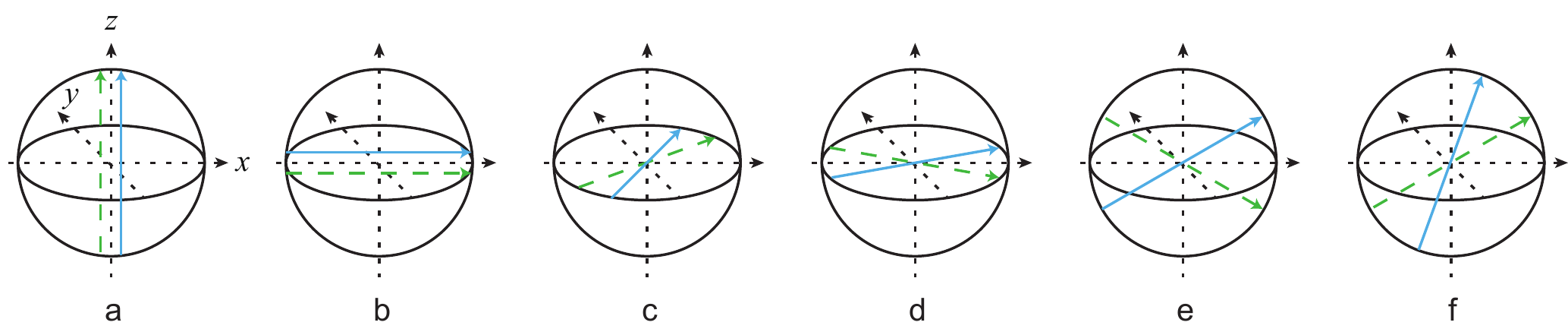}\\
  \caption{Illustration of the nonadiabatic initialization of the conditional eigenstates on Bloch sphere. (a) Assume that the qubit $1$ is originally in $\ket{0}$. The blue lines with arrows denote the Bloch vectors of qubit $1$ state when qubit $2$ is in $\ket{0}$ while the green dashed lines denote the Bloch vectors of qubit $1$ state when  qubit $2$ is in $\ket{1}$. (b)-(f) shows the five operations in the $S$ control sequence, \eqref{S_sequence}. Finally, the two state vectors are in the $xz$-plane, with  nutation angles $\theta_{0}=\pi/2-Jt_c-\varphi'$ (blue line) and $\theta_{1}=\pi/2+Jt_c-\varphi'$ (green dashed line).}\label{fig:initialization}
\end{figure}
Let us consider the NMR Hamiltonian 
\begin{equation}\label{eq:hnmr}
    H(t)=\frac{\omega_0}{2}\sigma_z+\frac{\omega_1}{2}[\cos(\omega t)\sigma_x+\sin(\omega t)\sigma_y],
\end{equation}
{\it i.e.}, \eqref{eq:hNMR} with $\phi = 0$.
The corresponding evolution operator takes the form $U(t)=e^{-i\sigma_z\frac{\omega t}{2}}e^{-i\tilde{H}t}$,
where $\tilde{H}=(\omega_0-\omega)\sigma_z/2+\omega_1\sigma_x/2$.
It turns out that there are two ways to exactly control the system evolution \cite{wang2001nonadiabatic}. First, we prepare the initial state $\ket{\tilde{\psi}(0)}$ to be one of the eigenstates of $\tilde{H}$, which completes a cyclic evolution at $t=2\pi/\omega$. Second, we add an extra magnetic field $\omega\sigma_z/2$ to $H(t)$, and prepare the initial state to be an eigenstate of $H(0)$\footnote{The latter setting allows us to use a time-dependent angular speed $\omega(t)$ of the rf field,  by imposing the cyclic condition $\int_0^T\omega(t)\mathrm{d}t=2\pi$.}.

\begin{figure}[tb]
  \centering
  \includegraphics[width=0.8\textwidth]{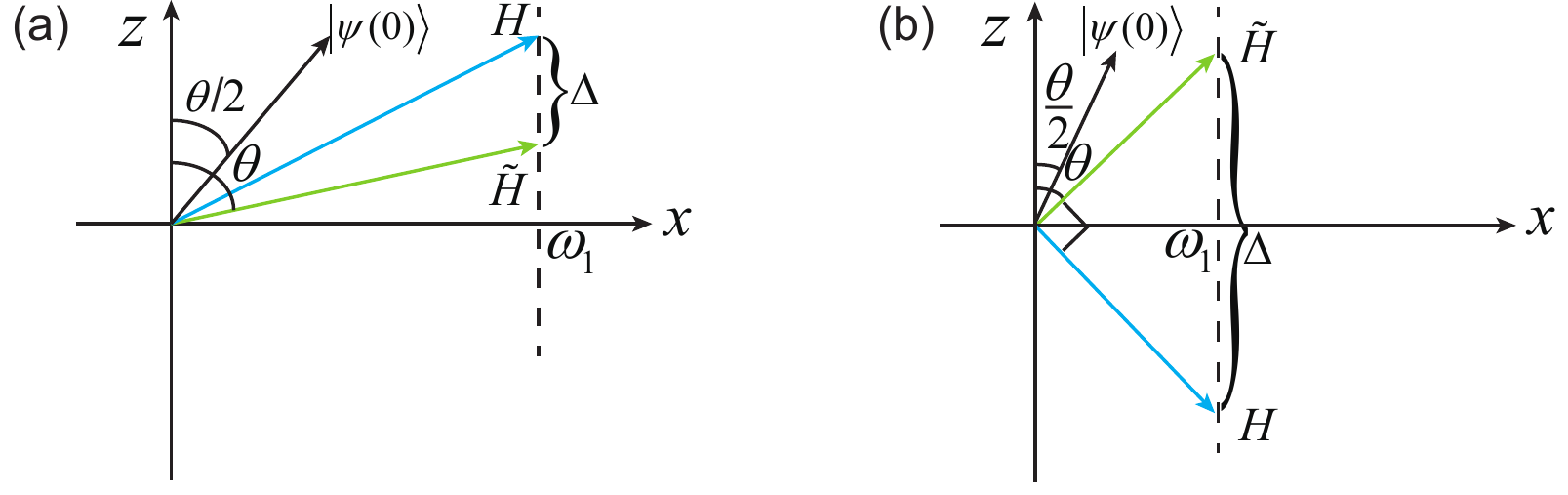}\\
  \caption{Schematic diagram of the Hamiltonians ($H(t)$ shown in blue lines and $\tilde{H}$ shown in green lines) in the parameter space at time $t=0$. The $x(z)$ axis shows the $\sigma_{x(z)}$ component of the Hamiltonians; thus, each Hamiltonian corresponds to a vector in the diagram. The lengths of the vectors are not scaled properly. (a) When $\omega>0$, an extra Hamiltonian $\Delta=\omega\sigma_z/2$ is added along $z$ direction. In this case, the nutation angle of $H$ is smaller than that for $\tilde{H}$. (b) When $\omega<0$, an extra Hamiltonian $\Delta=-\omega\sigma_z/2$ is added against $z$ direction. In this opposite case, the nutation angle of $H$ is larger than that for $\tilde{H}$. When $\omega=-(\omega_0^2+\omega_1^2)/\omega_0$, the vector associated with $H(0)$ is vertical to that of $\tilde{H}$ for a set of given $\omega_0$ and $\omega_1$, making $\bra{\psi(t)}H(t)\ket{\psi(t)}=0$ all the time during the evolution.}\label{fig:HandS}
\end{figure}

To realize a single-qubit phase-shift gate, let us initialize the spin in the state $\ket{0}$. After a rotation $e^{-i\sigma_y\theta/2}$ ($\tan\theta=\omega_1/\omega_0$), the state is prepared in $\ket{\psi(0)} = \cos(\theta/2)\ket{0} + \sin(\theta/2)\ket{1}$. The above initialization is a rotation around the $y$-axis by an angle $\theta$ on Bloch sphere\footnote{Note that the Bloch sphere representation is used as the state space for the nonadiabatic geometric phases.}. Since such rotation is along a geodesic, no phase factor is accumulated. 
By switching on the rf field for time $T=2\pi/\omega$, $\ket{\psi(0)}$ would undergo a cyclic evolution. The related dynamical and geometric phases can be calculated by using \eqref{gamma}, yielding $-\pi[\cos\theta+(\omega_0^2+\omega_1^2)^{1/2}/\omega]$ and $-\pi(1-\cos\theta)$, respectively. On the other hand, if the initial state is $\ket{1}$, the corresponding dynamical and geometric phases are 
$\pi[\cos\theta+(\omega_0^2+\omega_1^2)^{1/2}/\omega]$ and $\pi(1-\cos\theta)$, respectively.
The dynamical phases can be set to zero by choosing $\omega=-(\omega_0^2+\omega_1^2)/\omega_0$ so that only the geometric ones remain.
This condition amounts to making the total field instantaneously `perpendicular' to the
state during the evolution [see \figref{fig:HandS} (b)], ensuring that $\bra{\psi(t)}H(t)\ket{\psi(t)}=0$.
Then, the geometric phase-shift gate in the basis of $\{\ket{0}, \ket{1}\}$ reads
\begin{equation}
    U(C)=\left(\begin{array}{cc}
       e^{-i\pi(1-\cos\theta)}  & 0 \\
        0 & e^{i\pi(1-\cos\theta)}
    \end{array}\right).
\end{equation}

Conditional nonadiabatic phase-shift gates can be constructed with an interacting spin pair in NMR. After adding the horizontal field, the Hamiltonian for qubit 1 takes the form 
\begin{equation}
    H_1(t)=\frac{1}{2}(\omega_0^1\pm J)\sigma_z^1+\frac{1}{2}\omega_1[\cos(\omega t)\sigma_x^1+\sin(\omega t)\sigma_y^1],
\end{equation}
where the $+$ and $-$ sign in front of $J$ depends on the state of qubit 2, be it either $\ket{0}$ or $\ket{1}$, respectively.
Here, we assume the driving frequency $\omega$ is close to $\omega_0^1$ but away from $\omega_0^2$.
In the rotating frame $W(t)=e^{i\omega t\sigma_z^1/2}$, the effective Hamiltonian is
\begin{equation}
    \tilde{H}_1=WH(t)W^{-1}+i \frac{\mathrm{d} W(t)}{\mathrm{d} t}W^{-1}(t) = \frac{1}{2}(\omega_0^1 -\omega \pm J)\sigma_z^1+\frac{1}{2}\omega_1\sigma_x^1.
\end{equation}

Before evolving the eigenstate of $\tilde{H}_1$ around $z$-axis, one should prepare the eigenstate from $\ket{0}$. Notably, the eigenstate must be created conditionally. A control sequence $S$ for the nonadiabatic evolution has been shown in Fig.~\ref{fig:initialization}, adapted from \cite{wang2001nonadiabatic}, and can be summarized as:
\begin{equation}\label{S_sequence}
    S:\left[\frac{\pi}{2}\right]^y\rightarrow J'[\varphi_{\pm}(t_c)]\rightarrow [-\delta_{\pm}\cdot t_c]^z\rightarrow \left[\frac{\pi}{2}\right]^x\rightarrow [-\varphi']^y,
\end{equation}
where $[\beta]^\alpha$ represents the rotation angles $\beta$ in the Bloch sphere around the $\alpha$-axis ($\alpha=x,y,z$), and $J'[\varphi_{\pm}(t_c)]$ is the evolution over time $t_c$ by the Hamiltonian $\delta_{\pm}\sigma_z/2$ ($\delta_{\pm}=\omega_0^1-\omega\pm J$). In this setting, $\varphi_{\pm}=\delta_{\pm}t_c$.
After the operation $S$, the angle between the qubit $1$ and the $z$-axis is $\theta_{0(1)}=\pi/2\mp Jt_c-\varphi'$, where $\theta_{0(1)}$ is for qubit 2 in $\ket{0} (\ket{1})$.
To make sure that the prepared state is the eigenstate of the Hamiltonian $\tilde{H}_1$, the following two  conditions must be satisfied:
\begin{equation}
        \tan(\varphi'+Jt_c)=\delta_+/\omega_1, \quad
        \tan(\varphi'-Jt_c)=\delta_-/\omega_1.
\end{equation}
It follows that, given specific values of $\delta_{\pm}$ and $\omega_1$, it is easy to obtain the evolution time $t_c$ and the angle $\varphi'$.

The initialization process prepares the conditional state $\ket{\psi(0)}_{0(1)}=\cos(\theta_{0(1)}/2)\ket{0}+\sin(\theta_{0(1)}/2)\ket{1}$, where $\ket{\psi(0)}_{0(1)}$ is the state when qubit 2 is in $\ket{0} (\ket{1})$. A cyclic path can be carried out by applying the rf field for a time interval $\omega t=2\pi$. The corresponding dynamical and geometric phases for $\ket{\psi(0)}_{0(1)}$ are $\gamma_{+0(1)}^d=-\pi[\cos\theta_{0(1)}+(\delta_{\pm}^2+\omega_1^2)^{1/2}/\omega]$ and $\gamma_{+0(1)}^g=-\pi(1-\cos\theta_{0(1)})$, respectively. On the other hand, the state $\ket{1}$ will be transformed by the $S$ operation to the conditional eigenstates  $\ket{\psi_{\perp}(0)}_{0(1)}=\sin(\theta_{0(1)}/2)\ket{0}-\cos(\theta_{0(1)}/2)\ket{1}$, perpendicular to $\ket{\psi(0)}_{0(1)}$. During the cyclic evolution, the dynamical and geometric phases acquired by  $\ket{\psi_{\perp}(0)}_{0(1)}$ are $\gamma_{-0(1)}^d=\pi[\cos\theta_{0(1)}+(\delta_{\pm}^2+\omega_1^2)^{1/2}/\omega]$ and $\gamma_{-0(1)}^g=\pi(1-\cos\theta_{0(1)})$, respectively. 
To let the state evolve in a `dynamical phase free' path, the following two conditions must be satisfied:
\begin{equation}
        \cos\theta_{0}=-\sqrt{\delta_+^2+\omega_1^2} / \omega, \quad
        \cos\theta_{1}=-\sqrt{\delta_-^2+\omega_1^2}/\omega,
\end{equation}
which yields the final constraint $\omega_1^2=\delta_+\delta_-$. In this situation, the final conditional phase-shift gate can be shown, in the ordered basis $\{\ket{00},\ket{01},\ket{10},\ket{11}\}$, to take the form 
\begin{equation}
    U_2(C)=\left(\begin{array}{cccc}
       e^{i\gamma_{+0}^g}  & 0 & 0 & 0 \\
        0 & e^{i\gamma_{+1}^g} & 0 & 0 \\
        0 & 0 & e^{i\gamma_{-0}^g} & 0 \\
        0 & 0 & 0 & e^{i\gamma_{-1}^g}
    \end{array}\right).
\end{equation}
In addition, the dynamical phases in both the single-qubit and conditional gates are to be removed by multi-loop schemes \cite{wang2001nonadiabatic} similar to the ones used in the adiabatic case.  The multi-loop schemes are useful when the conditions for the parameter are not easy to satisfy in some quantum systems.
Also, they can be eliminated with a modified two-loop method (see Fig.~\ref{fig:twoloop}) \cite{zhu03universal} or a one-loop method \cite{zhang2005nonadiabatic}. 
Other proposals based on NMR include CNOT-like gates \cite{wang2001NMR} and geometric gates in fictitious spin-$1/2$ subspaces  \cite{gopinath2006geo}.
Another method to avoid the dynamical phases for those systems with nonrotating Hamiltonian is to let the state evolve along geodesic lines (so-called `orange slice scheme', see Fig.~\ref{fig:orangeslice}) \cite{li02non,zhao17rydberg}.

The NMR Hamiltonian used in the adiabatic and nonadiabatic schemes is obtained under the assumption that the spins are in the weak coupling regime where the couplings are mediated via the covalent bonds. This is the case for liquid samples \cite{jones2000geometric}. However, when the spins are strongly coupled with one another through dipolar couplings, such as in liquid crystals, more interaction terms will appear in the system Hamiltonian and thus make it complicated for a gate design. This problem can be addressed by treating the $2^n$ nondegenerate energy levels as an $n$-qubit system and using transition selective pulses to control them. As an example, procedures for building nonadiabatic geometric phase gates can be found in such a system in \cite{gopinath2006geo}.

\subsubsection{Universal single-qubit geometric gates}\label{sec:universalgeo}
Phase-shift gates based on a certain set of bases  commute with each other, so they can not form an arbitrary single-qubit gate. 
The implementation of universal quantum computation \cite{lloyd95almost} requires the capability to realize an arbitrary single-qubit gate as well as an entangling two-qubit gate. The conditional phase-shift gate that we have demonstrated is a proper entangling two-qubit gate.
It has been shown that any single-qubit gate can be obtained with geometric phase-shift gates \cite{zhu02,zhu03universal}.

We have so far considered the process in which a pair of orthogonal states $\ket{\eta_{\pm}}$ evolves cyclically under an evolution operator $U(T)$. A phase difference between the two states is introduced in this process. By removing the dynamical phases, the geometric phase-shift gate in the ordered basis $\{\ket{\eta_+},\ket{\eta_-}\}$ reads $\mathrm{diag}\{e^{i\gamma_+^g},e^{i\gamma_-^g}\}$. This means that if an initial state is $\ket{\psi_i}=a_+\ket{\eta_+}+a_-\ket{\eta_-}$ with $a_{\pm}=\braket{\eta_{\pm}}{\psi_i}$, the final state after the cyclic evolution can be written as 
$\ket{\psi_f}=a_+e^{i\gamma_+^g}\ket{\eta_+}+a_-e^{i\gamma_-^g}\ket{\eta_-}$. Assuming that $\ket{\eta_+}=\cos(\chi/2)\ket{0}+\sin(\chi/2)\ket{1}$ and $\ket{\eta_-}=\sin(\chi/2)\ket{0}-\cos(\chi/2)\ket{1}$, the phase-shift in the ordered basis $\{\ket{0},\ket{1}\}$ takes the form
\begin{equation}\label{eq:uchi}
    U(\gamma^g,\chi)=\left(\begin{array}{cc}
       e^{i\gamma^g}\cos^2\frac{\chi}{2}+e^{-i\gamma^g}\sin^2\frac{\chi}{2}  & i\sin\chi\sin\gamma^g \\
        i\sin\chi\sin\gamma^g & e^{i\gamma^g}\sin^2\frac{\chi}{2}+e^{-i\gamma^g}\cos^2\frac{\chi}{2}
    \end{array}\right),
\end{equation}
where we have used the relation $\gamma^g=\gamma_+^g=-\gamma_-^g$.

When another cyclic evolution based on a different pair of eigenstates $\ket{\eta'_{\pm}}$ is completed, two new geometric phases ${\gamma_{\pm}^g}'$ can be obtained, forming another phase-shift gate $U({\gamma^g}',\chi')$. It is easy to check that the two operators $U(\gamma^g,\chi)$ and $U({\gamma^g}',\chi')$ are noncommuting unless $\sin\gamma^g\sin{\gamma^g}'\sin(\chi-\chi')=0$. In fact, $U(\gamma^g,\chi)$ represents a kind of rotation around an axis related to $\chi$ by an angle $\gamma^g$. Thus, by choosing two different $\chi$ for any phase $\gamma^g$, an arbitrary single-qubit gate can be achieved. 
For example, the rotation $U=\exp(i\gamma\sigma_x)$ is obtained when $\chi=\pi/2$, which produces a spin flip (NOT gate) when $\gamma=\pi/2$ and an equal-weight superposition of spin states when $\gamma=\pi/4$.

\begin{figure}[tb]
  \centering
  \includegraphics[width=1.0\textwidth]{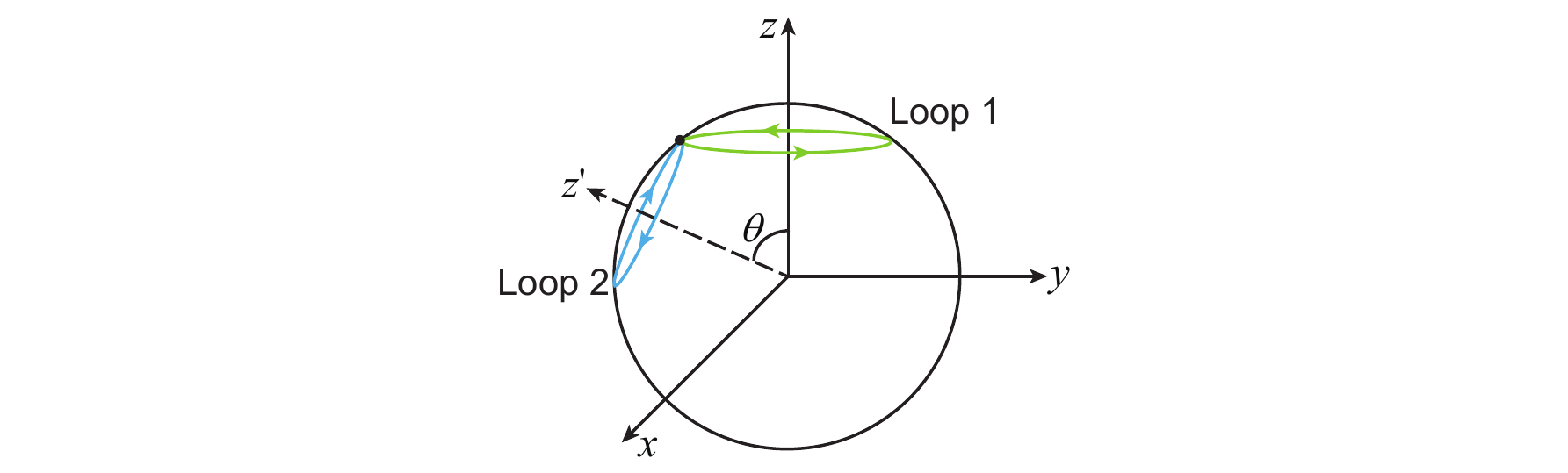}\\
  \caption{Schematic diagram of the two-loop method to cancel out the dynamical phases. The evolution loops are shown in the Bloch sphere. The black dot shows the position of the initial state for the two loops. The symmetric axis of $Loop$ $1$ is $z$-axis, while the symmetric axis for $Loop$ $2$ is $z'$ which is obtained by rotating $z$-axis around $y$-axis for an angle $\theta=\alpha+\alpha'$.}\label{fig:twoloop}
\end{figure}

It is remarkable that there is no more constraint on the universal scheme than the nonadiabatic one. Therefore, dynamical phases can be avoided by evolving the eigenstates along the `dynamical-phase free' paths \cite{wang2001nonadiabatic,zhu02geometric} or the geodesic lines \cite{li02non},  otherwise they can be removed by means of the two-loop technique \cite{zhu03universal}. 

To give an explicit example of the two-loop method, we consider the single-qubit NMR Hamiltonian in \eqref{eq:hnmr}. The initial state is the eigenstate of $\tilde{H}$ with eigenvalue $+1$, and the evolution path for these loops on the Bloch sphere is shown in Fig.~\ref{fig:twoloop}. \\
$Loop$ $1$:\\
The vector of the Hamiltonian is $\Vec{B}=(\omega_1\cos\omega t,\omega_1\sin\omega t,\omega_0)$, where $t\in[0,\tau)$, and $\tau=2\pi/\omega$.  This indicates that the first loop is rotated around $z$-axis with a nutation angle $\alpha=\arctan\omega_1/(\omega_0-\omega)$. \\
$Loop$ $2$:\\
The vector of the Hamiltonian is $\Vec{B}'=R_y(\alpha'+\alpha)(\omega_1^\prime\cos\omega t,\omega_1^\prime\sin\omega t,-\omega_0^\prime)$, where $t\in[\tau,2\tau]$.
$R_y(\alpha'+\alpha)$ represents the rotation of the Hamiltonian vector around $y$-axis for the angle $\alpha'+\alpha$, where $\tan\alpha'=\omega_1^\prime/(\omega_0^\prime+\omega)$. This rotation ensures that the initial states of the two loops are the same at $t=2\pi/\omega$.

By assuming that the dynamical and geometric phases corresponding to $Loop$ $1(2)$ are $\gamma^d_{1(2)}$ and $\gamma^g_{1(2)}$, respectively, the total phase for the two loops is  $\gamma=\gamma^d_1+\gamma^d_2+\gamma^g_1+\gamma^g_2$. For our proposal, it is required that 
\begin{equation}
    \gamma_1^d+\gamma_2^d=0, \quad \gamma_1^g+\gamma_2^g=\Gamma\pi,
\end{equation}
where $\Gamma\pi$ is the nontrivial geometric phase that determines the geometric gate.
Note that the initial state is the positive eigenstate of the Hamiltonian in $Loop$ $1$, but the negative eigenstate of the Hamiltonian is $Loop$ $2$.
This requires the parameters to satisfy the following equations:
\begin{equation}
    \frac{\omega_0^2+\omega_1^2-\omega_0\omega}{\sqrt{(\omega_0-\omega)^2+\omega_1^2}}=\frac{(\omega_0^\prime)^2+(\omega_1^\prime)^2+\omega_0^\prime\omega}{\sqrt{(\omega_0^\prime+\omega)^2+(\omega_1^\prime)^2}}, \quad \frac{\omega_0-\omega}{\sqrt{(\omega_0-\omega)^2+\omega_1^2}}-\frac{\omega_0^\prime+\omega}{\sqrt{(\omega_0^\prime+\omega)^2+(\omega_1^\prime)^2}}=\Gamma.
\end{equation}
For a given $\Gamma$, it is always possible to find such parameters because there are five variables in two equations. An example is provided in \cite{zhu03universal}. Recall that the two-qubit gate is obtained by rotating the Hamiltonian; the two-loop method is feasible for this case too \cite{zhu03universal}.

The NMR system considered in \cite{zhu03universal} is a good example of the two-loop scheme since the three components of Hamiltonians are available, and the evolution path can be precisely controlled. It has been experimentally demonstrated in liquid-state NMR that, using a spin-echo technique, a dynamical phase can be successfully eliminated, and a single-qubit gate can be realized \cite{ota09geo}. In contrast, it could be a better choice for other systems, such as trapped ions \cite{li02non}, to let the states evolve along geodesic lines.

\begin{figure}[tb]
  \centering
  \includegraphics[width=1.0\textwidth]{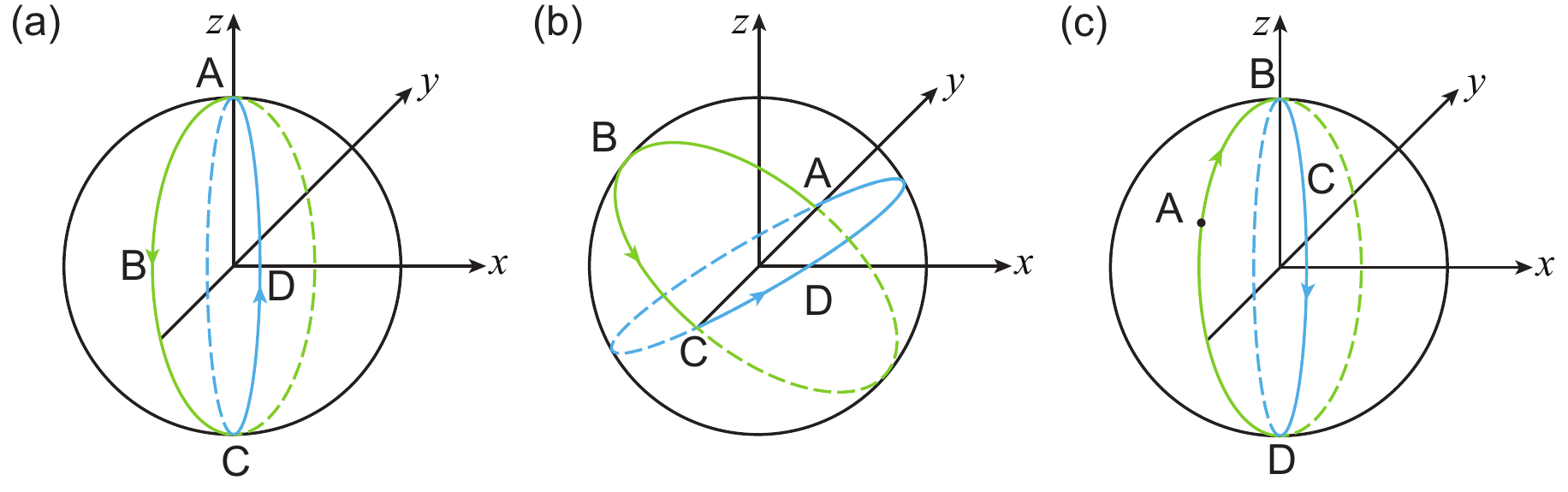}\\
  \caption{Illustration of evolution paths consisting of geodesic lines. Each loop has two semi-circles; the initial state evolves on the green and blue ones successively. (a) Evolution loop to realize a rotation around $z$-axis. The initial state is $\ket{0}$. (b) Evolution loop to construct a rotation around $y$-axis. The initial state is $\ket{+}_y$. (c) Evolution loop to build a rotation around $\Vec{n}$ for an angle $\gamma$ where $\Vec{n}$ is decided by the initial state $\ket{d}$ labelled as the black dot located at point A.}\label{fig:orangeslice}
\end{figure}

Consider a set of trapped ions, each of which  has two internal states $\ket{0}$ and $\ket{1}$ (with energy gap $\omega_0$) and can be selectively addressed by lasers.
When the $j$th ion is exposed to a traveling-wave laser with frequency $\omega_L$ and phase $\phi$, the effective Hamiltonian becomes 
\begin{equation}
    H^{j}(t)=\frac{\omega_0}{2}\sigma_z^j+\omega_j[\sigma^+_j e^{i\eta(a+a^\dag)-i\omega_Lt+i\phi}+\mathrm{H.c.}],
\end{equation}
where $\omega_j$ is the Rabi frequency, $\sigma^+_j=\ket{0}_j\bra{1}$, $\eta$ is the Lamb-Dicke parameter that accounts for the coupling strength between internal and motional states, and $a$ ($a^\dag$) is the phonon (lowering) raising operator \cite{li02non}.
In the rotating frame $e^{-i\omega_L t\sigma_z/2}$, $H^j(t)$ can be recast as
\begin{equation}
    \tilde{H}^j=\omega_j\cos\phi\sigma_x^j+\omega_j\sin\phi\sigma_y^j+\frac{\omega_0-\omega_L}{2}\sigma_z^j.
\end{equation}
In the resonant regime ($\omega_L=\omega_0$), only the $x$ and $y$ components are relevant. If the initial state is $\ket{0}$, it can be rotated  along the geodesic line ABC [see Fig.~\ref{fig:orangeslice}(a)] to $\ket{1}$ by turning on the Hamiltonian $\tilde{H}^j(\phi_0)=\omega_j\cos(\phi_0)\sigma_x^j+\omega_j\sin(\phi_0)\sigma_y^j$ for a $\pi$ pulse. Then change the laser phase to $-\phi_0$ [{\it i.e.}, $\tilde{H}^j(-\phi_0)=\omega_j\cos(\phi_0)\sigma_x^j-\omega_j\sin(\phi_0)\sigma_y^j$] and apply the other $\pi$ pulse after which state $\ket{1}$ will return to $\ket{0}$ along the geodesic line CDA on the Bloch sphere [see Fig.~\ref{fig:orangeslice}(a)]. The total evolution operator for the two connected paths is
\begin{equation}
    \tilde{U}_\pi=e^{-i\tilde{H}^j(-\phi_0)\pi/2}e^{-i\tilde{H}^j(\phi_0)\pi/2}.
\end{equation}
After the evolution, $\ket{0}$ is transformed to $\tilde{U}_\pi\ket{0}=e^{2i\phi_0}\ket{0}$. It is easy to check that the dynamical phase during each $\pi$ pulse is zero ($\bra{0}\tilde{U}^\dag(t)\tilde{H}(t)\tilde{U}(t)\ket{0}=0$), thus the total phase factor $2\phi_0$ is purely geometric. If the initial state is $\ket{1}$, it will acquire a geometric phase of $e^{-2i\phi_0}$. By changing $\phi_0$, an arbitrary phase-shift gate [$\chi=0$ in \eqref{eq:uchi}] can be implemented. 

In the nonresonant regime, with $\phi=0$, $\tilde{H^j}$ lies in the $xz$ plane in which the nutation angle is $\theta_0=\arctan[2\omega_j/(\omega_0-\omega_L)]$. 
The initial state $\ket{+}_y$ ($\sigma_y\ket{+}_y=\ket{+}_y$) will be moved along the geodesic line ABC to $\ket{-}_y$ ($\sigma_y\ket{-}_y=-\ket{-}_y$)  [see Fig.~\ref{fig:orangeslice}(b)]. Then $\ket{-}_y$ can be transformed back to $\ket{+}_y$ along CDA by setting $\phi=\pi$.
Similar to the resonant case, the total phase factor $e^{2i\theta_0}$ acquired by $\ket{+}_y$ is  purely geometric. Therefore, an arbitrary phase-shift gate around $y$-axis can be achieved by choosing different $\theta_0$. Together with the rotations around $z$-axis, any single-qubit gate is available \cite{li02non}.  

In the above scheme, two loops and three components of system Hamiltonians are required to satisfy the universality. In fact, only one loop and two components are needed for the same purpose \cite{zhao17rydberg}. As shown in Fig.~\ref{fig:orangeslice}(c), an arbitrary single-qubit geometric gate $U(\gamma)=e^{i\gamma\Vec{n}\cdot\Vec{\sigma}}$ where $\Vec{n}=(\sin\theta\cos\varphi,\sin\theta\sin\varphi,\cos\theta)$ can be realized by means of the following Hamiltonian:  
\begin{equation}
H(t) = \begin{cases}
          h(t)[\cos(\varphi-\frac{\pi}{2})\sigma_x+\sin(\varphi-\frac{\pi}{2})\sigma_y], & 0\leq t\leq \tau_1,\\
          h(t)[\cos(\gamma+\varphi+\frac{\pi}{2})\sigma_x+\sin(\gamma+\varphi+\frac{\pi}{2})\sigma_y]  , & \tau_1\leq t\leq\tau_2,\\
          h(t)[\cos(\varphi-\frac{\pi}{2})\sigma_x+\sin(\varphi-\frac{\pi}{2})\sigma_y], & \tau_2\leq t\leq\tau_3 
\end{cases}
\end{equation}
under the conditions
\begin{equation}
    \int_0^{\tau_1}h(t)\mathrm{d}t=\frac{\theta}{2}, \quad \int_{\tau_1}^{\tau_2}h(t)\mathrm{d}t=\frac{\pi}{2}, \quad  \int_{\tau_2}^{\tau_3}h(t)\mathrm{d}t=\frac{\pi-\theta}{2}.
\end{equation}
The initial state $\ket{d}=\cos(\theta/2)\ket{0}+\sin(\theta/2)e^{i\phi}\ket{1}$ is evolved along the geodesic lines AB ($0\leq t\leq \tau_1$), BCD ($\tau_1\leq t\leq\tau_2$), and DA ($\tau_2\leq t\leq\tau_3$). When it returns to A, $\ket{d}$ acquires a geometric (total) phase factor $e^{i\gamma}$. Accordingly, its orthogonal state $\ket{b}=\sin(\theta/2)e^{-i\phi}\ket{0}-\cos(\theta/2)\ket{1}$ transforms into  $e^{-i\gamma}\ket{b}$. Hence, the evolution operator reads 
\begin{equation}
    U(\tau_3)=e^{i\gamma}\ket{d}\bra{d}+e^{-i\gamma}\ket{b}\bra{b}=U(\gamma).
\end{equation}
This indicates that the axis $\Vec{n}$ in $U(\gamma)$ is determined by the initial state $\ket{d}$ ($\ket{b}$), and the rotation angle $\gamma$ around $\Vec{n}$ is the angle between the two planes related to BAD and BCD [see Fig.~\ref{fig:orangeslice}(c)]. 
Besides the Rydberg atoms \cite{zhao17rydberg}, this one loop scheme is also suitable for transmon qubits \cite{chen2018non} and silicon-based qubits \cite{zhang20high}.

Until now, many schemes have been introduced to remove or avoid dynamical phases. We stress that the geometric and dynamical phases should be calculated in the same frame (usually not in the lab frame). Therefore, one only needs to remove the dynamical phases in the relevant frame.

\subsubsection{Unconventional geometric gates}\label{sec:unconventional}
In all the aforementioned proposals, dynamical phases have to be avoided or removed. 
The reason to get rid of the dynamical phases is that they usually depend on experimental details rather than global geometric features.
So, it is interesting to ask whether there is a kind of dynamical phase that possesses the same feature as geometric ones. 
The answer is yes, and when the dynamical phase $\gamma^d$ is proportional to the geometric phase $\gamma^g$ with a constant ratio $a$, the total phase $\gamma=\gamma^d+\gamma^g=(1+a)\gamma^g$ can be treated as a geometric one. 
As an example, the geometric phase for a qubit state is the negative half of the solid angle ($-\Omega/2$) associated with the evolution path. If the corresponding dynamical phase turns out to be $a\Omega$ with $a$ a constant independent of $\Omega$, the total phase $(a-1/2)\Omega$ would have the same geometric property and thus could be used to construct geometric phase-shift gates (when $a\neq1/2$).
Due to this nature, this kind of geometric gate is referred to as unconventional geometric quantum gates \cite{leibfried2003experi,zhu03unconventional,du06experimental}.
Compared with conventional geometric gates, unconventional geometric gates can simplify experimental operations because they do not require the removal of dynamical phases in order to ensure global geometric features.

To get into some details, let us look at a one-dimensional harmonic oscillator: 
\begin{equation}
    H_0=\frac{1}{2m}\hat{p}^2+\frac{m\omega^2}{2}\hat{x}^2,
\end{equation}
where $m$ is the oscillator's mass, $\omega$ is its natural frequency, $\hat{p}$ is the momentum operator, and $\hat{x}$ is the position operator. Let us perturb the system by adding a term proportional to the position, {\it i.e.}, $H_1=\epsilon \hat{x}$ where $\epsilon$ is a constant, yielding the total Hamiltonian 
\begin{equation}
    H=H_0+H_1=\frac{1}{2m}\hat{p}^2+\frac{m\omega^2}{2}\hat{x}^2+\epsilon \hat{x}.
\end{equation}
In terms of the standard creation and annihilation operators 
\begin{equation}
    a^\dag=\frac{1}{\sqrt{2 m\omega}}(m\omega\hat{x}-i\hat{p}), \quad a=\frac{1}{\sqrt{2 m\omega}}(m\omega\hat{x}+i\hat{p}).
\end{equation}
the total Hamiltonian $H$ takes the form $(\hbar=1)$
\begin{equation}
    H=\omega(a^\dag a+\frac{1}{2})+\beta(a^\dag+a),
\end{equation}
where $\beta=\epsilon/\sqrt{2m\omega}$.
By transforming to the rotating frame by applying the rotation operator $R(t)=\exp[-i\omega(a^\dag a+1/2)t]$, we find
\begin{equation}\label{eq:heff}
    \tilde{H}(t)=R^{\dagger}(t)HR(t)+i \frac{\mathrm{d}R^{\dagger}(t)}{\mathrm{d}t}R(t)
    =\beta(a^\dag e^{i\omega t}+a e^{-i\omega t}).
\end{equation}
Since $\tilde{H}(t)$ is time-dependent, we cannot obtain $\tilde{U}(t)=T\exp[-i\int_0^t\tilde{H}(t')\mathrm{d}t']$ directly. However, the commutation of $\tilde{H}(t)$ at different time is proportional to the identity, {\it i.e.}, $[\tilde{H}(t_1), \tilde{H}(t_2)]=2i\beta^2\sin\omega(t_2-t_1) \hat{1}$, which makes it possible to find the analytical form of $\tilde{U}(t)$ as 
\begin{equation}
    \tilde{U}(t)=e^{i\frac{\beta^2}{\omega^2}(\omega t-\sin\omega t)} e^{-i\int_0^t\tilde{H}(t')\mathrm{d}t'}.
\end{equation}
The second factor on the right-hand side can be evaluated as
\begin{equation}
    e^{-i\int_0^t\tilde{H}(t')\mathrm{d}t'} = e^{a^\dag\frac{\beta}{\omega} (1-e^{i\omega t}) - a\frac{\beta}{\omega} (1-e^{-i\omega t})} = D\left[\frac{\beta}{\omega}(1-e^{i\omega t}) \right] 
\end{equation}
with $D(z)$ the displacement operator acting as $D(z)\ket{0}_c=\ket{z}_c$, $\ket{z}_c$ being a coherent state localized at $z \in \mathbb{C}$ in phase space.
Accordingly, the evolution operator takes the form $\tilde{U}(t)=e^{i\beta^2(\omega t-\sin\omega t)/\omega^2}D[\frac{\beta}{\omega}(1-e^{i\omega t})]$.

Suppose now that the system is initially in the vacuum state $\ket{0}_c$. Then, the instantaneous state can be written as 
\begin{equation}
    \ket{\psi(t)}=e^{i\frac{\beta^2}{\omega^2}(\omega t-\sin\omega t)}\ket{\frac{\beta}{\omega}(1-e^{i\omega t})}_c.
\end{equation}
It turns out that the time-dependent coherent state $\ket{z(t)}_c=D(\frac{\beta}{\omega}(1-e^{i\omega t}))\ket{0}_c$ serves properly as the auxiliary state $\ket{\varphi(t)}$ in \eqref{gammat}, since $\ket{z(\frac{2\pi}{\omega})}_c=\ket{z(0)}_c=\ket{0}_c$. The state parameter $z(t)$ traverses a circle with radius $\beta/\omega$ and centered at the point $(\beta/\omega,0)$ in phase space.
The total phase angle accumulated during the evolution is $\gamma(t)=\beta^2(\omega t-\sin\omega t)/\omega^2$.
When $t=2\pi/\omega$, a cyclic evolution of $\ket{z(t)}_c$ is accomplished and $\ket{\psi(t)}$ arrives at  $\ket{\psi(T)}=e^{2i\pi\beta^2/\omega^2}\ket{0}_c$.
The total phase $\gamma(T)$ for the cyclic evolution is $2\pi\beta^2/\omega^2$, which depends on the parameter $\beta$ and the angular frequency $\omega$; in fact it is the phase space area enclosed by the circle $[0,2\pi/\omega] \ni t \mapsto z(t)$.
The related dynamical phase can be calculated as 
\begin{equation}
    \gamma^d(t)=-\int_0^t {}_c\bra{z(t')}\tilde{H}(t')\ket{z(t')}_c\mathrm{d}t'=\frac{2\beta^2}{\omega^2}(\omega t-\sin\omega t).
\end{equation}
When $\omega t=2\pi$, the dynamical phase is $4\pi\beta^2/\omega^2$.
Therefore, the geometric phase during the evolution becomes $\gamma^g(T)=\gamma(T)-\gamma^d(T)=-2\pi\beta^2/\omega^2=-\gamma^d/2$.
Note that both the dynamical and total phases do not depend on the evolution time but on the system parameter $\beta$ and $\omega$, just as the geometric phase.
Note also that the unconventional relation $\gamma^d(t)=-2\gamma^g(t)$ is always valid during the evolution \cite{zhu03unconventional}.

The ratio between the dynamical and geometric phases in the harmonic oscillator system is fixed and equal to $-2$, but this number can change in NMR or other systems. Recall that the dynamical and geometric phases for the cyclic path of an NMR qubit read $\gamma^d_+=-\pi[\cos\theta+(\omega_0^2+\omega_1^2)^{1/2}/\omega]$
and $\gamma^g_+=-\pi(1-\cos\theta)$, respectively. If we ask $\gamma_+^d=a\gamma_+^g$, it is equal to set $a=[\cos\theta+(\omega_0^2+\omega_1^2)^{1/2}/\omega]/(1-\cos\theta)$, where $\tan\theta=\omega_1/\omega_0$ \cite{wang07non}. For a given $\gamma_+^g$ ($\omega_0$ and $\omega_1$ are fixed), there are many choices of $a$, each of which corresponds to a constant $\omega$. For example, $a=0$, we obtain $\omega\omega_0=-\omega_0^2+\omega_1^2$ coinciding with the condition to  the `dynamical-free path' \cite{wang01multibit}.

Such a proposal can also be realized in a trapped ion system, where ions interact with laser light and move in a linear trap. It 
provides a good platform for  implementing a quantum computer \cite{cirac95quantum,sorensen99quantum,milburn00ion,sorensen00entangl} (for reviews, see Refs. \cite{duan10collo,haffner08quantum}).  The significant features of this system  
are that it allows the realization of multi-qubit quantum gates between any set of ions (not necessarily neighboring) and that decoherence can be made negligible during the whole control process \cite{cirac95quantum}. 
In \cite{zhu03unconventional}, the authors consider a system in which two ions are confined in a harmonic trap potential and interact with laser radiation. Each ion has two internal states, denoted by $\ket{\!\!\downarrow}$ and $\ket{\!\!\uparrow}$, which are used to define the qubit states $\ket{0}$ and $\ket{1}$, respectively. The quantum state of the harmonic oscillator can be described as $\ket{\Psi}$.
The motion of the ions is strongly coupled by their mutual Coulomb repulsion and can be described with normal modes. 
Along the direction in which the two ions are aligned, there are two normal modes: the center-of-mass mode where the displacements of both ions from equilibrium are the same, and the stretch mode where the displacements are equal but in opposite direction  \cite{leibfried2003experi}. The center-of-mass mode has been used to realize dynamical gates for the ions \cite{cirac95quantum}.
When appropriate laser beams are chosen, the trap potential can excite a stretch mode with the frequency $\omega_s$ when the ions are in different internal states, while nothing will happen when the ions are in the same internal state \cite{milburn00ion,sorensen00entangl,wang01multibit}. 
This means that when the initial state of the system is  $\ket{\!\!\downarrow\downarrow}\ket{\Psi}$ or $\ket{\!\!\uparrow\uparrow}\ket{\Psi}$, it will be left unchanged. On the other hand, when the state is $\ket{\!\!\downarrow\uparrow}\ket{\Psi}$ or $\ket{\!\!\uparrow\downarrow}\ket{\Psi}$, there will be a driving Hamiltonian acting on $\ket{\Psi}$.
With the rotating wave approximation (RWA), the Hamiltonian for the harmonic oscillator in the rotating frame reads
\begin{equation}\label{eq:hd}
    H(t)=i\Omega_D(a^\dag e^{-i\delta t+i\phi}-ae^{i\delta t-i\phi}),
\end{equation}
where $a^\dag$ and $a$ are the usual harmonic oscillator raising and lowering operators, $\delta$ is the detuning, $\phi$ is the phase of the driving field, and $\Omega_D=(F_{0\uparrow}-F_{0\downarrow})z_{0s}/2$ with $z_{0s}$ being the spread of the ground state wave function of the stretch mode and $F_{0\uparrow}$ ($F_{0\downarrow}$) being the dipole force acting on the $\ket{\!\!\uparrow}$ ($\ket{\!\!\downarrow}$) state.
The quantum state $\ket{\Psi}$ under this force can be coherently displaced in position-momentum phase space.

Comparing with \eqref{eq:heff}, $\delta$ in \eqref{eq:hd} corresponds to $-\omega$ in \eqref{eq:heff}, while $\Omega_D$ in \eqref{eq:hd} resembles $\Omega$ in \eqref{eq:heff}. 
Although there is an additional phase difference $ie^{i\phi}$ ($-ie^{-i\phi}$) for $a^\dag$ ($a$), they do not affect the commutation relation. Explicitly, the commutation of $H(t)$ at different time is $[H(t_1),H(t_2)]=2i\Omega_D^2\sin\delta(t_1-t_2)$.
Therefore, the evolution operator $U(t)$ for $H(t)$ in \eqref{eq:hd} is
\begin{equation}
    U(t)=e^{i\frac{\Omega_D^2}{\delta^2}(\sin\delta t-\delta t)}D[\frac{\Omega_D}{\delta}e^{i(\phi+\frac{\pi}{2})}(e^{-i\delta t}-1)].
\end{equation}
It follows that the state $\ket{\Psi}$ will complete a cyclic evolution at $t=2\pi/\delta$. The total phase is $\gamma=-\gamma^g=-2\pi(\Omega_D/\delta)^2$.
Thus the corresponding two-qubit gate is
\begin{equation}
    U(\gamma)=\left(\begin{array}{cccc}
       1  & 0 & 0 & 0 \\
        0 & e^{i\gamma} & 0 & 0 \\
        0 & 0 & e^{i\gamma} & 0 \\
        0 & 0 & 0 & 1
    \end{array}\right),
\end{equation}
in the basis $\{\ket{00},\ket{01},\ket{10},\ket{11}\}$ \cite{zhu03unconventional}. 
The broadly used control-phase gate (CZ) can be realized with $U(\gamma)$ (by setting $\gamma=-\pi/2$) and single-qubit phase gates ($S$), {\it i.e.}, 
\begin{equation}
    \mathrm{CZ}=\left(\begin{array}{cccc}
       1  & 0 & 0 & 0 \\
        0 & 1 & 0 & 0 \\
        0 & 0 & 1 & 0 \\
        0 & 0 & 0 & -1
    \end{array}\right)= U(-\pi/2)(S_1\otimes S_2)=\left(\begin{array}{cccc}
       1  & 0 & 0 & 0 \\
        0 & -i & 0 & 0 \\
        0 & 0 & -i & 0 \\
        0 & 0 & 0 & 1
    \end{array}\right)\left(\begin{array}{cccc}
       1  & 0 & 0 & 0 \\
        0 & i & 0 & 0 \\
        0 & 0 & i & 0 \\
        0 & 0 & 0 & -1
    \end{array}\right),
\end{equation}
where $S_{1(2)}$ is the $S$ gate for qubit 1 (2).

In Ref. \cite{zhu03unconventional}, the evolution operator $U(t)$ is derived by means of the coherent-state path integral formulation in phase space \cite{gazeau09coherent}. The geometric and dynamical phases obtained coincide with ours. As pointed out in Ref. \cite{leibfried2003experi}, the geometric phase, which is the negative value of the total phase, can be shown to be  $\gamma^g=-\mathrm{Im}\{\oint_Cz^*\mathrm{d}z\}$, where $C$ is the path in the phase space. With 
$z=x\sqrt{m\omega/2\hbar}+ip/\sqrt{2\hbar m\omega}$ and by using Green's formula, we can show that ($\hbar$ is included here)
\begin{equation}
    \gamma^g=-\frac{1}{\hbar}\int\!\!\int_S\mathrm{d}x\mathrm{d}p=-\frac{S}{\hbar},
\end{equation}
where $x$ and $p$ are the position and momentum in the phase space and $S$ is the area surrounded by $C$. 
This equation reveals the geometric property of the unconventional geometric phase.

The experimental technique used in \cite{leibfried2003experi} can be improved in various aspects. For example, additional Stark shifts can be efficiently suppressed by choosing particular laser beams \cite{seidelin06micro}. A version of the geometric gate where the laser intensities impinging on the ions are controlled by transporting the ion crystal through the laser fields \cite{leibfried07transport}. A more detailed discussion on the experimental progress can be found in \cite{haffner08quantum}.

This scheme uses only stretch mode to displace coherent states in phase space. In practice, other modes exist, and all the vibrational modes cannot simultaneously return to their original point because they have different frequencies. To suppress the displacement of other modes, the relative detuning between the two lasers is close to the frequency of the stretch mode, and the strength of the displacement for other modes should be much smaller than the vibrational frequency. This constraint makes the gate rate much smaller than the vibrational frequencies. 
This problem could be overcome by driving the ions with a single off-resonant standing-wave laser so that several vibrational modes are simultaneously displaced. Under certain conditions, all the vibrational modes can return to the original point in the phase space at the same time \cite{zheng06high}.

The unconventional geometric gates can also be realized in cavity QED  where identical three-level atoms couple simultaneously to a highly detuned cavity mode. Each atom has one excited state $\ket{e}$ and two ground state $\ket{g}$ and $\ket{s}$ which encode the quantum information. The transition between $\ket{e}$ and $\ket{s}$ is driven by the cavity mode and a classical laser field with detuning \cite{zheng04un}. During the interaction, the atoms remain in their ground states with no transitions while the cavity mode is displaced along a closed loop in phase space. Thus, a phase factor proportional to the geometric phase is conditional upon the atomic state. Furthermore, the atoms can be disentangled with the cavity mode, and thus the geometric gate is insensitive to the cavity decay.   
When the transition between $\ket{e}$ and $\ket{s}$ is resonant with the classical field, the qubit states can be defined as $\ket{g}$ and the dressed state $\ket{-}=(\ket{s}-\ket{e})/\sqrt{2}$. This allows the atoms to be driven in the strong-driving regime, in which the two states are not affected, but the cavity mode will undergo cyclic evolution with a faster speed \cite{chen06strong}. 
The spontaneous emission from $\ket{e}$ can be avoided by adiabatically eliminating the excited state from the evolution with a two-channel Raman interaction with a small cavity detuning \cite{feng07scheme,wu2007unconventional}.

\subsection{Experimental realizations and gate robustness}

\begin{table}[tbh]   
\begin{center}   
\caption{Summary of reported geometric gate fidelities in various quantum systems. For the fidelity obtained for a particular gate, the gate is indicated in the following parentheses; otherwise, it is an average fidelity. In the column for nonadiabatic two-qubit gates, CZ is an abbreviation for control-NOT, while CCZ is for control-control-NOT (a three-qubit entangling gate) and so on.}
\label{tb:Gfidelity} 
\begin{tabular}{c|cc|ccc}
\hline\hline
 & \multicolumn{2}{c|}{Adiabatic} & \multicolumn{3}{c}{Nonadiabatic} \\
   & Single-qubit & Two-qubit & Single-qubit & Two-qubit & Unconventional \\ \hline
NMR &  &  &  &  & \makecell{$98.5\%$ \cite{du06experimental}\\$93.4\%$\cite{du06experimental}}\\
Trapped ions &  &  &  &  & $97\%$ \cite{leibfried2003experi} \\
Cold atoms &  &  & $90\%$ \cite{imai08demonstration} &  &  \\
Quantum dots &  &  & $90\%$ \cite{wang16experimental} &  &  \\
Superconducting &  &  & \makecell{$99.8\%$ \cite{xu2020experimental}\\$99.6\%$ \cite{zhao2021experimental}} & \makecell{$97.7\%$ (CZ,\cite{xu2020experimental})\\$94\%$ (CZ,\cite{song17continuous})\\$86.8\%$ (CCZ,\cite{song17continuous})\\$81.7\%$ (CCCZ,\cite{song17continuous})} &  \\
NV centers & \makecell{$98\%$ ($Z_{\pi/2}$,\cite{huang19experimental})\\$98.8\%$ ($Z_{\pi/8}$,\cite{huang19experimental})\\$97\%$ ($X_{\pi/2}$,\cite{huang19experimental})} & $94\%$ \cite{huang19experimental} &  \makecell{$99.4\%$ ($X$,\cite{kleissler2018universal})\\$99.5\%$ ($Z$,\cite{kleissler2018universal})\\$99.2\%$ ($H$,\cite{kleissler2018universal})} &  &  \\
\hline\hline
\end{tabular}   
\end{center}   
\end{table}

\textit{Adiabatic geometric gates.}--A universal set of adiabatic geometric gates was realized in solid-state spins in a diamond defect, in which $\ket{m_s=0}$ and $\ket{m_s=-1}$ were selected as $\ket{0}$ and $\ket{1}$, respectively \cite{huang19experimental}. The experiment demonstrated that the adiabatic scheme offers a unique advantage of inherent robustness to parameter variations in the coupling rate and the frequency detuning. The  gates achieved and their fidelities can be found in Table \ref{tb:Gfidelity}.  

\textit{Nonadiabatic geometric gates.}--Nonadiabatic geometric gates have been experimentally demonstrated in various quantum systems, including cold atoms \cite{imai08demonstration}, quantum dots \cite{wang16experimental}, superconducting systems \cite{xu2020experimental,zhao2021experimental,song17continuous}, and NV centers \cite{kleissler2018universal}. The highest gate fidelity up to now is achieved by Xu {\it et al.} \cite{xu2020experimental} in a superconducting qubit chain. Remarkably, multi-qubit geometric gates, up to a four-qubit CCCZ gate (see Table \ref{tb:Gfidelity}), were built in a superconducting circuit, where five qubits were controllably coupled to a resonator \cite{song17continuous}.

\textit{Unconventional geometric gates.}--As we mentioned in Section \ref{sec:unconventional}, the unconventional geometric gates were first realized in trapped ions \cite{leibfried2003experi} with the ratio between dynamical and geometric phases being $-2$. Then, a more general idea in which $\gamma_d=\alpha_g+\eta\gamma_g$ with $\alpha_g$ being a coefficient dependent only on the geometric feature of the evolution path and $\eta$ a constant was implemented in NMR \cite{du06experimental}. Notably, this scheme allows the achievement of a universal set of unconventional geometric gates.

We have summarized the geometric gate fidelities reported in the existing literature in Table \ref{tb:Gfidelity}. The data are obtained from various quantum systems, most of which are among the most promising platforms for scalable quantum computation. It is evident that high-fidelity geometric gates have been available in superconducting qubits and NV centers. More importantly, the record of fidelity is renewed rapidly, approaching the modest threshold required for quantum error correction.

At the end of this section, we would like to discuss the robustness of the geometric gates. Building quantum gates with geometric phases is motivated by the belief that geometric phases depend only on the path of evolution rather than the dynamical details and thus may have built-in error-resilient features. In fact, the geometric feature of the Berry phase has been studied extensively and experimentally verified in various quantum systems ({\it e.g.}, in an electronic harmonic oscillator \cite{pechal2012geometric} and in a superconducting phase qubit \cite{zhang2017measuring}). However, this belief is challenged in the context of quantum information processing. In Ref.~\cite{nazir02decoherence}, Nazir {\it et al.} showed that geometric gates are more sensitive to decoherence than dynamical ones.
Moreover, Blais and Tremblay studied the effect of fluctuations of the control parameters on geometric gates and found that dynamical gates outperform geometric ones \cite{blais03effect}.
In contrast to these observations from numerical simulations, analytical results showed that the Berry phase is robust against dephasing \cite{carollo2003geometric} as well as stochastic fluctuations of control parameters \cite{de03berry}. It was found in Ref.~\cite{de03berry} that the geometric phase contributions to dephasing are path-dependent and will vanish in the adiabatic limit. These two facts have been experimentally verified in \cite{leek07observation,berger2013exploring,filipp2009experimental}. It is pointed out that the geometric gates studied in the numerical approach are realized with three rotations, while a dynamical gate is implemented with only one. In this sense, the error model used in Ref.~\cite{nazir02decoherence} is not suitable to compare the performance of the geometric and dynamical gates because the gate fidelity in the model is determined only by the gate time, irrelevant to the geometric or dynamical nature.

To provide a direct and fair comparison, Zhu and Zanardi proposed a scheme in which the gate fidelity and dynamical phase depend on the same parameters \cite{zhu05geo}. It was found that, in the presence of stochastic control errors, the maximum gate fidelity appears when the dynamical phase vanishes, showing clear evidence for the robustness of GQC. In addition, evidence of the robustness is also found in the presence of systematic errors both theoretically \cite{thomas2011} and experimentally \cite{wu13geo}. 

There are more concrete examples for certain quantum systems. For instance, in solid-state spins, the average probability of error per gate for geometric gates is found to be one order of magnitude better than that for the dynamical gates, suggesting that the geometric property is significantly more resilient to the error and parameter imperfections present in the system \cite{kleissler2018universal}. As another piece of evidence shown 
for solid-state spins, adiabatic geometric gates remain unchanged when the driving field amplitude varies by a factor of 2 or the detuning fluctuates in a range comparable to the inverse of the gate time \cite{huang19experimental}. The same robustness feature is also found in a superconducting circuit \cite{xu2020experimental,zhao2021experimental} where geometric gates are demonstrated to be more immune to the control amplitude error and qubit frequency shift-induced error than conventional dynamical gates.

The noise effect on the geometric gates can be suppressed by many other techniques, such as composite pulses \cite{ota2009composite} and dynamical decoupling sequences \cite{qin2017suppressing}. For now, many new-type high-fidelity geometric gates have been proposed for various systems \cite{chen2020high,liu2021super,li2021superrobust,chen2022fault,chen2022enhanced,kang2022nonadiabatic}.

\section{Holonomic quantum computation} 

\label{sec:nonabelianHQC} 

\subsection{Adiabatic holonomic gates}
The adiabatic quantum holonomy emerges from cyclic evolution of a parameter dependent subspace $\mathcal{V}(R)$ of Hilbert space $\mathcal{H}$ corresponding to a degenerate eigensubspace of the system Hamiltonian $H(R)$. 
The holonomic transformation on the subspace is described by the path-ordered integral $U(C)=\mathcal{P}e^{i\oint_CA(R)\cdot\mathrm{d}R}$ with  $A_{pk}(R)=i\bra{\varphi_p(R)}\nabla_R\ket{\varphi_k(R)}$ the matrix-valued Wilczek-Zee connection \cite{wilczek1984appearance}, $C$ a loop in parameter space, and the set $\{ \ket{\varphi_k (R)} \}$ spans $\mathcal{V}(R)$. 
In contrast, the additional dynamical contribution is merely a global phase factor and can be ignored. 
Since $A(R)$ for different loops generally do not commute, the use of holonomies to build quantum gates naturally provides universality.   
These holonomies have a richer underlying geometric structure as compared to the Berry phase, and thus may offer more control feasibility from a practical point of view. 

Before proceeding to concrete models, it is essential to ask the following question: are the holonomic transformations capable of building an arbitrary gate in the subspace? 
In fact, all holonomies generated by the connection $A$ form a subgroup of $U(N)$ ($N$ is the dimension of the degenerate subspace $\mathcal{V}$) known as the holonomy group $\mathrm{Hol}(A)$.  It is clear that universality requires that $\mathrm{Hol}(A)=U(N)$. 
Whether this condition can be satisfied is determined by the curvature 2-form  $F=\sum_{\mu\nu}F_{\mu\nu}\mathrm{d}x_{\mu}\wedge \mathrm{d}x_{\nu}$, where
\begin{equation}
    F_{\mu\nu}=\partial_\mu A_\mu-\partial_\nu A_\nu-[A_\mu,A_\nu].
\end{equation}
If $F_{\mu\nu}$ linearly spans the whole Lie algebra $U(N)$ for each $\mu$ and $\nu$, then $A$ is irreducible (for the details to the mathematical concepts used here, we refer to  \cite{nakahara2003geometry,fujii2001mathematical}).
Zanardi and Rasetti argued in their seminal paper \cite{zanardi1999holonomic} that the connections associated with holonomies are actually irreducible. 
As an example, they considered an $(N+1)$-dimensional Hilbert space that comprises an $N$-dimensional degenerate subspace. The related curvature was explicitly calculated and they showed that the components span the whole $U(N)$. In addition, a specific model was proposed in Ref. \cite{pachos1999non} to determine the loops that generate any required gate, thus forming a universal set of quantum gates. 

The basic property of quantum holonomies is that they can be combined according to the underlying loops, {\it i.e.}, 
\begin{equation}
    U_n(C_2\cdot C_1)=U_n(C_2)U_n(C_1).
\end{equation}
As such, each loop can generate a generic holonomy. Now the situation is very similar to the more general universality condition studied in Ref. \cite{lloyd95almost}. Therefore, for practical purposes, two generic loops suffice to generate a universal set of gates acting on the subspace \cite{zanardi1999holonomic}. A formal description of the underlying mathematical foundation of HQC together with an example in an optical system can be found in Ref. \cite{fujii2001mathematical}. 

Another essential ingredient in HQC is the scalability, based on which the above approach can be generalized to many-qubit situations. 
One possible method is to use the tensor product structure of composite quantum systems, {\it i.e.}, $\mathcal{H}=\otimes_{j=1}^n\mathcal{H}_j$, where $\mathcal{H}_j$ is the Hilbert space for $j$th system, and $n$ is the number of the systems. In this case, cyclic evolution of the subspace $\mathcal{V}_i\otimes\mathcal{V}_j$ can be realized in $\mathcal{H}_i\otimes\mathcal{H}_j$ via a varying Hamiltonian of the form $H_{i,j}=H_i+H_j+H_{I}$, where $H_{i(j)}$ is the Hamiltonian acting on $\mathcal{H}_{i(j)}$ and $H_I$ is the interaction between the two systems, as long as $H_I$ does not perturb the structure of $H_i$ and $H_j$.
Another method is to add an auxiliary system, {\it e.g.}, a qubit, to the target systems. The total Hilbert space is of the form $\mathcal{H}=\otimes_{j=1}^n\mathcal{H}_j\otimes\mathcal{H}_a$, where $\mathcal{H}_a$ is the Hilbert space of the auxiliary system \cite{pachos1999non}.
The cyclic evolution of the subspace $\mathcal{V}_i\otimes\mathcal{V}_j$ can be performed in $\mathcal{V}_i\otimes\mathcal{V}_j\otimes\mathcal{H}_a$ by slowly changing the Hamiltonian of the combined system.
Some control aspects of quantum gates with non-Abelian holonomy are provided in Ref.~\cite{lucarelli2005control}.

\subsubsection{Tripod scheme}\label{sec:tripod}

\begin{figure}
  \centering
  \includegraphics[width=0.8\textwidth]{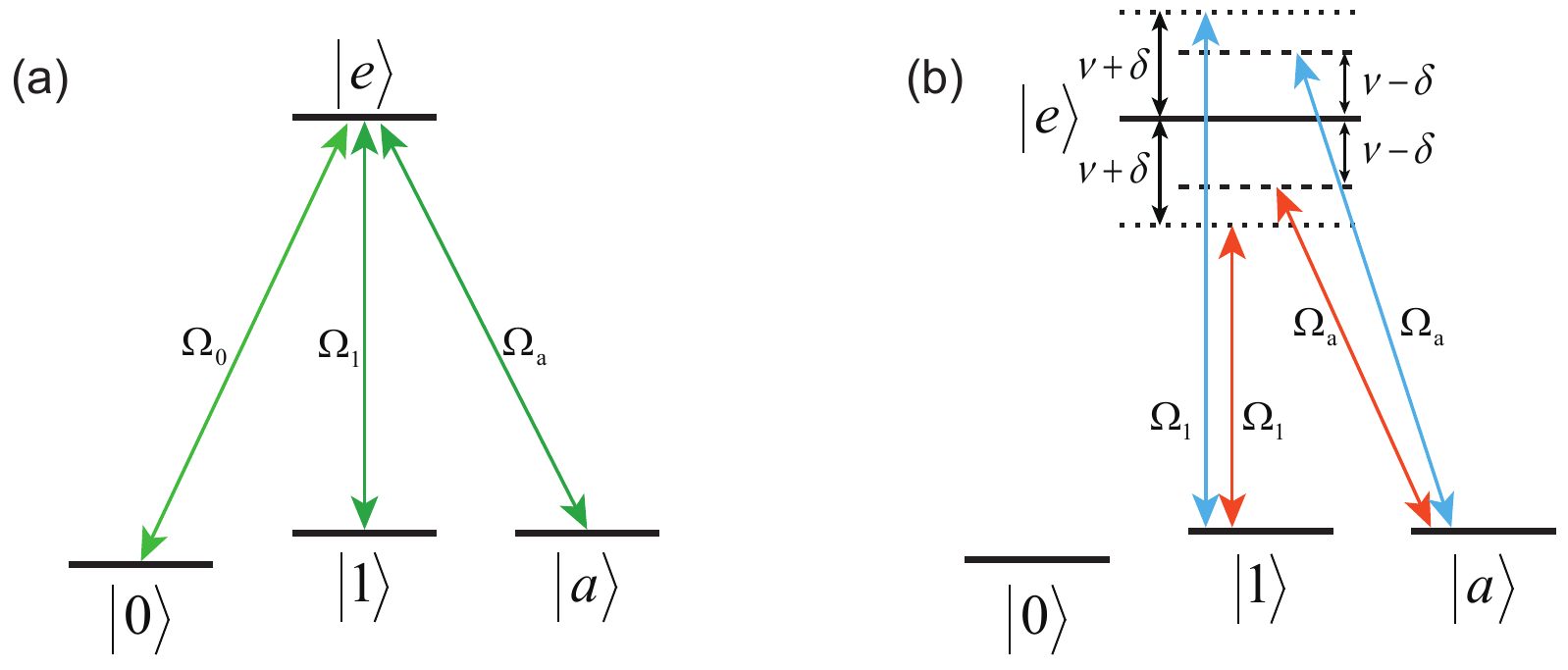}\\
  \caption{Level structures and laser configurations for trapped ions. (a) Three ground or metastable states $\ket{0}$, $\ket{1}$, and $\ket{a}$ coupled to an excited state $\ket{e}$ forming a tripod are used for each ion, where $\ket{0}$ and $\ket{1}$ are defined as qubit states. A possible way to choose the three ground states is that $\ket{1}$ and $\ket{a}$ are two degenerate Zeeman sub-levels addressed by lasers with different polarizations, and $\ket{0}$ is the ground state with slightly different energy so that it can be addressed by a laser with a different frequency. The same structure can be realized in trapped ions with lasers that couple the vibrational mode with two internal energy levels of the atom \cite{pachos2002topological}, neutral atoms in a cavity QED \cite{recati2002holonomic}, or Ising chain \cite{ota2008implementation}. Interestingly, in a semiconductor-based system \cite{solinas2003holonomic,solinas2003semiconductor}, the ground state of the system is used as the excited state $\ket{e}$, while the three states of the light holes are depicted as $\ket{0}$, $\ket{1}$, and $\ket{a}$. (b) Two-color laser method to build conditional phase-shift gate between two ions. The same configuration  is used on both ions.}\label{fig:tripod}
\end{figure}
The ion-trap setup, theoretically described in Refs. \cite{cirac95quantum,sorensen99quantum,molmer1999multiparticle,cirac2000scalable} and experimentally demonstrated in Refs.  \cite{monroe1995demonstration,roos1999quantum,sackett2000experimental}, is an ideal platform for tripod-based HQC. The tripod system has also been studied in atoms \cite{coulston1992population,unanyan1998robust,unanyan1999laser}, the latter of these works delineated a non-Abelian gauge structure of the tripod. 
Duan {\it et al.} used this structure to achieve a universal set of holonomic quantum gates \cite{duan2001geometric}. In the latter system, an array of ions are confined in a linear Paul trap. Each ion has three ground (or metastable) states ($\ket{0}$, $\ket{1}$, and $\ket{a}$), and one excited state $\ket{e}$ [see Fig.~\ref{fig:tripod} (a)]. The states $\ket{0}$ and $\ket{1}$ are used as qubit states, while the state $\ket{a}$ is an auxiliary level for gate realization. All three ground states are coupled to $\ket{e}$ separately through a resonant classical laser field, each with a suitable polarization and frequency. In the rotating frame, the Hamiltonian for each ion takes the form
\begin{equation}\label{eq:htripod}
    H_j=\Omega_0\ket{e}_j\bra{0}+\Omega_1\ket{e}_j\bra{1}+\Omega_a\ket{e}_j\bra{a}+\mathrm{H.c.},
\end{equation}
where $\Omega_k$ ($k=0,1,a$) are controllable Rabi frequencies. With this Hamiltonian, two kinds of single-qubit gates acting on the $j$th ion can be realized. 

The first kind of single-qubit operations is $U^j_z(\phi_1)=e^{i\phi_1\ket{1}_j\bra{1}}$, which is equivalent to $e^{-i\phi_1\sigma_z^j/2}$, {\it i.e.}, a qubit rotation around $z$-axis by the angle $\phi_1$. To this end, $\Omega_0$ is set to zero during the adiabatic process so that $\ket{0}$ is decoupled, while $\Omega_1=-\Omega\sin(\theta/2)e^{i\varphi}$ 
and  $\Omega_a=\Omega\cos(\theta/2)$. Here, $\theta$ and $\varphi$ serve as control parameters. The absolute magnitude $\Omega$ should be sufficiently large to satisfy the adiabatic condition. In this setting, the Hamiltonian has a dark state  $\ket{d}_j=\cos(\theta/2)\ket{1}_j+\sin(\theta/2)e^{i\varphi}\ket{a}_j$ with vanishing energy and thereby trivial dynamical phase factor. Initially, $\theta = 0$ so as the dark state starts in $\ket{1}_j$ and  evolves in a cyclic manner by changing $\theta$ and $\varphi$ so as to end up at $\theta = 0$. This essentially reduces to a calculation of the Berry phase acquired by $\ket{d}_j$, yielding $\phi_1=\oint\sin\theta\mathrm{d}\theta\mathrm{d}\varphi$. Hence, $U^j_z(\phi_1)$ with an arbitrary $\phi_1$ can be obtained by choosing a proper loop of $\theta$ and $\varphi$.

The second kind of single-qubit gate is $U^j_y(\phi_2)=e^{i\phi_2\sigma_y^j}$, which implements a rotation around the $y$-axis. For this purpose, the Rabi frequencies are parameterized as $\Omega_0=\Omega\sin\theta\cos\varphi$, $\Omega_1=\Omega\sin\theta\sin\varphi$, and $\Omega_a=\Omega\cos\theta$. The resulting Hamiltonian has two degenerate dark states $\ket{d_1}=\cos\theta(\cos\varphi\ket{0}_j+\sin\varphi\ket{1}_j)-\sin\theta\ket{a}_j$ and $\ket{d_2}=\cos\varphi\ket{1}_j-\sin\varphi\ket{0}_j$, which form a 2-D subspace. Similarly, an adiabatic cyclic evolution of the subspace, starting and ending at ${\rm Span} \{ \ket{0},\ket{1} \}$, can be realized by taking the parameters $\theta,\varphi$ around a loop based at $\theta =0$. The Wilczek-Zee connection becomes
\begin{equation}
    A=\cos\theta\left(
    \begin{array}{cc}
        0 & -i \\
        i & 0
    \end{array}
    \right)\Vec{\varphi},
\end{equation}
where $\Vec{\varphi}$ is a unit vector along the direction of increasing $\varphi$. By using Stokes' theorem, one finds that the holonomy $U_y^j(\phi_2)$ with $\phi_2=\iint\sin\theta\mathrm{d}\theta\mathrm{d}\varphi$ the solid angle enclosed by the loop in parameter space. The combination of $U_z^j(\phi_1)$ and $U_y^j(\phi_2)$ allows one to implement any single-qubit holonomic gate \cite{lloyd95almost}.

To complete a set of universal gates, a conditional geometric phase-shift gate $U^{jk}(\phi_3)=e^{i\phi_3\ket{11}_{jk}\bra{11}}$ between the $j$th and $k$th ion qubits is required. The geometric phase $\phi_3$ can be realized in a two-ion system coupled through Coulomb  interactions. Based on the two-color control proposal in Ref. \cite{molmer1999multiparticle}, the transition $\ket{1}\leftrightarrow\ket{e}$ of each ion is driven by a red and blue detuned laser, respectively [see Fig.~\ref{fig:tripod}(b)]. The red and blue lasers have a detuning $-(\nu+\delta)$ and  $\nu+\delta$, respectively, where $\nu$ is the phonon frequency of one oscillation mode ({\it e.g.}, the center of mass mode) and $\delta$ is an extra detuning. Similarly, the transition $\ket{a}\leftrightarrow\ket{e}$ is driven by a red and blue detuned laser, but with the detuning $-(\nu-\delta)$ and $\nu-\delta$, respectively. When the Lamb-Dicke criterion is satisfied, {\it i.e.}, $\eta^2\ll1$, where $\eta$ is defined by the ratio of the ion oscillation amplitude to the optical wavelength, the interaction Hamiltonian takes the form
\begin{equation}\label{eq:hjk}
    H_{jk}=\frac{\eta^2}{\delta}\left(-|\Omega'|\sin\frac{\theta}{2}e^{i\varphi}\ket{ee}_{jk}\bra{11}+|\Omega'|\cos\frac{\theta}{2}\ket{ee}_{jk}\bra{aa}+\mathrm{H.c.}\right),
\end{equation}
where $|\Omega'|^2=|\Omega_1|^4+|\Omega_a|^4$, $|\Omega_1|^2/|\Omega_a|^2=\tan(\theta/2)$, and $\varphi/2$ is the phase difference between the two lasers. In this situation, the basis vectors $\ket{00}_{jk}$, $\ket{01}_{jk}$, and $\ket{10}_{jk}$ are decoupled, leaving only $\ket{11}_{jk}$ relevant. The dark state of $H_{jk}$ is $\ket{d}_{jk}=\cos(\theta/2)\ket{11}+\sin(\theta/2)e^{i\varphi}\ket{aa}$. It follows that $\theta$ and $\varphi$ can be used as control parameters, undergoing a cyclic adiabatic evolution from $\theta=0$. The Berry phase acquired by $\ket{d}_{jk}$ is the solid angle $\phi_3$ swept by the vector characterized by spherical polar angles $(\theta, \varphi)$ in parameter space. 

When demonstrating these holonomic gates experimentally, several practical issues should be addressed \cite{duan2001geometric}. First, the adiabatic condition must be satisfied, demanding that the gate operation time must be significantly longer than the inverse of the energy gap between the dark states and other states (the adiabatic condition). The energy gap for $H_j$ is characterized by $\Delta_1=|\Omega|$, while for $H_{jk}$, it is given by $\Delta_2=\eta^2|\Omega'|/\delta$. Thus, it is required that $1/(t_1\Delta_1)^2$ and $1/(t_2\Delta_2)^2$, $t_1$ and $t_2$ being the run time of the single- and two-qubit holonomic gates, respectively, should be small. Second, since the excited state $\ket{e}$ is coupled by resonant lasers, one should avoid spontaneous emission (assuming the rate $\gamma_s$). Due to the fact that $\ket{e}$ is only weakly populated during the adiabatic evolution, the effective spontaneous emission rate is reduced by the leakage probability $1/(\Delta_it_i)^2$, $i=1,2$. In this case, the condition $\gamma_s/(\Delta_i^2t_i)\ll1$ can guarantee the spontaneous emission to be negligible. Finally, the influence of the heating of ion motion should be small. This requires $\gamma_h\ll\delta$, where $\gamma_h$ is the heating rate. It is worth noting that all the conditions mentioned above are equal to those in the dynamical schemes using the off-resonant Raman transitions.

As a promising platform for quantum computing, superconducting systems (for reviews, see Refs. \cite{you2006superconducting,huang2020superconducting}) are shown to be suitable for designing holonomic gates. Faoro {\it et al.} put forward a scheme in which the four levels in Fig.~\ref{fig:tripod}(a) are represented by states encoded in four superconducting islands \cite{faoro2003non}. The related Josephson network operates in the charge regime, {\it i.e.}, the Josephson energies of the junctions are much smaller than the charging energy of the setup. Under some parameter conditions \cite{faoro2003non}, the charge state $\ket{1000}$ of the four superconducting qubits is an excited state, and thus can be used as an encoded $\ket{e}_L$ in Fig.~\ref{fig:tripod}(a); three other charge states $\ket{0100}$, $\ket{0001}$, and $\ket{0010}$ are degenerate and can be used as logical $\ket{0}_L$, $\ket{1}_L$, and auxiliary $\ket{a}_L$, respectively. With this structure and the three fluxes (through the latter three islands) as control parameters, $U_z(\phi_1)$ and $U_y(\phi_2)$ can be realized. By coupling two such Josephson networks together via Josephson junctions, the effective Hamiltonian $H_{\mathrm{eff}}=J_{ea}\ket{ee}_L\bra{ea}+J_{11}\ket{ee}_L\bra{11}+\mathrm{H.c.}$ is obtained. Similar to $H_{jk}$ in \eqref{eq:hjk}, the conditional phase-shift gate $U^{jk}(\phi_3)$ can be implemented. This idea has been simplified to cases where three \cite{choi2003geometric,kamleitner2011geometric} and also two \cite{cholascinski2004quantum} superconducting qubits are utilized for one logical qubit. In Ref.~\cite{cholascinski2004quantum}, a setup similar to that depicted in Fig.~\ref{fig:SQUID}(c) was employed. The difference is that the islands in this setup were coupled with dc SQUID \cite{cholascinski2004quantum} instead of a capacitor \cite{falci00detection} because of the need for controllable Josephson couplings.

Another way to find the tripod structure in a superconducting system is to use the energy levels of a device in a microwave cavity. Zhang {\it et al.} studied the energy levels in a SQUID coupled to a single-mode cavity field with constant coupling strength $g$ and two microwave pulses \cite{zhang2005holonomic}. The three lowest levels in the SQUID are used as $\ket{0}$, $\ket{1}$, and $\ket{a}$ while an excited level is used as $\ket{e}$. In this system, the $\ket{a}\!\leftrightarrow\!\ket{e}$ transition is coupled to the cavity, and the $\ket{0}\!\leftrightarrow\!\ket{e}$, $\ket{1}\!\leftrightarrow\!\ket{e}$ transitions are coupled to the microwave pulses. When two such SQUIDs are in a microwave cavity, each $\ket{a}\!\leftrightarrow\!\ket{e}$ transition is coupled to the cavity field. In this setting, the system Hamiltonian contains a degenerate invariant subspace, the adiabatic cyclic evolution of which implements a controlled phase-shift gate on the two qubits. A charge-phase qubit in a cavity \cite{feng2008holonomic} and two phase qubits in a cavity \cite{peng2008implementation} were used for implementing holonomic gates. In Ref.~\cite{lin2009robust}, $N+1$ atoms (the first one acts as a control system and the others act as target systems) interact with a single-mode cavity, and  multi-qubit phase gates can be performed on the target atoms.

Adiabatic holonomic gates have also been proposed in P-type cubic symmetry semiconductors \cite{bernevig2005holonomic}, electron spins in quantum dots \cite{golovach2010holonomic}, heavy holes in semiconductors \cite{budich2012all}, systems carrying electric dipole moment \cite{bakke2011quantum,bakke2012holonomic}, and ensembles of neutral atoms \cite{zheng2012geometric}. Apart from building quantum gates, adiabatic holonomy has been shown to be useful for quantum information storage \cite{li2004non} and for charge pumping devices \cite{brosco2008non,pirkkalainen2010non,solinas2010ground,Sugawa2021wilson}.

\subsection{Nonadiabatic holonomic gates}

Nontrivial nonadiabatic quantum holonomies may emerge from cyclic evolution of a proper subspace $\mathcal{V}(t)$ of Hilbert space $\mathcal{H}$. As shown in Section \ref{subsec:non_quant_holonomy} above, the unitary transformation $U(T,0) \equiv U(T)$ acting on $\mathcal{V}(0)$ contains both dynamical and geometric contributions: $U(T)=\mathcal{T}e^{i\int_0^T[A(t)-K(t)]\mathrm{d}t}$. Unlike adiabatic holonomies, the dynamical component $K(t)$ in $U(T)$ is a Hermitian matrix and generally does not commute with the geometric one $A(t)$.

To build quantum gates with nonadiabatic holonomies, one needs to deal with the dynamical contribution. One possible method is to find some special system Hamiltonians $H(t)$ with $L$-dimensional subspace $\mathcal{V}(t)$ such that the corresponding $K(t)$ on $\mathcal{V}(t)$ vanishes, {\it i.e.}, $K(t)=0$ during the evolution. This choice results in trivial (global) dynamical phases. Two conditions to realize a holonomic gate were identified in Refs.~\cite{sjoqvist2012non,xu2012non}: 
\begin{align}\label{eq:2condition}
   & \text{(i) } \sum_{k=1}^{L}\ket{\varphi_k(T)}\bra{\varphi_k(T)}=\sum_{k=1}^{L}\ket{\varphi_k(0)}\bra{\varphi_k(0)}, \quad \text{{\it i.e.}, } \mathcal{V}(T)=\mathcal{V}(0);\\ \nonumber
    & \text{(ii) } \bra{\varphi_k(t)}H(t)\ket{\varphi_l(t)}=0, \quad k,l=1,2,\cdots,L ,\quad \text{{\it i.e.}, } K(t)=0, \\ \nonumber
\end{align}
$L$ being the dimension of $\mathcal{V}(t)$. 
The first condition guarantees that the subspace  $\mathcal{V}(0)$ undergoes a cyclic evolution during time $T$ (the `run time'), while the second condition ensures that there is no dynamical contribution during the evolution.

It follows from the two conditions that the unitary transformation acting on $\mathcal{V}(T)=\mathcal{V}(0)$ is\footnote{Note that a more general formulation of condition (ii) is $K_{kl}(t) = \eta \delta_{kl}$ \cite{sjoqvist2016conceptual}, $\eta$ being real-valued and possible time-dependent, which allows for a global phase factor in $U(T)$. This phase factor will have no physical influence on the gate and will therefore be ignored in the following.} $U(T)=\mathcal{T}e^{i\int_0^TA(t)\mathrm{d}t}$. Since $A(t)$ serves as a proper gauge potential under gauge transformations, $U(T)$ is the holonomy acting on the subspace $\mathcal{V}(0)$ \cite{sjoqvist2012non,xu2012non}.
In the following, a realization of holonomic gates based on the three- and four-level settings uses Hamiltonians that satisfy the above two conditions.

Another possible method involves the design of proper Hamiltonians such that $[A(t),K(t')]=0$, $t,t'\in [0,T]$. In this manner, the evolution $U(T)$ can be separated into the geometric part $U_g(T)=\mathcal{T}e^{i\int_0^TA(t)\mathrm{d}t}$ and the dynamical part $U_d(T)=\mathcal{T}e^{-i\int_0^TK(t)\mathrm{d}t}$. The latter can be compensated for by another purely dynamical evolution resulting in $U'(T)=U_d^{\dagger}(T)$.

\subsubsection{Three-level schemes}\label{sec:3level}
\begin{figure}
  \centering
  \includegraphics[width=1.0\textwidth]{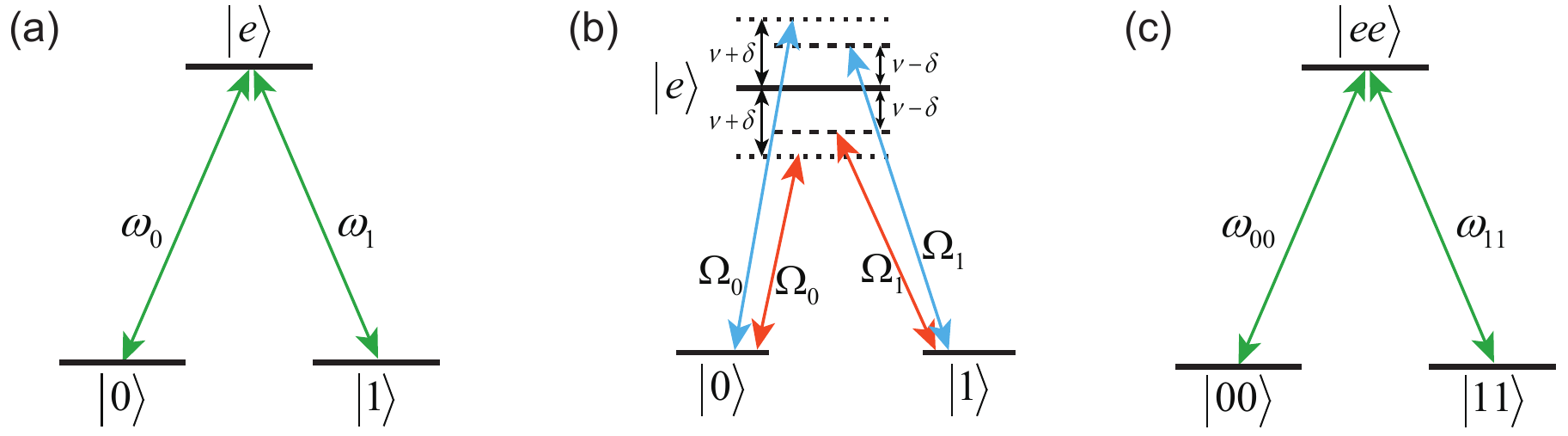}\\
  \caption{Level structures for nonadiabatic holonomic gates in a $\Lambda$ configuration. (a) The two ground state levels $\ket{0}$ and $\ket{1}$ are resonantly coupled to an excited state level $\ket{e}$ via a pair of pulses. Each pulse's relative strength and phase are characterized by the parameters $\omega_k$ ($k=0,1$) satisfying $|\omega_0|^2+|\omega_1|^2=1$. (b) Two-color laser method to realize nonadiabatic two-qubit holonomic gates. Compared with the adiabatic tripod scheme shown in Fig.~\ref{fig:tripod}(b), the $\Lambda$-like structure used here does not need the auxiliary level $\ket{a}$. The ratio between $|\Omega_0(t)|^2$ and $|\Omega_1(t)|^2$ is constant, {\it i.e.}, $|\Omega_0(t)|^2/|\Omega_1(t)|^2=\tan(\theta/2)$. (c) Level structure for the effective Hamiltonian $H_e$ in which $\Omega_{00}=\sin(\theta/2)e^{i\phi/2}$ and $\Omega_{11}=-\cos(\theta/2)e^{-i\phi/2}$.}\label{fig:three-level}
\end{figure}
Consider a three-level quantum system consisting of the `bare' energy eigenstates $\ket{0}$, $\ket{1}$, and $\ket{e}$ with energies $E_0$, $E_1$, and $E_e$, respectively. While the transition $\ket{0}\!\leftrightarrow\!\ket{1}$ is forbidden, each transition $\ket{k}\!\leftrightarrow\!\ket{e}$, $k=0,1$, is driven separately by a laser or microwave pulse with frequency $\nu_k$. The Hamiltonian describing the interaction is written in the rotating frame by means of RWA as
\begin{equation}\label{eq:hdelta01}
    H(t)=\Delta_0\ket{0}\bra{0}+\Delta_1\ket{1}\bra{1}+\Omega(t)\left(\omega_0\ket{e}\bra{0}+\omega_1\ket{e}\bra{1}+\mathrm{H.c.}\right),
\end{equation}
where $\Delta_k=2\pi\nu_k-(E_e-E_k)$, $k=0,1$, are detunings controlled by $\nu_k$. The parameters $\omega_0$ and $\omega_1$, which satisfy $|\omega_0|^2+|\omega_1|^2=1$, describe the relative strength and relative phase of the two pulses. The Hamiltonian can be turned on and off by controlling the pulse envelope $\Omega(t)$. 
This $\Lambda$-type configuration can be found in a variety of quantum systems such as superconducting transmons \cite{abdumalikov2013experimental} or Xmons \cite{zhang2019single}, NV-centers \cite{arroyo2014room,zu2014experimental}, NMR \cite{feng2013experimental}, and trapped ions \cite{toyoda2013realization}.

To realize single-qubit nonadiabatic holonomic gates, it is convenient to work in the resonant regime, where $\Delta_0=\Delta_1=0$ \cite{sjoqvist2012non,xu2012non}. The states $\ket{0}$ and $\ket{1}$ are used to define the single-qubit state space $\mathcal{V}(0)$. Here, the parameters $\omega_0$ and $\omega_1$ are required to be time-independent over the duration of the pulse implementation. In this case, the Hamiltonian reduces to
\begin{equation}\label{eq:h1omega}
H_1(t)=\Omega(t) \left( \omega_0 \ket{e} \bra{0} + \omega_1 \ket{e} \bra{1} + \mathrm{H.c.} \right),
\end{equation}
the level structure of which is shown in Fig.~\ref{fig:three-level}(a).
$H_1(t)$ has a dark eigenstate $\ket{d}=-\omega_1\ket{0}+\omega_0\ket{1}$, which is decoupled from the dynamics, as well as two other eigenstates $\ket{E_{\pm}}=(\pm \ket{e}+\omega_0^*\ket{0}+\omega_1^*\ket{1})/\sqrt{2} \equiv (\pm \ket{e}+\ket{b})/\sqrt{2}$ with eigenvalues $\pm\Omega(t)$ and `bright state' $\ket{b} = \omega_0^{\ast} \ket{0} + \omega_1^{\ast} \ket{1}$. The evolution $U_1(t)$ generated by $H_1(t)$ takes the form 
\begin{equation}
    U_1(t)= \ket{d} \bra{d} + e^{-i\int_0^t\Omega(t')\mathrm{d}t'} \ket{E_+} \bra{E_+} + e^{i\int_0^t\Omega(t')\mathrm{d}t'} \ket{E_-} \bra{E_-}.
\end{equation}
The subspace $\mathcal{V}(0)$ evolves into $\mathcal{V}(t)$ according to $\ket{k}\rightarrow \ket{\psi_k(t)}=U_1(t)\ket{k}$, $k=0,1$. $\mathcal{V}(0)$ undergoes cyclic evolution if the pulse envelope satisfies the condition $\int_0^T\Omega(t)\mathrm{d}t=\pi$. 
The evolution is purely geometric because $K_{kl}(t)=\bra{\psi_k(t)}H_1(t)\ket{\psi_l(t)}=\bra{k}H_1(t)\ket{l}=0$ for $t\in[0,T]$. 
Therefore, the evolution operator projected onto the subspace $\mathcal{V}(0)$ is the traceless holonomic single-qubit gate 
\begin{equation}\label{eq:u1g}
    U_1^g(C_n) = \ket{d}\bra{d} - \ket{b}\bra{b} = \Vec{n}\cdot\Vec{\sigma},
\end{equation}
where $\Vec{n}$ is a unit vector in $\mathbb{R}^3$ and $\Vec{\sigma}=(\sigma_x,\sigma_y,\sigma_z)$ are the Pauli operators. The transformation induced by $\ket{d}\bra{d} - \ket{b}\bra{b}$ is a complex-valued Householder reflection of a state through a plane orthogonal to $\ket{b}$; a transformation that has been extensively studied in the context of multilevel atoms and ions \cite{kyoseva2006,ivanov2006,ivanov2008}. Explicitly, by setting $\omega_0=\sin(\theta/2)e^{i\phi}$ and $\omega_1=-\cos(\theta/2)$, the unit vector takes the form $\Vec{n}=(\sin\theta\cos\phi,\sin\theta\sin\phi,\cos\theta)$. In this way, $U_1^g(C)$ can be regarded as a single-qubit rotation around $\Vec{n}$ by an angle 
$\pi$. An arbitrary single-qubit holonomic gate can be obtained by combining two such $U_1^g(C)$ corresponding to different unit vectors  {\it i.e.},
\begin{equation}
    U_1^g(C)=U_1^g(C_m)U_1^g(C_n)=\Vec{n}\cdot\Vec{m}-i\Vec{\sigma}\cdot(\Vec{n}\times\Vec{m}), 
\end{equation}
where $C=C_m\cdot C_n$, $C_m$ and $C_n$ are the two loops for the two cyclic evolutions with $\Vec{m}$ and $\Vec{n}$ being their associated unit vectors. 
$U_1^g(C)$ is a rotation by an angle $2\arccos(\Vec{n}\cdot\Vec{m})$ around the normal of the plane spanned by $\Vec{n}$ and $\Vec{m}$.
Hence, any desired single-qubit gate can be realized by a suitable choice of $\Vec{n}$ and $\Vec{m}$.
For example, the Hadamard gate $\ket{k}\mapsto[(-1)^k\ket{k}+\ket{k\oplus1}]/\sqrt{2}$ ($k=0,1$) can be achieved by a single loop with $\Vec{n}=(1,0,1)/\sqrt{2}$. The phase-shift gate $\ket{k}\mapsto e^{2ik(\phi'-\phi)}\ket{k}$ corresponds to two successive loops with $\Vec{n}=(\cos\phi,\sin\phi,0)$ and $\Vec{m}=(\cos\phi',\sin\phi',0)$.

Nontrivial two-qubit gates can be realized with the S\o{}rensen-M\o{}lmer setting \cite{sorensen99quantum,molmer1999multiparticle}. In the nonadiabatic version [see Fig.~\ref{fig:three-level}(b)], the transitions $\ket{k}\!\leftrightarrow\!\ket{e}$ ($k=0,1$) for a pair of ions are addressed by lasers with detunings $\pm\nu\pm\delta$ and $\pm\nu\mp\delta$, respectively. 
The off-resonant coupling between the ground states and the four singly excited states ($\ket{e0},\ket{0e},\ket{e1},\ket{1e}$) can be suppressed by setting the Rabi frequencies $|\Omega_0(t)|$ and $|\Omega_1(t)|$ smaller than $\nu$ \cite{sorensen99quantum}. In the Lamb-Dicke regime ($\eta^2\ll1$), the effective two-ion Hamiltonian reads 
\begin{equation}
    H_{2}=\frac{\eta^2}{\delta}\left( |\Omega_0(t)|^2\sigma_0(\phi)\otimes\sigma_0(\phi)-|\Omega_1(t)|^2\sigma_1(-\phi)\otimes\sigma_1(-\phi) \right),
\end{equation}
where $\eta$ is the Lamb-Dicke parameter, $\sigma_0(\phi)=e^{i\phi/4\ket{e}\bra{0}}+\mathrm{H.c.}$, and $\sigma_1(-\phi)=e^{-i\phi/4\ket{e}\bra{1}}+\mathrm{H.c.}$. The Hamiltonian $H_{2}$ can be rearranged so as to take the form  
\begin{equation}
    H_{2}(t)=\frac{\eta^2}{\delta}\Omega'(t) \left( H_e+H_a \right),
\end{equation}
where $\Omega'(t)=\sqrt{|\Omega_0(t)|^4+|\Omega_1(t)|^4}$.
The two terms $H_e$ and $H_a$ read
\begin{equation}
    H_e=\sin\frac{\theta}{2}e^{i\frac{\phi}{2}}\ket{ee}\bra{00}-\cos\frac{\theta}{2}e^{-i\frac{\phi}{2}}\ket{ee}\bra{11} +\mathrm{H.c.}, \ \ \ \ 
    H_a=\sin\frac{\theta}{2}\ket{e0}\bra{0e}-\cos\frac{\theta}{2}\ket{e1}\bra{1e} +\mathrm{H.c.},
\end{equation}
where  $\tan(\theta/2)=|\Omega_0(t)|^2/|\Omega_1(t)|^2$.

Due to the fact that $[H_e,H_a]=0$, the evolution operator can be written as
\begin{equation}\label{eq:u2heha}
    U_2(t)=e^{-i\int_0^tH_{2}(t')\mathrm{d}t'}=e^{-i\int_0^t\frac{\eta^2}{\delta}\Omega'(t')\mathrm{d}t'H_e}e^{-i\int_0^t\frac{\eta^2}{\delta}\Omega'(t')\mathrm{d}t'H_a}.
\end{equation}
When the cyclic condition $\int_0^T\frac{\eta^2}{\delta}\Omega'(t)\mathrm{d}t=\pi$ is satisfied, the second term on the right-hand side of \eqref{eq:u2heha} acts trivially on the computational subspace. 
Therefore, $H_{2}$ reduces to the $\Lambda$-like configuration $H_e$ shown in Fig.~\ref{fig:three-level}(c). 
The corresponding evolution operator at time $T$ reads
\begin{equation}
    U_2^g(C)=\left(
    \begin{array}{cccc}
        \cos\theta & 0 & 0 & \sin\theta e^{-i\phi} \\
        0 & 1 & 0 & 0 \\
        0 & 0 & 1 & 0 \\
        \sin\theta e^{i\phi} & 0 & 0 &-\cos\theta
    \end{array}
    \right)
\end{equation}
in the ordered basis $\{\ket{00}, \ket{01},\ket{10},\ket{11}\}$ spanning the computational subspace of the qubit pair. The purely holonomic nature of $U_2^g(C)$ follows from the triviality of $H_a$ on this subspace and the $\Lambda$ structure of $H_e$. Furthermore, $U_2^g(C)$ can entangle the two qubits, as is, {\it e.g.}, evident by noting that $\theta=0$ yields the conditional phase-shift gate $\ket{kl}\mapsto e^{ikl\pi}\ket{kl}$ ($k,l=0,1$). 

The geometric interpretation of $U_1^g(C_n)$ and $U_2^g(C)$ in terms of paths in the Grassmannian $\mathcal{G} (3;2)$ was provided in Ref. \cite{sjoqvist2012non}. Numerical simulation shows that  the nonadiabatic holonomic gates are robust against control error and decay \cite{sjoqvist2012non,johansson2012robustness}. Although the nonadiabatic scheme can be performed at high speed, there is a lower limit at which the RWA breaks down, severely affecting the fidelity of the holonomic gates \cite{spiegelberg2013rwa}. This opens up a trade-off between decay and RWA validity that can be used to optimize the performance of the holonomic gates with respect to run time \cite{alves2022rwa}.

The required $\Lambda$ model can be applied to many other quantum systems. For example, in a two-dimensional honeycomb lattice consisting of 
TLR (transmission line resonators) transmons units \cite{xue2017nonadiabatic,hong2018implementing}, the lowest three energy levels of each unit can be used to represent $\ket{0}$, $\ket{1}$, and $\ket{e}$, respectively. 
Furthermore, the $\Lambda$ structure can be realized by encoding the three states into physical qubits. One possible approach to this end is to use the `collective encoding' for three qubits: $\ket{0}\rightarrow\ket{100}$, $\ket{1}\rightarrow\ket{010}$, $\ket{e}\rightarrow\ket{001}$ \cite{xu2012non}.
The above encoding is suitable for a three-quantum-dot device containing only a single electron \cite{mousolou2017universal}, where there are only three one-electron states ($\ket{100}$, $\ket{010}$, and $\ket{001}$) in the Fock space. The transitions between the encoded states are performed by hopping between different sites induced by an external magnetic field. More encoding schemes are  used in combining  HQC with environmental-noise suppressing techniques, which will be addressed in section \ref{sec:combineHQC}.

\subsubsection{Single-shot scheme}\label{sec:1shot}
\begin{figure}
  \centering
  \includegraphics[width=1.0\textwidth]{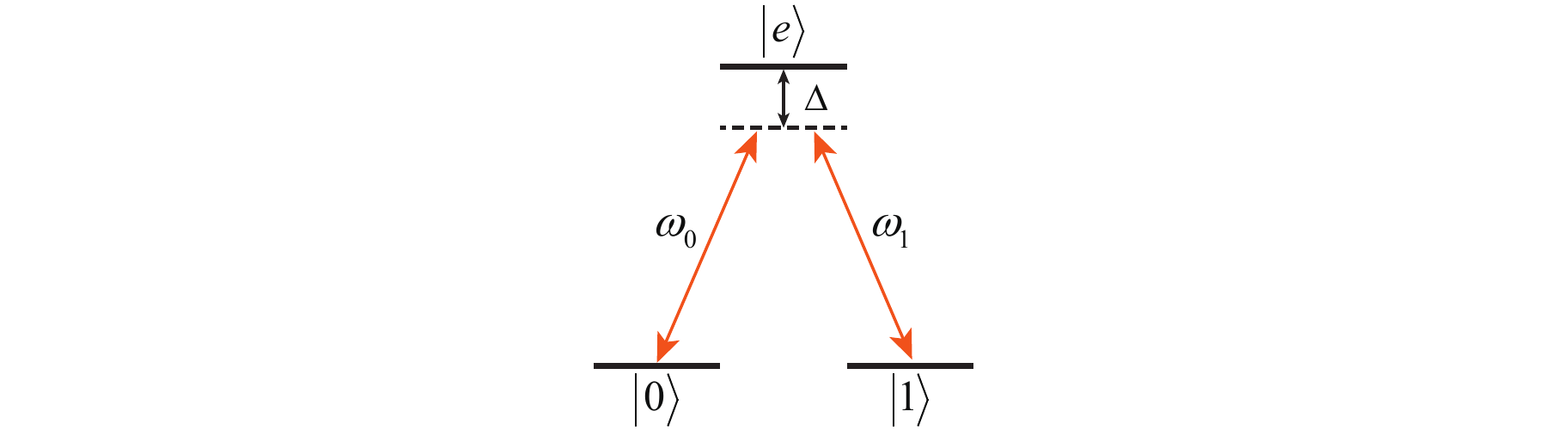}\\
  \caption{Level structure for a off-resonant $\Lambda$ configuration. The energy levels $\ket{0}$ and $\ket{1}$ are coupled to the excited state $\ket{e}$ by two pulses with detuning $\Delta$ and associated with frequencies $\omega_0$ and $\omega_1$, respectively.}\label{fig:singleshot}
\end{figure}

As shown in \eqref{eq:u1g}, a holonomic gate based on the  $\Lambda$ configuration in the resonant regime is a $\pi$ rotation around an axis dependent on the laser parameters $\theta$ and $\phi$. In this case, the universality of single-qubit gates requires a combination of two such holonomic gates. In the `single-shot' (or `off-resonant') scheme \cite{xu2015non,sjoqvist2016non}, on the other hand, the rotation angle can take any value by introducing a detuning in the control fields, thereby resulting in an arbitrary single-qubit gate by applying a single pair of laser pulses. 

To understand the idea of the single-shot scheme, we recall the Hamiltonian in \eqref{eq:hdelta01}. By setting the detunings as $\Delta_0=\Delta_1=\Delta$, the effective Hamiltonian can be written as 
\begin{equation}
    H_s(t)=-\Delta\ket{e}\bra{e}+\Omega(t)\left(\omega_0\ket{e}\bra{0}+\omega_1\ket{e}\bra{1}+\mathrm{H.c.}\right),
\end{equation}
where we have off-set the zero-point energy by subtracting $\Delta \hat{1}$. The level setting is illustrated in Fig.~\ref{fig:singleshot}.
We next assume that the envelope  $\Omega(t)$ is a square pulse, {\it i.e.}, $\Omega(t)=\Omega$ when $t\in[0,T]$ and $\Omega(t)=0$ otherwise. 
By choosing
$\Delta=-2\Omega\sin\gamma$, $\omega_0=\cos\alpha\cos\gamma$, and $\omega_1=\sin\alpha\cos\gamma e^{i\beta}$ during the pulse, the effective Hamiltonian takes the form  
\begin{equation}
    H_s=2\Omega\sin\gamma\ket{e}\bra{e}+\Omega\cos\gamma \left( \ket{b}\bra{e}+\ket{e}\bra{b} \right),
\end{equation}
where $\ket{b}=\cos\alpha\ket{0}+e^{i\beta}\sin\alpha\ket{1}$. $H_s$ describes a two-level Hamiltonian because the dark state $\ket{d}=\sin\alpha\ket{0}-e^{i\beta}\cos\alpha\ket{1}$ decouples from the dynamics, just as in the resonant case. The choice of a square pulse ensures that no dynamical phases are acquired during the evolution, {\it i.e.}, that condition (i) above is satisfied\footnote{It should be noted that this can also be achieved by choosing $\Delta \equiv \Delta (t)$ and $\Omega(t)$ to have precisely the same shape \cite{zhang2019single}.}. 

When the run time $T$ is set to satisfy $\Omega T=\pi$, the evolution operator $U_s(T)$ generated by $H_s$ takes the form:
\begin{equation}
    U_s(T)=\left(
    \begin{array}{ccc}
        e^{-i\phi} & 0 & 0  \\
        0 & e^{-i\phi} & 0  \\
        0 & 0 & 1   
    \end{array}
    \right)
\end{equation}
in the basis $\ket{e}, \ket{b}, \ket{d}$. Here, the phase angle $\phi=\pi(1+\sin\gamma)$ can be designed by adjusting the detuning $\Delta$.
Since the basis $\{\ket{b},\ket{d}\}$ spans the same subspace as $\{\ket{0},\ket{1}\}$, the qubit subspace returns at time $T$. Hence, the operation acting on the qubit reads
\begin{equation}
    U_s(T)=e^{-i\frac{\phi}{2}\Vec{n}\cdot\Vec{\sigma}},
\end{equation}
where $\Vec{n}=(\sin2\alpha\cos\beta,\sin2\alpha\sin\beta,\cos2\alpha)$ is the qubit rotation axis and $\phi$ is the rotation angle.
This is an explicit expression of $SU(2)$. Moreover, this $U_s(T)$ is holonomic since the condition $\bra{k}U_s^{-1}(t)H_s(t)U_s(t)\ket{j}=0$, $k,j=0,1$, is satisfied for $t\in[0,T]$.

Compared with the resonant three-level scheme in which only the rotation axis $\Vec{n}$ can be adjusted, the single-shot scheme introduces an extra parameter, {\it i.e.}, the detuning $\Delta$, so that both the rotation axis $\Vec{n}$ and angle $\phi$ can be controlled independently. As a result, an arbitrary single-qubit gate can be realized in one step by properly designing $\Vec{n}$ and $\phi$. However, it is noteworthy that a small $\phi$ is achieved for a large $\Delta/\Omega$, which can be reached either by using a large $\Delta$ or a small $\Omega$, making the gate potentially unstable to small perturbations.
Implementation of this scheme has been proposed for  NV-center in diamond \cite{zhou2018fast} and Rydberg atoms \cite{zhao2018nonadiabatic}, where collective states are employed. 

\subsubsection{Multi-pulse schemes}\label{sec:multipulse}
\begin{figure}
  \centering
  \includegraphics[width=1.0\textwidth]{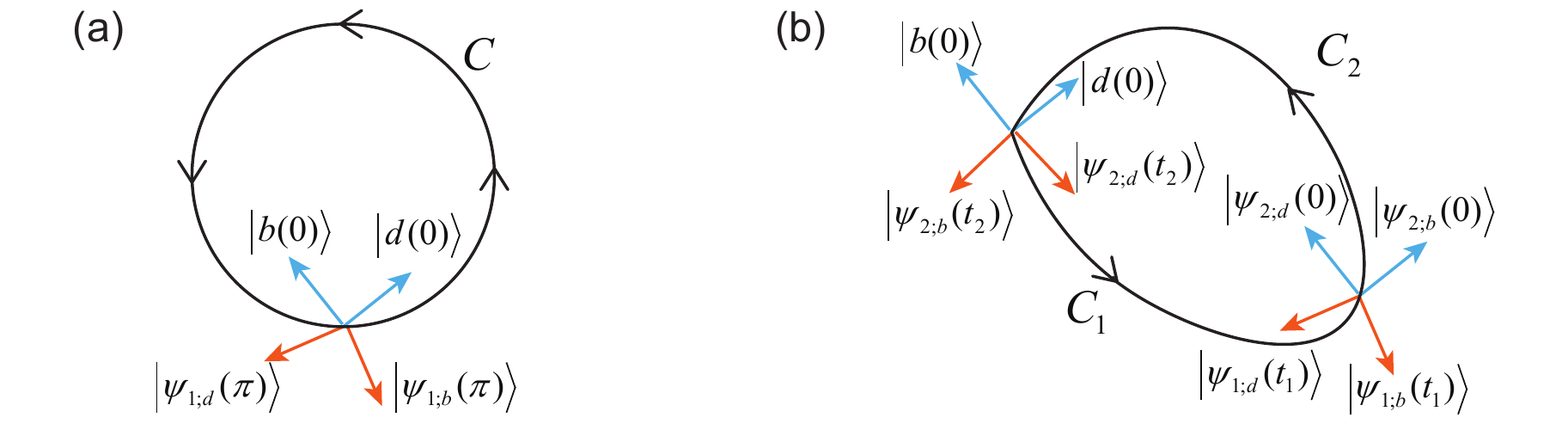}\\
  \caption{The multi-pulse scheme adapted to the resonant $\Lambda$-model. (a) In the standard scheme \cite{sjoqvist2012non}, a loop $C$ in $\mathcal{G}(3;2)$ is generated by a single resonant pulse pair. The initial subspace $\mathcal{V}(0)$ spanned by $\{\ket{b(0)},\ket{d(0)}\}$ is evolved by the $\Lambda$ Hamiltonian $H(t)=\Omega(t)(\ket{e}\bra{b}+\ket{b}\bra{e})$ around a loop when $\int_0^T\Omega(t)\mathrm{d}t=\pi$. This induces a traceless holonomic gate which transforms $\ket{b(0)}$ to $\ket{\psi_{1;b}(\pi)}$ and $\ket{d(0)}$ to $\ket{\psi_{1;d}(\pi)}$. (b) A loop generated by the multi-pulse scheme ($L=2$). Here, the initial subspace $\mathcal{V}_1(0)$ moves along the first path segment $C_1$ generated by $H_1(t)$ to $\mathcal{V}_1(t_1)=\mathcal{V}_2(0)$ spanned by $\ket{\psi_{1;b}(t_1)}$ and $\ket{\psi_{1;d}(t_1)}$. The second segment is generated by $H_2(t)$, which evolves $\mathcal{V}_2(0)$ to $\mathcal{V}_2(t_2)$. Since $H_1(t)$ and $H_2(t)$ are generally obtained by means of different external laser field pairs, a unitary basis transformation $V_2$ should act on the final state, {\it i.e.}, $\ket{\psi_{2;k}(0)}=V_2\ket{\psi_{1,k}(t_1)}$ ($k=b,d$). The quantum holonomy for the loop reads $U=U_2\cdot U_1$, where $U_1$ and $U_2$ are the evolution operators for $C_1$ and $C_2$, respectively. Contrary to the single-pulse holonomy in (a), $U$ is no longer traceless and can in fact sweep over all SU(2) by varying the laser parameters defining $\ket{d(0)}$ and $\ket{b(0)}$, as well as the relative phase $\eta$ between the two pulse pairs.}\label{fig:multi-pulse}
\end{figure}

A `single-loop, multi-pulse' scheme \cite{herterich2016single} can be used to generate arbitrary holonomic gates in a resonant $\Lambda$ system by dividing the loop into open path segments $C_1$, $C_2$, $\cdots$, $C_L$ in $\mathcal{G}(3;2)$, driven by a sequence of tailored laser pulses. Figure \ref{fig:multi-pulse} illustrates the idea for $L=2$. 

The scheme can be described in terms of the following procedure:

(i) The first path segment $C_1$ starts at the computational subspace, {\it i.e.},  $\mathcal{V}_1(0)=\mathrm{Span}\{\ket{0},\ket{1}\}=\mathrm{Span}\{\ket{b},\ket{d}\}$ with  $\ket{b}=\omega_0^*\ket{0}+\omega_1^*\ket{1}$ and $\ket{d}=-\omega_1\ket{0}+\omega_0\ket{1}$. $\mathcal{V}_1(0)$ is subsequently driven by the resonant Hamiltonian  $H_1=\Omega_1(t)(\ket{e}\bra{b}+\ket{b}\bra{e})$ to end up at $\mathcal{V}_1(t_1)=\mathrm{Span}\{\ket{\psi_{1;b}(t_1)},\ket{\psi_{1;d}(t_1)}\}$ at $t=t_1$.

(ii) The initial point ($\mathcal{V}_2 (0) = \mathrm{Span}\{\ket{\psi_{2;b}(0)},\ket{\psi_{2;d}(0)}\}$) of the second path segment $C_2$ coincides with the final point $\mathcal{V}_1(t_1)$ of the first segment. The driving Hamiltonian $H_2=\Omega_2(t)(\ket{\psi_{2;e}}\bra{\psi_{2;b}}+\ket{\psi_{2;b}}\bra{\psi_{2;e}})$ takes $\mathcal{V}_2 (0)$ along $C_2$. Since $H_1(t)$ and $H_2(t)$ are generally obtained by means of different external laser field pairs, a unitary basis transformation $V_2$ should act on the final state, {\it i.e.}, $\ket{\psi_{2;k}(0)}=V_2\ket{\psi_{1,k}(t_1)}$ ($k=b,d$). In addition, to maintain the geometric nature of the  gate, one must require that $V_2\ket{\psi_{1,e}(t_1)}=\ket{\psi_{1,e}(t_1)}$. 

(iii) The procedure is iterated from $C_2$ up to $C_{L}$ so that the subspace at the end-point of $C_L$ coincides with the computational subspace, {\it i.e.},   $\mathrm{Span}\{\ket{\psi_{n;b}(t_n)},\ket{\psi_{n;d}(t_n)}\}=\mathrm{Span}\{\ket{b},\ket{d}\}$, thereby closing the loop. 

By using the evolution operator $U_n=e^{-i\int_{0}^{t_n}H_n(t)\mathrm{d}t}$ for each segment $n$, we may write the resulting quantum holonomy as  
\begin{equation}
    U(C_1\cdots C_L)=U_L\cdots U_1.
\end{equation}
This shows that for a loop with $L$ segments, $U(C_1\cdots C_L)$ is a proper single-qubit gate acting on $\mathrm{Span}\{\ket{0},\ket{1}\}$.
It is demonstrated \cite{herterich2016single} that two segments ($L=2$) with $\int_{t_0}^{t_1}\Omega_1(t)\mathrm{d}t=\int_{t_1}^{t_2}\Omega_2(t)\mathrm{d}t=\pi/2$ are sufficient to construct an arbitrary single-qubit gate of the form $e^{-i\frac{1}{2} (\pi - \eta) {\bf n} \cdot \boldsymbol{\sigma}}$, $\eta$ being the relative phase of the two pulse pairs. Thus, similar to the single-shot scheme, the multi-pulse scheme can build an arbitrary single-qubit gate with one loop. An early experimental demonstration of the $L=2$ multi-pulse scheme in a solid state qubit can be found in Ref. \cite{rippe2008solidstate}.

In the ideal situation, all the three above-mentioned settings, {\it i.e.}, two-loop, single-shot, and multi-pulse schemes, are equivalent in the terms of universality. However, they may behave differently in the presence of errors due to the fact that distinctly different paths are traversed. The robustness of the three schemes against the systematic errors in the Rabi frequencies was studied in Ref. \cite{xu2017robust}, where it is shown  that the two-loop scheme is more immune to such kind of errors. Notably, the systematic error can be suppressed with composite-pulse scheme \cite{xu2017composite} or optimal control method \cite{zhang2019searching}. 

In the above schemes, the cyclic condition $\int_0^T\Omega(t)\mathrm{d}t=\pi$ must be satisfied regardless of the size of the geometric phase angle. A practical question is: are there paths shorter than $\pi$ for a given gate? A positive answer will provide a new possibility for realizing holonomic gates that are exposed to decoherence for a shorter time. Based on the off-resonant $\Lambda$ model, Xu {\it et al.} \cite{xu2018path} proposed a path-shortening scheme in which the evolution paths were separated into several (at least 2) `head-to-tail' segments, the total length of which can be less than $\pi$. An additional merit of this setting is that  decoherence can be eliminated by combining it with dynamical decoupling in a nonredundant fashion. The multi-pulse and path-shortening schemes look similar in the sense that they both divide a loop into several segments, but they are essentially different because the multi-pulse scheme is based on the resonant $\Lambda$ model while the path-shortening scheme is based on the off-resonant model, and the length of the loop in the multi-pulse scheme is still $\pi$ but can be shorter than $\pi$ in the path-shortening scheme.

\subsubsection{Four-level schemes}
\begin{figure}
  \centering
  \includegraphics[width=1.0\textwidth]{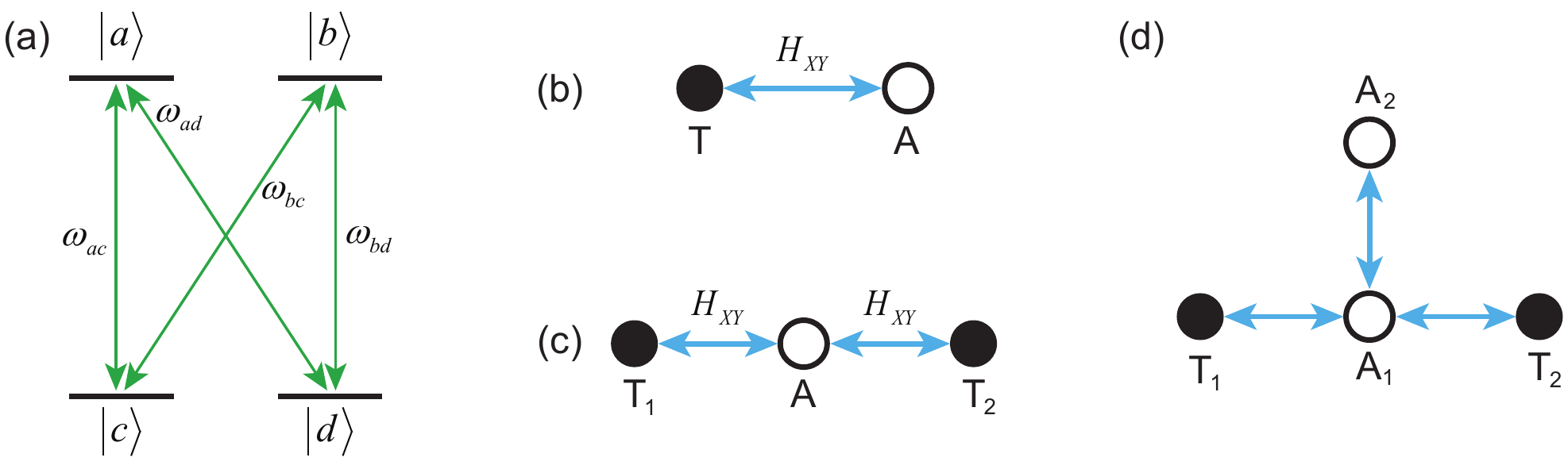}\\
  \caption{Level structure and qubit arrangement for the four-level scheme. (a) The quantum system employed here has four energy levels, in which $\ket{a}$ and $\ket{b}$ ($\ket{c}$ and $\ket{d}$) constitute the subspace $\mathcal{V}_0$ ($\mathcal{V}_1$). The arrows indicate the transitions between different levels. (b) The four-level structure can be provided with a target qubit T (solid circle) interacting with an auxiliary qubit A (open circle) through the $XY$-type interaction. (c) Two-qubit holonomic gates can be applied to a pair of target qubits $\mathrm{T}_1$ and $\mathrm{T}_2$ with the help of an auxiliary qubit A. $\mathrm{T}_1$ and $\mathrm{T}_2$ interact with qubit A through the $XY$-type interaction with different coupling strengths. (d) With an extra auxiliary qubit, the frequently used two-qubit gates acting on $\mathrm{T}_1$ and $\mathrm{T}_2$, such as SWAP, controlled-Z, and CNOT, can be realized.} \label{fig:fourlevel}
\end{figure}

In an attempt to go beyond the standard $\Lambda$ setting, Azimi Mousolou {\it et al.} \cite{mousolou2014universal}
considered the pulsed four-level system shown in Fig.~\ref{fig:fourlevel}(a).  This corresponds to the Hamiltonian 
\begin{equation}\label{eq:hT}
    H(t)=\Omega(t)\left( 
    \begin{array}{cc}
        0 & S \\
        S^\dag & 0
    \end{array}
    \right),
\end{equation}
expressed with respect to the ordered basis $\{\ket{a},\ket{b}$ $\ket{c},\ket{d}\}$, and where 
\begin{equation}
S = \left( \begin{array}{cc}
\omega_{ac} & \omega_{ad} \\ 
\omega_{bc} & \omega_{bd}
\end{array} \right) .
\end{equation}
The form of $H(t)$ naturally splits the  four-dimensional Hilbert space into two subspaces  $\mathcal{V}_0 = {\rm Span}\{\ket{a},\ket{b}\}$ and $\mathcal{V}_1 = {\rm Span} \{\ket{c},\ket{d}\}$. This kind of level structure can be realized in settings involving four coupled quantum dots arranged in closed ring configuration \cite{trif2010spin} and in a class of single-molecule magnet comprised of antiferromagnetic spin rings \cite{gatteschi2006molecular}. 

For invertible $S$ ({\it i.e.}, $\det S \neq 0$), there is a unique singular value decomposition $S=U_lDU_r^\dag$ with $U_l$ and $U_r$ unitary $2\times 2$ matrices and $D=\mathrm{diag}\{\alpha,\beta\} \geq 0$ \cite{nielsen2002quantum} that can be used to evaluate the time evolution operator $U(t)$ as 
\begin{equation}\label{eq:ufour}
    U(t)=e^{-i\int_0^tH(s)\mathrm{d}s}=\left(
    \begin{array}{cc}
        U_l\cos(a_tD)U_l^\dag & -iU_l\sin(a_tD)U_r^\dag \\
        -iU_r\sin(a_tD)U_l^\dag & U_r\cos(a_tD)U_r^\dag
    \end{array}
    \right).
\end{equation}
Here, $a_t=\int_0^t\Omega(s)\mathrm{d}s$ and we have taken $S$ to be time-independent over the duration of the pulse.
By assuming that there exists a time $t=T$ such that $\sin(a_TD)=\mathrm{diag}\{\sin(a_T\alpha),\sin(a_T\beta)\}=\mathrm{diag}\{0,0\}$ and $\cos(a_TD)=\mathrm{diag}\{\cos(a_T\alpha),\cos(a_T\beta)\}=\mathrm{diag}\{1,-1\}=Z$ , $U(T)$ is block diagonal, implying that the evolution of $\mathcal{V}_0$ and $\mathcal{V}_1$ is cyclic at time $T$.
With this condition, the transformation $U_k(C_k)$ acting on $\mathcal{V}_k$, $k=0,1$, reads
\begin{equation}
    U_0(C_0)=U_l ZU_l^{\dagger}, \quad U_1(C_1)=U_r ZU_r^{\dagger},
\end{equation}
where $C_0$ and $C_1$ are the paths of $\mathcal{V}_0$ and $\mathcal{V}_1$, respectively.

To realize nonadiabatic holonomic gates with $H(t)$, the two conditions (i) and (ii) in Eq. (\ref{eq:2condition}) must be satisfied. 
For condition (i), the cyclic evolution of $\mathcal{V}_0$ and $\mathcal{V}_1$ is  accomplished provided $T$ exists\footnote{A necessary and sufficient condition for the existence of $T$ is that the singular values $\alpha$ and $\beta$ are commensurate, {\it i.e.}, that $\alpha/\beta$ is a rational number.}. 
Furthermore, condition (ii) is satisfied since $\bra{l(t)}H(t)\ket{k(t)}=\bra{l}U^{\dagger}(t)H(t)U(t)\ket{k}=0$ for $l,k=a,b \ {\rm or} \  c,d$, where the commutation relation $[U(t),H(t)]=0$ has been used.
Hence, $U_0(C_0)$ and $U_1(C_1)$ can under these circumstances be regarded as holonomic gates acting on $\mathcal{V}_0$ and $\mathcal{V}_1$, respectively.

This four-level structure can be used to represent a two-qubit system.
By encoding the four basis states into the two-qubit states according to  $\ket{a}\rightarrow\ket{00}$, $\ket{b}\rightarrow\ket{01}$, $\ket{c}\rightarrow\ket{10}$, $\ket{d}\rightarrow\ket{11}$, the evolution operator $U(T)$ can be rewritten as 
\begin{equation}\label{eq:uc0c1}
    U(C_0,C_1)=\ket{0}\bra{0}\otimes U_0(C_0)+ \ket{1}\bra{1}\otimes U_1(C_1).
\end{equation}
When $U_l\neq U_r$, $U(C_0,C_1)$ is a conditional gate with the first qubit as control.
On the other hand, when $U_l=U_r$, $U(C_0,C_1)$ is reduced to $I\otimes U_0$, which is a single-qubit gate acting on the second qubit.
An arbitrary single-qubit gate is available since $U_l$ ($U_r$) can be chosen as long as $\det S\neq 0$, implying that a universal set of gates can be realized with  $U(C_0,C_1)$. 
The definition of the first and second qubits can be switched by swapping the encoding of $\ket{b}$ and $\ket{c}$. Therefore, an arbitrary single-qubit gate and an arbitrary conditional gate can be performed on the first qubit too \cite{mousolou2014universal}.
The four-level holonomic scheme can be viewed as a non-Abelian counterpart of the geometric phase-shift gate proposed in Ref.  \cite{zhu02,zhu03universal,wang01multibit}. However, the non-Abelian scheme possesses the advantage that there is no dynamical contribution to the evolution, and thus there is no need to remove it.

A key difference between the three- and four-level systems is in which Grassmannian paths reside. While the paths in the $\Lambda$ system are traced out in $\mathcal{G} (3;2)$, the relevant Grassmannian in the four-level case is, as pointed out in Ref. \cite{karle2003}, $\mathcal{G} (4;2)$. In other words, $C_0$ and $C_1$ combined is a path that resides in the space of two-dimensional subspaces of a four-dimensional Hilbert space\footnote{In some cases, the full $\mathcal{G} (4;2)$ can be reduced to $\mathcal{G} (3;2)$ when interpreting the paths in the four-level system. This happens precisely when $\det S = 0$. An example of this is the NMR experiment in Ref.  \cite{feng2013experimental}, which utilizes a pair of spin$-\frac{1}{2}$ nuclei designed to have vanishing $\det S$ in order to realize the original $\Lambda$ scheme. Conversely, when $\det S \neq 0$ the system is irreducible and will genuinely trace out a path in $\mathcal{G} (4;2)$.}. 

Next, let us reconsider the evolution operator $U(t)$ in \eqref{eq:ufour}. When $\cos(a_TD)=0$ and $\sin(a_TD)=\mathrm{diag}\{1,-1\}=Z$ or $I$ for a certain time $T$, $U(T)$ takes the form 
\begin{equation}
    U(T)=\left(
    \begin{array}{cc}
        0 & -iU_lZ^{P_1}U_r^\dag \\
        -iU_rZ^{P_1}U_l^\dag & 0
    \end{array}
    \right),
\end{equation}
where $P_1=0,1$. One can see that the subspaces $\mathcal{V}_0$ and $\mathcal{V}_1$ exchange their positions at time $T$. 
In the same way,  we can have an evolution operator with the same form for another time duration $T'$, i.e.
\begin{equation}
    U(T')=\left(
    \begin{array}{cc}
        0 & iU_l^\prime Z^{P_2}U_r^{\prime\dag} \\
        iU_r^\prime Z^{P_2}U_l^{\prime\dag} & 0
    \end{array}
    \right),
\end{equation}
where $P_2=0,1$. By sequentially combining $U(T)$ and $U(T')$, we obtain
\begin{equation}
    U(T+T')=\left(
    \begin{array}{cc}
        U_l^\prime Z^{P_2}U_r^{\prime\dag}U_rZ^{P_1}U_l^\dag & 0 \\
        0 & U_r^\prime Z^{P_2}U_l^{\prime\dag}U_lZ^{P_1}U_r^\dag
    \end{array}
    \right),
\end{equation}
which shows that each subspace has been transferred back to its original position, and attains an arbitrary single-qubit holonomic gate acting on each subspace \cite{mousolou2014non}. This process can be treated as a non-Abelian  generalization of the orange-slice-shaped path used in measurements of the Abelian geometric phase \cite{kwiat1991,allman1997} as well as in schemes for GQC \cite{zhao17rydberg,thomas2011}.

As shown in Ref.  \cite{zhang2018holonomic} (see also Ref. \cite{wang2020dephasing}), the four-level structure can be provided by an auxiliary qubit interacting with the target qubit via the $XY$-type interaction  $H_{XY}^{12}=\sigma_x^1\sigma_x^2+\sigma_y^1\sigma_y^2$, which is available in a variety of quantum system such as superconducting circuits \cite{majer2007coupling}, quantum dots \cite{imamog1999quantum}, and solid-state spins \cite{mozyrsky2001indirect}. Thus, the scheme using auxiliary qubit is scalable in state-of-the-art two-dimensional quantum chips \cite{arute2019quantum,gong2021quantum,yan2019strongly} and therefore constitutes an ideal platform for the realization of the four-level holonomies and thereby large-scale holonomic quantum computers. To see how this scheme works, we denote the auxiliary (target) qubit as 1 (2) [see Fig.~\ref{fig:fourlevel}(b)] and consider the Hamiltonian that generates holonomic transformations on the target qubit:
\begin{equation}\label{eq:h1xy}
    H_1(t)=\frac{J(t)}{2}\sin\theta(\cos\beta\sigma_x^1+\sin\beta\sigma_y^1)+\frac{J(t)}{2}\cos\theta(\sigma_x^1\sigma_x^2+\sigma_y^1\sigma_y^2),
\end{equation}
where $J(t)$ is the control envelope, $\sigma_{x(y)}^k$ is the Pauli-$x(y)$ matrix acting on qubit $k=1,2$, and $\theta$ ($\beta$) is a control parameter that is constant during each gate operation.
Naturally, the 4-dimensional Hilbert space spanned by $H_1(t)$ in \eqref{eq:h1xy} can be separated into two 2-dimensional subspaces: $\mathcal{V}_0$ spanned by $\{\ket{00}, \ket{01}\}$ and   $\mathcal{V}_1$ spanned by $\{\ket{10}, \ket{11}\}$. By defining $a_t=\int_0^tJ(s)\mathrm{d}s$ and $D={\rm diag} \{ \cos^2(\theta/2),\sin^2(\theta/2) \}$, and by assuming a time $\tau$, such that $\sin(a_\tau D)=\mathrm{diag}\{0,0\}$ and  $\cos(a_\tau D)=\mathrm{diag}\{-1,1\}$, one finds
\begin{equation}
    U_1(\tau)=\ket{0}\bra{0}\otimes U_1^h(C_0)+\ket{1}\bra{1}\otimes U_1^h(C_1).
\end{equation}
Such a $\tau$ exists provided $\tan^2 (\theta/2)$ is a rational number.
If the auxiliary qubit (qubit 1) is initialized in $\ket{0}$, then the  gate on the target qubit (qubit 2) is the single-qubit gate $U_1^h(C_0)=\cos\theta\sigma_z-\sin\theta(\cos\beta\sigma_x+\sin\beta\sigma_y)$. In fact, an arbitrary single-qubit gate can be obtained with $U_1^h(C_0)$ \cite{zhang2018holonomic}, {\it e.g.}, a Hadamard gate is available when $\theta=\pi/4$ and $\beta=\pi$. Also, it is easy to show that $U_1^h(C_0)$ is holonomic \cite{zhang2018holonomic}.   

Following the same logic, two-qubit holonomic gates on a pair of target qubits ($\mathrm{T}_1$ and $\mathrm{T}_2$ shown in Fig.~\ref{fig:fourlevel}(c), denoted as qubit 1 and 2, respectively) can be implemented with the help of an auxiliary qubit (denoted as qubit 3). To this end, the control Hamiltonian for the three qubits comprises only  $XY$-type interaction and reads 
\begin{equation}
    H_2=J_{13}(\sigma_x^1\sigma_x^3+\sigma_y^1\sigma_y^3)+J_{23}(\sigma_x^2\sigma_x^3+\sigma_y^2\sigma_y^3),
\end{equation}
where $J_{13}$ ($J_{23}$) is the coupling strength between the qubits 1 and 3 (2 and 3). Under the action of $H_2$, the Hilbert space for the three qubits has four invariant subspaces: $\mathcal{V}_1$ spanned by $\{\ket{000}\}$, $\mathcal{V}_2$ spanned by $\{\ket{001},\ket{010},\ket{100}\}$, $\mathcal{V}_3$ spanned by $\{\ket{110},\ket{101},\ket{011}\}$, and $\mathcal{V}_4$ spanned by $\{\ket{111}\}$. 
While the the action of $H_2$ on $\mathcal{V}_1$ and $\mathcal{V}_4$ is trivial ($H_{\mathcal{V}_1}=H_{\mathcal{V}_4}=0$), the reduced Hamiltonians for $\mathcal{V}_2$ and $\mathcal{V}_3$ take the same form,
\begin{equation}
    H_{\mathcal{V}_2}=H_{\mathcal{V}_3}=\left(
                                               \begin{array}{ccc}
                                                  0 & J_{23} & J_{13} \\
                                                  J_{23} & 0 & 0 \\
                                                  J_{13} & 0 & 0
                                               \end{array}\right).
\end{equation}
It is evident that $H_{\mathcal{V}_2}$ and $H_{\mathcal{V}_3}$ both represent a $\Lambda$-type Hamiltonian, implying that proper holonomic gates can be performed on a logical qubit encoded within $\mathcal{V}_2$ and $\mathcal{V}_3$. For example, the states $\ket{001}$, $\ket{010}$, and $\ket{100}$ can be used as the levels $\ket{a}$, $\ket{0}$, and $\ket{1}$, respectively, and by setting $J_{13}=\Omega\sin(\theta/2)$ and $J_{23}=\Omega\cos(\theta/2)$, the evolution operator $U_{\mathcal{V}_2}$ generated with $H_{\mathcal{V}_2}$ at time $\tau=\pi/\Omega$ becomes
\begin{equation}
    U_{\mathcal{V}_2}(\tau)=\left(
                            \begin{array}{ccc}
                                -1 & 0 & 0 \\
                                0 & \cos\theta & -\sin\theta \\
                                0 & -\sin\theta & -\cos\theta
                            \end{array}\right)=-\ket{001} \bra{001} \oplus U_{\mathcal{V}_2}^h.
\end{equation}
Here, $U_{\mathcal{V}_2^h}$ is acting on the logical qubit ${\rm Span} \{ \ket{010},\ket{100} \}$.
In this case, if the auxiliary qubit is initially in $\ket{0}$ and the two target qubits are in an arbitrary two-qubit state $\ket{\psi}_a=a_{00}\ket{00}+a_{01}\ket{01}+a_{10}\ket{10}+a_{11}\ket{11}$, the total initial state $\ket{\psi}_i$ will be transferred to the final state $\ket{\psi}_f=\left[ a_{00}\ket{00}+U_{\mathcal{V}_2}^h\left(a_{01}\ket{01}+a_{10}\ket{10}\right)-a_{11}\ket{11}\right]\otimes \ket{0}$,
demonstrating that the corresponding holonomic two-qubit gate on the two target qubits is
\begin{equation}
    U_2^h=\ket{00} \bra{00}\oplus U_{\mathcal{V}_2}^h\oplus(-\ket{11} \bra{11}) . 
\end{equation}
It is clear that $U_2^h$ is an entangling gate and thus a universal set of holonomic gates can be achieved with this method.

A similar structure is proposed in  Ref.~\cite{gurkan2015realization} for a chain of three-level systems. Essentially, this idea is shared by Refs. \cite{gurkan2015realization,zhang2018holonomic,mousolou2018scalable}. Building two-qubit holonomic gates with the help of an auxiliary qubit is demonstrated in Ref.  \cite{mousolou2017electric}, where the holonomic gates are shown to possess arbitrary entangling power. The difference between \cite{mousolou2017electric} and \cite{zhang2018holonomic,mousolou2018scalable} is that in Ref.~\cite{mousolou2017electric} both XY-type interaction and Dzyalozhinsky-Moriya (DM) interaction are required, while in Refs.~\cite{zhang2018holonomic,mousolou2018scalable} only single-qubit Hamiltonian and XY-type interaction are utilized. The latter proposal has been demonstrated explicitly with transmons \cite{chen2018nonadiabatic,xue2017nonadiabatic} and Rabi Lattice \cite{wang2020dephasing}, where the XY-type interaction is effectively generated by coupling two transmons with a capacitor and by ultrastrong couplings between quantum Rabi models, respectively. Interestingly, due to the intrinsic characteristics of the ultrastrong coupling and the quantum Rabi model, the gates in Ref.  \cite{wang2020dephasing} are immune to dephasing noise.

A key advantage of using an auxiliary qubit is that the auxiliary qubit is in its ground state ($\ket{0}$) before and after each gate operation, {\it i.e.}, an ideal final state can always be written as $\ket{\psi}_f=\ket{\psi}_T\otimes\ket{0}$, where $\ket{\psi}_T$ is the final state for the target system and $\ket{0}$ is the state of the auxiliary qubit. Therefore, the auxiliary qubit can be used repeatedly without any further operation. In addition, in practice, when the system is perturbed by noise, the final state may be $\ket{\psi}'_f=a_T\ket{\psi}_T \otimes \ket{0}+a_\perp\ket{\psi}_\perp\otimes\ket{1}$, where $a_T$ and  $a_\perp$ are complex-valued normalization factors and $\ket{\psi}_\perp$ is another state of the target system. In this case, by measuring $\sigma_z$ of the target qubit and the result $\ket{0}$ is obtained, the final state is again reset to $\ket{\psi}_f$ and the fidelity of the operation is increased; on the other hand, if the result $\ket{1}$ appears, the final state is changed to $\ket{\psi}_\perp$ which heralds that an error occurs \cite{zhang2018holonomic,kang2020heralded}.  
Another advantage is that this method is scalable on a square lattice, where qubits are arranged on the corners \cite{zhang2018holonomic}. This property allows the application of the method to the implementation of surface codes, which will be discussed in detail in the following section.

\subsubsection{Reverse-engineering schemes}\label{sec:reverseengin}

In all the  previous geometric and holonomic schemes, the underlying strategies to build a quantum gate are the same: start with a Hamiltonian $H(\Vec{\lambda}(t);t)$ ($\Vec{\lambda}$ is a set of possibly time dependent control parameters) of a certain system; find its evolution operator $U(\Vec{\lambda}(t);t)=\sum_k\ket{\psi_k(t)}\bra{\psi_k(0)}$, where $\ket{\psi_k(0)}$ is the initial state and $\ket{\psi_k(t)}$ is the corresponding instantaneous state satisfying the time dependent Schr\"{o}dinger equation; choose a particular set of $\Vec{\lambda}_1 (t)$ and run time $\tau$ such that $U(\Vec{\lambda}_1(t);t)$ satisfies the geometric or holonomic conditions, thereby ensuring that $U(\Vec{\lambda}_1(\tau);\tau)$ arises from purely geometric evolution. For quantum computation purposes, one only needs to pick a target $U(\Vec{\lambda}_1(\tau);\tau)$ and then immediately obtains the corresponding $H(\Vec{\lambda};t)$ based on the chosen parameters. The advantage of this strategy is that the geometric phases are calculated by the difference between the total phase and the dynamical phase, where the latter two phases can be obtained with $\ket{\psi_k(t)}$, thereby eliminating the need to find a set of auxiliary instantaneous basis states $\ket{\varphi_k(t)}$ [as shown in \eqref{gamma} for the Abelian case and in \eqref{ut} for the non-Abelian case] to calculate the geometric phases and quantum holonomies.

In fact, one can do this in a reverse-engineering way: start from a target gate $U_{\mathrm{tar}}$; associate $U_{\mathrm{tar}}$ with a loop in state space (projective space)
generated by a unitary operator $U(\Vec{\lambda};t)$ ($t\in[0,\tau]$); find a Hamiltonian that can realize the cyclic evolution and make sure that there is no dynamical contribution picked up during the evolution. The advantage of this approach is that there are normally infinitely many Hamiltonians satisfying the geometric or holonomic conditions so that one can choose a robust one against certain types of errors in practice. However, since the Hamiltonian is unknown, one has to choose a set of auxiliary bases to describe the geometric phases or quantum holonomies based on which the Hamiltonian can be found. 

The reverse-engineering idea has been adopted in many other quantum control theories to design fast and robust procedures. Shortcut to adiabaticity (STA) (for a review, see Ref. \cite{guery2019shortcuts}), which has received increasing attention recently, is one of them. 
STA aims to speed up adiabatic evolution so that quantum control tasks can be completed in a shorter time. This theory is useful in GQC and HQC since the adiabatic schemes \cite{jones2000geometric,duan2001geometric} in both can be accelerated. A thorough review of applying STA in GQC and HQC has been provided in Ref. \cite{du2019geometric}. Below, we shall give a brief summary of this topic.

Although STA is called by many other names, such as counterdiabatic field algorithm \cite{demirplak2003adiabatic,demirplak2005assisted}, transitionless quantum driving \cite{berry2009transitionless}, and invariant-based inverse engineering \cite{muga2009frictionless}, their underlying ideas are the same \cite{guery2019shortcuts}. According to the spectrum of the adiabatic Hamiltonian $H_0(t)$, STA theory has two categories, related to whether the energy spectrum is nondegenerate or degenerate.

In the nondegenerate case, a  Hamiltonian $H_0(t)$ changes adiabatically from $t=0$ to $\tau$ along a path in  parameter space, so that the corresponding evolution operator takes the approximate form $U(t)=\sum_ke^{i\gamma_k(t)}\ket{\psi_k(t)}\bra{\psi_k(0)}$, where $e^{i\gamma_k(t)}$ are phase factors and $\ket{\psi_k(t)}$ are nondegenerate eigenstates of $H_0(t)$.  By using the reverse-engineering method, a Hamiltonian $H(t) = i\dot{U}(t)U^{\dagger}(t)$ that accomplishes the same evolution without approximation and at arbitrary speed can be obtained as 
\begin{equation}
    H(t)=-\sum_k\dot{\gamma_k}(t)\ket{\psi_k(t)}\bra{\psi_k(t)}+i\sum_k\ket{\dot{\psi}_k(t)}\bra{\psi_k(t)}.
\end{equation}
The above equation shows that $H(t)$ can be written as $H(t)=H_0(t)+H_a(t)$ with $H_a(t)$ an additional Hamiltonian that suppresses the transitions between different eigenstates of $H_0(t)$ caused by the fast change of them.  By choosing $\gamma_k(t)=-\int_0^tE_k(t')\mathrm{d}t'$, $E_k(t)$ the $k$th eigenvalue of $H_0(t)$, {\it i.e.}, $\gamma_k(t)$ taken to be the dynamical phase of $\ket{\psi_k(t)}$), one arrives at $H_a(t)=i\sum_k\ket{\dot{\psi}_k(t)}\bra{\psi_k(t)}$, which is the counterdiabatic Hamiltonian. It is clear that this protocol is only suitable for noncyclic evolution, where geometric phases are irrelevant. For a more general case, one should set $\gamma_k(t)=-\int_0^tE_k(t')\mathrm{d}t'+i\int_0^t\braket{\psi_k(t)}{\dot{\psi}_k(t)}\mathrm{d}t'$, where both the dynamical and geometric contributions are taken into account \cite{berry2009transitionless}. Under this choice, one retains the transitionless Hamiltonian
\begin{equation}
    H_a(t)=i\sum_k\ket{\dot{\psi}_k(t)}\bra{\psi_k(t)}-i\sum_k\braket{\psi_k(t)}{\dot{\psi}_k(t)}\ket{\psi_k(t)}\bra{\psi_k(t)}.
\end{equation}
In the following, we focus on cyclic evolution. With the help of $H_a(t)$, an adiabatic geometric phase-shift gate can be realized at arbitrary speed. Now it is clear that the price that STA pays for a faster speed is the higher energy involved with a more complicated control Hamiltonian than $H_0(t)$. As shown in $H_a(t)$, the implementation of STA demands extra couplings which may be infeasible for practical quantum systems. This problem becomes even more serious when multi-level and many-body systems are considered.

In Ref.~\cite{liang2016proposal}, a GQC scheme realized with STA in the NV-center system was proposed. The qubit states were chosen to be the Zeeman levels $\ket{m_s=0}$ and $\ket{m_s=-1}$ from the NV spin-triplet ground states. The $\sigma_x$ and $\sigma_z$ rotations were achieved by driving the initial states along orange-slice paths in state space, just as in the adiabatic evolution scheme. However, the path of the control Hamiltonian was distorted owing to the presence of the time-dependent $H_a(t)$. The dynamical phases were canceled out by flipping $H(t)$ during the evolution, with no contribution to the dynamical phases from $H_a(t)$.   

In the degenerate case, the eigensubspace $\mathcal{V}_k$ of an $n_k$-fold degenerate eigenenergy $E_k(t)$ of $H_0(t)$, acquires a quantum holonomy after completing an adiabatic cyclic change  \cite{wilczek1984appearance}. 
The Hamiltonian that can accelerate this process was given by Zhang {\it et al.} \cite{zhang2015fast}, and takes the general form
\begin{equation}
    H(t)=H_0(t)+H_a(t)=\sum_{k,n}E_k(t)\ket{\psi^k_{n}(t)}\bra{\psi^k_{n}(t)}+i\sum_{k,n,m}\left( \ket{\dot{\psi}_n^k(t)}\bra{\psi^k_n(t)}-A^k_{nm}(t)\ket{\psi^k_{n}(t)}\bra{\psi^k_{m}(t)}\right),
\end{equation}
where $A^k_{nm}(t)=\braket{\psi^k_{n}(t)}{\dot{\psi}^k_m(t)}$ are the elements of the connection matrix $A(t)$. Compared to the STA Hamiltonian in the nondegenerate case, the above $H(t)$ includes new couplings between eigenstates within each eigensubspace $\mathcal{V}_k$, implying a fast but potentially more complicated control configuration. As an application, a fast version of the tripod scheme discussed in Sec.~\ref{sec:tripod} was developed in Ref.~\cite{zhang2015fast}. It was also explicitly shown that the coupling structure induced by $H_a(t)$ can be substantially simplified by evolving the dark subspace along an orange-slice-shaped path in parameter space. Moreover, the authors demonstrated that the scheme is feasible in a tunable transmon system. In addition to picking a specific loop, one can modify the amplitudes of the control pulses to cancel  unwanted couplings \cite{liu2017superadiabatic}.  

Besides speeding up adiabatic schemes with STA, one can also design geometric and holonomic gates via reverse engineering directly. For the Abelian case, the reverse-engineering scheme can be explicitly explained as follows. Any target quantum gate has a decomposition as $U_{\mathrm{tar}}=\sum_ke^{i\gamma_k^g}\ket{\psi_k(0)}\bra{\psi_k(0)}$. To realize $U_{\mathrm{tar}}$ with geometric phases, $\ket{\psi_k(0)}$ should be taken as initial states of a cyclic evolution with corresponding  geometric phase factors $e^{i\gamma_k^g}$. Thus, $U_{\mathrm{tar}}$ relates to  a unitary operator $U(t)$ reading $U(t) = \sum_k \ket{\psi_k(t)} \bra{\psi_k(0)}$, where $\ket{\psi_k(t)}$ are the instantaneous states and indicate the evolution path. At $t=\tau$, $\ket{\psi_k(t)}$ completes a cyclic evolution and returns to $\ket{\psi_k(0)}$ up to a total phase $e^{i\gamma_k(\tau)}$, {\it i.e.}, $U(\tau) = \sum_{k} e^{i\gamma_k(\tau)} \ket{\psi_k(0)}\bra{\psi_k(0)}$. As shown in \eqref{gamma}, the total phase is $\gamma_k(\tau) = -\int_0^\tau\bra{\varphi(t)}H(t)\ket{\varphi(t)}\mathrm{d}t+i\int_0^\tau \bra{\varphi(t)}\frac{\mathrm{d}}{\mathrm{d} t}\ket{\varphi(t)}\mathrm{d}t$, where $\ket{\varphi_k(t)}=e^{-i\gamma_k(t)}\ket{\psi_k(t)}$. Therefore, the required Hamiltonian that can accomplish the cyclic evolution takes the general form
\begin{equation}\label{}
    H(t)=i\sum_k\ket{\dot{\psi}_k(t)}\bra{\psi_k(t)}=\sum_k-\dot{\gamma}_k(t)\ket{\varphi_k(t)}\bra{\varphi_k(t)}+i\ket{\dot{\varphi}_k(t)}\bra{\varphi_k(t)}.
\end{equation}
The above equation shows that once one can find such a set of $\ket{\varphi_k(t)}$, $H(t)$ is easily obtained. However, to ensure that $U(\tau)$ is a purely geometric gate, the dynamical phases accumulating along the loop should be made to vanish. 

To satisfy the geometric condition, one basically has two choices. The first one is to require the total dynamical phases to be zero, {\it i.e.}, $\int_0^\tau\bra{\varphi_k(t)}H(t)\ket{\varphi_k(t)}\mathrm{d}t=0$. In Ref. \cite{liu2019plug}, Liu {\it et al.} used this choice in order to realize robust geometric gates with extensible flexibility gained by the many candidates of the system Hamiltonian (Theorem 1 stated in this work is equivalent to the relation $\ket{\varphi_k(t)}=e^{-i\gamma_k(t)}\ket{\psi_k(t)}$). It is explicitly shown that an arbitrary single-qubit gate based on an NV center can be built up by separating a loop into two equal parts, the dynamical phases of which are of opposite values and thus compensating each other, while the geometric phases are added up. Besides the robustness against slow Rabi errors and random quasistatic noise, this method is compatible with various error suppressing techniques, such as single-shot shaped pulse and dynamical decoupling, so that the gate fidelity can be further improved \cite{liu2019plug}.

The other choice is to let the dynamical phases vanish locally along the path by taking $\bra{\varphi_k(t)}H(t)\ket{\varphi_k(t)}=0$, $k=1,\cdots,N$. This option was used in Ref. \cite{li2020approach}, where the authors gave a general form of Hamiltonians that can drive a quantum system from an initial state $\ket{\psi_k(0)}$ to a final state  $\ket{\psi_k(\tau)}$ along a desired path prescribed by $\ket{\psi_k(t)}$, such that $\ket{\psi_k(\tau)}=e^{i\gamma_k^g(\tau)}\ket{\psi_k(0)}$, $\gamma_k^g(\tau)$ being the geometric phase. By constraining the Hamiltonian so as to give rise to dynamical phases that vanish along the paths, one obtains
\begin{equation}
    H(t)=i\sum_{k\neq l}\braket{\varphi_k(t)}{\dot{\varphi}_l(t)}\ket{\varphi_k(t)}\bra{\varphi_l(t)}.
\end{equation}
Consequently, the geometric gate can be written as $U(\tau)=\sum_ke^{i\gamma_k^g(\tau)}\ket{\psi_k(0)}\bra{\psi_k(0)}$. In the two-level case, $U(\tau)$ corresponds to a single-qubit gate. As shown before, any single-qubit gate can be achieved by properly choosing $\ket{\psi_k(0)}$ and $\gamma_k^g(\tau)$. A merit provided by this scheme is that one can choose a much shorter path for a target geometric gate, getting rid of the need to evolve along the orange-slice path (the path is long) or rotation around a fixed axis (the two-loop scheme is needed to eliminate the dynamical phases). In this way, the evolution time can be minimized, decreasing the exposure of the system to decoherence.

In the non-Abelian case, we do not require each $\ket{\psi_k(0)}$ to return to itself, but demand each $\ket{\varphi_k(0)}$ returns. A target gate is associated to the matrix-value difference $C(\tau)$ between $\ket{\psi_k(\tau)}$ and $\ket{\varphi_l(\tau)}$, {\it i.e.}, $\ket{\psi_k(t)}=\sum_lC_{lk}(t)\ket{\varphi_l(t)}$, $k,l=1,\cdots,L$, where  $C(\tau)=\mathcal{T}e^{i\int_0^\tau A(t)-K(t)\mathrm{d}t}$ with $A_{kl}(t)=i\braket{\varphi_k(t)}{\dot{\varphi}_l(t)}$ and $K_{kl}(t)=\bra{\varphi_k(t)}H(t)\ket{\varphi_l(t)}$. To achieve a nontrivial $C(\tau)$, at least one extra auxiliary state should be added. For simplicity, one can choose  $\ket{\psi_{L+1}(t)} = e^{i\gamma(t)} \ket{\varphi_{L+1}(t)}$, where $\gamma(t)$ is a real function of $t$ with initial condition $\gamma(0)=0$. To let the evolution be purely geometric, the condition $K_{kl}=0$, $\forall k,l$, must be satisfied. Under this condition, the general form of the required Hamiltonian $H(t)=i\sum_{k=1}^{L+1}\ket{\dot{\psi}_k(t)}\bra{\psi_k(t)}$ can be written in the representation of $\ket{\varphi_k(t)}$ as \cite{zhao2020general}
\begin{equation}
    H(t)=\left[ i\sum_{k=1}^{L}\braket{\varphi_k(t)}{\dot{\varphi}_{L+1}}\ket{\varphi_k(t)}\bra{\varphi_{L+1}(t)}+\mathrm{H.c.}\right]+\left[ i\braket{\varphi_{L+1}(t)}{\dot{\varphi}_{L+1}(t)}-\dot{\gamma}(t) \right]\ket{\varphi_{L+1}(t)}\bra{\varphi_{L+1}(t)}.
\end{equation}
The Hamiltonian can be directly generalized to the case where more auxiliary states are used. As shown in Ref. \cite{zhao2020general}, the above Hamiltonian can be used to shorten the evolution path for a given holonomic gate. With this method, the minimum time needed to realize a general single-qubit holonomic gate can be estimated. 

In the above, the Hamiltonian required for a target gate is obtained by putting the associated evolution operator describing the evolution path into the Schr\"{o}dinger equation. Alternatively, the evolution path can be characterized by a dynamical invariant operator $I(t)$ \cite{lewis1969an} whose dynamics is given by solving the von Neumann-like equation $\dot{I}(t)=-i[H(t),I(t)]$. The idea of using dynamical invariant theory to develop geometric gates is first introduced in Ref. \cite{teo2005geometric}, and it has been applied to both geometric \cite{shao2007implementation,wang2009geometric,chen2020high} and holonomic \cite{kang2020flexible} schemes. When a dynamical invariant operator is associated with a given evolution loop, the corresponding Hamiltonian can be obtained by means of the above von Neumann equation. As in the Schr\"{o}dinger equation framework, the reverse-engineering method based on dynamical invariant operators provides a flexible choice of Hamiltonian. This method is shown to be useful in building heralded geometric phase \cite{kang2020heralded} and suppressing leakage errors in multi-level systems \cite{liu2020leakage}.

Another example to achieve a degenerate eigensubspace is the $\Delta$-model Hamiltonian $H_0(t)$ \cite{du2017degenerate}. When the control parameters in $H_0(t)$ satisfy some conditions, two degenerate ground states and one excited state exist. Since all the couplings between the three levels are already employed, the additional Hamiltonian only reshapes the control pulse. As a result, the evolution can be made 20 times faster with perfect fidelity with only a modest increase of the Rabi frequency \cite{du2017degenerate}. These examples clearly show that the STA+GQC/HQC scheme opens a new horizon for implementing geometric quantum gates. 

\subsection{Experimental realizations and gate robustness}

\begin{table}[tbh]   
\begin{center}   
\caption{Summary of reported holonomic gate fidelities in various quantum systems. For the fidelity obtained for a particular gate, the gate is indicated in the following parentheses; otherwise, it is an average fidelity. The gate $A$ measured in Ref.~\cite{zu2014experimental} is a specially defined rotation gate. It is worth noting that holonomic gates are also built in cold atoms \cite{leroux2018non} and in NMR \cite{zhu2019single}, but the gate performance is characterized by distance. Thus they are not included here.}
\label{tb:Hfidelity} 
\begin{tabular}{c|cc|cc}
\hline\hline
 & \multicolumn{2}{c|}{Adiabatic} & \multicolumn{2}{c}{Nonadiabatic} \\
   & Single-qubit & Two-qubit & Single-qubit & Two-qubit \\ \hline
NMR &  &  &  \makecell{$95.9\%$($Z_{\pi/2}$,\cite{feng2013experimental}), $95.9\%$($Z_{\pi}$,\cite{feng2013experimental})\\$98.1\%$($X_{\pi/2}$,\cite{feng2013experimental}), $96.3\%$($X_{\pi}$,\cite{feng2013experimental})\\$98.29\%$($Z_{\pi/2}$,\cite{li2017experimental}), $99.75\%$($Z_{\pi}$,\cite{li2017experimental})\\$98.07\%$($X_{\pi/2}$,\cite{li2017experimental}), $99.68\%$($X_{\pi}$,\cite{li2017experimental})}& $91.43\%$(CNOT,\cite{feng2013experimental}) \\ \hline
Trapped ions & \makecell{$96.5\%$($X$,\cite{toyoda2013realization})\\ $93.1\%$($Z$,\cite{toyoda2013realization})\\$96.5\%$($H$,\cite{toyoda2013realization})}&  & $99\%$ \cite{ai2020experimental} &   \\ \hline
Superconducting &  &  & \makecell{$95.4\%$($H$,\cite{abdumalikov2013experimental}), $97.5\%$($X$,\cite{abdumalikov2013experimental})\\$99.4\%$\cite{zhao2021experimental}, $99.6\%$\cite{xu2018single}\\$96.6\%$($X$,\cite{yan2019exp}), $97.6\%$($H$,\cite{yan2019exp})\\$96.4\%$($X_{\pi/2}$,\cite{yan2019exp}), $99.3\%$\cite{zhang2019single}} &   \\ \hline
NV centers &  &  & \makecell{$98.0\%$($X$,\cite{arroyo2014room}), $98.0\%$($Y$,\cite{arroyo2014room})\\$97.5\%$($Z$,\cite{arroyo2014room}), $97.8\%$($H$,\cite{arroyo2014room})\\$96.5\%$($X$,\cite{zu2014experimental}), $96.9\%$($A$,\cite{zu2014experimental})\\$92.1\%$($H$,\cite{zu2014experimental}), $92\%$($X$,\cite{sekiguchi2017optical})\\$89\%$($Y$,\cite{sekiguchi2017optical}), $90\%$($Z$,\cite{sekiguchi2017optical}) \\$75\%$($X$,\cite{zhou2017holonomic}), $78\%$($Y$,\cite{zhou2017holonomic})\\$74\%$($H$,\cite{zhou2017holonomic}), $93\%$,\cite{ishida2018universal}\\$99\%$($X$,\cite{nagata2018universal}), $93\%$($Y$,\cite{nagata2018universal})\\$93\%$($Z$,\cite{nagata2018universal}), $93\%$($H$,\cite{nagata2018universal})}  &  \makecell{$90.2\%$(CNOT,\cite{zu2014experimental})\\$90\%$(CZ,\cite{nagata2018universal})}\\
\hline\hline
\end{tabular}   
\end{center}   
\end{table}

\textit{Tripod scheme.}--The tripod scheme described in Section \ref{sec:tripod} has been experimentally realized for a single $^{40}\mathrm{Ca}^+$ ion trapped in vacuum using a linear Paul trap with an operating frequency of 23 MHz and secular frequencies of $(\omega_x,\omega_y,\omega_z)/2\pi=(2.4,2.2,0.69)$ MHz \cite{toyoda2013realization}.  A bias magnetic field of $2.9\times 10^{-4}$ T was applied to cause a Zeeman splitting of $4.9$ MHz between $D_{5/2}$ sublevels. The $S_{1/2}-D_{5/2}$
transition was chosen for the holonomic gates. The state $S_{1/2} (m_s=-1/2)$ was encoding $\ket{e}$, and the three Zeeman sublevels in $D_{5/2}$ ($m_s=-3/2,1/2,-5/2$) were used as $\ket{0}, \ket{1}, \ket{a}$, respectively. By using the measured populations, the fidelities of $\sigma_x$, $\sigma_y$, and Hadamard (${\rm H}$) were estimated as $(96.5 \pm 3.8)\%$, $(93.1 \pm 3.8)\%$ and $(96.5\pm3.8)\%$, respectively. The robustness of holonomic gates against variations in Rabi frequencies was demonstrated in Ref. \cite{toyoda2013realization}. Furthermore, a laser-cooled gas of strontium-87 atoms was coupled to laser fields to form a tripod structure, in terms of which holonomic gates were realized \cite{leroux2018non}. In this setting, three sublevels ($m_g=5/2$, $7/2$, $9/2$) of $F_g=9/2$ were used as $\ket{0}$, $\ket{1}$, and $\ket{a}$; while one sublevel ($m_e=7/2$) of $F_e=9/2$ represented $\ket{e}$. 

\textit{Three-level scheme.}--The single-qubit nonadiabatic holonomic gates shown in Section \ref{sec:3level} were experimentally demonstrated by Abdumalikov {\it et al.} with a superconducting transmon embedded in a three-dimensional cavity \cite{abdumalikov2013experimental}. The lowest three energy levels of the transmon were taken to represent $\ket{0}$, $\ket{e}$, and $\ket{1}$ (from the bottom up), respectively. The reason for choosing the second level as $\ket{e}$ (rather than the third level as in the standard $\Lambda$ setting) was that the direct transition between the ground state and the third level is forbidden. Two driving microwave signals with the same truncated Gaussian envelop $\Omega(t)$ but different amplitudes $\omega_0$ and $\omega_1$ coupled the states $\ket{0}$ and $\ket{1}$ to the state $\ket{e}$, respectively. The transition frequencies measured by Ramsey spectroscopy were 8.086 GHz ($\ket{0}\!\leftrightarrow\!\ket{e}$) and 7.776 GHz ($\ket{e}\!\leftrightarrow\!\ket{1}$). After executing the holonomic gate, the state of the transmon was read out through the cavity. A holonomic Hadamard and NOT ($\sigma_x$) gates  were demonstrated. Using quantum process tomography (QPT), the gate fidelities for the two gates were found to be $F_H=(95.4\pm0.6)\%$ (Hadamard) and $F_{\mathrm{NOT}}=(97.5\pm0.9)\%$ (NOT), which were shown to be in good agreement with numerical simulations of a master equation including dissipation  \cite{abdumalikov2013experimental}. Another experiment with transmon can be found in Ref.~\cite{danilin2018experimental}.

The single-qubit holonomic gates were performed by Feng {\it et al.} in NMR system \cite{feng2013experimental}. A $^{13}C$ spin and a $^1H$ spin were used for the single-qubit gates. 
The spin states $\ket{10}$ and $\ket{11}$ were used to encode logical $\ket{0}$ and $\ket{1}$, respectively, while $\ket{01}$ was used as $\ket{e}$. 
By using the interaction between the two spins, four rotations [$R_z(\pi/2)$, $R_z(\pi)$, $R_x(\pi/2)$, $R_x(\pi)$] were demonstrated with average fidelity over $96\%$ estimated by QPT.

Indeed, the $\Lambda$ structure naturally exists and is well-studied in nitrogen-vacancy (NV) centers in diamond, where it emerges as a suitable platform for HQC \cite{arroyo2014room,zu2014experimental,sekiguchi2017optical,nagata2018universal}. The ground states of an NV center are the three degenerate electron spin states (corresponding to $S=1$). The $m_s=0$, $m_s=-1$, and $m_s=+1$ levels are denoted by $\ket{0_s}$, $\ket{-_s}$, and $\ket{+_s}$, respectively. As the $\ket{+_s} \! \leftrightarrow \! \ket{-_s}$ transition is forbidden, the two states are well suited to form the qubit subspace while $\ket{0_s}$ is used as $\ket{e}$. This structure was adopted in Refs.~\cite{arroyo2014room,zu2014experimental,nagata2018universal}. As shown in Ref.~\cite{arroyo2014room}, at an external magnetic field of about 404 G, the transitions $\ket{0_s}\!\leftrightarrow\!\ket{+_s}$ with $\omega_{0+}=2\pi\times4.003$ GHz and $\ket{0_s}\!\leftrightarrow\!\ket{-_s}$ with $\omega_{0-}=2\pi\times1.738$ GHz can be coherently driven by microwave radiation. Four holonomic single-qubit gates were tested, and their fidelities were determined by performing QPT. For the three Pauli gates and the Hadamard gate, each process fidelity was found to be around $98\%$. When the infidelity caused by errors in preparation and projective readout was taken into account, a fidelity close to unity could be achieved for each gate \cite{arroyo2014room}. The same set of gates were conducted in Ref.~\cite{nagata2018universal}, with fidelities $99\%$ ($\sigma_x$), $93\%$ ($\sigma_y$), $93\%$ ($\sigma_z$), and $93\%$ (${\rm H}$) for electron spin as well as $97\%$ ($\sigma_x$), $93\%$ ($\sigma_y$), $98\%$ ($\sigma_z$), and $95\%$ (${\rm H}$) for nuclear spin. A universal set of single-qubit gates was realized in Ref.~\cite{zu2014experimental}, where the NOT,  $\pi/8$, and ${\rm H}$ gates were reported with fidelities of $(96.5 \pm 1.9)\%$, $(96.9 \pm 1.5)\%$, and $(92.1 \pm 1.8)\%$, respectively. 
Another approach to construct the $\Lambda$ model is to use the $\ket{\pm_s}$ states for the qubit, but an orbital excited state (labelled as $\ket{A_2}$) to implement $\ket{e}$ \cite{ishida2018universal}. The holonomic gates $\sigma_x$, $\sigma_y$, and ${\rm H}$ were demonstrated in this setting with fidelities $95\%$, $94\%$, and $91\%$, respectively.

As an important component of a set of universal gates, the holonomic CNOT gate was experimentally demonstrated in NMR \cite{feng2013experimental} and NV centers
\cite{zu2014experimental}. 
The NMR experiment was performed by using three qubits with the encoding $\ket{100}\rightarrow\ket{00}$, $\ket{101}\rightarrow\ket{01}$,  $\ket{110}\rightarrow\ket{10}$, $\ket{111}\rightarrow\ket{11}$, where the first qubit works as an auxiliary qubit. The CNOT gate was obtained with a fidelity of $91.43\%$.
In the NV center experiment, the electron spin was used as the target qubit and a nearby $C^{13}$ nuclear spin as the control qubit (with the qubit states $\ket{\!\!\uparrow}$ and $\ket{\!\!\downarrow}$) \cite{zu2014experimental}. Both the electron and the nuclear spins were  polarized through optical pumping under a 451 G magnetic field, and made to interact with each other through hyperfine and dipole coupling. The effective Hamiltonian $H(t)=\Omega(t)\left[\big(\ket{0\!\!\uparrow}-\ket{1\!\!\uparrow}\big)\bra{e\uparrow\!\!}+\mathrm{H.c.}\right]/\sqrt{2}$ in the presence of state-selective microwave and rf pulses was implemented. Under the cyclic condition, a CNOT gate $\ket{\!\!\uparrow}\bra{\uparrow\!\!}\otimes\sigma_x+\ket{\!\!\downarrow}\bra{\downarrow\!\!}\otimes I$ with fidelity of $(90.2\pm2.5)\%$ was demonstrated.

\textit{Single-shot scheme.}--The single-shot scheme introduced in Section \ref{sec:1shot} has been realized experimentally with NMR \cite{li2017experimental}, NV center  \cite{zhou2017holonomic,sekiguchi2017optical}, and superconducting Xmon \cite{zhang2019single}. Ref.  \cite{li2017experimental} used the same setup as in Ref.  \cite{feng2013experimental} with two interacting nuclear spins residing at $^{13}\mathrm{C}$ and $^1\mathrm{H}$, respectively. Explicitly, the spin states $\ket{10}$ and $\ket{11}$ were encoding logical $\ket{0}$ and $\ket{1}$, while $\ket{01}$ was used as $\ket{e}$. The detuning and transitions were introduced in the single- and two-qubit Hamiltonian for the two qubits. Four single-qubit rotation gates $R_x(\pi/2)$, $R_x(\pi)$, $R_z(\pi/2)$, and $R_z(\pi)$ were realized with fidelities $99.82\%$, $99.86\%$, $99.73\%$, and $99.88\%$, respectively. 

In the NV center experiments \cite{sekiguchi2017optical,zhou2017holonomic}, the two ground states $\ket{\pm_s}$ (used as qubit states) were coupled to $\ket{A_2}$ (used as the excited $\ket{e}$) by a two-tone optical field with one-photon detuning $\Delta$. A set of single-qubit gates was demonstrated with the resultant effective Hamiltonian. For example,   $R_x(\pi/2)$, $R_x(-\pi/2)$, $R_y(\pi/2)$, and $R_y(-\pi/2)$ with fidelities of $83\%$, $80\%$, $82\%$, and $80\%$, respectively, were obtained in Ref. \cite{zhou2017holonomic}. The three Pauli gates $\sigma_x$, $\sigma_y$, and $\sigma_z$ were reported with fidelities $92\%$, $89\%$, and $90\%$, respectively, in Ref.  \cite{sekiguchi2017optical}. As a comparison, the gates $\sigma_x$, $\sigma_y$, and ${\rm H}$ were realized in Ref. \cite{zhou2017holonomic} with resonant control ($\Delta=0$). It turns out that the fidelity of these three resonant gates were a bit lower, showing evidence that the single-shot scheme performs better than the three-level scheme in this particular setup \cite{zhou2017holonomic}.

To demonstrate the capability of the single-shot scheme, single-qubit Clifford gates were implemented in a superconducting Xmon \cite{zhang2019single}. In this experiment, the lowest three energy levels of the Xmon were utilized to implement $\ket{0}$, $\ket{e}$, and $\ket{1}$ with relevant transition frequencies $\omega_{0e}=2\pi\times 4.849$ GHz and $\omega_{e1}=2\pi\times 4.597$ GHz. All the Clifford gates were performed with fidelities above $99\%$. 

\textit{Multi-pulse schemes.}--The first experimental demonstration of the multi-pulse scheme (see Section \ref{sec:multipulse}) was performed for both a superconducting transmon qubit and a microwave cavity in Ref.  \cite{xu2018single}. The encoding of the transmon was the same as in Ref.  \cite{abdumalikov2013experimental}, but the total evolution was divided into two equal halves. Both QPT and random benchmarking (RB) were used for four different single-qubit gates, both yielding gate fidelities over $99.6\%$ in the transmon case. The highest gate fidelity for the cavity was found to be over $99.0\%$.

Another experiment \cite{ai2020experimental} combined the single-loop scheme with optimal control. It was performed with a trapped $^{171}\mathrm{Yb}^+$ ion, where $m_F=0$ ($F=1$), $m_F=1$ ($F=1$), and $m_F=0$ ($F=0$) were used to implement $\ket{0}$, $\ket{1}$, and $\ket{e}$, respectively. The four gates $\sigma_x$, ${\rm H}$, $S$, and $T$ were found by means of QPT to have fidelities above $97\%$.
Moreover, an NMR experiment \cite{zhu2019single} demonstrated that the composite scheme is more robust against the pulse errors than the single-loop scheme when performed in a decoherence-free subspace.

\textit{Reverse-engineering schemes.}--The reverse-engineering technique proposed in Ref.~\cite{liang2016proposal} (see Section \ref{sec:reverseengin}) was experimentally demonstrated in Ref. \cite{kleissler2018universal}, where gate fidelities $F_x=99.4\%$ for $\sigma_x$ and $F_z=99.5\%$ for $\sigma_z$ were reported. Since the main resource of errors in this system is parameter noise, the authors compared the geometric gates based on STA with its dynamical counterpart and concluded that the fidelity for the geometric gate is much higher than that of the dynamical one \cite{kleissler2018universal}.
Another reverse-engineering scheme uses a $\Xi$-type level structure with a single-photon detuning \cite{yan2019exp}. The obtained effective Hamiltonian is similar to the one used in the single-shot scheme, but due to the adiabatic nature, control parameters can change gradually now. This scheme was realized for a superconducting Xmon qutrit \cite{yan2019exp}, and three holonomic gates ($X$, $H$, and $X_{\pi/2}$) were constructed with fidelities $96.6\%$, $97.6\%$, and $96.4\%$, respectively.

We have summarized the holonomic gate fidelities reported in the existing literature in Table \ref{tb:Hfidelity}. Compared with the adiabatic schemes, nonadiabatic ones certainly attract more attention due to their modest requirement for level structure. 
Among the systems, most experiments have been  performed in NV center settings, due to their built-in $\Lambda$-type level structure. By comparing Tables \ref{tb:Gfidelity} and \ref{tb:Hfidelity}, one can find that the highest fidelity for the holonomic gates is at the same level as the geometric ones.

\textit{Dynamical gate fidelities.}--One interesting question is to compare the performance of geometric and holonomic gates with that of the dynamical ones. 
To the best of our knowledge, the highest fidelity reported for single-qubit gates in NV centers is 99.9952\% \cite{rong2015experimental}, while that for two-qubit gates is  99.92\% \cite{xie202399}. For trapped ions, the highest single-qubit gate fidelity reported is 99.9999\% \cite{harty2014high}, and that for two-qubit gates is 99.92\% \cite{ballance2016high,gaebler2016high}. For superconducting qubits, the highest recorded single-qubit gate fidelity is above 99.99\% \cite{somoroff2021millisecond}, and that for two-qubit gates is 99.92\% \cite{acharya2022suppressing}. It is clear that the dynamical gates are superior to the geometric ones found so far. However, we remark that these outstanding dynamical results are obtained with error-suppressing techniques. Hence, we can expect that the fidelities for the geometric and holonomic gates can be improved further by combining them with such techniques. 

\textit{Robustness.}--At the end of this section, we discuss the robustness of holonomic gates. Just like geometric phases, quantum holonomies depend only on the geometric properties of parameters or state space. Therefore, holonomic gates should be robust to any kind of errors that keep these properties unchanged \cite{kowarsky2014non}. However, this assumption may not hold when environmental noise and control errors are present  \cite{ellinas2001universal,cen2003evaluation}. In this sense, instead of a theoretic proof, case-by-case studies on the error-resilience of holonomic gates should be performed. 

In the presence of control errors, Kuvshinov and Kuzmin developed a general expression for the fidelity of adiabatic holonomic gates \cite{kuvshinov2003stability}. A numerical study of adiabatic holonomic gates with stochastic errors in control parameters was carried out in Ref. \cite{solinas2004robustness}. In this latter work, it was shown that for fast and slowly random varying fluctuations, the holonomic gates are robust because the loops in parameter space turn out to be simply shifted rather than deformed, thus essentially preserving the swept solid angles. A similar result was obtained in finite-time gates with parametric errors, in which the holonomic gates present some ability to cancel the errors \cite{lupo2007robustness}.

For the effect of environmental noise, the gate performance has been studied via the master equation \cite{sarandy2006abelian}. Employing the quantum jump approach, holonomic gates exhibit robustness in the no-jump trajectory. However, a single jump can ruin the gate completely \cite{fuentes2005holo}. Nevertheless, in the environment modeled as an ensemble of harmonic oscillators, a fidelity revival behavior is observed, and an optimal finite operation time can be determined accordingly \cite{florio2006robust}. This revival behavior was also found in Ref.~\cite{parodi2007environmental} in which it is shown that the performance of the gate increases over a long time. Remarkably, the freedom in the choice of loop enables one to realize holonomic gates that perform better than the STIRAP gates \cite{parodi2007environmental}.

A comparison of the robustness of adiabatic and nonadiabatic holonomic gates was provided in Ref. \cite{johansson2012robustness}. It turned out that the adiabatic gates are robust to decay and mean detuning in the limit of large run time, while the nonadiabatic gate can be made resilient to decay of the excited state and to constant mean and relative detunings by using pulses that are sufficiently shorter than the time scales of the decay and detuning.
In addition to the built-in robustness, both the control errors \cite{kuvshinov2005robust,kuvshinov2006decoherence} and environmental noise \cite{cen2004refocusing,parodi2006fidelity} can be suppressed by combining holonomic gates with other quantum control technologies.

\section{Combining GQC and HQC with other quantum technologies} 
\label{sec:combineHQC} 
In this section, we explain how GQC and HQC can be combined with other techniques to eliminate environmental or control errors that cannot be handled solely via the geometric method. 
The interplay amongst different mechanisms provides robustness to errors. However, it is not straightforward to achieve the desired outcomes since each mechanism imposes its own constraints and conditions, just as in the application of GQC and HQC.
For example, the environment-error eliminating mechanisms, such as Decoherence-Free Subspace (DFS), Noiseless Subsystem (NS), Dynamical Decoupling (DD), and quantum error correction (QEC), require encoding the computational space in a larger Hilbert space spanned by a set of qubits.
To perform geometric or holonomic gates with these mechanisms, one needs to find a proper logical space permitted by the physical system, to embed the GQC or HQC level structure into the logical space, and to find the corresponding Hamiltonian that induces the geometric or holonomic gates acting on the logical state space.

A common feature of the logical spaces is that they are orthogonal to the rest of the corresponding Hilbert space. This means that logical space occupies only a part of the total Hilbert space which in turn takes the form of $\mathcal{H}=\mathcal{H}_l\oplus \mathcal{H}_o$ where $\mathcal{H}_{l(o)}$ are the Hilbert space for the logical (orthogonal) subspace, respectively.  A logical space $\mathcal{H}_l$ is usually an eigenspace of a system Hamiltonian, and $\mathcal{H}_o$ contains other eigenspaces. When an `excitation' happens, $\mathcal{H}_l$ is transformed into a part of $\mathcal{H}_o$. In some cases, such as DFS and NS in a collective environment, the dimensions of the eigenspaces are different, while for others, such as DD and QEC, the dimensions of the spaces are the same. Therefore, the proportions of $\mathcal{H}_l$ to $\mathcal{H}$ are different for different mechanisms.

\begin{table}[tbh]   
\begin{center}   
\caption{A summary of the quantum technologies and their various features. In principle, dynamical decoupling with ideal control can eliminate any type of qubit error with ideal control, but it is linear interactions that are usually considered in practice. Also, it is worth noting that there are optimal control theories with measurement feedback control, but this is not the case we are going to study here.}  
\label{tb:combine} 
\begin{tabular}{|c|c|c|c|c|}   
\hline   Quantum technology       & error type   & prior knowledge & error strength & feedback control \\   
\hline   DFS/NS                   & collective   & $\surd$         & arbitrary      & $\times$         \\ 
\hline   Dynamical decoupling     & arbitrary    & $\times$        & limited        & $\times$         \\ 
\hline   Quantum error correction & local        & $\times$        &  low           & $\surd$          \\   
\hline   Optimal control          & quasi-static & $\surd$         &  limited       & $\times$         \\
\hline   
\end{tabular}   
\end{center}   
\end{table}

\subsection{Combining GQC/HQC with decoherence-free subspace and noiseless subsystem}
Decoherence-free subspaces and noiseless subsystems protect quantum information by encoding it into a certain subspace or subsystem that is immune to decoherence (for reviews, see Refs.  \cite{lidar2014review,schlosshauer2019quantum}). In this way, although the physical systems used for encoding are exposed to environmental noise, the evolution in the DFS or NS remains unitary. The concept of DFS was first proposed for qubits in a collective dephasing environment (by `collective', we mean that the physical systems interact with their environment in the same way or with permutation symmetry) \cite{zanardi1997noiseless,duan1997preserving,lidar1998decoherence}. A single decoherence-free state of two photons and a DFS of two trapped ions have been demonstrated experimentally in Refs. \cite{kwiat2000experimental} and \cite{kielpinski2001decoherence}, respectively. NSs \cite{knill2000theory} are a natural generalization of the concept of DFS and exist when the decohering interaction possesses some nontrivial algebra of symmetries. For instance, when a set of qubits interacts with their environment in an arbitrary collective manner, nontrivial NS can be found in the total Hilbert space spanned by the qubits \cite{viola2001experimental}.

The concept of DFSs and NSs can be unified with representation theory \cite{knill2000theory,lidar2014review}. The general form of the system-environment interaction Hamiltonian is $H_{I}=\sum_\alpha S_\alpha\otimes E_\alpha$, where the $S_\alpha$ are called error operators and $E_\alpha$ are the corresponding environmental Hamiltonians. The error operators $S_\alpha$ generate the algebra $\mathcal{A}=\{S_\alpha\}$. Assume that $\mathcal{A}$ is $\dag$-closed, {\it i.e.}, that $A\in \mathcal{A}\Rightarrow A^\dag\in\mathcal{A}$, and that $I\in \mathcal{A}$ ($I$ is the identity), then the orthogonal decomposition of $\mathcal{A}$ is given by $\mathcal{A}\cong\oplus_J I_{n_J}\otimes\mathcal{M}_{d_J}$, where $J$ is the label of an irreducible representation (irrep.) of $\mathcal{A}$, $n_J$ is the degeneracy of the $J$th irrep., and $d_J$ is the dimension of the $J$th full matrix algebra $\mathcal{M}_{d_J}$. 
Accordingly, the system Hilbert space can be decomposed as $\mathcal{H}=\oplus_J\mathbb{C}^{n_J}\otimes\mathbb{C}^{d_J}$, where each left factor $\mathbb{C}^{n_J}$ is called a `subsystem' and the corresponding right factor $\mathbb{C}^{d_J}$ is called a `gauge'. The tensor product $\mathbb{C}^{n_J}\otimes\mathbb{C}^{d_J}$ forms a `subspace' of the system Hilbert space. An immediate consequence of the above space decomposition is that the subsystem factors $\mathbb{C}^{n_J}$ are unaffected by the error operators (and thus decoherence). Therefore, it is possible to safely store quantum information in each of the subsystem factors. In this sense, the $\mathbb{C}^{n_J}$ factors are called noiseless subsystems. When $d_J=1$, one recovers the concept of DFS: $\mathbb{C}^1$ is just a scalar, and $\mathbb{C}^{n_J}\otimes\mathbb{C}=\mathbb{C}^{n_J}$ reduces to a proper subspace that is decoherence-free. 

\subsubsection{DFS arising in collective environment}
\begin{figure}
  \centering
  \includegraphics[width=1.0\textwidth]{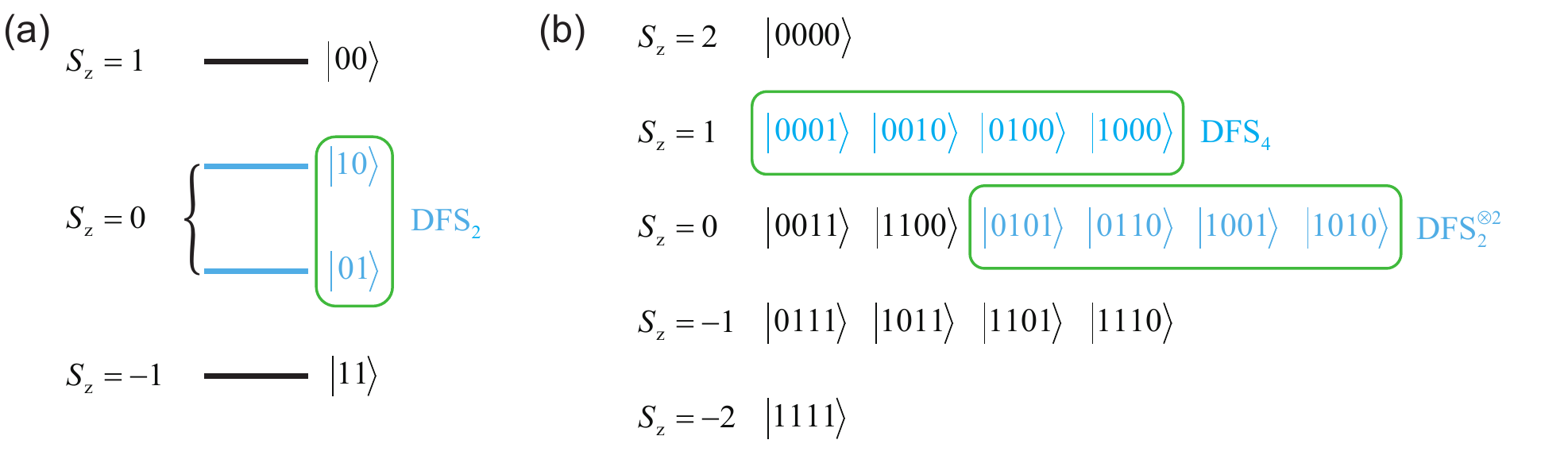}\\
  \caption{Encoding of DFS$_2$ and DFS$_4$. (a) A DFS$_2$ is embedded in a two-qubit space. The two logical bases are in the subspace with 0 eigenvalue of $S_z$ ($S_z=0$). (b) The DFS$_2$ encoding is scalable. A $\mathrm{DFS}_2\otimes\mathrm{DFS}_2$ ($\mathrm{DFS}_2^{\otimes 2}$) can be found in the $S_z=0$ subspace of a four-qubit space, representing two logical qubits. Since the four logical bases are in the same DFS,  $\mathrm{DFS}_2^{\otimes 2}$ is also a DFS. On the other hand, a DFS$_4$ is in the $S_z=2$ subspace.}\label{fig:DFS2}
\end{figure}
Generally, DFSs/NSs emerge when the system-environment interaction possesses symmetry. One prototypical example is the case of the collective environment, where a set of qubits couples to the environment with permutation symmetry. In this case, the interaction Hamiltonian can be written as $H_I=\sum_\alpha S_\alpha\otimes E_\alpha$, where $S_\alpha=\frac{1}{2}\sum_k\sigma_\alpha^k$ with $\alpha=x,y,z$, and $k$ being the physical qubit index. 
Here, the Hilbert space can be decomposed into $\mathcal{H} = \oplus_{J} \mathbb{C}^{n_J} \otimes \mathbb{C}^{d_J}$, where $J$ denotes the total spin, and the sum is from $0$ or $\frac{1}{2}$ dependent on the number of qubits $N$ being even or odd, respectively. It is clear that for a given $J$, $d_J=2J+1$.  On the other hand, $n_J$ can be obtained by using angular momentum addition rules or the Bratteli diagram \cite{lidar2014review}, which gives $n_J=(2J+1)N!/(N/2+1+J)!(N/2-J)!$. 

In the collective dephasing case where $H_I=S_z\otimes E_z$, two qubits are enough to encode a single logical qubit. The two states $\ket{01}$ and $\ket{10}$ are annihilated by $S_z$, {\it i.e.}, $S_z\ket{01}=S_z\ket{10}=0$, and thus span a decoherence-free logical qubit: $\ket{0}_L=\ket{01}$, $\ket{1}_L=\ket{10}$, labeled as $\mathrm{DFS}_2$ in Fig.~\ref{fig:DFS2} (a)  \cite{feng2009geometric}. An arbitrary geometric phase-shift gate can be performed on the logical qubit by using only single-qubit and two-body XXZ interaction terms in the Hamiltonian \cite{xu2014universal}. This scheme is straightforwardly scalable: when another pair of qubits is encoded as a logical qubit, the four qubits represent two logical qubits with the logical states $\ket{00}_L,\ket{01}_L,\ket{10}_L,\ket{11}_L$, labeled as $\mathrm{DFS}_2^{\otimes 2}$ in Fig.~\ref{fig:DFS2} (b). The four states form a larger DFS that is invariant under the action of certain nearest-neighbor two-body interactions. It is shown that a set of geometric entangling gates, both conventional \cite{xu2014universal} and unconventional \cite{feng2009geometric}, can be realized for the two logical qubits, and thus a universal set of gates is available.

The just discussed encoding can be generalized to multi-level logical systems. In fact, it was first used by Wu {\it et al.} for realizing the adiabatic holonomic gates in a four-qubit DFS, which is spanned by $\ket{0}_L = \ket{0001},\ket{1}_L = \ket{0010},\ket{a}_L = \ket{0100},\ket{e}_L = \ket{1000}$, labeled as $\mathrm{DFS}_4$ in Fig.~\ref{fig:DFS2} (b) \cite{wu2005holonomic}. 
Note that the four logical states do not vanish under the action of $S_z$, but all share the same eigenvalue $+1$. In the total Hilbert space, there is a 6-dimensional DFS (labeled as $\mathrm{DFS}_6$) spanned by the four-qubit states that have two 1s and two 0s, {\it e.g.}, $\ket{0011}$. The logical states in this DFS are all annihilated by $S_z$. Adiabatic holonomic gates acting on the 4-dimensional DFS can be built with two- and four-body interactions \cite{wu2005holonomic}. 
It was later shown that the adiabatic process can be accelerated by means of STA \cite{zhang2015fast,pyshkin2016expedited}. Notably, the scheme proposed in Ref.~\cite{song2016shortcuts} is not only fast but also more suitable for physical implementation since there are only two-body interactions involved in the gate construction. 

\begin{figure}
  \centering
  \includegraphics[width=1.0\textwidth]{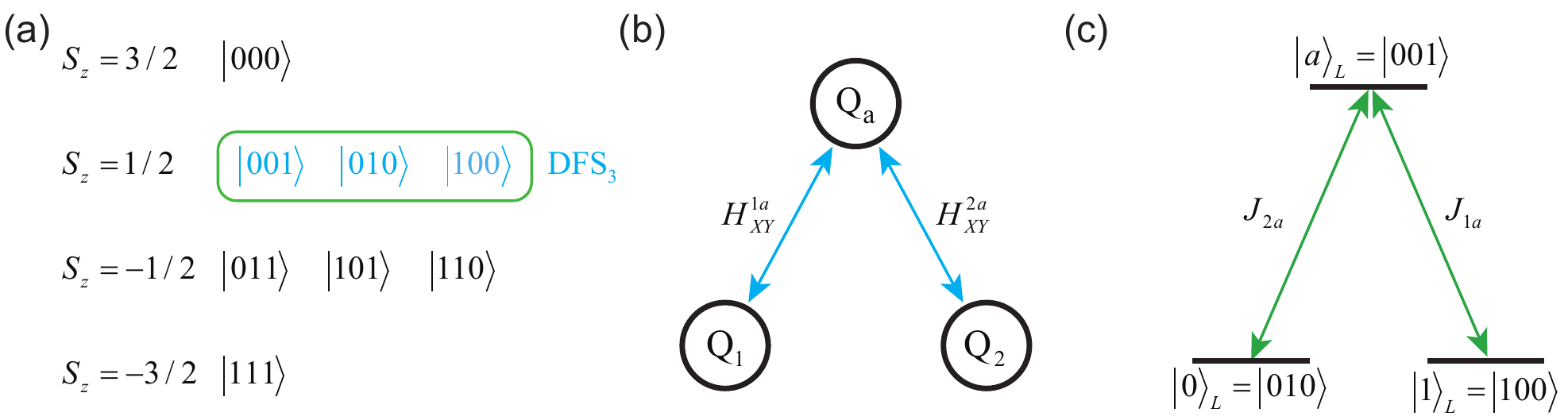}\\
  \caption{Encoding of DFS$_3$ and a physical realization with three qubits. (a) A DFS$_3$ is embedded in the $S_z=1/2$ subspace spanned by three qubits. (b) An experimental feasible realization of DFS$_3$ with three qubits, labeled as Q$_1$, Q$_2$, and Q$_a$, respectively. Q$_1$ and Q$_2$ are used as data qubits while Q$_a$ is the auxiliary qubit. Q$_{1(2)}$ couples with Q$_a$ through $XY$-type interaction $H_{XY}$ with coupling strength $J_{1a}$ ($J_{2a}$). (c) The three logical states $\ket{0}_L$, $\ket{1}_L$, and $\ket{a}_L$ construct an invariant subspace under the action of $H_{XY}^{1a}$ and $H_{XY}^{2a}$. $\ket{0(1)}_L$ is coupled to $\ket{a}_L$ with strength $J_{2a(1a)}$.}\label{fig:DFS3}
\end{figure}

It is clear that while the adiabatic holonomic gates require at least four qubits to encode a logical qubit, the nonadiabatic holonomic gates in the three-level scheme use only three: $\ket{0}_L=\ket{010}$, $\ket{1}_L=\ket{001}$, and $\ket{a}_L=\ket{100}$, labeled as $\mathrm{DFS}_3$ in Fig.~\ref{fig:DFS3} \cite{xu2012non}. 
This encoding and method to implement holonomic gates have been applied in different systems, such as trapped ions with two-body interaction \cite{liang2014nonadiabatic}, NV centers \cite{zhou2015cavity}, and transmons \cite{xue2015universal,xue2016nonadiabatic}. 
Moreover, the encoding is used to realize quantum gates with different schemes, such as unconventional geometric gates \cite{zhao2016nonadiabatic} and single-shot scheme \cite{zhao2017single}.  

$\mathrm{DFS}_2$ and $\mathrm{DFS}_3$ encoding can be combined to implement nonadiabatic holonomic gates \cite{mousolou2018realization}. The key idea is to encode the logical qubit in three physical qubits as $\ket{0}_L=\ket{01}\otimes\ket{0}$ and $\ket{1}_L=\ket{10}\otimes\ket{0}$ with the third qubit playing the role of an auxiliary system, as illustrated  Fig.~\ref{fig:DFS3} (b). Together with the auxiliary state $\ket{a}_L=\ket{00}\otimes\ket{1}$, a $\Lambda$ system can be realized by means of two-body interactions between the three physical qubits, see Fig.~\ref{fig:DFS3} (c). The initial logical state is stored in $\mathrm{DFS}_2$, and after a cyclic evolution, it will return to  $\mathrm{DFS}_2$, thereby realizing a holonomic gate. To perform an entangling logical gate on two such logical qubits, we first note that the four logical states $\ket{00}_L=\ket{0101}$, $\ket{01}_L=\ket{0110}$, $\ket{10}_L=\ket{1001}$, and $\ket{11}_L=\ket{1010}$ are four states in $\mathrm{DFS}_6$.
The remaining two logical states in $\mathrm{DFS}_6$ can be used as auxiliary states, such that the six states form a double-$\Lambda$ structure. This encoding is also suitable for single-loop scheme \cite{wang2018single} and has been experimentally demonstrated in NMR \cite{zhu2019single}.

\subsubsection{DFS arising in atom-cavity couplings}
When a set of multilevel systems couples to a common `bus', the main decoherence is caused by the bus since it is usually strongly perturbed by its environment. In the case of cavity QED, the bus is a cavity mode, which may leak into the environment.
In addition, the cavity couples to an excited state of the atom that shows spontaneous emission. These two issues can be addressed with DFS encoding when some symmetry emerges.

The first example is the system comprising two four-level atoms inside an optical cavity. Each atom has three lower states, $\ket{0}$, $\ket{1}$ and $\ket{a}$, which could be different hyperfine levels or Zeeman levels, and an excited state $\ket{e}$ coupled to each ground state by laser radiation. The lasers act commonly on both atoms and the two atoms interact with each other via the cavity radiation field. An adiabatic procedure is employed such that the cavity is kept empty and the excited state $\ket{e}$ is depopulated. When the empty cavity tuned along the $\ket{1}\leftrightarrow\ket{e}$ transition with equal atom-cavity coupling, a DFS against cavity emission is spanned by $\{\ket{aa}, \ket{a1}, \ket{1a}, \ket{11}\}$ (labeled as $\mathrm{DFS}_{ac}$) because they are annihilated by the atom-cavity interaction Hamiltonian and immune to the atomic emission \cite{pachos2002quantum}. As an application, adiabatic conditional phase-shift gates can be constructed by means of the tripod scheme with the four states in $\mathrm{DFS}_{ac}$. 
A more complicated version of this setup was demonstrated in \cite{zheng2008deterministic}, in which each of the atoms has two excited states and four ground states.

Remarkably, besides finite-dimensional atoms, DFS emerges from continuous variable systems, such as oscillators \cite{albert2016holonomic} and many other systems \cite{you2005correlation}. The decoherence of a single mode oscillator is described with the Lindblad equation $\dot{\rho}=F\rho F^\dag-\{F^\dag F,\rho\}/2$ with $F=\sqrt{\kappa}(\hat{a}^d-a_0^d)$, $[\hat{a},\hat{a}^\dag]=1$, $\kappa$ being a real number, and $a_0$ being a complex number. A DFS is spanned by the $d$ well-separated coherence states $\ket{a_0^d}$, which are annihilated by $F$. It is explicitly demonstrated that universal control in this DFS can be achieved via two simple gate families, loop gates and collision gates, together with an entangling two-oscillator gate \cite{albert2016holonomic}.
Moreover, GQC can be implemented with other encodings, such as the binomial codes, in coupled atom-cavity systems \cite{chen2021fast}. 

\subsubsection{NS arising in collective environment}

For the general collective decoherence case, {\it i.e.}, where all three $S_\alpha$:s  are present, it is more convenient to encode the logical states in an NS since it usually requires fewer qubits than in a DFS. To see why, we list the space decomposition of the 4-, 5-, and 6-qubit Hilbert space, respectively, according to $S_\alpha$ as follows 
\begin{align}\label{eq:NS}
    \mathcal{H}_{1/2}^{\otimes 4} =& \mathbb{C}^2\otimes\mathbb{C}\oplus\mathbb{C}^3\otimes\mathbb{C}^3\oplus\mathbb{C}\otimes\mathbb{C}^5, \nonumber \\
    \mathcal{H}_{1/2}^{\otimes 5} =& \mathbb{C}^5\otimes\mathbb{C}^2\oplus\mathbb{C}^4\otimes\mathbb{C}^4\oplus\mathbb{C}\otimes\mathbb{C}^6, \nonumber \\
    \mathcal{H}_{1/2}^{\otimes 6} =& \mathbb{C}^5\otimes\mathbb{C}\oplus\mathbb{C}^9\otimes\mathbb{C}^3\oplus\mathbb{C}^5\otimes\mathbb{C}^5\oplus\mathbb{C}\otimes\mathbb{C}^7.
\end{align}
Each section of a space decomposition is related to a particular $J$, containing a $\mathbb{C}^{n_J}$ as well as a $\mathbb{C}^{d_J}$ subsystems. The error operators $S_\alpha$ act only on $\mathbb{C}^{d_J}$, and thus information stored in $\mathbb{C}^{n_J}$ remains unchanged. 

\begin{figure}
  \centering
  \includegraphics[width=1.0\textwidth]{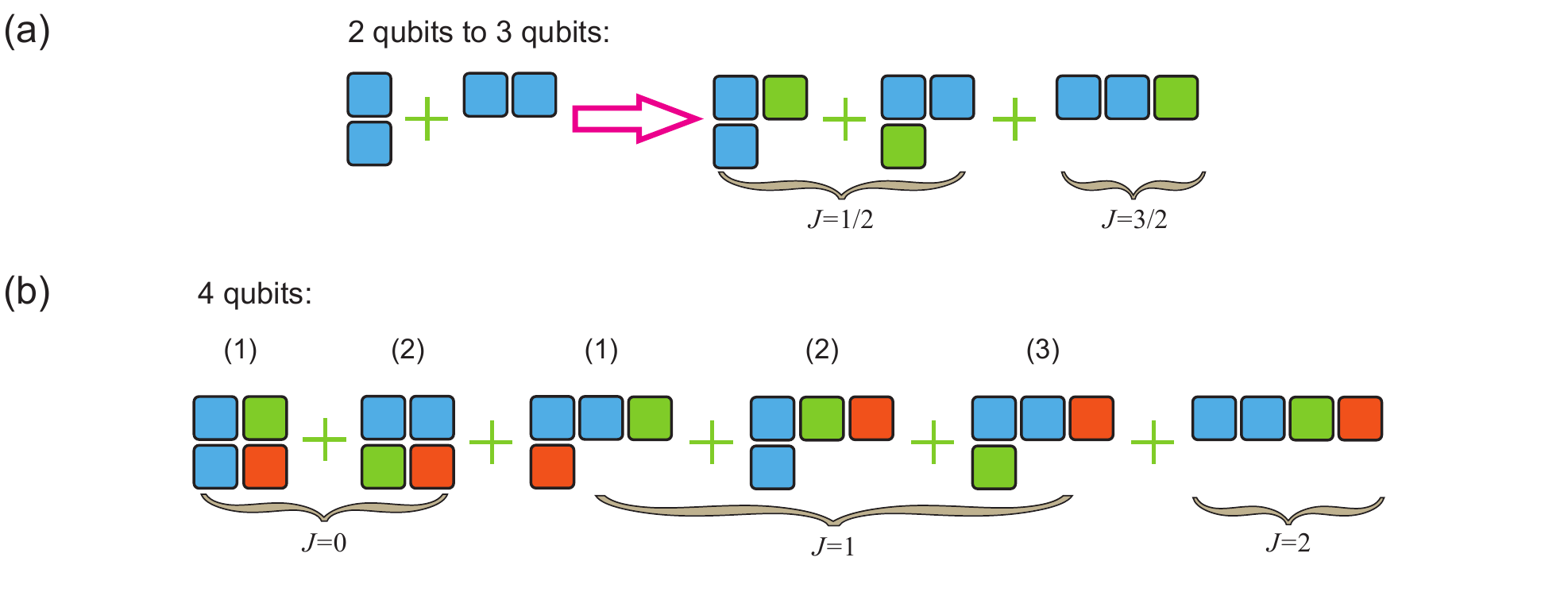}\\
  \caption{Young diagrams and tableaux for the qubit-space decomposition. (a) When a third qubit (the green one) is attached to the Young diagrams of the first two qubits (the blue ones), the $J=1/2$ Young diagram can be filled in two different ways. Here, one obtains an NS of dimension $n_{1/2}=2$. (b) After attaching the fourth qubit (the red one), one obtains a 2-dimensional NS ($n_{0}=2$) and a 3-dimensional one ($n_{1}=3$).} \label{fig:ns}
\end{figure}

According to angular momentum theory, each $J$ corresponds to a Young diagram (for
examples, see Fig.~\ref{fig:ns}). 
The dimension of $\mathbb{C}^{n_J}$ is determined by the number of Young tableaux for a given Young diagram.
For instance, there are three different ways to fill the $J=1$ Young diagram, and hence $n_1=3$.
It is evident from Eq.~(\ref{eq:NS}) that a 3-dimensional NS, which can be used for nonadiabatic holonomic gates, is supported by at least 4 qubits \cite{zhang2014quantum}, while a 4-dimensional NS, which can be used for adiabatic holonomic gates, is supported by at least 5 qubits \cite{wu2005holonomic}. 
The DFSs only emerge when the number of qubits is even since only in this case $J$ starts from $0$.
From Eq.~(\ref{eq:NS}), one finds that the 4-qubit Hilbert space only supports a 2-dimensional DFS while the 6-qubit Hilbert space can support a 5-dimensional DFS.
In contrast, a 2-dimensional NS can be found in the 3-qubit case and a 5-dimensional NS requires only five qubits.

Now, we explain how to encode in a given NS by taking the $\mathbb{C}^3\otimes\mathbb{C}^3$ section in the 4-qubit space as an example. 
As shown in Fig.~\ref{fig:ns}(b), there are three different ways to construct a $J=1$ space. 
The corresponding three bases in $\mathbb{C}^3$ subsystem are labeled as $\ket{k}\ket{1}$, $\ket{k}\ket{2}$, and $\ket{k}\ket{3}$, $k=1,2,3$; for details, see appendix of Ref. \cite{zhang2014quantum}. There are two key points: the error operators $S_\alpha$ act as  $S_\alpha\ket{k}\ket{j}=\ket{k}\sum_m a_m\ket{m}$ ($j=1,2,3$), with $a_m$ being complex numbers, {\it e.g.}, $S_z\ket{3}\ket{j}=(4-2j)\ket{3}\ket{1}$, and the error operators act on the three different bases in the same way, {\it i.e.}, the coefficients $a_m$ do not depend on the index $k$. 
In this sense, an encoded state can be defined on an arbitrary superposition of the three bases for a given $k$, {\it e.g.}, $\ket{k}_L=(\ket{k}\ket{1}+\ket{k}\ket{2}+\ket{k}\ket{3})/\sqrt{3}$. Then, we have a 3-dimensional NS, in which a logical state takes the form of
\begin{equation}
    \ket{\psi}_L=b_1\ket{1}_L+b_2\ket{2}_L+b_3\ket{3}_L 
\end{equation}
with $b_m$, $m=1,2,3$, being complex-valued.
It is easy to check that, although the logical states are changed by $S_\alpha$, $b_m$ are left unchanged.

To perform holonomic gates with the four-qubit encoding, a set of control Hamiltonians is needed to achieve the appropriate transitions so that the computation stays within the $J=1$ subspace. In fact, all the Gell-Mann matrices on the NS can be found in terms of qubit permutation operators \cite{zhang2014quantum}, thus the $\Lambda$ type logical Hamiltonian is available in this setting.

\subsection{Combining GQC with dynamical decoupling}
Dynamical decoupling (DD) is the strategy to protect coherence of quantum systems that are surrounded by an environment. In a typical DD scheme, a sequence of operators is applied to qubits to average the qubit-environment couplings to zero. The idea of DD originates from the Hahn echo \cite{hahn1950spin} and was developed for high-precision magnetic resonance spectroscopy \cite{mehring2012principles}. The first implementation of DD in the field of quantum computation was developed in order to suppress dephasing on a single qubit by successive application of Pauli operators \cite{viola1998dynamical}. It was subsequently  extended to a general framework based on
identifying which independent linear interactions between the qubits and their environment can be decoupled \cite{viola1999dynamical,viola2000dynamical}. Many other DD schemes and their experimental demonstration have been reported since then (for a review, see Ref. \cite{yang2011preserving}).

\begin{figure}
  \centering
  \includegraphics[width=1.0\textwidth]{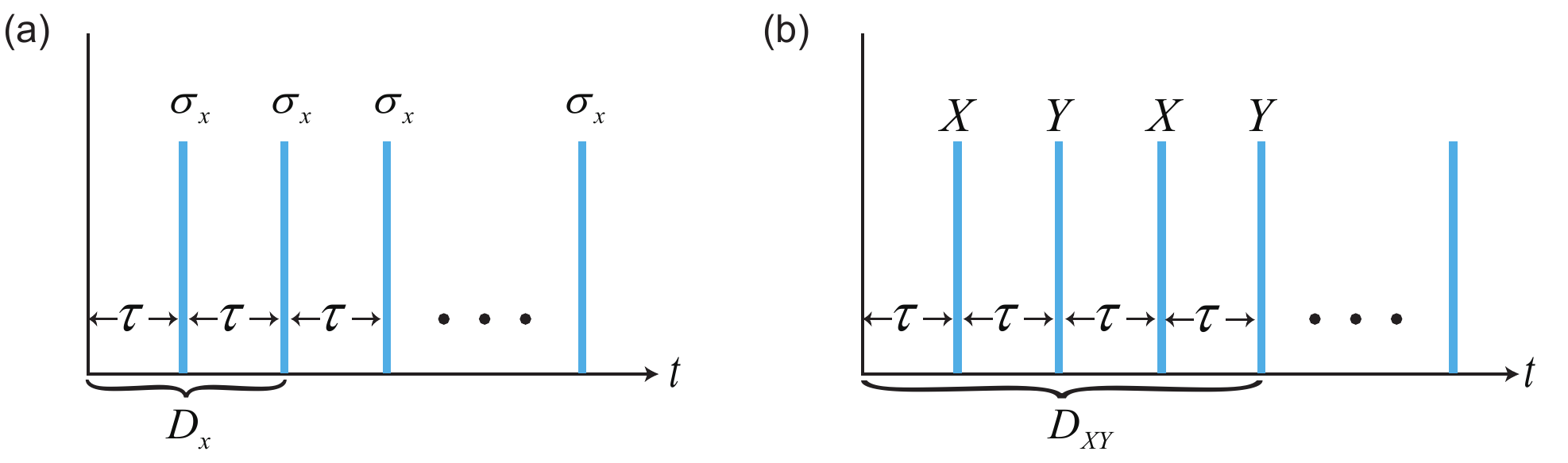}\\
  \caption{Schematic diagrams for decoupling procedures $D_x$ and $D_{XY}$. (a) Let the system evolves freely for a time period $\tau$ followed by a $\sigma_x$ on the qubit instantaneously. Repeat once more and $D_x$ is done. (b) Let the system evolves freely for a time period $\tau$, then use the decoupling operator $X$. Let the system evolves freely for another time period $\tau$ followed by the decoupling operator $Y$. Repeat it and $D_{XY}$ is accomplished.} \label{fig:dd}
\end{figure}
While there have been many approaches to the DD method, we here consider only the first-order average Hamiltonian obtained through the Baker-Cambell-Hausdorff (BCH) formula in the simplest cases. More thorough analyses of the efficiency of the schemes can be found in \cite{lidar2013quantum}. The starting point is a total Hamiltonian of the generic form  
\begin{equation}
    H_t=H_S+H_E+H_I
\end{equation}
with $H_S$ the qubit system Hamiltonian, $H_E$ the environment Hamiltonian, and $H_I$ the system-environment interaction. The aim of DD is to seek some proper $H_S$ acting on the qubits periodically so that the effect of $H_I$ is removed. 

We start with the pure dephasing system-environment interaction $H_I=\sigma_z\otimes E_z$. To decouple this interaction, the system Hamiltonian is set to $H_S=\lambda(t)\sigma_x$, where $\lambda(t)$ is a square-shaped bang-bang controllable pulse of duration $\delta$. When the pulse lasts for a time period $\delta$ such that $\delta\lambda=\pi/2$, the evolution of the qubit is described by $U_c(\delta,0)=\sigma_x$. The ideal case corresponds to $\delta\rightarrow0$ and $\lambda\rightarrow\infty$, implying that the pulses should be Dirac delta-like. In the ideal pulse scenario, the influence of $H_E$ and $H_I$ during the realization of $U_c$ is negligible as long as their norm is bounded. The decoupling procedure can be shown as 
\begin{equation}
    D_x=[\tau,\sigma_x,\tau,\sigma_x],
\end{equation}
where $\tau$ denotes a free evolution of $H_t$ for time $\tau$, and the  $\sigma_x$:s are decoupling operators each applied instantaneously.
The total evolution describing the procedure $D_x$ reads
\begin{equation}
    U_{x}=\sigma_x e^{-i(H_E+H_I)\tau}\sigma_x e^{-i(H_E+H_I)\tau}=e^{-i(H_E+\sigma_x H_I\sigma_x)\tau}e^{-i(H_E+H_I)\tau}=e^{-i(H_E-H_I)\tau}e^{-i(H_E+H_I)\tau}.
\end{equation}
When $\tau$ is small enough, $U_{x}$ can be rewritten by using the BCH formula to the first order as
\begin{equation}\label{eq:ux}
    U_{x}=e^{-i(H_E-H_I)\tau-i(H_E+H_I)\tau}=e^{-i2H_E\tau},
\end{equation}
where higher order terms are smaller quantities and neglected. 
Since $H_E$ affects only the environment, the qubit is decoupled at $t=2\tau$.

In a more general setting, a set of qubits may independently interact with the environment in a linear fashion as described by
\begin{equation}
    H_I=\sum_{k,\alpha}\sigma_\alpha^k\otimes E_\alpha^k,
\end{equation}
where $k$ is the qubit index and $\alpha=x,y,z$ correspond to the three Pauli operators. Note that, unlike the collective interaction, we do not assume any symmetry in this general case. To decouple the qubits from $H_I$, we need four decoupling operators $\{I,X,Y,Z\}$ in which $I$ is the identity for all the qubits and
\begin{equation}
    X=\otimes_k\sigma_{x}^k, \quad Y=\otimes_k\sigma_{y}^k, \quad 
    Z=\otimes_k\sigma_{z}^k.
\end{equation}
The latter three decoupling operators can be realized by setting $H_S=\lambda(t)\sum_k\sigma_\alpha^k$ with $\delta\lambda=\pi/2$.
The decoupling procedure based on the four decoupling operators is
\begin{equation}
    D_{XY}=[\tau, X, \tau, X, Z, \tau, Z, Y, \tau, Y]=[\tau,X, \tau,Y,\tau,X,\tau,Y],
\end{equation}
in which the equation $Y=XZ$ (up to a global phase) is used.
Similar to Eq.~(\ref{eq:ux}), it is easy to check that the evolution operator for $D_{XY}$ is
\begin{equation}
    U_{XY}=ZU_\tau Z YU_\tau Y XU_\tau X U_\tau ,
\end{equation}
where $U_\tau=e^{-i(H_E+H_I)\tau}$.
Again, for a small $\tau$, $U_{XY}$ can be reformulated to the first order as 
\begin{equation}
    U_{XY}=e^{-i4H_E\tau},
\end{equation}
which shows that $H_I$ is effectively removed from the dynamics of the qubits at $t=4\tau$.

DD can protect quantum states and gates from environmental noise \cite{khodjasteh2009dynamically}. The combination of DD and geometric quantum gates leads to schemes that are robust against both environmental noise and control error. However, to achieve this, one must make sure that the desired gates are not spoiled by the decoupling operators. 
One method to address this issue is to employ a modified Eulerian circle \cite{khodjasteh2009dynamically}.  In \cite{xu2014protecting}, Xu and Long proposed another method in which a logical qubit is encoded into two physical three-qubit states: $\ket{0}_L=\ket{+++}, \ket{1}_L=\ket{--+}$ with $\ket{+}$ and $\ket{-}$ being eigenstates of $\sigma_x$. To realize the geometric gates, the control Hamiltonian takes the form  $\sigma_{kl}(\phi)=e^{i\phi}\ket{1}_{kl}\bra{0}+e^{-i\phi}\ket{0}_{kl}\bra{1}$, where $k,l$ are qubit indexes. The reason to choose this encoding is that $\sigma_{kl}(\phi)$ commutes with the decoupling operators and thus quantum control and DD can be performed simultaneously. A logical $\sigma_x$ and logical rotation $e^{i\theta\sigma_y}$ can be achieved with this setting with geometric phases. As discussed in the DFS schemes, the above encoding can be directly generalized to the two-qubit case, and a logical entangling gate $e^{i\phi\sigma_y\otimes\sigma_z}$ can be generated.  

The encoding proposed in Ref.~\cite{xu2014protecting} can be used to construct a universal set of logical gates by utilizing the two-qubit interaction $J_{x}^{kl}\sigma_x^k\sigma_x^l+J_{y}^{kl}\sigma_y^k\sigma_y^l$ as control Hamiltonian \cite{wu2020universal}.
A similar encoding was proposed in order to combine DD with nonadiabatic holonomic schemes, in which the logical-qubit states were $\ket{0}_L=\ket{001}$ and $\ket{1}_L=\ket{010}$, and the auxiliary state  $\ket{a}_L=\ket{100}$.

\subsection{Combining HQC with Quantum Error Correction}

\begin{figure}
  \centering
  \includegraphics[width=1.0\textwidth]{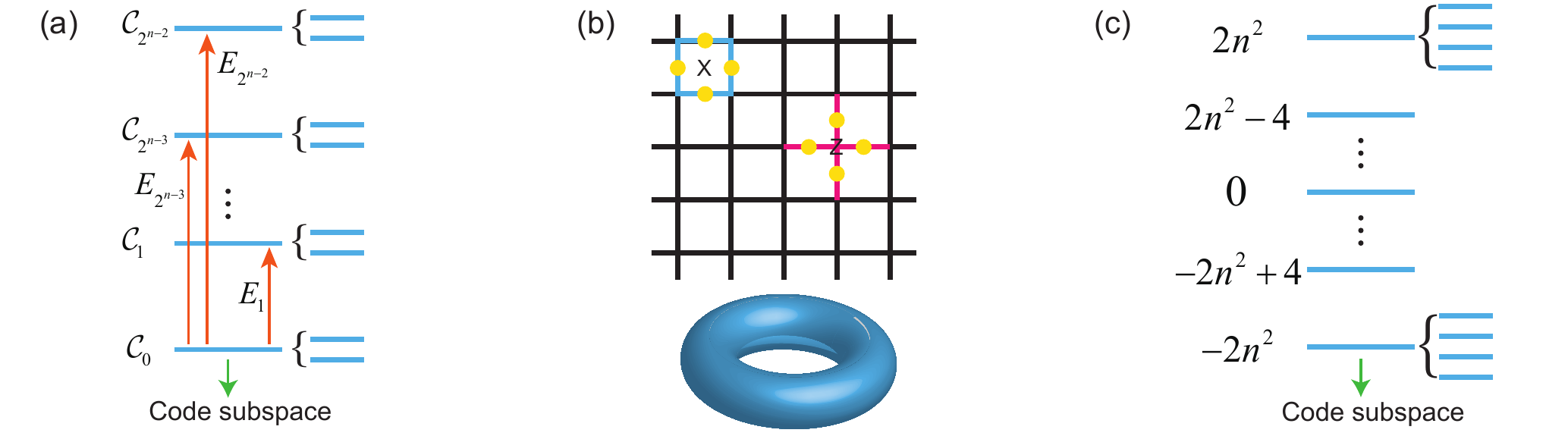}\\
  \caption{(a) Space structure of a quantum code used in QEC. Here, one logical qubit is encoded in $n$ physical qubits. Thus the dimension of the code subspace $\mathcal{C}_0$ is two. When an error $E_1$ occurs, $\mathcal{C}_0$ is excited to $\mathcal{C}_1$ with the same space structure. Such an encoding allows at most $2^{n-2}$ different correctable errors. (b) Square lattice attached to the surface of a torus. Consider an $n\times n$ lattice, a qubit (shown by a solid yellow circle) is attached to each edge. Hence there are a total of $2n^2$ qubits on the lattice. An $X$-type stabilizer is defined on four qubits that surround a square while a $Z$-type stabilizer is defined on four qubits that form a cross. (c) Energy spectrum and code subspace for the toric codes.}\label{fig:qec}
\end{figure}

Quantum error correction (QEC) is an active strategy for realizing quantum computation in a fault-tolerant manner (for a review, see Ref. \cite{terhal2015quantum}; for books, see Refs. \cite{nielsen2002quantum,lidar2013quantum}).
A typical QEC procedure consists of encoding logical states in a set of physical qubits, 
performing a syndrome measurement to detect any single-qubit error, and recovering the logical states with unitary operations.

In QEC, logical code states are the common eigenstates of the stabilizers, which are commuting strings of Pauli operators defined on a set of physical qubits. By assuming we have $n$ physical qubits, spanning a $2^n$-dimensional Hilbert space $\mathcal{H}$, each stabilizer has two eigenvalues $\pm1$, each of which corresponds to $2^{n-1}$ orthogonal basis vectors. Apparently, $m$ linearly independent stabilizers have $2^{n-m}$ common eigenstates with the same eigenvalue. Hence, to define $k$ logical qubits with these $n$ physical qubits, we need $(n-k)$ stabilizers.  

The aim of QEC is to find good codes that can detect more errors with the same overhead. A necessary and sufficient condition under which a set of errors can be corrected by a quantum code is provided in \cite{nielsen2002quantum}. Here, we explain this in a more intuitive way (see Fig.~\ref{fig:qec} for an example). Note that the $k$ logical qubits span a $2^{k}$-dimensional subspace $\mathcal{C}$ of $\mathcal{H}$. When an error occurs, {\it e.g.}, a phase-flip $\sigma_z$ on one of the physical qubits, it transforms each logical state in $\mathcal{C}$ to a new state in $\mathcal{H}$. We ask whether the $2^{k}$ new states form another subspace $\mathcal{C}'$ orthogonal to $\mathcal{C}$, but with the same structure so that the vectors in $\mathcal{C}'$ are related to a different set of eigenvalues of the stabilizers. In this way, by measuring the logical states affected by the error, we can determine which subspace they have moved into.
Accordingly, the error that happened can be identified with the measurement results and then corrected for.
It is shown by threshold theorems that if the errors of each type are local and their rates are below a certain threshold, large-scale quantum computation with arbitrarily small errors can be implemented \cite{gottesman1997stabilizer,lidar2013quantum}.

The quantum error correction scheme that has attracted the most interest recently is based on surface codes \cite{bravyi1998quantum,dennis2002topological,fowler2012surface}, which are defined on a 2-dimensional qubit lattice. Surface codes are similar to toric codes \cite{kitaev2003fault}, {\it i.e.}, topological codes on a torus, see Fig.~\ref{fig:qec} (b). 
By attaching qubits to
the toroidal lattice, two sorts of stabilizers are defined: $X$-type stabilizers defined on the squares and $Z$-type stabilizer defined on the crosses.
For an $n\times n$ square lattice, there are a total of $n^2$ $X(Z)-$type stabilizers, in terms of 
which the Hamiltonian 
\begin{equation}
    H=-\sum_{k=1}^{n^2}X_k-\sum_{j=1}^{n^2}Z_j 
\end{equation}
can be defined.
The ground states is four-fold degenerate with energy eigenvalue $-2n^2$, determined by the genus of the torus \cite{kitaev2003fault}. The ground-state manifold is used as the code subspace. When a single-qubit $\sigma_x$ or $\sigma_z$ error occurs, the code subspace is transformed to the first excited subspace with eigenvalue ($-2n^2+4$). 

Surface codes possess the same code subspace as toric codes but avoid the trouble of arranging the qubits on the surface of a 3-dimension object. One appealing feature of surface codes is that only nearest-neighbor interactions are involved, facilitating experimental implementation. Another one is the appreciable tolerance to local errors due to their topological nature. As a result, surface codes are considered a promising approach to realizing QEC.


One can realize logical transformations acting on the logical qubits by means of quantum holonomies. This is possible since in QEC, the code space and error spaces are subspaces or subsystems of the total Hilbert space. By moving the subspaces or subsystem \cite{oreshkov2009holonomic} to complete cyclic evolution, nontrivial holonomies can be obtained.
Furthermore, one can use HQC to build the required gates in the circuit to implement QEC. In this case, the holonomic gates act on the physical qubits and need not be universal. In both cases, the advantages of HQC and QEC are combined.

The motivation for combining HQC with QEC is that HQC alone is universal but not fault-tolerant since it is solely robust to certain kinds of errors.  
The fault tolerance can be gained by protecting holonomic gates by means of QEC techniques on codes \cite{oreshkov2009fault,oreshkov2009scheme}, {\it i.e.}, finding a holonomic realization of a universal set of encoded gates.  This can be done by adiabatically transporting the code space along suitable loops via sequences of elementary fault-tolerant transformation. 
In the subspace stabilizer code case, an initial  Hamiltonian is chosen as an element of the stabilizer and varied in an adiabatic way along proper paths so that the vectors in different eigenspaces of the Hamiltonian undergo the same transformation. It is worth noting that each eigenspace will acquire a dynamical phase. However, since the dynamical phases are global phases for the eigenspaces, they will be projected out by the syndrome measurements. A distinction between the combined HQC-QEC scheme and original adiabatic HQC is that the holonomic gates in the former are performed in several subspaces at the same time, so as to be compatible with error correction.  
An alternative scheme \cite{zheng2014fault} to construct fault-tolerant HQC via QEC is proposed with the merit that the energy gap between the ground space and other excited spaces does not change with the problem size. This enhances the capability to prevent low-weight thermal excitation.
An explicit application of HQC on surface codes is provided in \cite{zheng2015fault}, showing that the frequency of error correction and the physical resources needed can be greatly reduced by the constant energy gap.

On the other hand, the quantum gates needed to form circuits on surface codes can be realized in a holonomic manner. This task is not straightforward because the required projective measurement on a multi-level quantum system can collapse the state to the auxiliary level. To overcome this problem, a modified lattice has been proposed in which the holonomic gates on the data and measurement qubits are built with auxiliary qubits (the scheme shown in Fig.~\ref{fig:fourlevel}) \cite{zhang2018holonomic}. Owing to the use of auxiliary qubits, the proposed scheme consumes more physical resources than the original surface code lattice. The limitation can be substantially reduced by using an experimental achievable lattice based on the quantum spin model \cite{wu2020holonomic}.

\subsection{Combining HQC with quantum optimal control theory}
There have been a load of toolboxes available to study  quantum dynamics and quantum control, {\it e.g.}, QuTiP \cite{johansson2012qutip}. A holonomic control procedure can also be searched via a classical computer through quantum optimal control theory (QOCT) \cite{khaneja2005optimal,montangero2007robust,motzoi2009simple,kelly2014optimal}. 
The QOCT method \cite{zhang2019searching} introduced here is based on the gradient ascent pulse engineering (GRAPE) algorithm \cite{khaneja2005optimal}, which exploits the gradient of a fidelity function to update the control field in an iterative manner. The combination of HQC and QOTC not only allows finding proper control pulses that generate a target holonomic gate numerically via GRAPE, but also enables one to suppress various kinds of errors, such as thermal noise and amplitude errors, by taking the errors into account in a target function.

\begin{figure}
  \centering
  \includegraphics[width=1.0\textwidth]{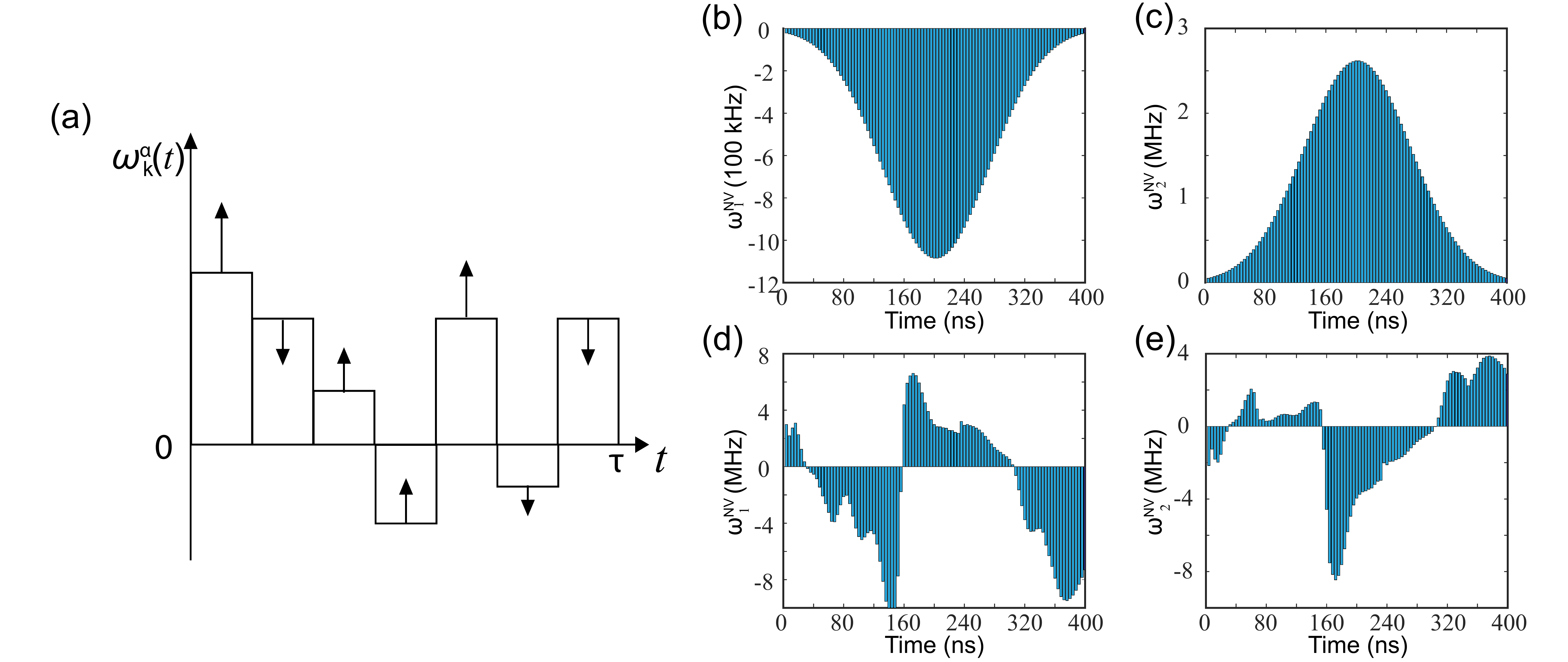}\\
  \caption{An illustration of the gradient ascent optimization and the control sequences for the holonomic Hadamard gate. (a) The evolution time $\tau$ is divided into $N$ (here, $N=7$) pieces, in each of which the control amplitudes $\omega_k^\alpha(t)$ is a constant. The sign and magnitude of the gradient $\partial O/\partial \omega_k^\alpha$ are represented as the direction (up or down) and length of the related vertical arrow, respectively. (b) and (c) show the original control sequences of $\omega_1^{NV}$ and $\omega_2^{NV}$ for the Hadamard gate, respectively. (d) and (e) show the control sequences of $\omega_1^{NV}$ and $\omega_2^{NV}$ obtained by QOCT for the Hadamard gate, respectively. Both processes have the same time $\tau=400$ ns and are divided into $N=100$ pieces. Adapted with permission from \cite{zhang2019searching}. Copyright 2019 by the American Physical Society.} \label{fig:QOCT}
\end{figure}

The main objective here is to find a proper control sequence of the system Hamiltonian to realize a target holonomic gate. The time-dependent Hamiltonian of the system is taken to be $H(t) = H_s + \sum_{k,\alpha}\omega_k^\alpha(t)H_k^\alpha$,
where $H_s$ is the static system Hamiltonian, $H_k^\alpha$ is the time independent $k$th control Hamiltonian for the $\alpha$th quantum system, and $\omega_k^\alpha(t)$ is the corresponding control parameter that describes the amplitude of $H_k^\alpha$. The system is manipulated by modulating the $\omega_k^\alpha(t)$:s. To obtain an analytic evolution operator, the total evolution time $\tau$ is divided equally into $N$ segments $\Delta t=\tau/N$. In each segment $j$, the amplitudes $\omega_k^\alpha(t)$ are set to be constant and labeled as $\omega_k^\alpha(j)$. Thus, the total Hamiltonian $H(t)$ is piecewise time-independent. It follows that the evolution operator takes the form $U(t)=e^{-iH(k)(t-t_k)}\cdots e^{-iH(2)\Delta t}e^{-iH(1)\Delta t}$, where $t_k$ is the initial time of the $k$th segments and $t\in[t_k,t_{k+1}]$. 

\begin{figure}
  \centering
  \includegraphics[width=1.0\textwidth]{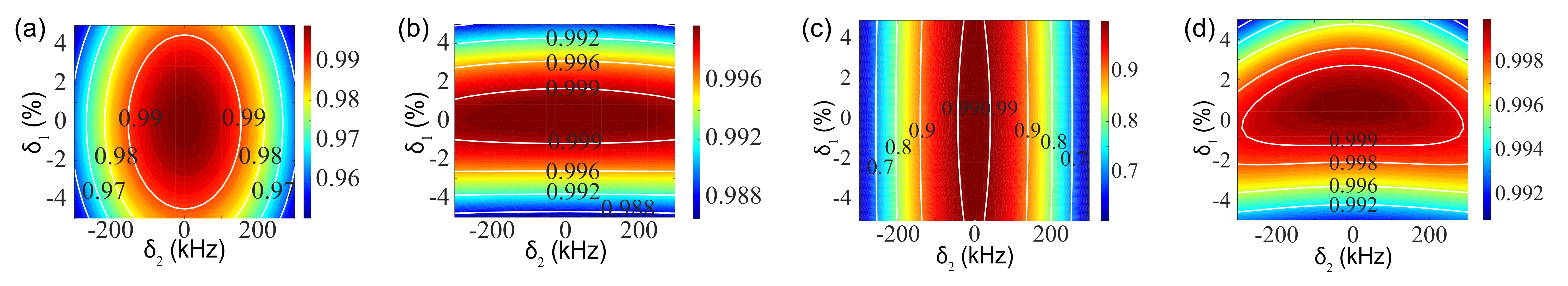}\\
  \caption{The fidelities of the Hadamard and CNOT gates under the thermal noise and the amplitude error. The $x$ axis represents the thermal noise $\delta_2$ and the $y$ axis represents the amplitude error $\delta_1$. The gate fidelities are indicated with the colors shown in the color bars. (a) (b) The fidelities of the Hadamard gate generated by the control sequences from the original NHQC scheme and obtained with the averaged target function $\Bar{O}(\omega_k^\alpha)$, respectively. (c) (d) The fidelities of the CNOT gate generated by the control sequences from the original NHQC scheme and obtained with the averaged target function $\Bar{O}(\omega_k^\alpha)$, respectively. Adapted with permission from \cite{zhang2019searching}. Copyright 2019 by the American Physical Society.} \label{fig:robustness}
\end{figure}

In addition to realizing the target operator $U_T$, the evolution driven by the obtained control sequence must satisfy the two holonomic conditions in Eq. (\ref{eq:2condition}). To this end, the authors of Ref.  \cite{zhang2019searching} used the target function 
\begin{equation}\label{eq:O}
    O(\omega_k^\alpha)=|\mathrm{Tr}[U_T^\dag U(\tau)P(0)]|^2/L^2-\eta\mathrm{Tr}[\int_0^\tau U^\dag(t)H(t)S(t)H(t)U(t)\mathrm{d}tS(0)],
\end{equation}
where $P(0)$ is the projection operator to the $L$-dimensional computational subspace $S(0)$ and $\eta$ is a penalty parameter. The first term in $O(\omega_k^\alpha)$ aims to compare the difference of the obtained evolution operator $U(\tau)$ with $U_T$, while the second term guarantees that there is no dynamical contribution during the evolution.
Based on the gradient of $O(\omega_k^\alpha)$, the explicit optimization procedure is as follows:

Step (1): Set the computational subspace $S(0)$, the target holonomic gate $U_T$, and the parameter $\eta$; guess a set of initial $\omega_k^\alpha$.

Step (2): Compute $U(t_k)$ ($k$ ranges from 1 to $N$); check whether $O(\omega_k^\alpha)$ reaches the predetermined value $O_p$ (the maximum value of $O(\omega_k^\alpha)$ is 1, but one should set $O_p$ a little smaller than 1). 

Step (3): If not, evaluate the gradient $\partial O(\omega_k^\alpha)/\partial\omega_k^\alpha$ and update the control amplitudes $\omega_k^\alpha$ with $\omega_k^\alpha+\varepsilon\partial O(\omega_k^\alpha)/\partial\omega_k^\alpha$, where $\varepsilon$ is a small step size.

Step (4): By using the new $\omega_k^\alpha$, return to Step (2) until $O(\omega_k^\alpha)\geqslant O_p$. 

In experiments, the main type of errors are quasi-static, {\it i.e.}, those errors that fluctuate so slowly that they can be taken as static in each gate operation. Some quasi-static errors, such as thermal noise and amplitude errors, are well-studied experimentally. In the presence of thermal and amplitude errors, the system Hamiltonian can be recast as $H(t,\delta)=H_s+\sum_l\delta_lE_l+(1+\delta_1)\sum_{k,\alpha}\omega_k^\alpha(t)H_k^\alpha$, where $E_l$ are the thermal error operators, $\delta_l$ are the error amplitudes, and $\delta_1$ describe the amplitude error strength.  To make the control sequence robust against the errors, $O(\omega_k^\alpha)$ in \eqref{eq:O} should be replaced by the averaged target function 
$\Bar{O}(\omega_k^\alpha)=\int\cdots\int \mathrm{d}\delta_l\prod _lP(\delta_l)O(\omega_k^\alpha,\delta)$,
where $p(\delta_l)$ is the error distribution and the integral is performed for all the $\delta_l$.

As an example, the general theory is explicitly demonstrated in the case of NV centers, where the thermal noise term is taken to be $\delta_2S_z$, and the amplitude error strength is $\delta_1$. By setting $\eta=1\times10^{-6}$ and total control time $\tau=400$ ns with $N=100$, the control sequences for the Hadamard gate are obtained and shown in Fig.~\ref{fig:QOCT}(d) and (e). The performances of the original control sequences and the searched ones are illustrated in Fig.~\ref{fig:robustness}(a) and (b). For a typical NV center in a type-II diamond, the thermal noise $\delta_2$ has a standard deviation of 130 kHz, and the $\delta_1$ is about $1\%$. Fig.~\ref{fig:robustness}(a) and (b) show that the fidelity for the former scheme is about $95\%$ while that for the optimization scheme is over $99.9\%$.   
In the same way, the performance of a CNOT gate built between the NV center and an adjacent $^{13}C$ are shown in Fig.~\ref{fig:robustness}(c) and (d). With the same $\delta_1$ and $\delta_2$, the fidelity for the former scheme is about $91\%$, and that for the optimization scheme is still over $99.9\%$. Note that the fidelities evaluated for the Hadamard and CNOT gate based on the original scheme coincide with the average gate fidelity reported in Ref.~\cite{zu2014experimental}, implying that the numerical simulation is faithful. Therefore, the improvement in the fidelity guided by the QOCT scheme is promising for future quantum experiments.

\section{Conclusions and outlook}
\label{sec:conclude}

It is reasonable to expect that the quantum computing community is decades away from building general-purpose fully error-corrected quantum computers, which can perform daunting tasks such as efficient prime number factorization. Existing quantum computers have at most fifty to seventy noisy qubits \cite{arute2019quantum,zhong2020quantum,wu2021strong}. 
Such noisy intermediate-scale quantum (NISQ) machines can, however, still execute tasks that are intractable with classical devices. 
This contemporary quantum computing paradigm is often referred to as the `NISQ era' \cite{preskill2018quantum}, in which the design of algorithms \cite{bharti2021noisy,lau2022nisq,buluta2009quantum,georgescu2014quantum} requires a thorough consideration of the limited accessible resources to execute classically demanding tasks.
In order to push forward the performance of such NISQ devices, one way to advance in this direction, we believe, is to combine pulse-level control \cite{shevchenko2010landau,alexander2020qiskit} of traditional dynamical quantum gates with geometric quantum computation (GQC) and holonomic quantum computation (HQC). 
We hope this review article will play an important role in achieving this goal.

We have presented this review article in a pedagogical way to benefit both novices and experts in the field of GQC and HQC. 
Both GQC and HQC possess universality, fast speed, and compatibility for robust quantum computing. Notably, different flavors of GQC and HQC can be adopted and combined together as ingredients of a universal scheme for qubit computation. Novel holonomic schemes may also be proposed based on quantum holonomies for qudits \cite{xu2021realizing,andre2022qudit} or  measurement-induced quantum holonomies \cite{ashhab2010control,kofman2012nonperturbative,oi2014unitary,mommers2021universal}. 

Geometric phases and quantum holonomies are deeply connected with topological concepts (such as the Chern number, for a recent paper, see Ref. \cite{weisbrich2021second}) and phenomena (such as the quantum Hall effect \cite{semenoff1986non,zhang2020topological}). Therefore, extending GQC and HQC to the topological regime could be a possible future research direction, with the motivation that topological quantities are highly insensitive to local errors. On the other hand, geometric phases are helpful in constructing universal gates in topological quantum computation \cite{karzig2016universal}. In this sense, researchers working in both research fields can benefit from each other.

Enhancing the robustness of geometric and holonomic gates is another direction that requires further research. A possible approach is to combine them with other quantum technologies. Some works have been done in implementing geometric or holonomic gates with protection from decoherence-free subspaces, noiseless subsystems, dynamical decoupling, quantum error correction, and dynamically correct approach \cite{dong2021doubly}. The aim of these efforts is to selectively bring together the advantages of the error resilience of geometric and holonomic gates and the environmental noise resilience of the aforementioned techniques. For systematic errors, one can resort to the use of quantum feedback control \cite{zhang2017quantum} and quantum optimal control \cite{werschnik2007quantum} theory. 

Due to its experimental feasibility, nonadiabatic HQC has attracted much attention recently. Almost every family of nonadiabatic holonomic schemes has been experimentally demonstrated, and some of them have been implemented on different experimental platforms. 
However, the gate fidelities in these experimental realizations are still below the threshold required by fault-tolerant quantum error correction.
Since the fidelity of holonomic gates keeps increasing and improving, further progress in experiments could be made by choosing suitable schemes for certain kinds of errors in practice. 
In this manner, we believe that GQC and HQC could pave the way towards semi-fault-tolerant quantum computation in the NISQ era in the near future.

\section*{Acknowledgement}
We acknowledge fruitful discussions with Al\'an Aspuru-Guzik, Xiao-Dong Yu, Bao-Jie Liu, Guofu Xu, and Peizi Zhao.
J.Z. is supported by the National Natural Science Foundation of China (Grant No. 12004206). 
L.-C.K. acknowledges funding from the National Research Foundation and the Ministry of Education, Singapore.
E.S. acknowledges financial support from the Swedish Research Council (VR) through Grant No. 2017-03832. 
D.M.T. is supported by the National Natural Science Foundation of China (Grant No. 12174224).

\bibliographystyle{unsrt}

\bibliography{references}

\begin{thebibliography}{100}

\bibitem{berry90}
M.~V. Berry.
\newblock Anticipations of the geometric phase.
\newblock {\em Phys. Today}, 43:34--40, 1990.

\bibitem{wilczek1989geometric}
F.~Wilczek and A.~Shapere.
\newblock {\em Geometric phases in physics}.
\newblock World Scientific, 1989.

\bibitem{bohm2003the}
A.~Bohm, A.~Mostafazadeh, H.~Koizumi, Q.~Niu, and J.~Zwanziger.
\newblock {\em The Geometric Phase in Quantum Systems: Foundations,
  Mathematical Concepts, and Applications in Molecular and Condensed Matter
  Physics}.
\newblock Springer Verlag, 2003.

\bibitem{chruscinski2004}
D.~Chru\'sci\'nski and A.~Jamio{\l}kowski.
\newblock {\em Geometric Phases in Classical and Quantum Mechanics}.
\newblock Springer Verlag, 2004.

\bibitem{berry1984quantal}
M.~V. Berry.
\newblock Quantal phase factors accompanying adiabatic changes.
\newblock {\em Proc. R. Soc. A}, 392:45--57, 1984.

\bibitem{longuet1975intersection}
H.~C. Longuet-Higgins.
\newblock The intersection of potential energy surfaces in polyatomic
  molecules.
\newblock {\em Proc. R. Soc. A}, 344:147--156, 1975.

\bibitem{stone1976spin}
A.~J. Stone.
\newblock Spin-orbit coupling and the intersection of potential energy surfaces
  in polyatomic molecules.
\newblock {\em Proc. R. Soc. A}, 351:141--150, 1976.

\bibitem{rytov1938transmitting}
S.~M. Rytov.
\newblock On the transmitting from wave to geometric optics.
\newblock {\em Dokl. Akad. Nauk. SSSR}, 18:263–265, 1938.

\bibitem{vladimirskii1941rotation}
V.~V. Vladimirskii.
\newblock The rotation of a polarization plane for curved light ray.
\newblock {\em Dokl. Akad. Nauk. SSSR}, 31:222--224, 1941.

\bibitem{pancharatnam1956}
S.~Pancharatnam.
\newblock Generalized theory of interference and its applications.
\newblock {\em Proc. Indian Acad. Sci.}, 44:398--417, 1956.

\bibitem{bhandari1997polarization}
R.~Bhandari.
\newblock Polarization of light and topological phases.
\newblock {\em Physics Reports}, 281:1--64, 1997.

\bibitem{bliokh2015spin}
K.~Y. Bliokh, F.~J. Rodr{\'\i}guez-Fortu{\~n}o, F.~Nori, and A.~V. Zayats.
\newblock Spin-orbit interactions of light.
\newblock {\em Nat. Photonics}, 9:796--808, 2015.

\bibitem{cisowski2022geometric}
C.~Cisowski, J.~B. G\"otte, and S.~Franke-Arnold.
\newblock Geometric phases of light: Insights from fiber bundle theory.
\newblock {\em Rev. Mod. Phys.}, 94:031001, 2022.

\bibitem{aharonov1959significance}
Y.~Aharonov and D.~Bohm.
\newblock Significance of electromagnetic potentials in the quantum theory.
\newblock {\em Phys. Rev.}, 115:485--491, 1959.

\bibitem{mead1979on}
C.~A. Mead and D.~G. Truhlar.
\newblock On the determination of {B}orn–{O}ppenheimer nuclear motion wave
  functions including complications due to conical intersections and identical
  nuclei.
\newblock {\em J. Chem. Phys.}, 70:2284--2296, 1979.

\bibitem{longuet1958studies}
H.~C. Longuet-Higgins, U.~\"Opik, M.~H.~L. Pryce, and R.~A. Sack.
\newblock Studies of the {Jahn-Teller} effect {.II.} {T}he dynamical problem.
\newblock {\em Proc. R. Soc. Lond. A}, 244:1--16, 1958.

\bibitem{aharonov1987phase}
Y.~Aharonov and J.~Anandan.
\newblock Phase change during a cyclic quantum evolution.
\newblock {\em Phys. Rev. Lett.}, 58:1593--1596, 1987.

\bibitem{Samuel1988general}
J.~Samuel and R.~Bhandari.
\newblock General setting for {Berry's} phase.
\newblock {\em Phys. Rev. Lett.}, 60:2339--2342, 1988.

\bibitem{simon1983holonomy}
B.~Simon.
\newblock Holonomy, the quantum adiabatic theorem, and {B}erry's phase.
\newblock {\em Phys. Rev. Lett.}, 51:2167--2170, 1983.

\bibitem{wilczek1984appearance}
F.~Wilczek and A.~Zee.
\newblock Appearance of gauge structure in simple dynamical systems.
\newblock {\em Phys. Rev. Lett.}, 52:2111--2114, 1984.

\bibitem{anadan1988non}
J.~Anandan.
\newblock Non-adiabatic non-{A}belian geometric phase.
\newblock {\em Phys. Lett. A}, 133:171--175, 1988.

\bibitem{mostafazadeh99}
A.~Mostafazadeh.
\newblock Noncyclic geometric phase and its non-{A}belian generalization.
\newblock {\em J. Phys. A: Math. Gen.}, 32:8157--8171, 1999.

\bibitem{kult2006noncyclic}
D.~Kult, J.~\AA{}berg, and E.~Sj\"oqvist.
\newblock Noncyclic geometric changes of quantum states.
\newblock {\em Phys. Rev. A}, 74:022106, 2006.

\bibitem{moore1991calculation}
D.~J. Moore.
\newblock The calculation of nonadiabatic {B}erry phases.
\newblock {\em Phys. Rep.}, 210:1--43, 1991.

\bibitem{kolodrubetz2017geometry}
M.~Kolodrubetz, D.~Sels, P.~Mehta, and A.~Polkovnikov.
\newblock {Geometry and non-adiabatic response in quantum and classical
  systems}.
\newblock {\em Phys. Rep.}, 697:1--87, 2017.

\bibitem{von1927wahrscheinlichkeitstheoretischer}
J.~Von~Neumann.
\newblock Wahrscheinlichkeitstheoretischer aufbau der quantenmechanik.
\newblock {\em Nachrichten von der Gesellschaft der Wissenschaften zu
  G{\"o}ttingen, Mathematisch-Physikalische Klasse}, 1927:245--272, 1927.

\bibitem{uhlmann1986parallel}
A.~Uhlmann.
\newblock Parallel transport and “quantum holonomy” along density
  operators.
\newblock {\em Rep. Math. Phys.}, 24:229--240, 1986.

\bibitem{sjoqvist2000geometric}
E.~Sj\"oqvist, A.~K. Pati, A.~Ekert, J.~S. Anandan, M.~Ericsson, D.~K.~L. Oi,
  and V.~Vedral.
\newblock Geometric phases for mixed states in interferometry.
\newblock {\em Phys. Rev. Lett.}, 85:2845--2849, 2000.

\bibitem{tong2004kinematic}
D.~M. Tong, E.~Sj\"oqvist, L.~C. Kwek, and C.~H. Oh.
\newblock Kinematic approach to the mixed state geometric phase in nonunitary
  evolution.
\newblock {\em Phys. Rev. Lett.}, 93:080405, 2004.

\bibitem{singh2003geometric}
K.~Singh, D.~M. Tong, K.~Basu, J.~L. Chen, and J.~F. Du.
\newblock Geometric phases for nondegenerate and degenerate mixed states.
\newblock {\em Phys. Rev. A}, 67:032106, 2003.

\bibitem{filipp2003off}
S.~Filipp and E.~Sj\"oqvist.
\newblock Off-diagonal geometric phase for mixed states.
\newblock {\em Phys. Rev. Lett.}, 90:050403, 2003.

\bibitem{ericsson2003mixed}
M.~Ericsson, A.~K. Pati, E.~Sj\"oqvist, J.~Br\"annlund, and D.~K.~L. Oi.
\newblock Mixed state geometric phases, entangled systems, and local unitary
  transformations.
\newblock {\em Phys. Rev. Lett.}, 91:090405, 2003.

\bibitem{chaturvedi04}
S.~Chaturvedi, E.~Ercolessi, G.~Marmo, G.~Morandi, N.~Mukunda, and R.~Simon.
\newblock Geometric phase for mixed states: a differential geometric approach.
\newblock {\em Eur. Phys. J. C}, 35:413--423, 2004.

\bibitem{marzlin2004geometric}
K.-P. Marzlin, S.~Ghose, and B.~C. Sanders.
\newblock Geometric phase distributions for open quantum systems.
\newblock {\em Phys. Rev. Lett.}, 93:260402, 2004.

\bibitem{mikael2005mixed}
M.~Nordling and E.~Sj\"oqvist.
\newblock Mixed-state non-{A}belian holonomy for subsystems.
\newblock {\em Phys. Rev. A}, 71:012110, 2005.

\bibitem{carollo2003geometric}
A.~Carollo, I.~Fuentes-Guridi, M.~Fran\ifmmode\mbox{\c{c}}\else\c{c}\fi{}a
  Santos, and V.~Vedral.
\newblock Geometric phase in open systems.
\newblock {\em Phys. Rev. Lett.}, 90:160402, 2003.

\bibitem{fuentes2005holo}
I.~Fuentes-Guridi, F.~Girelli, and E.~Livine.
\newblock Holonomic quantum computation in the presence of decoherence.
\newblock {\em Phys. Rev. Lett.}, 94:020503, 2005.

\bibitem{ericsson2003general}
M.~Ericsson, E.~Sj\"oqvist, J.~Br\"annlund, D.~K.~L. Oi, and A.~K. Pati.
\newblock Generalization of the geometric phase to completely positive maps.
\newblock {\em Phys. Rev. A}, 67:020101, 2003.

\bibitem{ramberg2019environment}
N.~Ramberg and E.~Sj{\"o}qvist.
\newblock Environment-assisted holonomic quantum maps.
\newblock {\em Phys. Rev. Lett.}, 122:140501, 2019.

\bibitem{thunstrom2005weak}
P.~Thunstr\"om, J.~{\AA}berg, and E.~Sj\"oqvist.
\newblock Adiabatic approximation for weakly open systems.
\newblock {\em Phys. Rev. A}, 72:022328, 2005.

\bibitem{sarandy2006abelian}
M.~S. Sarandy and D.~A. Lidar.
\newblock {A}belian and non-{A}belian geometric phases in adiabatic open
  quantum systems.
\newblock {\em Phys. Rev. A}, 73:062101, 2006.

\bibitem{oreshkov2010adiabatic}
O.~Oreshkov and J.~Calsamiglia.
\newblock Adiabatic {M}arkovian dynamics.
\newblock {\em Phys. Rev. Lett.}, 105:050503, 2010.

\bibitem{carollo2006coherent}
A.~Carollo, M.~F. Santos, and V.~Vedral.
\newblock Coherent quantum evolution via reservoir driven holonomies.
\newblock {\em Phys. Rev. Lett.}, 96:020403, 2006.

\bibitem{zheng2012dissipation}
S.-B. Zheng.
\newblock Dissipation-induced geometric phase for an atom trapped in an optical
  cavity.
\newblock {\em Phys. Rev. A}, 85:052106, 2012.

\bibitem{dasgupta2007decoherence}
S.~Dasgupta and D.~A. Lidar.
\newblock Decoherence-induced geometric phase in a multilevel atomic system.
\newblock {\em J. Phys. B: At. Mol. Opt. Phys.}, 40:S127, 2007.

\bibitem{kult2008holonomy}
D.~Kult, J.~{\AA}berg, and E.~Sj{\"o}qvist.
\newblock Holonomy for quantum channels.
\newblock {\em Phys. Rev. A}, 77:012114, 2008.

\bibitem{khosla2013quantum}
K.~E. Khosla, M.~R. Vanner, W.~P. Bowen, and G.~J. Milburn.
\newblock {Quantum state preparation of a mechanical resonator using an
  optomechanical geometric phase}.
\newblock {\em New J. Phys.}, 15:043025, 2013.

\bibitem{fu2019geometric}
H.~Fu, Z.-C. Gong, T.-H. Mao, C.-Y. Shen, C.-P. Sun, S.~Yi, Y.~Li, and G.-Y.
  Cao.
\newblock Geometric energy transfer in a {S}t{\"u}ckelberg interferometer of
  two parametrically coupled mechanical modes.
\newblock {\em Phys. Rev. Applied}, 11:034010, 2019.

\bibitem{kippenberg2008cavity}
T.~J. Kippenberg and K.~J. Vahala.
\newblock {Cavity Optomechanics: Back-Action at the Mesoscale}.
\newblock {\em Science}, 321:1172--1176, 2008.

\bibitem{aspelmeyer2014cavity}
M.~Aspelmeyer, T.~J. Kippenberg, and F.~Marquardt.
\newblock Cavity optomechanics.
\newblock {\em Rev. Mod. Phys.}, 86:1391, 2014.

\bibitem{xiong2022higher}
W.~Xiong, Z.~Li, G.-Q. Zhang, M.~Wang, H.-C. Li, X.-Q. Luo, and J.~Chen.
\newblock Higher-order exceptional point in a blue-detuned non-hermitian cavity
  optomechanical system.
\newblock {\em Phys. Rev. A}, 106:033518, 2022.

\bibitem{vedral2003geometric}
V.~Vedral.
\newblock Geometric phases and topological quantum computation.
\newblock {\em Int. J. Quantum Inf.}, 1:1--23, 2003.

\bibitem{sjoqvist2008trend}
E.~Sj{\"o}qvist.
\newblock A new phase in quantum computation.
\newblock {\em Physics}, 1:35, 2008.

\bibitem{sjoqvist15}
E.~Sj{\"o}qvist.
\newblock Geometric phases in quantum information.
\newblock {\em Int. J. Quantum Chem.}, 115:1311--1326, 2015.

\bibitem{preskill2018quantum}
J.~Preskill.
\newblock Quantum computing in the {NISQ} era and beyond.
\newblock {\em Quantum}, 2:79, 2018.

\bibitem{jones2000geometric}
J.~A. Jones, V.~Vedral, A.~Ekert, and G.~Castagnoli.
\newblock Geometric quantum computation using nuclear magnetic resonance.
\newblock {\em Nature}, 403:869--871, 2000.

\bibitem{ekert2000geometric}
A.~Ekert, M.~Ericsson, P.~Hayden, H.~Inamori, J.~A. Jones, D.~K.~L. Oi, and
  V.~Vedral.
\newblock Geometric quantum computation.
\newblock {\em J. Mod. Opt.}, 47:2501--2513, 2000.

\bibitem{zanardi1999holonomic}
P.~Zanardi and M.~Rasetti.
\newblock Holonomic quantum computation.
\newblock {\em Phys. Lett. A}, 264:94--99, 1999.

\bibitem{falci00detection}
G.~Falci, R.~Fazio, G.~M. Palma, J.~Siewert, and V.~Vedral.
\newblock Detection of geometric phases in superconducting nanocircuits.
\newblock {\em Nature}, 407:355--358, 2000.

\bibitem{wang2001nonadiabatic}
X.-B. Wang and K.~Matsumoto.
\newblock Nonadiabatic conditional geometric phase shift with {NMR}.
\newblock {\em Phys. Rev. Lett.}, 87:097901, 2001.

\bibitem{zhu02}
S.-L. Zhu and Z.~D. Wang.
\newblock Implementation of universal quantum gates based on nonadiabatic
  geometric phases.
\newblock {\em Phys. Rev. Lett.}, 89:097902, 2002.

\bibitem{li02non}
X.-Q. Li, L.-X. Cen, G.~Huang, L.~Ma, and Y.-J. Yan.
\newblock Nonadiabatic geometric quantum computation with trapped ions.
\newblock {\em Phys. Rev. A}, 66:042320, 2002.

\bibitem{li2002ultrafast}
X.-Q. Li, C.-Y. Hu, L.-X. Cen, H.-Z. Zheng, and Y.~J. Yan.
\newblock Ultrafast geometric manipulation of electron spin and detection of
  the geometric phase via {F}araday rotation spectroscopy.
\newblock {\em Phys. Rev. B}, 66:235207, 2002.

\bibitem{zhu02geometric}
S.-L. Zhu and Z.~D. Wang.
\newblock Geometric phase shift in quantum computation using superconducting
  nanocircuits: Nonadiabatic effects.
\newblock {\em Phys. Rev. A}, 66:042322, 2002.

\bibitem{solinas2003nonadiabatic}
P.~Solinas, P.~Zanardi, N.~Zangh{\`\i}, and F.~Rossi.
\newblock Nonadiabatic geometrical quantum gates in semiconductor quantum dots.
\newblock {\em Phys. Rev. A}, 67:052309, 2003.

\bibitem{zhao17rydberg}
P.~Z. Zhao, X.-D. Cui, G.~F. Xu, E.~Sj\"oqvist, and D.~M. Tong.
\newblock Rydberg-atom-based scheme of nonadiabatic geometric quantum
  computation.
\newblock {\em Phys. Rev. A}, 96:052316, 2017.

\bibitem{chen2018non}
T.~Chen and Z.-Y. Xue.
\newblock Nonadiabatic geometric quantum computation with parametrically
  tunable coupling.
\newblock {\em Phys. Rev. Appl.}, 10:054051, 2018.

\bibitem{zhang20high}
C.~Zhang, T.~Chen, S.~Li, X.~Wang, and Z.-Y. Xue.
\newblock High-fidelity geometric gate for silicon-based spin qubits.
\newblock {\em Phys. Rev. A}, 101:052302, 2020.

\bibitem{teo2005geometric}
J.~C.~Y. Teo and Z.~D. Wang.
\newblock Geometric phase in eigenspace evolution of invariant and adiabatic
  action operators.
\newblock {\em Phys. Rev. Lett.}, 95:050406, 2005.

\bibitem{shao2007implementation}
L.B. Shao, Z.D. Wang, and D.~Y. Xing.
\newblock Implementation of quantum gates based on geometric phases accumulated
  in the eigenstates of periodic invariant operators.
\newblock {\em Phys. Rev. A}, 75:014301, 2007.

\bibitem{wang2009noncyclic}
Z.~S. Wang, G.~Q. Liu, and Y.~H. Ji.
\newblock Noncyclic geometric quantum computation in a
  nuclear-magnetic-resonance system.
\newblock {\em Phys. Rev. A}, 79:054301, 2009.

\bibitem{wang2009geometric}
Z.~S. Wang.
\newblock Geometric quantum computation and dynamical invariant operators.
\newblock {\em Phys. Rev. A}, 79:024304, 2009.

\bibitem{leibfried2003experi}
D.~Leibfried, B.~DeMarco, V.~Meyer, D.~Lucas, M.~Barrett, J.~Britton, W.~M.
  Itano, B.~Jelenkovi\'{c}, C.~Langer, T.~Rosenband, and D.~J. Wineland.
\newblock Experimental demonstration of a robust, high-fidelity geometric two
  ion-qubit phase gate.
\newblock {\em Nature}, 422:412--415, 2003.

\bibitem{zhu03unconventional}
S.-L. Zhu and Z.~D. Wang.
\newblock Unconventional geometric quantum computation.
\newblock {\em Phys. Rev. Lett.}, 91:187902, 2003.

\bibitem{pachos1999non}
J.~K. Pachos, P.~Zanardi, and M.~Rasetti.
\newblock Non-{A}belian {B}erry connections for quantum computation.
\newblock {\em Phys. Rev. A}, 61:010305, 1999.

\bibitem{duan2001geometric}
L.-M. Duan, J.~I. Cirac, and P.~Zoller.
\newblock Geometric manipulation of trapped ions for quantum computation.
\newblock {\em Science}, 292:1695--1697, 2001.

\bibitem{faoro2003non}
L.~Faoro, J.~Siewert, and R.~Fazio.
\newblock Non-{A}belian holonomies, charge pumping, and quantum computation
  with {J}osephson junctions.
\newblock {\em Phys. Rev. Lett.}, 90:028301, 2003.

\bibitem{solinas2003semiconductor}
P.~Solinas, P.~Zanardi, N.~Zangh{\`\i}, and F.~Rossi.
\newblock Semiconductor-based geometrical quantum gates.
\newblock {\em Phys. Rev. B}, 67:121307, 2003.

\bibitem{toyoda2013realization}
K.~Toyoda, K.~Uchida, A.~Noguchi, S.~Haze, and S.~Urabe.
\newblock Realization of holonomic single-qubit operations.
\newblock {\em Phys. Rev. A}, 87:052307, 2013.

\bibitem{sjoqvist2012non}
E.~Sj{\"o}qvist, D.-M. Tong, L.~M. Andersson, B.~Hessmo, M.~Johansson, and
  K.~Singh.
\newblock Non-adiabatic holonomic quantum computation.
\newblock {\em New J. Phys.}, 14:103035, 2012.

\bibitem{xu2012non}
G.~F. Xu, J.~Zhang, D.~M. Tong, E.~Sj\"oqvist, and L.~C. Kwek.
\newblock Nonadiabatic holonomic quantum computation in decoherence-free
  subspaces.
\newblock {\em Phys. Rev. Lett.}, 109:170501, 2012.

\bibitem{abdumalikov2013experimental}
A.~A. Abdumalikov~Jr, J.~M. Fink, K.~Juliusson, M.~Pechal, S.~Berger,
  A.~Wallraff, and S.~Filipp.
\newblock Experimental realization of non-{A}belian non-adiabatic geometric
  gates.
\newblock {\em Nature}, 496:482--485, 2013.

\bibitem{feng2013experimental}
G.~R. Feng, G.~F. Xu, and G.~L. Long.
\newblock Experimental realization of nonadiabatic holonomic quantum
  computation.
\newblock {\em Phys. Rev. Lett.}, 110:190501, 2013.

\bibitem{arroyo2014room}
S.~Arroyo-Camejo, A.~Lazariev, S.~W. Hell, and G.~Balasubramanian.
\newblock Room temperature high-fidelity holonomic single-qubit gate on a
  solid-state spin.
\newblock {\em Nat. Commun.}, 5:4870, 2014.

\bibitem{zu2014experimental}
C.~Zu, W.-B. Wang, L.~He, W.-G. Zhang, C.-Y. Dai, F.~Wang, and L.-M. Duan.
\newblock Experimental realization of universal geometric quantum gates with
  solid-state spins.
\newblock {\em Nature}, 514:72--75, 2014.

\bibitem{xu2015non}
G.~F. Xu, C.~L. Liu, P.~Z. Zhao, and D.~M. Tong.
\newblock Nonadiabatic holonomic gates realized by a single-shot
  implementation.
\newblock {\em Phys. Rev. A}, 92:052302, 2015.

\bibitem{sjoqvist2016non}
E.~Sj{\"o}qvist.
\newblock Nonadiabatic holonomic single-qubit gates in off-resonant {$\Lambda$}
  systems.
\newblock {\em Phys. Lett. A}, 380:65--67, 2016.

\bibitem{xu2018path}
G.~F. Xu, D.~M. Tong, and E.~Sj{\"o}qvist.
\newblock Path-shortening realizations of nonadiabatic holonomic gates.
\newblock {\em Phys. Rev. A}, 98:052315, 2018.

\bibitem{herterich2016single}
E.~Herterich and E.~Sj{\"o}qvist.
\newblock Single-loop multiple-pulse nonadiabatic holonomic quantum gates.
\newblock {\em Phys. Rev. A}, 94:052310, 2016.

\bibitem{mousolou2014universal}
V.~Azimi~Mousolou, C.~M. Canali, and E.~Sj{\"o}qvist.
\newblock Universal non-adiabatic holonomic gates in quantum dots and
  single-molecule magnets.
\newblock {\em New J. Phys.}, 16:013029, 2014.

\bibitem{gurkan2015realization}
Z.~N. G{\"u}rkan and E.~Sj{\"o}qvist.
\newblock Realization of a holonomic quantum computer in a chain of three-level
  systems.
\newblock {\em Phys. Lett. A}, 379:3050--3053, 2015.

\bibitem{zhang2018holonomic}
J.~Zhang, S.~J. Devitt, J.~Q. You, and F.~Nori.
\newblock Holonomic surface codes for fault-tolerant quantum computation.
\newblock {\em Phys. Rev. A}, 97:022335, 2018.

\bibitem{wang2020dephasing}
Y.~Wang, Y.~Su, X.~Chen, and C.~Wu.
\newblock Dephasing-protected scalable holonomic quantum computation on a
  {R}abi lattice.
\newblock {\em Phys. Rev. Appl.}, 14:044043, 2020.

\bibitem{zhang2015fast}
J.~Zhang, T.~H. Kyaw, D.~M. Tong, E.~Sj{\"o}qvist, and L.-C. Kwek.
\newblock Fast non-{A}belian geometric gates via transitionless quantum
  driving.
\newblock {\em Sci. Rep.}, 5:1--7, 2015.

\bibitem{liu2019plug}
B.-J. Liu, X.-K. Song, Z.-Y. Xue, X.~Wang, and M.-H. Yung.
\newblock Plug-and-play approach to nonadiabatic geometric quantum gates.
\newblock {\em Phys. Rev. Lett.}, 123:100501, 2019.

\bibitem{li2020approach}
K.~Z. Li, P.~Z. Zhao, and D.~M. Tong.
\newblock Approach to realizing nonadiabatic geometric gates with prescribed
  evolution paths.
\newblock {\em Phys. Rev. Research}, 2:023295, 2020.

\bibitem{zhao2020general}
P.~Z. Zhao, K.~Z. Li, G.~F. Xu, and D.~M. Tong.
\newblock General approach for constructing {H}amiltonians for nonadiabatic
  holonomic quantum computation.
\newblock {\em Phys. Rev. A}, 101:062306, 2020.

\bibitem{liu2021super}
B.-J. Liu, Y.-S. Wang, and M.-H. Yung.
\newblock Super-robust nonadiabatic geometric quantum control.
\newblock {\em Phys. Rev. Research}, 3:L032066, 2021.

\bibitem{werschnik2007quantum}
J.~Werschnik and E.~K.~U. Gross.
\newblock Quantum optimal control theory.
\newblock {\em J. Phys. B: At. Mol. Opt. Phys.}, 40:R175--R211, 2007.

\bibitem{clerk2010introduction}
A.~A. Clerk, M.~H. Devoret, S.~M. Girvin, F.~Marquardt, and R.~J. Schoelkopf.
\newblock Introduction to quantum noise, measurement, and amplification.
\newblock {\em Rev. Mod. Phys.}, 82:1155, 2010.

\bibitem{schlosshauer2019quantum}
M.~Schlosshauer.
\newblock Quantum decoherence.
\newblock {\em Phys. Rep.}, 831:1--57, 2019.

\bibitem{zanardi1997noiseless}
P.~Zanardi and M.~Rasetti.
\newblock Noiseless quantum codes.
\newblock {\em Phys. Rev. Lett.}, 79:3306--3309, 1997.

\bibitem{duan1997preserving}
L.-M. Duan and G.-C. Guo.
\newblock Preserving coherence in quantum computation by pairing quantum bits.
\newblock {\em Phys. Rev. Lett.}, 79:1953--1956, 1997.

\bibitem{lidar1998decoherence}
D.~A Lidar, I.~L. Chuang, and K.~B. Whaley.
\newblock Decoherence-free subspaces for quantum computation.
\newblock {\em Phys. Rev. Lett.}, 81:2594--2597, 1998.

\bibitem{knill2000theory}
E.~Knill, R.~Laflamme, and L.~Viola.
\newblock Theory of quantum error correction for general noise.
\newblock {\em Phys. Rev. Lett.}, 84:2525--2528, 2000.

\bibitem{yang2011preserving}
W.~Yang, Z.-Y. Wang, and R.-B. Liu.
\newblock Preserving qubit coherence by dynamical decoupling.
\newblock {\em Front. Phys.}, 6:2--14, 2011.

\bibitem{terhal2015quantum}
B.~M. Terhal.
\newblock Quantum error correction for quantum memories.
\newblock {\em Rev. Mod. Phys.}, 87:307--346, 2015.

\bibitem{lidar2013quantum}
D.~A. Lidar and T.~A. Brun.
\newblock {\em Quantum error correction}.
\newblock Cambridge university press, 2013.

\bibitem{feng2009geometric}
X.-L. Feng, C.~Wu, H.~Sun, and C.~H. Oh.
\newblock Geometric entangling gates in decoherence-free subspaces with minimal
  requirements.
\newblock {\em Phys. Rev. Lett.}, 103:200501, 2009.

\bibitem{wu2005holonomic}
L-A Wu, P.~Zanardi, and D.~A. Lidar.
\newblock Holonomic quantum computation in decoherence-free subspaces.
\newblock {\em Phys. Rev. Lett.}, 95:130501, 2005.

\bibitem{zhang2014quantum}
J.~Zhang, L.-C. Kwek, E.~Sj{\"o}qvist, D.~M. Tong, and P.~Zanardi.
\newblock Quantum computation in noiseless subsystems with fast non-{A}belian
  holonomies.
\newblock {\em Phys. Rev. A}, 89:042302, 2014.

\bibitem{xu2014protecting}
G.~Xu and G.~Long.
\newblock Protecting geometric gates by dynamical decoupling.
\newblock {\em Phys. Rev. A}, 90:022323, 2014.

\bibitem{wu2020universal}
X.~Wu and P.~Z. Zhao.
\newblock Universal nonadiabatic geometric gates protected by dynamical
  decoupling.
\newblock {\em Phys. Rev. A}, 102:032627, 2020.

\bibitem{zhao2021dynamical}
P.~Z. Zhao, X.~Wu, and D.~M. Tong.
\newblock Dynamical-decoupling-protected nonadiabatic holonomic quantum
  computation.
\newblock {\em Phys. Rev. A}, 103:012205, 2021.

\bibitem{oreshkov2009fault}
O.~Oreshkov, T.~A. Brun, and D.~A. Lidar.
\newblock Fault-tolerant holonomic quantum computation.
\newblock {\em Phys. Rev. Lett.}, 102:070502, 2009.

\bibitem{oreshkov2009scheme}
O.~Oreshkov, T.~A. Brun, and D.~A. Lidar.
\newblock Scheme for fault-tolerant holonomic computation on stabilizer codes.
\newblock {\em Phys. Rev. A}, 80:022325, 2009.

\bibitem{yin2007implementation}
Z.-Q. Yin, F.-l. Li, and P.~Peng.
\newblock Implementation of holonomic quantum computation through engineering
  and manipulating the environment.
\newblock {\em Phys. Rev. A}, 76:062311, 2007.

\bibitem{marzlin2004incosistency}
K.-P. Marzlin and B.~C. Sanders.
\newblock Inconsistency in the application of the adiabatic theorem.
\newblock {\em Phys. Rev. Lett.}, 93:160408, 2004.

\bibitem{tong2005quantitative}
D.~M. Tong, K.~Singh, L.~C. Kwek, and C.~H. Oh.
\newblock Quantitative conditions do not guarantee the validity of the
  adiabatic approximation.
\newblock {\em Phys. Rev. Lett.}, 95:110407, 2005.

\bibitem{tong2010quantitative}
D.~M. Tong.
\newblock Quantitative condition is necessary in guaranteeing the validity of
  the adiabatic approximation.
\newblock {\em Phys. Rev. Lett.}, 104:120401, 2010.

\bibitem{liu2005optical}
Y.-X. Liu, J.~Q. You, L.~F. Wei, C.~P. Sun, and F.~Nori.
\newblock {Optical Selection Rules and Phase-Dependent Adiabatic State Control
  in a Superconducting Quantum Circuit}.
\newblock {\em Phys. Rev. Lett.}, 95:087001, 2005.

\bibitem{mousolou2016nano}
V.~Azimi~Mousolou, C.~M. Canali, and E.~Sj{\"o}qvist.
\newblock Spin-electric {B}erry phase shift in triangular molecular magnets.
\newblock {\em Phys. Rev. B}, 94:60011, 2016.

\bibitem{vandersypen2005NMR}
L.~M.~K. Vandersypen and I.~L. Chuang.
\newblock {NMR} techniques for quantum control and computation.
\newblock {\em Rev. Mod. Phys.}, 76:1037--1069, 2005.

\bibitem{buluta2011natural}
I.~Buluta, S.~Ashhab, and F.~Nori.
\newblock Natural and artificial atoms for quantum computation.
\newblock {\em Rep. Prog. Phys.}, 74:104401, 2011.

\bibitem{gu2017microwave}
X.~Gu, A.~F. Kockum, A.~Miranowicz, Y.-X. Liu, and F.~Nori.
\newblock Microwave photonics with superconducting quantum circuits.
\newblock {\em Phys. Rep.}, 718:1--102, 2017.

\bibitem{kockum2019quantum}
A.~F. Kockum and F.~Nori.
\newblock Quantum bits with {J}osephson junctions.
\newblock In {\em Fundamentals and Frontiers of the Josephson Effect}, pages
  703--741. Springer, 2019.

\bibitem{shnirman1997quantum}
A.~Shnirman, G.~Sch\"on, and Z.~Hermon.
\newblock Quantum manipulations of small {J}osephson junctions.
\newblock {\em Phys. Rev. Lett.}, 79:2371--2374, 1997.

\bibitem{makhlin1999josephson}
Y.~Makhlin, G.~Sc\"ohn, and A.~Shnirman.
\newblock Josephson-junction qubits with controlled couplings.
\newblock {\em Nature}, 398:305--307, 1999.

\bibitem{wei2008controllable}
L.~F. Wei, J.~R. Johansson, L.~X. Cen, S.~Ashhab, and F.~Nori.
\newblock Controllable coherent population transfers in superconducting qubits
  for quantum computing.
\newblock {\em Phys. Rev. Lett.}, 100:113601, 2008.

\bibitem{you2011atomic}
J.~Q. You and F.~Nori.
\newblock Atomic physics and quantum optics using superconducting circuits.
\newblock {\em Nature}, 474:589--597, 2011.

\bibitem{kockum2019ultrastrong}
A.~F. Kockum, A.~Miranowicz, S.~De~Liberato, S.~Savasta, and F.~Nori.
\newblock Ultrastrong coupling between light and matter.
\newblock {\em Nat. Rev. Phys.}, 1:19--40, 2019.

\bibitem{tinkham2004introduction}
M.~Tinkham.
\newblock {\em Introduction to superconductivity}.
\newblock Dover Publication, Mineola, N.Y., 2004.

\bibitem{kyaw2019towards}
T.~H. Kyaw.
\newblock {\em Towards a Scalable Quantum Computing Platform in the Ultrastrong
  Coupling Regime}.
\newblock Springer, 2019.

\bibitem{stassi2020scalable}
R.~Stassi, M.~Cirio, and F.~Nori.
\newblock {Scalable quantum computer with superconducting circuits in the
  ultrastrong coupling regime}.
\newblock {\em npj Quantum Inf.}, 6:1--6, 2020.

\bibitem{zhu03universal}
S.-L. Zhu and Z.~D. Wang.
\newblock Universal quantum gates based on a pair of orthogonal cyclic states:
  Application to {NMR} systems.
\newblock {\em Phys. Rev. A}, 67:022319, 2003.

\bibitem{zhang2005nonadiabatic}
X.-D. Zhang, S.-L. Zhu, L.~Hu, and Z.~D. Wang.
\newblock Nonadiabatic geometric quantum computation using a single-loop
  scenario.
\newblock {\em Phys. Rev. A}, 71:014302, 2005.

\bibitem{wang2001NMR}
X.~Wang and K.~Matsumoto.
\newblock {NMR} {C}-{NOT} gate through the {A}haranov-{A}nandan phase shift.
\newblock {\em J. Phys. A: Math. Gen.}, 34:L631--L634, 2001.

\bibitem{gopinath2006geo}
T.~Gopinath and A.~Kumar.
\newblock Geometric quantum computation using fictitious spin-$\frac{1}{2}$
  subspaces of strongly dipolar coupled nuclear spins.
\newblock {\em Phys. Rev. A}, 73:022326, 2006.

\bibitem{lloyd95almost}
S.~Lloyd.
\newblock Almost any quantum logic gate is universal.
\newblock {\em Phys. Rev. Lett.}, 75:346--349, 1995.

\bibitem{ota09geo}
Y.~Ota, Y.~Goto, Y.~Kondo, and M.~Nakahara.
\newblock Geometric quantum gates in liquid-state {NMR} based on a cancellation
  of dynamical phases.
\newblock {\em Phys. Rev. A}, 80:052311, 2009.

\bibitem{du06experimental}
J.~Du, P.~Zou, and Z.~D. Wang.
\newblock Experimental implementation of high-fidelity unconventional geometric
  quantum gates using an {NMR} interferometer.
\newblock {\em Phys. Rev. A}, 74:020302, 2006.

\bibitem{wang07non}
Z.~S. Wang, C.~Wu, X.-L. Feng, L.~C. Kwek, C.~H. Lai, C.~H. Oh, and V.~Vedral.
\newblock Nonadiabatic geometric quantum computation.
\newblock {\em Phys. Rev. A}, 76:044303, 2007.

\bibitem{wang01multibit}
X.~Wang, A.~S\o{}rensen, and K.~M\o{}lmer.
\newblock Multibit gates for quantum computing.
\newblock {\em Phys. Rev. Lett.}, 86:3907--3910, 2001.

\bibitem{cirac95quantum}
J.~I. Cirac and P.~Zoller.
\newblock Quantum computations with cold trapped ions.
\newblock {\em Phys. Rev. Lett.}, 74:4091--4094, 1995.

\bibitem{sorensen99quantum}
A.~S\o{}rensen and K.~M\o{}lmer.
\newblock Quantum computation with ions in thermal motion.
\newblock {\em Phys. Rev. Lett.}, 82:1971--1974, 1999.

\bibitem{milburn00ion}
G.~J. Milburn, S.~Schneider, and D.~F.~V. James.
\newblock Ion trap quantum computing with warm ions.
\newblock {\em Fortschritte der Phys.}, 48:801--810, 2000.

\bibitem{sorensen00entangl}
A.~S{\o}rensen and K.~M{\o}lmer.
\newblock Entanglement and quantum computation with ions in thermal motion.
\newblock {\em Phys. Rev. A}, 62:022311, 2000.

\bibitem{duan10collo}
L.-M. Duan and C.~Monroe.
\newblock Colloquium: Quantum networks with trapped ions.
\newblock {\em Rev. Mod. Phys.}, 82:1209--1224, 2010.

\bibitem{haffner08quantum}
H.~H{\"{a}}ffner, C.~F. Roos, and R.~Blatt.
\newblock Quantum computing with trapped ions.
\newblock {\em Phys. Rep.}, 469:155--203, 2008.

\bibitem{gazeau09coherent}
J.-P. Gazeau.
\newblock {\em Coherent states in quantum physics}.
\newblock Wiley, Weinheim, 2009.

\bibitem{seidelin06micro}
S.~Seidelin, J.~Chiaverini, R.~Reichle, J.~J. Bollinger, D.~Leibfried,
  J.~Britton, J.~H. Wesenberg, R.~B. Blakestad, R.~J. Epstein, D.~B. Hume,
  W.~M. Itano, J.~D. Jost, C.~Langer, R.~Ozeri, N.~Shiga, and D.~J. Wineland.
\newblock Microfabricated surface-electrode ion trap for scalable quantum
  information processing.
\newblock {\em Phys. Rev. Lett.}, 96:253003, 2006.

\bibitem{leibfried07transport}
D.~Leibfried, E.~Knill, C.~Ospelkaus, and D.~J. Wineland.
\newblock Transport quantum logic gates for trapped ions.
\newblock {\em Phys. Rev. A}, 76:032324, 2007.

\bibitem{zheng06high}
S.-B. Zheng.
\newblock High-speed geometric quantum phase gates for trapped ions in thermal
  motion.
\newblock {\em Phys. Rev. A}, 74:032322, 2006.

\bibitem{zheng04un}
S.-B. Zheng.
\newblock Unconventional geometric quantum phase gates with a cavity {QED}
  system.
\newblock {\em Phys. Rev. A}, 70:052320, 2004.

\bibitem{chen06strong}
C.-Y. Chen, M.~Feng, X.-L. Zhang, and K.-L. Gao.
\newblock Strong-driving-assisted unconventional geometric logic gate in cavity
  {QED}.
\newblock {\em Phys. Rev. A}, 73:032344, 2006.

\bibitem{feng07scheme}
X.-L. Feng, Z.~Wang, C.~Wu, L.~C. Kwek, C.~H. Lai, and C.~H. Oh.
\newblock Scheme for unconventional geometric quantum computation in cavity
  {QED}.
\newblock {\em Phys. Rev. A}, 75:052312, 2007.

\bibitem{wu2007unconventional}
C.~Wu, Z.~Wang, X.-L. Feng, H.-S. Goan, L.~C. Kwek, C.~H. Lai, and C.~H. Oh.
\newblock Unconventional geometric quantum computation in a two-mode cavity.
\newblock {\em Phys. Rev. A}, 76:024302, 2007.

\bibitem{imai08demonstration}
H.~Imai and A.~Morinaga.
\newblock Demonstration of pure geometric universal single-qubit operation on
  two-level atoms.
\newblock {\em Phys. Rev. A}, 78:010302, 2008.

\bibitem{wang16experimental}
L.~Wang, T.~Tu, B.~Gong, C.~Zhou, and G.-C. Guo.
\newblock Experimental realization of non-adiabatic universal quantum gates
  using geometric {L}andau-{Z}ener-{S}tückelberg interferometry.
\newblock {\em Sci. Rep.}, 6:19048, 2016.

\bibitem{xu2020experimental}
Y.~Xu, Z.~Hua, T.~Chen, X.~Pan, X.~Li, J.~Han, W.~Cai, Y.~Ma, H.~Wang, Y.~P.
  Song, Z.-Y. Xue, and L.~Sun.
\newblock Experimental implementation of universal nonadiabatic geometric
  quantum gates in a superconducting circuit.
\newblock {\em Phys. Rev. Lett.}, 124:230503, 2020.

\bibitem{zhao2021experimental}
P.~Z. Zhao, Z.~J.~Z. Dong, Z.~X. Zhang, G.~P. Guo, D.~M. Tong, and Y.~Yin.
\newblock Experimental realization of nonadiabatic geometric gates with a
  superconducting {X}mon qubit.
\newblock {\em Sci. China-Phys. Mech. Astron.}, 64:250362, 2021.

\bibitem{song17continuous}
C.~Song, S.-B. Zheng, P.~Zhang, K.~Xu, L.~Zhang, Q.~Guo, W.~Liu, D.~Xu,
  H.~Deng, K.~Huang, D.~Zheng, X.~Zhu, and H.~Wang.
\newblock Continuous-variable geometric phase and its manipulation for quantum
  computation in a superconducting circuit.
\newblock {\em Nat. Commun.}, 8:1--7, 2017.

\bibitem{huang19experimental}
Y.-Y. Huang, Y.-K. Wu, F.~Wang, P.-Y. Hou, W.-B. Wang, W.-G. Zhang, W.-Q. Lian,
  Y.-Q. Liu, H.-Y. Wang, H.-Y. Zhang, L.~He, X.-Y. Chang, Y.~Xu, and L.-M.
  Duan.
\newblock Experimental realization of robust geometric quantum gates with
  solid-state spins.
\newblock {\em Phys. Rev. Lett.}, 122:010503, 2019.

\bibitem{kleissler2018universal}
F.~Klei{\ss}ler, A.~Lazariev, and S.~Arroyo-Camejo.
\newblock Universal, high-fidelity quantum gates based on superadiabatic,
  geometric phases on a solid-state spin-qubit at room temperature.
\newblock {\em npj Quantum Information}, 4:49, 2018.

\bibitem{pechal2012geometric}
M.~Pechal, S.~Berger, A.~A. Abdumalikov~Jr, J.~M. Fink, J.~A. Mlynek,
  L.~Steffen, A.~Wallraff, and S.~Filipp.
\newblock Geometric phase and nonadiabatic effects in an electronic harmonic
  oscillator.
\newblock {\em Phys. Rev. Lett.}, 108:170401, 2012.

\bibitem{zhang2017measuring}
Z.~Zhang, T.~Wang, L.~Xiang, J.~Yao, J.~Wu, and Y.~Yin.
\newblock Measuring the {B}erry phase in a superconducting phase qubit by a
  shortcut to adiabaticity.
\newblock {\em Phys. Rev. A}, 95:042345, 2017.

\bibitem{nazir02decoherence}
A.~Nazir, T.~P. Spiller, and W.~J. Munro.
\newblock Decoherence of geometric phase gates.
\newblock {\em Phys. Rev. A}, 65:042303, 2002.

\bibitem{blais03effect}
A.~Blais and A.-M.~S. Tremblay.
\newblock Effect of noise on geometric logic gates for quantum computation.
\newblock {\em Phys. Rev. A}, 67:012308, 2003.

\bibitem{de03berry}
G.~De~Chiara and G.~M. Palma.
\newblock Berry phase for a spin $1/2$ particle in a classical fluctuating
  field.
\newblock {\em Phys. Rev. Lett.}, 91:090404, 2003.

\bibitem{leek07observation}
P.~J. Leek, J.~M. Fink, A.~Blais, R.~Bianchetti, M.~G{\"o}ppl, J.~M. Gambetta,
  D.~I. Schuster, L.~Frunzio, R.~J. Schoelkopf, and A.~Wallraff.
\newblock Observation of {B}erry's phase in a solid-state qubit.
\newblock {\em Science}, 318:1889--1892, 2007.

\bibitem{berger2013exploring}
S.~Berger, M.~Pechal, A.~A. Abdumalikov~Jr, C.~Eichler, L.~Steffen, A.~Fedorov,
  A.~Wallraff, and S.~Filipp.
\newblock Exploring the effect of noise on the {B}erry phase.
\newblock {\em Phys. Rev. A}, 87:060303, 2013.

\bibitem{filipp2009experimental}
S.~Filipp, J.~Klepp, Y.~Hasegawa, C.~Plonka-Spehr, U.~Schmidt, P.~Geltenbort,
  and H.~Rauch.
\newblock Experimental demonstration of the stability of {B}erry’s phase for
  a spin-1/2 particle.
\newblock {\em Phys. Rev. Lett.}, 102:030404, 2009.

\bibitem{zhu05geo}
S.-L. Zhu, Z.~D. Wang, and P.~Zanardi.
\newblock Geometric quantum computation and multiqubit entanglement with
  superconducting qubits inside a cavity.
\newblock {\em Phys. Rev. Lett.}, 94:100502, 2005.

\bibitem{thomas2011}
J.~T. Thomas, M.~Lababidi, and M.~Tian.
\newblock Robustness of single-qubit geometric gate against systematic error.
\newblock {\em Phys. Rev. A}, 84:042335, 2011.

\bibitem{wu13geo}
H.~Wu, E.~M. Gauger, R.~E. George, M.~M\"ott\"onen, H.~Riemann, N.~V.
  Abrosimov, P.~Becker, H.-J. Pohl, K.~M. Itoh, M.~L.~W. Thewalt, and J.~J.~L.
  Morton.
\newblock Geometric phase gates with adiabatic control in electron spin
  resonance.
\newblock {\em Phys. Rev. A}, 87:032326, 2013.

\bibitem{ota2009composite}
Y.~Ota and Y.~Kondo.
\newblock Composite pulses in {NMR} as nonadiabatic geometric quantum gates.
\newblock {\em Phys. Rev. A}, 80:024302, 2009.

\bibitem{qin2017suppressing}
X.~Qin, G.-C. Guo, and Z.-W. Zhou.
\newblock Suppressing the geometric dephasing of {B}erry phase by using
  modified dynamical decoupling sequences.
\newblock {\em New J. Phys.}, 19:013025, 2017.

\bibitem{chen2020high}
T.~Chen and Z.-Y. Xue.
\newblock High-fidelity and robust geometric quantum gates that outperform
  dynamical ones.
\newblock {\em Phys. Rev. Appl.}, 14:064009, 2020.

\bibitem{li2021superrobust}
S.~Li, B.-J. Liu, Z.~Ni, L.~Zhang, Z.-Y. Xue, J.~Li, F.~Yan, Y.~Chen, S.~Liu,
  M.-H. Yung, Y.~Xu, and D.~Yu.
\newblock Superrobust geometric control of a superconducting circuit.
\newblock {\em Phys. Rev. Applied}, 16:064003, 2021.

\bibitem{chen2022fault}
Y.-H. Chen, R.~Stassi, W.~Qin, A.~Miranowicz, and F.~Nori.
\newblock Fault-tolerant multiqubit geometric entangling gates using photonic
  cat-state qubits.
\newblock {\em Phys. Rev. Applied}, 18:024076, 2022.

\bibitem{chen2022enhanced}
Y.-H. Chen, A.~Miranowicz, X.~Chen, Y.~Xia, and F.~Nori.
\newblock Enhanced-fidelity ultrafast geometric quantum computation using
  strong classical drives.
\newblock {\em Phys. Rev. Applied}, 18:064059, 2022.

\bibitem{kang2022nonadiabatic}
Y.-H. Kang, Y.-H. Chen, X.~Wang, J.~Song, Y.~Xia, A.~Miranowicz, S.-B. Zheng,
  and F.~Nori.
\newblock Nonadiabatic geometric quantum computation with cat-state qubits via
  invariant-based reverse engineering.
\newblock {\em Phys. Rev. Research}, 4:013233, 2022.

\bibitem{nakahara2003geometry}
M.~Nakahara.
\newblock {\em Geometry, topology and physics}.
\newblock CRC press, 2003.

\bibitem{fujii2001mathematical}
K.~Fujii.
\newblock Mathematical foundations of holonomic quantum computer.
\newblock {\em Rep. Math. Phys.}, 48:75--82, 2001.

\bibitem{lucarelli2005control}
D.~Lucarelli.
\newblock Control aspects of holonomic quantum computation.
\newblock {\em J. Math. Phys.}, 46:052103, 2005.

\bibitem{pachos2002topological}
J.~Pachos.
\newblock Topological features in ion-trap holonomic computation.
\newblock {\em Phys. Rev. A}, 66:042318, 2002.

\bibitem{recati2002holonomic}
A.~Recati, T.~Calarco, P.~Zanardi, J.~I. Cirac, and P.~Zoller.
\newblock Holonomic quantum computation with neutral atoms.
\newblock {\em Phys. Rev. A}, 66:032309, 2002.

\bibitem{ota2008implementation}
Y.~Ota, M.~Bando, Y.~Kondo, and M.~Nakahara.
\newblock Implementation of holonomic quantum gates by an isospectral
  deformation of an ising dimer chain.
\newblock {\em Phys. Rev. A}, 78:052315, 2008.

\bibitem{solinas2003holonomic}
P.~Solinas, P.~Zanardi, N.~Zangh{\`\i}, and F.~Rossi.
\newblock Holonomic quantum gates: a semiconductor-based implementation.
\newblock {\em Phys. Rev. A}, 67:062315, 2003.

\bibitem{molmer1999multiparticle}
K.~M{\o}lmer and A.~S{\o}rensen.
\newblock Multiparticle entanglement of hot trapped ions.
\newblock {\em Phys. Rev. Lett.}, 82:1835, 1999.

\bibitem{cirac2000scalable}
J.~I. Cirac and P.~Zoller.
\newblock A scalable quantum computer with ions in an array of microtraps.
\newblock {\em Nature}, 404:579--581, 2000.

\bibitem{monroe1995demonstration}
C.~Monroe, D.~M. Meekhof, B.~E. King, W.~M. Itano, and D.~J. Wineland.
\newblock Demonstration of a fundamental quantum logic gate.
\newblock {\em Phys. Rev. Lett.}, 75:4714, 1995.

\bibitem{roos1999quantum}
Ch. Roos, Th. Zeiger, H.~Rohde, H.~C. N{\"a}gerl, J.~Eschner, D.~Leibfried,
  F.~Schmidt-Kaler, and R.~Blatt.
\newblock Quantum state engineering on an optical transition and decoherence in
  a {P}aul trap.
\newblock {\em Phys. Rev. Lett.}, 83:4713, 1999.

\bibitem{sackett2000experimental}
C.~A. Sackett, D.~Kielpinski, B.~E. King, C.~Langer, V.~Meyer, C.~J. Myatt,
  M.~Rowe, Q.~A. Turchette, W.~M. Itano, D.~J. Wineland, and C.~Monroe.
\newblock Experimental entanglement of four particles.
\newblock {\em Nature}, 404:256--259, 2000.

\bibitem{coulston1992population}
G.~W. Coulston and K.~Bergmann.
\newblock Population transfer by stimulated {R}aman scattering with delayed
  pulses: analytical results for multilevel systems.
\newblock {\em J. Chem. Phys.}, 96:3467--3475, 1992.

\bibitem{unanyan1998robust}
R.~G. Unanyan, M.~Fleischhauer, B.~W. Shore, and K.~Bergmann.
\newblock Robust creation and phase-sensitive probing of superposition states
  via stimulated {R}aman adiabatic passage ({STIRAP}) with degenerate dark
  states.
\newblock {\em Opt. Commun.}, 155:144--154, 1998.

\bibitem{unanyan1999laser}
R.~G. Unanyan, B.~W. Shore, and K.~Bergmann.
\newblock Laser-driven population transfer in four-level atoms: Consequences of
  non-{A}belian geometrical adiabatic phase factors.
\newblock {\em Phys. Rev. A}, 59:2910, 1999.

\bibitem{you2006superconducting}
J.~Q. You and F.~Nori.
\newblock Superconducting circuits and quantum information.
\newblock {\em Phys. Today}, 58:42--47, 2005.

\bibitem{huang2020superconducting}
H.-L. Huang, D.~Wu, D.~Fan, and X.~Zhu.
\newblock Superconducting quantum computing: a review.
\newblock {\em Sci. China Inf. Sci.}, 63:1--32, 2020.

\bibitem{choi2003geometric}
M.-S. Choi.
\newblock Geometric quantum computation on solid-state qubits.
\newblock {\em J. Phys.: Condens. Matter}, 15:7823--7833, 2003.

\bibitem{kamleitner2011geometric}
I.~Kamleitner, P.~Solinas, C.~M{\"u}ller, A.~Shnirman, and M.~M{\"o}tt{\"o}nen.
\newblock Geometric quantum gates with superconducting qubits.
\newblock {\em Phys. Rev. B}, 83:214518, 2011.

\bibitem{cholascinski2004quantum}
M.~Cholascinski.
\newblock Quantum holonomies with {J}osephson-junction devices.
\newblock {\em Phys. Rev. B}, 69:134516, 2004.

\bibitem{zhang2005holonomic}
P.~Zhang, Z.~D. Wang, J.~D. Sun, and C.~P. Sun.
\newblock Holonomic quantum computation using rf superconducting quantum
  interference devices coupled through a microwave cavity.
\newblock {\em Phys. Rev. A}, 71:042301, 2005.

\bibitem{feng2008holonomic}
Z.-B. Feng and X.-D. Zhang.
\newblock Holonomic quantum computation with superconducting charge-phase
  qubits in a cavity.
\newblock {\em Phys. Lett. A}, 372:1589--1594, 2008.

\bibitem{peng2008implementation}
Z.~H. Peng, H.~F. Chu, Z.~D. Wang, and D.~N. Zheng.
\newblock Implementation of adiabatic geometric gates with superconducting
  phase qubits.
\newblock {\em J. Phys.: Condens. Matter}, 21:045701, 2008.

\bibitem{lin2009robust}
G.-W. Lin, X.-B. Zou, X.-M. Lin, and G.-C. Guo.
\newblock Robust and fast geometric quantum computation with multiqubit gates
  in cavity {QED}.
\newblock {\em Phys. Rev. A}, 79:064303, 2009.

\bibitem{bernevig2005holonomic}
B.~A. Bernevig and S.-C. Zhang.
\newblock Holonomic quantum computing based on the {S}tark effect.
\newblock {\em Phys. Rev. B}, 71:035303, 2005.

\bibitem{golovach2010holonomic}
V.~N. Golovach, M.~Borhani, and D.~Loss.
\newblock Holonomic quantum computation with electron spins in quantum dots.
\newblock {\em Phys. Rev. A}, 81:022315, 2010.

\bibitem{budich2012all}
J.~C. Budich, D.~G. Rothe, E.~M. Hankiewicz, and B.~Trauzettel.
\newblock All-electric qubit control in heavy hole quantum dots via
  non-{A}belian geometric phases.
\newblock {\em Phys. Rev. B}, 85:205425, 2012.

\bibitem{bakke2011quantum}
K.~Bakke and C.~Furtado.
\newblock Quantum holonomies for an electric dipole moment.
\newblock {\em Phys. Lett. A}, 375:3956--3959, 2011.

\bibitem{bakke2012holonomic}
K.~Bakke and C.~Furtado.
\newblock Holonomic quantum computation based on the scalar {A}haronov-{B}ohm
  effect for neutral particles and linear topological defects.
\newblock {\em Ann. Phys.}, 327:376--385, 2012.

\bibitem{zheng2012geometric}
Y.-C. Zheng and T.~A. Brun.
\newblock Geometric manipulation of ensembles of atoms on an atom chip for
  quantum computation.
\newblock {\em Phys. Rev. A}, 86:032323, 2012.

\bibitem{li2004non}
Y.~Li, P.~Zhang, P.~Zanardi, and C.~P. Sun.
\newblock Non-{A}belian geometric quantum memory with an atomic ensemble.
\newblock {\em Phys. Rev. A}, 70:032330, 2004.

\bibitem{brosco2008non}
V.~Brosco, R.~Fazio, F.~W.~J. Hekking, and A.~Joye.
\newblock Non-{A}belian superconducting pumps.
\newblock {\em Phys. Rev. Lett.}, 100:027002, 2008.

\bibitem{pirkkalainen2010non}
J.-M. Pirkkalainen, P.~Solinas, J.~P. Pekola, and M.~M{\"o}tt{\"o}nen.
\newblock Non-{A}belian geometric phases in ground-state {J}osephson devices.
\newblock {\em Phys. Rev. B}, 81:174506, 2010.

\bibitem{solinas2010ground}
P.~Solinas, J.-.M Pirkkalainen, and M.~M{\"o}tt{\"o}nen.
\newblock Ground-state geometric quantum computing in superconducting systems.
\newblock {\em Phys. Rev. A}, 82:052304, 2010.

\bibitem{Sugawa2021wilson}
S.~Sugawa, F.~Salces-Carcoba, Y.~Yue, A.~Putra, and I.~B. Spielman.
\newblock {Wilson loop and Wilczek-Zee phase from a non-{A}belian gauge field}.
\newblock {\em npj Quantum Inf.}, 7:1--9, 2021.

\bibitem{sjoqvist2016conceptual}
E.~Sj{\"o}qvist, V.~Azimi~Mousolou, and C.~M. Canali.
\newblock Conceptual aspects of geometric quantum computation.
\newblock {\em Quantum Inf. Process.}, 15:3995--4011, 2016.

\bibitem{zhang2019single}
Z.~Zhang, P.~Z. Zhao, T.~Wang, L.~Xiang, Z.~Jia, P.~Duan, D.~M. Tong, Y.~Yin,
  and G.-C Guo.
\newblock Single-shot realization of nonadiabatic holonomic gates with a
  superconducting {X}mon qutrit.
\newblock {\em New J. Phys.}, 21:073024, 2019.

\bibitem{kyoseva2006}
E.~S. Kyoseva and N.~V. Vitanov.
\newblock Coherent pulsed excitation of degenerate multistate systems: Exact
  analytic solutions.
\newblock {\em Phys. Rev. A}, 73:023420, 2006.

\bibitem{ivanov2006}
A.~Ivanov, E.~S. Kyoseva, and N.~V. Vitanov.
\newblock Engineering of arbitrary u($n$) transformations by quantum
  householder reflections.
\newblock {\em Phys. Rev. A}, 74:022323, 2006.

\bibitem{ivanov2008}
A.~Ivanov and N.~V. Vitanov.
\newblock Synthesis of arbitrary unitary transformations of collective states
  of trapped ions by quantum householder reflections.
\newblock {\em Phys. Rev. A}, 77:012335, 2008.

\bibitem{johansson2012robustness}
M.~Johansson, E.~Sj{\"o}qvist, L.~M. Andersson, M.~Ericsson, B.~Hessmo,
  K.~Singh, and D.~M. Tong.
\newblock Robustness of nonadiabatic holonomic gates.
\newblock {\em Phys. Rev. A}, 86:062322, 2012.

\bibitem{spiegelberg2013rwa}
J.~Spiegelberg and E.~Sj\"oqvist.
\newblock Validity of the rotating-wave approximation in nonadiabatic holonomic
  quantum computation.
\newblock {\em Phys. Rev. A}, 88:054301, 2013.

\bibitem{alves2022rwa}
G.~O. Alves and E.~Sj\"oqvist.
\newblock Time-optimal holonomic quantum computation.
\newblock {\em Phys. Rev. A}, 106:032406, 2022.

\bibitem{xue2017nonadiabatic}
Z.-Y. Xue, F.-L. Gu, Z.-P. Hong, Z.-H. Yang, D.-W. Zhang, Y.~Hu, and J.~Q. You.
\newblock Nonadiabatic holonomic quantum computation with dressed-state qubits.
\newblock {\em Phys. Rev. Appl.}, 7:054022, 2017.

\bibitem{hong2018implementing}
Z.-P. Hong, B.-J. Liu, J.-Q. Cai, X.-D. Zhang, Y.~Hu, Z.~D. Wang, and Z.-Y.
  Xue.
\newblock Implementing universal nonadiabatic holonomic quantum gates with
  transmons.
\newblock {\em Phys. Rev. A}, 97:022332, 2018.

\bibitem{mousolou2017universal}
V.~Azimi~Mousolou.
\newblock Universal non-adiabatic geometric manipulation of pseudo-spin charge
  qubits.
\newblock {\em Europhys. Lett.}, 117:10006, 2017.

\bibitem{zhou2018fast}
J.~Zhou, B.-J. Liu, Z.-P. Hong, and Z.-Y. Xue.
\newblock Fast holonomic quantum computation based on solid-state spins with
  all-optical control.
\newblock {\em Sci. China: Phys. Mech. Astron.}, 61:1--7, 2018.

\bibitem{zhao2018nonadiabatic}
P.~Z. Zhao, X.~Wu, T.~H. Xing, G.~F. Xu, and D.~M. Tong.
\newblock Nonadiabatic holonomic quantum computation with {R}ydberg superatoms.
\newblock {\em Phys. Rev. A}, 98:032313, 2018.

\bibitem{rippe2008solidstate}
L.~Rippe, B.~Julsgaard, A.~Walther, Y.~Ying, and S.~Kr\"oll.
\newblock Experimental quantum-state tomography of a solid-state qubit.
\newblock {\em Phys. Rev. A}, 77:022307, 2008.

\bibitem{xu2017robust}
G.~F. Xu, P.~Z. Zhao, D.~M. Tong, and E.~Sj{\"o}qvist.
\newblock Robust paths to realize nonadiabatic holonomic gates.
\newblock {\em Phys. Rev. A}, 95:052349, 2017.

\bibitem{xu2017composite}
G.~F. Xu, P.~Z. Zhao, T.~H. Xing, E.~Sj\"oqvist, and D.~M. Tong.
\newblock Composite nonadiabatic holonomic quantum computation.
\newblock {\em Phys. Rev. A}, 95:032311, 2017.

\bibitem{zhang2019searching}
F.~Zhang, J.~Zhang, P.~Gao, and G.~L. Long.
\newblock Searching nonadiabatic holonomic quantum gates via an optimization
  algorithm.
\newblock {\em Phys. Rev. A}, 100:012329, 2019.

\bibitem{trif2010spin}
M.~Trif, F.~Troiani, D.~Stepanenko, and D.~Loss.
\newblock Spin electric effects in molecular antiferromagnets.
\newblock {\em Phys. Rev. B}, 82:045429, 2010.

\bibitem{gatteschi2006molecular}
D.~Gatteschi, R.~Sessoli, and J.~Villain.
\newblock {\em Molecular nanomagnets}.
\newblock Oxford University Press, 2006.

\bibitem{nielsen2002quantum}
M.~A. Nielsen and I.~L. Chuang.
\newblock {\em Quantum computation and quantum information}.
\newblock Cambridge University Press, 2002.

\bibitem{karle2003}
R.~Karle and J.~Pachos.
\newblock Geometrical phases for the $g(4,2)$ grassmannian manifold.
\newblock {\em J. Math. Phys.}, 72:2463--2470, 2003.

\bibitem{mousolou2014non}
V.~Azimi~Mousolou and E.~Sj{\"o}qvist.
\newblock Non-{A}belian geometric phases in a system of coupled quantum bits.
\newblock {\em Phys. Rev. A}, 89:022117, 2014.

\bibitem{kwiat1991}
P.~G. Kwiat and R.~Y. Chiao.
\newblock Observation of a nonclassical {B}erry’s phase for the photon.
\newblock {\em Phys. Rev. Lett.}, 66:588, 1991.

\bibitem{allman1997}
B.~E. Allman, H.~Kaiser, S.~A. Werner, A.~G. Wagh, V.~C. Rakhecha, and
  J.~Summhammer.
\newblock Observation of geometric and dynamical phases by neutron
  interferometry.
\newblock {\em Phys. Rev. A}, 56:4420--4439, 1997.

\bibitem{majer2007coupling}
J.~Majer, J.~M. Chow, J.~M. Gambetta, J.~Koch, B.~R. Johnson, J.~A. Schreier,
  L.~Frunzio, D.~I. Schuster, A.~A. Houck, A.~Wallraff, A.~Blais, M.~H.
  Devoret, S.~M. Girvin, and R.~J. Schoelkopf.
\newblock Coupling superconducting qubits via a cavity bus.
\newblock {\em Nature}, 449:443--447, 2007.

\bibitem{imamog1999quantum}
A.~Imamoglu, D.~D. Awschalom, G.~Burkard, D.~P. DiVincenzo, D.~Loss,
  M.~Sherwin, and A.~Small.
\newblock Quantum information processing using quantum dot spins and cavity
  {QED}.
\newblock {\em Phys. Rev. Lett.}, 83:4204--4207, 1999.

\bibitem{mozyrsky2001indirect}
D.~Mozyrsky, V.~Privman, and M.~L. Glasser.
\newblock Indirect interaction of solid-state qubits via two-dimensional
  electron gas.
\newblock {\em Phys. Rev. Lett.}, 86:5112--5115, 2001.

\bibitem{arute2019quantum}
F.~Arute, K.~Arya, R.~Babbush, D.~Bacon, J.~C. Bardin, R.~Barends, R.~Biswas,
  S.~Boixo, F.~G. S.~L. Brandao, D.~A. Buell, et~al.
\newblock Quantum supremacy using a programmable superconducting processor.
\newblock {\em Nature}, 574:505--510, 2019.

\bibitem{gong2021quantum}
M.~Gong, S.~Wang, C.~Zha, M.-C. Chen, H.-L. Huang, Y.~Wu, Q.~Zhu, Y.~Zhao,
  S.~Li, S.~Guo, et~al.
\newblock Quantum walks on a programmable two-dimensional 62-qubit
  superconducting processor.
\newblock {\em Science}, 372:948--952, 2021.

\bibitem{yan2019strongly}
Z.~Yan, Y.-R. Zhang, M.~Gong, Y.~Wu, Y.~Zheng, S.~Li, C.~Wang, F.~Liang,
  J.~Lin, Y.~Xu, et~al.
\newblock Strongly correlated quantum walks with a 12-qubit superconducting
  processor.
\newblock {\em Science}, 364:753--756, 2019.

\bibitem{mousolou2018scalable}
V.~Azimi~Mousolou.
\newblock Scalable star-shaped architecture for universal spin-based
  nonadiabatic holonomic quantum computation.
\newblock {\em Phys. Rev. A}, 98:062340, 2018.

\bibitem{mousolou2017electric}
V.~Azimi~Mousolou.
\newblock Electric nonadiabatic geometric entangling gates on spin qubits.
\newblock {\em Phys. Rev. A}, 96:012307, 2017.

\bibitem{chen2018nonadiabatic}
T.~Chen, J.~Zhang, and Z.-Y. Xue.
\newblock Nonadiabatic holonomic quantum computation on coupled transmons with
  ancillaries.
\newblock {\em Phys. Rev. A}, 98:052314, 2018.

\bibitem{kang2020heralded}
Y.-H. Kang, Z.-C. Shi, J.~Song, and Y.~Xia.
\newblock Heralded atomic nonadiabatic holonomic quantum computation with
  {R}ydberg blockade.
\newblock {\em Phys. Rev. A}, 102:022617, 2020.

\bibitem{guery2019shortcuts}
D.~Gu{\'e}ry-Odelin, A.~Ruschhaupt, A.~Kiely, E.~Torrontegui,
  S.~Mart{\'\i}nez-Garaot, and J.~G. Muga.
\newblock Shortcuts to adiabaticity: concepts, methods, and applications.
\newblock {\em Rev. Mod. Phys.}, 91:045001, 2019.

\bibitem{du2019geometric}
Y.~Du, Z.~Liang, H.~Yan, and S.-L. Zhu.
\newblock Geometric quantum computation with shortcuts to adiabaticity.
\newblock {\em Adv. Quantum Technol.}, 2:1900013, 2019.

\bibitem{demirplak2003adiabatic}
M.~Demirplak and S.~A. Rice.
\newblock Adiabatic population transfer with control fields.
\newblock {\em J. Phys. Chem. A}, 107:9937--9945, 2003.

\bibitem{demirplak2005assisted}
M.~Demirplak and S.~A. Rice.
\newblock Assisted adiabatic passage revisited.
\newblock {\em J. Phys. Chem. A}, 109:6838--6844, 2005.

\bibitem{berry2009transitionless}
M.~V. Berry.
\newblock Transitionless quantum driving.
\newblock {\em J. Phys. A: Math. Theor.}, 42:365303, 2009.

\bibitem{muga2009frictionless}
J.~G. Muga, X.~Chen, A.~Ruschhaupt, and D.~Gu{\'e}ry-Odelin.
\newblock Frictionless dynamics of {B}ose-{E}instein condensates under fast
  trap variations.
\newblock {\em J. Phys. B: At. Mol. Opt. Phys.}, 42:241001, 2009.

\bibitem{liang2016proposal}
Z.-T. Liang, X.~Yue, Q.~Lv, Y.-X. Du, W.~Huang, H.~Yan, and S.-L. Zhu.
\newblock Proposal for implementing universal superadiabatic geometric quantum
  gates in nitrogen-vacancy centers.
\newblock {\em Phys. Rev. A}, 93:040305, 2016.

\bibitem{liu2017superadiabatic}
B.-J. Liu, Z.-H. Huang, Z.-Y. Xue, and X.-D. Zhang.
\newblock Superadiabatic holonomic quantum computation in cavity {QED}.
\newblock {\em Phys. Rev. A}, 95:062308, 2017.

\bibitem{lewis1969an}
H.~R. Lewis and W.~R. Riesenfeld.
\newblock An exact quantum theory of the time‐dependent harmonic oscillator
  and of a charged particle in a time‐dependent electromagnetic field.
\newblock {\em J. Math. Phys.}, 10:1458--1473, 1969.

\bibitem{kang2020flexible}
Y.-H. Kang, Z.-C. Shi, B.-H. Huang, J.~Song, and Y.~Xia.
\newblock Flexible scheme for the implementation of nonadiabatic geometric
  quantum computation.
\newblock {\em Phys. Rev. A}, 101:032322, 2020.

\bibitem{liu2020leakage}
B.-J. Liu and M.-H. Yung.
\newblock Leakage suppression for holonomic quantum gates.
\newblock {\em Phys. Rev. Appl.}, 14:034003, 2020.

\bibitem{du2017degenerate}
Y.-X. Du, B.-J. Liu, Q.-X. Lv, X.-D. Zhang, H.~Yan, and S.-L. Zhu.
\newblock Degenerate eigensubspace in a triangle-level system and its geometric
  quantum control.
\newblock {\em Phys. Rev. A}, 96:012333, 2017.

\bibitem{leroux2018non}
F.~Leroux, K.~Pandey, R.~Rehbi, F.~Chevy, C.~Miniatura, B.~Gr{\'e}maud, and
  D.~Wilkowski.
\newblock Non-{A}belian adiabatic geometric transformations in a cold
  {S}trontium gas.
\newblock {\em Nat. Commun.}, 9:1--7, 2018.

\bibitem{zhu2019single}
Z.~Zhu, T.~Chen, X.~Yang, J~Bian, Z.-Y. Xue, and X.~Peng.
\newblock Single-loop and composite-loop realization of nonadiabatic holonomic
  quantum gates in a decoherence-free subspace.
\newblock {\em Phys. Rev. Appl.}, 12:024024, 2019.

\bibitem{li2017experimental}
H.~Li, Y.~Liu, and G.~L. Long.
\newblock Experimental realization of single-shot nonadiabatic holonomic gates
  in nuclear spins.
\newblock {\em Sci. China-Phys. Mech. Astron.}, 60:1--7, 2017.

\bibitem{ai2020experimental}
M.-Z. Ai, S.~Li, Z.~Hou, R.~He, Z.-H. Qian, Z.-Y. Xue, J.-M. Cui, Y.-F. Huang,
  C.-F. Li, and G.-C. Guo.
\newblock Experimental realization of nonadiabatic holonomic single-qubit
  quantum gates with optimal control in a trapped ion.
\newblock {\em Phys. Rev. Appl.}, 14:054062, 2020.

\bibitem{xu2018single}
Y.~Xu, W.~Cai, Y.~Ma, X.~Mu, L.~Hu, T.~Chen, H.~Wang, Y.~P. Song, Z.-Y. Xue,
  Z.-q. Yin, and L.~Sun.
\newblock Single-loop realization of arbitrary nonadiabatic holonomic
  single-qubit quantum gates in a superconducting circuit.
\newblock {\em Phys. Rev. Lett.}, 121:110501, 2018.

\bibitem{yan2019exp}
T.~Yan, B.-J. Liu, K.~Xu, C.~Song, S.~Liu, Z.~Zhang, H.~Deng, Z.~Yan, H.~Rong,
  K.~Huang, M.-H. Yung, Y.~Chen, and D.~Yu.
\newblock Experimental realization of nonadiabatic shortcut to non-{A}belian
  geometric gates.
\newblock {\em Phys. Rev. Lett.}, 122:080501, 2019.

\bibitem{sekiguchi2017optical}
Y.~Sekiguchi, N.~Niikura, R.~Kuroiwa, H.~Kano, and H.~Kosaka.
\newblock Optical holonomic single quantum gates with a geometric spin under a
  zero field.
\newblock {\em Nat. Photonics}, 11:309--314, 2017.

\bibitem{zhou2017holonomic}
B.~B. Zhou, P.~C. Jerger, V.~O. Shkolnikov, F.~J. Heremans, G.~Burkard, and
  D.~D. Awschalom.
\newblock Holonomic quantum control by coherent optical excitation in diamond.
\newblock {\em Phys. Rev. Lett.}, 119:140503, 2017.

\bibitem{ishida2018universal}
N.~Ishida, T.~Nakamura, T.~Tanaka, S.~Mishima, H.~Kano, R.~Kuroiwa,
  Y.~Sekiguchi, and H.~Kosaka.
\newblock Universal holonomic single quantum gates over a geometric spin with
  phase-modulated polarized light.
\newblock {\em Opt. Lett.}, 43:002380, 2018.

\bibitem{nagata2018universal}
K.~Nagata, K.~Kuramitani, Y.~Sekiguchi, and H.~Kosaka.
\newblock Universal holonomic quantum gates over geometric spin qubits with
  polarised microwaves.
\newblock {\em Nat. Commun.}, 9:1--10, 2018.

\bibitem{danilin2018experimental}
S.~Danilin, A.~Veps{\"a}l{\"a}inen, and G.~S. Paraoanu.
\newblock Experimental state control by fast non-{A}belian holonomic gates with
  a superconducting qutrit.
\newblock {\em Physica Scripta}, 93:055101, 2018.

\bibitem{rong2015experimental}
X.~Rong, J.~Geng, F.~Shi, Y.~Liu, K.~Xu, W.~Ma, F.~Kong, Z.~Jiang, Y.~Wu, and
  J.~Du.
\newblock Experimental fault-tolerant universal quantum gates with solid-state
  spins under ambient conditions.
\newblock {\em Nat. Commun.}, 6:8748, 2015.

\bibitem{xie202399}
T.~Xie, Z.~Zhao, S.~Xu, X.~Kong, Z.~Yang, M.~Wang, Y.~Wang, F.~Shi, and J.~Du.
\newblock 99.92\%-fidelity {CNOT} gates in solids by noise filtering.
\newblock {\em Phys. Rev. Lett.}, 130:030601, 2023.

\bibitem{harty2014high}
T.~P. Harty, D.~T.~C. Allcock, C.~J. Ballance, L.~Guidoni, H.~A. Janacek, N.~M.
  Linke, D.~N. Stacey, and D.~M. Lucas.
\newblock High-fidelity preparation, gates, memory, and readout of a
  trapped-ion quantum bit.
\newblock {\em Phys. Rev. Lett.}, 113:220501, 2014.

\bibitem{ballance2016high}
C.~J. Ballance, T.~P. Harty, N.~M. Linke, M.~A. Sepiol, and D.~M. Lucas.
\newblock High-fidelity quantum logic gates using trapped-ion hyperfine qubits.
\newblock {\em Phys. Rev. Lett.}, 117:060504, 2016.

\bibitem{gaebler2016high}
J.~P. Gaebler, T.~R. Tan, Y.~Lin, Y.~Wan, R.~Bowler, A.~C. Keith, S.~Glancy,
  K.~Coakley, E.~Knill, D.~Leibfried, and D.~J. Wineland.
\newblock High-fidelity universal gate set for be $^9${Be}$^+$ ion qubits.
\newblock {\em Phys. Rev. Lett.}, 117:060505, 2016.

\bibitem{somoroff2021millisecond}
A.~Somoroff, Q.~Ficheux, R.~A. Mencia, H.~Xiong, R.~V. Kuzmin, and V.~E.
  Manucharyan.
\newblock Millisecond coherence in a superconducting qubit.
\newblock {\em arXiv:2103.08578}, 2021.

\bibitem{acharya2022suppressing}
R.~Acharya, I.~Aleiner, R.~Allen, T.~I. Andersen, M.~Ansmann, F.~Arute,
  K.~Arya, A.~Asfaw, J.~Atalaya, R.~Babbush, et~al.
\newblock {Suppressing quantum errors by scaling a surface code logical qubit}.
\newblock {\em arXiv:2207.06431}, 2022.

\bibitem{kowarsky2014non}
M.~A. Kowarsky, L.~C.~L. Hollenberg, and A.~M. Martin.
\newblock Non-{A}belian geometric phase in the diamond nitrogen-vacancy center.
\newblock {\em Phys. Rev. A}, 90:042116, 2014.

\bibitem{ellinas2001universal}
D.~Ellinas and J.~Pachos.
\newblock Universal quantum computation by holonomic and nonlocal gates with
  imperfections.
\newblock {\em Phys. Rev. A}, 64:022310, 2001.

\bibitem{cen2003evaluation}
L.~Cen, X.~Li, Y.~Yan, H.~Zheng, and S.~Wang.
\newblock Evaluation of holonomic quantum computation: Adiabatic versus
  nonadiabatic.
\newblock {\em Phys. Rev. Lett.}, 90:147902, 2003.

\bibitem{kuvshinov2003stability}
V.~I. Kuvshinov and A.V. Kuzmin.
\newblock Stability of holonomic quantum computations.
\newblock {\em Phys. Lett. A}, 316:391--394, 2003.

\bibitem{solinas2004robustness}
P.~Solinas, P.~Zanardi, and N.~Zangh{\`\i}.
\newblock Robustness of non-{A}belian holonomic quantum gates against
  parametric noise.
\newblock {\em Phys. Rev. A}, 70:042316, 2004.

\bibitem{lupo2007robustness}
C.~Lupo, P.~Aniello, M.~Napolitano, and G.~Florio.
\newblock Robustness against parametric noise of nonideal holonomic gates.
\newblock {\em Phys. Rev. A}, 76:012309, 2007.

\bibitem{florio2006robust}
G.~Florio, P.~Facchi, R.~Fazio, V.~Giovannetti, and S.~Pascazio.
\newblock Robust gates for holonomic quantum computation.
\newblock {\em Phys. Rev. A}, 73:022327, 2006.

\bibitem{parodi2007environmental}
D.~Parodi, M.~Sassetti, P.~Solinas, and N.~Zangh{\`\i}.
\newblock Environmental noise reduction for holonomic quantum gates.
\newblock {\em Phys. Rev. A}, 76:012337, 2007.

\bibitem{kuvshinov2005robust}
V.~I. Kuvshinov and A.V. Kuzmin.
\newblock Robust {H}adamard gate for optical and ion trap holonomic quantum
  computers.
\newblock {\em Phys. Lett. A}, 341:450--453, 2005.

\bibitem{kuvshinov2006decoherence}
V.~I. Kuvshinov and A.V. Kuzmin.
\newblock Decoherence induced by squeezing control errors in optical and ion
  trap holonomic quantum computations.
\newblock {\em Phys. Rev. A}, 73:052305, 2006.

\bibitem{cen2004refocusing}
L.-X.. Cen and P.~Zanardi.
\newblock Refocusing schemes for holonomic quantum computation in the presence
  of dissipation.
\newblock {\em Physical Review A}, 70:052323, 2004.

\bibitem{parodi2006fidelity}
D.~Parodi, M.~Sassetti, P.~Solinas, P.~Zanardi, and N.~Zangh{\`\i}.
\newblock Fidelity optimization for holonomic quantum gates in dissipative
  environments.
\newblock {\em Phys. Rev. A}, 73:052304, 2006.

\bibitem{lidar2014review}
D.~A. Lidar.
\newblock Review of {D}ecoherence-{F}ree {S}ubspaces, {N}oiseless {S}ubsystems,
  and {D}ynamical {D}ecoupling.
\newblock {\em Adv. Chem. Phys.}, 154:295--354, 2014.

\bibitem{kwiat2000experimental}
P.~G. Kwiat, A.~J. Berglund, J.~B. Altepeter, and A.~G. White.
\newblock Experimental verification of decoherence-free subspaces.
\newblock {\em Science}, 290:498--501, 2000.

\bibitem{kielpinski2001decoherence}
D.~Kielpinski, V.~Meyer, M.~A. Rowe, C.~A. Sackett, W.~M. Itano, C.~Monroe, and
  D.~J. Wineland.
\newblock A decoherence-free quantum memory using trapped ions.
\newblock {\em Science}, 291:1013--1015, 2001.

\bibitem{viola2001experimental}
L.~Viola, E.~M. Fortunato, M.~A. Pravia, E.~Knill, R.~Laflamme, and D.~G. Cory.
\newblock Experimental realization of noiseless subsystems for quantum
  information processing.
\newblock {\em Science}, 293:2059--2063, 2001.

\bibitem{xu2014universal}
G.~F. Xu and G.~L. Long.
\newblock Universal nonadiabatic geometric gates in two-qubit decoherence-free
  subspaces.
\newblock {\em Sci. Rep.}, 4:1--5, 2014.

\bibitem{pyshkin2016expedited}
P.~V. Pyshkin, D.-W. Luo, J.~Jing, J.~Q. You, and L.-A. Wu.
\newblock Expedited holonomic quantum computation via net zero-energy-cost
  control in decoherence-free subspace.
\newblock {\em Sci. Rep.}, 6:1--6, 2016.

\bibitem{song2016shortcuts}
X.-K. Song, H.~Zhang, Q.~Ai, J.~Qiu, and F.-G. Deng.
\newblock Shortcuts to adiabatic holonomic quantum computation in
  decoherence-free subspace with transitionless quantum driving algorithm.
\newblock {\em New J. Phys.}, 18:023001, 2016.

\bibitem{liang2014nonadiabatic}
Z.-T. Liang, Y.-X. Du, W.~Huang, Z.-Y. Xue, and H.~Yan.
\newblock Nonadiabatic holonomic quantum computation in decoherence-free
  subspaces with trapped ions.
\newblock {\em Phys. Rev. A}, 89:062312, 2014.

\bibitem{zhou2015cavity}
J.~Zhou, W.-C. Yu, Y.-M. Gao, and Z.-Y. Xue.
\newblock Cavity {QED} implementation of non-adiabatic holonomies for universal
  quantum gates in decoherence-free subspaces with nitrogen-vacancy centers.
\newblock {\em Opt. Express}, 23:14027--14035, 2015.

\bibitem{xue2015universal}
Z.-Y. Xue, J.~Zhou, and Z.~D. Wang.
\newblock Universal holonomic quantum gates in decoherence-free subspace on
  superconducting circuits.
\newblock {\em Phys. Rev. A}, 92:022320, 2015.

\bibitem{xue2016nonadiabatic}
Z.-Y. Xue, J.~Zhou, Y.-M. Chu, and Y.~Hu.
\newblock Nonadiabatic holonomic quantum computation with all-resonant control.
\newblock {\em Phys. Rev. A}, 94:022331, 2016.

\bibitem{zhao2016nonadiabatic}
P.~Z. Zhao, G.~F. Xu, and D.~M. Tong.
\newblock Nonadiabatic geometric quantum computation in decoherence-free
  subspaces based on unconventional geometric phases.
\newblock {\em Phys. Rev. A}, 94:062327, 2016.

\bibitem{zhao2017single}
P.~Z. Zhao, G.~F. Xu, Q.~M. Ding, E.~Sj{\"o}qvist, and D.~M. Tong.
\newblock Single-shot realization of nonadiabatic holonomic quantum gates in
  decoherence-free subspaces.
\newblock {\em Phys. Rev. A}, 95:062310, 2017.

\bibitem{mousolou2018realization}
V.~Azimi~Mousolou.
\newblock Realization of universal nonadiabatic geometric control on
  decoherence-free qubits in the {XY} model.
\newblock {\em Europhys. Lett.}, 121:20004, 2018.

\bibitem{wang2018single}
C.~Wang and Z.~Guo.
\newblock Single-loop realization of universal nonadiabatic holonomic gates in
  decoherence-free subspaces.
\newblock {\em Europhys. Lett.}, 124:40003, 2018.

\bibitem{pachos2002quantum}
J.~Pachos and H.~Walther.
\newblock Quantum computation with trapped ions in an optical cavity.
\newblock {\em Phys. Rev. Lett.}, 89:187903, 2002.

\bibitem{zheng2008deterministic}
S.-B. Zheng.
\newblock Deterministic geometric quantum phase gates for two atoms in
  decoherence-free subspace.
\newblock {\em Phys. Lett. A}, 372:6584--6587, 2008.

\bibitem{albert2016holonomic}
V.~V. Albert, C.~Shu, S.~Krastanov, C.~Shen, R.-B. Liu, Z.-B. Yang, R.~J.
  Schoelkopf, M.~Mirrahimi, M.~H. Devoret, and L.~Jiang.
\newblock Holonomic quantum control with continuous variable systems.
\newblock {\em Phys. Rev. Lett.}, 116:140502, 2016.

\bibitem{you2005correlation}
J.~Q. You, X.~Hu, and F.~Nori.
\newblock Correlation-induced suppression of decoherence in capacitively
  coupled cooper-pair boxes.
\newblock {\em Phys. Rev. B}, 72:144529, 2005.

\bibitem{chen2021fast}
Y.-H. Chen, W.~Qin, R.~Stassi, X.~Wang, and F.~Nori.
\newblock Fast binomial-code holonomic quantum computation with ultrastrong
  light-matter coupling.
\newblock {\em Phys. Rev. Research}, 3:033275, 2021.

\bibitem{hahn1950spin}
E.~L. Hahn.
\newblock Spin echoes.
\newblock {\em Phys. Rev.}, 80:580--594, 1950.

\bibitem{mehring2012principles}
M.~Mehring.
\newblock {\em Principles of high resolution {NMR} in solids}.
\newblock Springer Science \& Business Media, 2012.

\bibitem{viola1998dynamical}
L.~Viola and S.~Lloyd.
\newblock Dynamical suppression of decoherence in two-state quantum systems.
\newblock {\em Phys. Rev. A}, 58:2733, 1998.

\bibitem{viola1999dynamical}
L.~Viola, E.~Knill, and S.~Lloyd.
\newblock Dynamical decoupling of open quantum systems.
\newblock {\em Phys. Rev. Lett.}, 82:2417, 1999.

\bibitem{viola2000dynamical}
L.~Viola, E.~Knill, and S.~Lloyd.
\newblock Dynamical generation of noiseless quantum subsystems.
\newblock {\em Phys. Rev. Lett.}, 85:3520, 2000.

\bibitem{khodjasteh2009dynamically}
K.~Khodjasteh and L.~Viola.
\newblock Dynamically error-corrected gates for universal quantum computation.
\newblock {\em Phys. Rev. Lett.}, 102:080501, 2009.

\bibitem{gottesman1997stabilizer}
D.~Gottesman.
\newblock {\em Stabilizer codes and quantum error correction}.
\newblock California Institute of Technology, 1997.

\bibitem{bravyi1998quantum}
S.~B. Bravyi and A.~Yu. Kitaev.
\newblock Quantum codes on a lattice with boundary.
\newblock {\em arXiv:quant-ph/9811052}, 1998.

\bibitem{dennis2002topological}
E.~Dennis, A.~Kitaev, A.~Landahl, and J.~Preskill.
\newblock Topological quantum memory.
\newblock {\em J. Math. Phys.}, 43:4452--4505, 2002.

\bibitem{fowler2012surface}
A.~G. Fowler, M.~Mariantoni, J.~M. Martinis, and A.~N. Cleland.
\newblock Surface codes: Towards practical large-scale quantum computation.
\newblock {\em Phys. Rev. A}, 86:032324, 2012.

\bibitem{kitaev2003fault}
A~Yu Kitaev.
\newblock Fault-tolerant quantum computation by anyons.
\newblock {\em Ann. Phys.}, 303:2--30, 2003.

\bibitem{oreshkov2009holonomic}
O.~Oreshkov.
\newblock Holonomic quantum computation in subsystems.
\newblock {\em Phys. Rev. Lett.}, 103:090502, 2009.

\bibitem{zheng2014fault}
Y.-C. Zheng and T.~A. Brun.
\newblock Fault-tolerant scheme of holonomic quantum computation on stabilizer
  codes with robustness to low-weight thermal noise.
\newblock {\em Phys. Rev. A}, 89:032317, 2014.

\bibitem{zheng2015fault}
Y.-C. Zheng and T.~A. Brun.
\newblock Fault-tolerant holonomic quantum computation in surface codes.
\newblock {\em Phys. Rev. A}, 91:022302, 2015.

\bibitem{wu2020holonomic}
C.~Wu, Y.~Wang, X.-L. Feng, and J.-L. Chen.
\newblock Holonomic quantum computation in surface codes.
\newblock {\em Phys. Rev. Appl.}, 13:014055, 2020.

\bibitem{johansson2012qutip}
J.~R. Johansson, P.~D. Nation, and F.~Nori.
\newblock Qutip: An open-source python framework for the dynamics of open
  quantum systems.
\newblock {\em Comput. Phys. Commun.}, 183:1760--1772, 2012.

\bibitem{khaneja2005optimal}
N.~Khaneja, T.~Reiss, C.~Kehlet, T.~Schulte-Herbr{\"u}ggen, and S.~J. Glaser.
\newblock Optimal control of coupled spin dynamics: design of {NMR} pulse
  sequences by gradient ascent algorithms.
\newblock {\em J. Magn. Reson.}, 172:296--305, 2005.

\bibitem{montangero2007robust}
S.~Montangero, T.~Calarco, and R.~Fazio.
\newblock Robust optimal quantum gates for {J}osephson charge qubits.
\newblock {\em Phys. Rev. Lett.}, 99:170501, 2007.

\bibitem{motzoi2009simple}
F.~Motzoi, J.~M. Gambetta, P.~Rebentrost, and F.~K. Wilhelm.
\newblock Simple pulses for elimination of leakage in weakly nonlinear qubits.
\newblock {\em Phys. Rev. Lett.}, 103:110501, 2009.

\bibitem{kelly2014optimal}
J.~Kelly, R.~Barends, B.~Campbell, Y.~Chen, Z.~Chen, B.~Chiaro, A.~Dunsworth,
  A.~G. Fowler, I.-C. Hoi, E.~Jeffrey, et~al.
\newblock Optimal quantum control using randomized benchmarking.
\newblock {\em Phys. Rev. Lett.}, 112:240504, 2014.

\bibitem{zhong2020quantum}
H.-S. Zhong, H.~Wang, Y.-H. Deng, M.-C. Chen, L.-C. Peng, Y.-H. Luo, J.~Qin,
  D.~Wu, X.~Ding, Y.~Hu, P.~Hu, X.-Y. Yang, W.-J. Zhang, H.~Li, et~al.
\newblock Quantum computational advantage using photons.
\newblock {\em Science}, 370:1460--1463, 2020.

\bibitem{wu2021strong}
Y.~Wu, W.-S. Bao, S.~Cao, F.~Chen, M.-C. Chen, X.~Chen, T.-H. Chung, H.~Deng,
  Y.~Du, D.~Fan, et~al.
\newblock {Strong Quantum Computational Advantage Using a Superconducting
  Quantum Processor}.
\newblock {\em Phys. Rev. Lett.}, 127:180501, 2021.

\bibitem{bharti2021noisy}
K.~Bharti, A.~Cervera-Lierta, T.~H. Kyaw, T.~Haug, S.~Alperin-Lea, A.~Anand,
  M.~Degroote, H.~Heimonen, J.~S. Kottmann, T.~Menke, et~al.
\newblock {Noisy intermediate-scale quantum algorithms}.
\newblock {\em Rev. Mod. Phys.}, 94:015004, 2022.

\bibitem{lau2022nisq}
J.~W.~Z. Lau, K.~H. Lim, H.~Shrotriya, and L.~C. Kwek.
\newblock {NISQ computing: where are we and where do we go?}
\newblock {\em AAPPS Bull.}, 32:27--30, 2022.

\bibitem{buluta2009quantum}
I.~Buluta and F.~Nori.
\newblock Quantum simulators.
\newblock {\em Science}, 326:108--111, 2009.

\bibitem{georgescu2014quantum}
I.~M. Georgescu, S.~Ashhab, and F.~Nori.
\newblock Quantum simulation.
\newblock {\em Rev. Mod. Phys.}, 86:153, 2014.

\bibitem{shevchenko2010landau}
S.~N. Shevchenko, S.~Ashhab, and F.~Nori.
\newblock Landau-{Z}ener-{S}t{\"u}ckelberg interferometry.
\newblock {\em Phys. Rep.}, 492:1--30, 2010.

\bibitem{alexander2020qiskit}
T.~Alexander, N.~Kanazawa, D.~J. Egger, L.~Capelluto, C.~J. Wood,
  A.~Javadi-Abhari, and D.~McKay.
\newblock {Qiskit Pulse}: {P}rogramming quantum computers through the cloud
  with pulses.
\newblock {\em Quantum Sci. Technol.}, 5:044006, 2020.

\bibitem{xu2021realizing}
G.~F. Xu, P.~Z. Zhao, E.~Sj{\"o}qvist, and D.~M. Tong.
\newblock Realizing nonadiabatic holonomic quantum computation beyond the
  three-level setting.
\newblock {\em Phys. Rev. A}, 103:052605, 2021.

\bibitem{andre2022qudit}
T.~Andr\'e and E.~Sj\"oqvist.
\newblock Dark path holonomic qudit computation.
\newblock {\em Phys. Rev. A}, 106:062402, 2022.

\bibitem{ashhab2010control}
S.~Ashhab and F.~Nori.
\newblock Control-free control: Manipulating a quantum system using only a
  limited set of measurements.
\newblock {\em Phys. Rev. A}, 82:062103, 2010.

\bibitem{kofman2012nonperturbative}
A.~G. Kofman, S.~Ashhab, and F.~Nori.
\newblock Nonperturbative theory of weak pre-and post-selected measurements.
\newblock {\em Phys. Rep.}, 520:43--133, 2012.

\bibitem{oi2014unitary}
D.~K.~L. Oi.
\newblock Unitary holonomies by direct degenerate projections.
\newblock {\em Phys. Rev. A}, 89:050102, 2014.

\bibitem{mommers2021universal}
C.~J.~G. Mommers and E.~Sj{\ifmmode\ddot{o}\else\"{o}\fi}qvist.
\newblock {Universal quantum computation and quantum error correction using
  discrete holonomies}.
\newblock {\em Phys. Rev. A}, 105:022402, 2022.

\bibitem{weisbrich2021second}
H.~Weisbrich, R.~L. Klees, G.~Rastelli, and W.~Belzig.
\newblock Second {C}hern numbers and non-{A}belian {B}erry phase in topological
  superconducting systems.
\newblock {\em PRX Quantum}, 2:010310, 2021.

\bibitem{semenoff1986non}
G.~W. Semenoff and P.~Sodano.
\newblock Non-{A}belian adiabatic phases and the fractional quantum {H}all
  effect.
\newblock {\em Phys. Rev. Lett.}, 57:1195, 1986.

\bibitem{zhang2020topological}
S.-B. Zhang, W.~B. Rui, A.~Calzona, S.-J. Choi, A.~P. Schnyder, and
  B.~Trauzettel.
\newblock Topological and holonomic quantum computation based on second-order
  topological superconductors.
\newblock {\em Phys. Rev. Research}, 2:043025, 2020.

\bibitem{karzig2016universal}
T.~Karzig, Y.~Oreg, G.~Refael, and M.~H. Freedman.
\newblock Universal geometric path to a robust {M}ajorana magic gate.
\newblock {\em Phys. Rev. X}, 6:031019, 2016.

\bibitem{dong2021doubly}
W.~Dong, F.~Zhuang, S.~E. Economou, and E.~Barnes.
\newblock Doubly geometric quantum control.
\newblock {\em PRX Quantum}, 2:030333, 2021.

\bibitem{zhang2017quantum}
J.~Zhang, Y.-X. Liu, R.-B. Wu, K.~Jacobs, and F.~Nori.
\newblock Quantum feedback: theory, experiments, and applications.
\newblock {\em Phys. Rep.}, 679:1--60, 2017.

\end{thebibliography}

\end{document}